\documentclass{ws-ijmpe}   
\usepackage{graphicx}
\usepackage[super,compress]{cite}

\newcommand{\ben}{\begin{eqnarray}}
\newcommand{\een}{\end{eqnarray}}
\newcommand{\nnu}{\nonumber\\}
\newcommand{\bef}{\begin{figure}[htpb]\centering}
\newcommand{\eef}{\end{figure}}

\newcommand{\Lb}{\left(}
\newcommand{\Rb}{\right)}

\def\beq{\begin{equation}}
\def\eeq{\end{equation}}
\def\bea{\begin{eqnarray}}
\def\eea{\end{eqnarray}}
\def\eq#1{{Eq.~(\ref{#1})}}
\def\fig#1{{Fig.~\ref{#1}}}

\begin{document} 

\markboth{Albacete {\it et al.}}{Predictions for $p+$Pb Collisions at 
$\sqrt{s_{_{NN}}} = 5$ TeV}

\catchline{}{}{}{}{}
 

\title{Predictions for $p+$Pb Collisions at $\sqrt{s_{_{NN}}} = 5$ TeV}

\author{JAVIER L. ALBACETE} 
\address{IPNO, Universit\'e Paris-Sud 11, CNRS/IN2P3, 91406 Orsay,
  France} 

\author{NESTOR ARMESTO}
\address{Departamento de F\'{\i}sica de Part\'{\i}culas and IGFAE, 
Universidade de Santiago de Compostela, 15706 Santiago de Compostela, 
Galicia, Spain}

\author{RUDOLF BAIER}
\address{Fakult\"at f\"ur Physik, Universit\"at
Bielefeld, D-33501, Bielefeld, Germany}

\author{GERGELY G. BARNAF\"OLDI} 
\address{Institute for Particle and Nuclear Physics, Wigner Research Centre for
Physics, Hungarian Academy of Sciences, P.O.Box 49, Budapest, 1525, Hungary}

\author{JEAN BARRETTE} 
\address{McGill University, Montreal, H3A 2T8, Canada}

\author{SOMNATH DE}
\address{Variable Energy Cyclotron Centre, 1/AF, Bidhan Nagar, Kolkata,
700064, India}

\author{WEI-TIAN DENG}
\address{Theory Center, IPNS, KEK, 1-1 Oho, Tsukuba, Ibaraki 305-0801, 
Japan}

\author{ADRIAN DUMITRU}
\address{Department of
    Natural Sciences, Baruch College, CUNY, 17 Lexington Avenue, New
    York, NY 10010, USA}
\address{RIKEN BNL Research Center,
    Brookhaven National Laboratory, Upton, NY 11973, USA}

\author{KEVIN DUSLING}
\address{Physics Department, North Carolina State University, Raleigh, NC 2
7695, USA}

\author{KARI J. ESKOLA}
\address{Department of Physics, P.O. Box 35, FI-40014 University of 
Jyv\"{a}skyl\"{a}, Finland}
\address{Helsinki Institute of Physics, P.O. Box 64, FIN-00014 University 
of Helsinki, Finland}

\author{RAINER FRIES}
\address{Cyclotron Institute and Department of Physics and Astronomy,
Texas A\&M University, College Station, TX 77843, USA}

\author{HIROTSUGU FUJII}
\address{Institute of Physics, University of Tokyo, Komaba,
    Tokyo 153-8902, Japan}

\author{FRANCOIS GELIS}
\address{Institut de Physique Th\'eorique,
   CEA, 91191 Gif-sur-Yvette Cedex, France}

\author{MIKLOS GYULASSY} 
\address{Department of Physics, Columbia University, New York, NY 10027, USA}
\address{WIGNER RCP, Institute for Particle 
and Nuclear Physics P.O.Box 49, Budapest, 1525, Hungary}

\author{YUNCUN HE}
\address{Key Laboratory of Quark $\&$ Lepton Physics (MOE) and
Institute of Particle Physics, Central China Normal University, Wuhan 430079, China}

\author{ILKKA HELENIUS}
\address{Department of Physics, P.O. Box 35, FI-40014 University of 
Jyv\"{a}skyl\"{a}, Finland}
\address{Helsinki Institute of Physics, P.O. Box 64, FIN-00014 University 
of Helsinki, Finland}

\author{ZHONG-BO KANG}  
\address{Theoretical Division, MS B283, Los Alamos National Laboratory, 
Los Alamos, NM 87545, USA}

\author{BORIS Z. KOPELIOVICH}
\address{Departamento de F\'{\i}sica,
Universidad T\'ecnica Federico Santa Mar\'{\i}a, Avda. Espa\~na 1680,
Casilla 110-V, Valpara\'iso, Chile}

\author{KRZYSZTOF KUTAK}
\address{Instytut Fizyki Jadrowej 
im. Henryka Niewodnicza\'nskiego,
Radzikowskiego 152, 31-342 Krak\'ow, Poland}

\author{PETER LEVAI} 
\address{Institute for Particle and Nuclear Physics, Wigner Research Centre for
Physics, Hungarian Academy of Sciences, P.O.Box 49, Budapest, 1525, Hungary}

\author{ZI-WEI LIN}
\address{C-209 Howell Science Complex, Department of Physics, East
  Carolina University, Greenville, NC 27858, USA} 

\author{ALFRED H. MUELLER}
\address{Department of Physics, Columbia University, New York, NY 10027, 
USA}

\author{YASUSHI NARA}
\address{Akita International
    University, Yuwa, Akita-city 010-1292, Japan}

\author{JAN NEMCHIK}
\address{Czech Technical University in Prague,
FNSPE, B\v rehov\'a 7,
11519 Prague, Czech Republic}

\author{G\' ABOR PAPP}
\address{E\"otv\"os Lor\'and University, P\'azm\'any P\'eter s\'et\'any 1/A,
H-1117, Budapest, Hungary}

\author{MIHAI PETROVICI} 
\address{National Institute for Physics and Nuclear Engineering,
Horia~Hulubei, R-077125, Bucharest, Romania}

\author{JIAN-WEI QIU}
\address{Physics Department, Brookhaven National Laboratory, Upton, NY 
11973, USA} 
\address{C.N. Yang Institute for Theoretical Physics, Stony Brook 
University, Stony Brook, NY 11794, USA}

\author{AMIR H. REZAEIAN}
\address{Departamento de F\'\i sica, Universidad T\'ecnica
Federico Santa Mar\'\i a, Avda. Espa\~na 1680,
Casilla 110-V, Valpara\'iso, Chile }

\author{PENG RU}
\address{School of Physics $\&$ Optoelectronic Technology, Dalian
University of Technology, Dalian, 116024 China}
\address{Institute of Particle Physics,
Central China Normal University, Wuhan, 430079 China}

\author{DOMINIQUE SCHIFF}
\address{LPT, Universit\'e Paris-Sud, B\^atiment 210, F-91405 Orsay, France}

\author{SEBASTIAN SAPETA}
\address{Institute for Particle Physics Phenomenology, Durham University,
South Rd, Durham DH1 3LE, UK}

\author{VASILE TOPOR POP} 
\address{McGill University, Montreal, H3A 2T8, Canada}

\author{PRITHWISH TRIBEDY}
\address{Variable Energy Cyclotron Centre, 1/AF Bidhan Nagar, Kolkata-70006
4, India}

\author{RAJU VENUGOPALAN}
\address{Physics Department, Brookhaven National Laboratory,
  Upton, NY 11973, USA}

\author{IVAN VITEV}  
\address{Theoretical Division, MS B283, Los Alamos National Laboratory, 
Los Alamos, NM 87545, USA}

\author{RAMONA VOGT}
\address{Physics Division, Lawrence Livermore National Laboratory, 
Livermore, CA 94551, USA}
\address{Physics Department, University of California at Davis, 
Davis, CA 95616, USA \\ vogt@physics.ucdavis.edu}

\author{ENKE WANG}
\address{Key Laboratory of Quark $\&$ Lepton Physics (MOE) and 
Institute of Particle Physics, Central China Normal University, Wuhan 430079, 
China}

\author{XIN-NIAN WANG}
\address{Key Laboratory of Quark and Lepton Physics (MOE) and Institute of 
Particle Physics, Central China  Normal University, Wuhan 430079, China}
\address{Nuclear Science Division, MS 70R0319, Lawrence Berkeley National 
Laboratory, Berkeley, CA 94720, USA}

\author{HONGXI XING}
\address{Institute of Particle Physics, 
Central China Normal University,
Wuhan 430079, China}

\author{RONG XU}
\address{Key Laboratory of Quark and Lepton Physics (MOE) and Institute of 
Particle Physics, Central China Normal University, Wuhan 430079, China}

\author{BEN-WEI ZHANG}
\address{Key Laboratory of Quark $\&$ Lepton Physics (MOE) and 
Institute of Particle Physics, Central China Normal University, Wuhan 430079, 
China}
\address{Nuclear Science Division, MS 70R0319, Lawrence Berkeley National 
Laboratory, Berkeley, CA 94720, USA}

\author{WEI-NING ZHANG}
\address{ School of Physics $\&$ Optoelectronic
Technology, Dalian University of
Technology, Dalian, 116024 China}
\address{Physics Department, Harbin
Institute of Technology, Harbin 150006, China}

\maketitle

\begin{abstract}
Predictions for charged hadron, identified light hadron, quarkonium, photon,
jet and gauge bosons in $p+$Pb collisions at $\sqrt{s_{_{NN}}} = 5$ TeV are
compiled and compared.  When test run data are available, they are compared
to the model predictions.  
\end{abstract}

\ccode{12.38.Bx, 25.75.Bh, 25.75.Cj, 13.87.-a}

\section{Introduction}
\label{sec:Intro}

Here predictions for the upcoming $p+$Pb LHC run at 
$\sqrt{s_{_{NN}}} = 5$~TeV compiled by members and friends of the JET
Collaboration \cite{JET} are presented.  
The test run data published by the ALICE
Collaboration \cite{ALICE_dndeta,ALICE_rpa} are compared to model
calculations available before the test run.  Most calculations are for 
midrapidity and minimum bias collisions.  Other results at different
rapidities and centralities are presented when available.

The predictions presented here, as well as the corresponding discussion, were 
made assuming that the proton circulated toward positive 
rapidity and the nucleus toward negative rapidity, $p+$Pb
collisions, similar to fixed-target 
configurations.  In the fixed-target configuration, the low $x$ nuclear parton 
distributions are probed at positive rapidity.
Throughout this paper, many of the results shown have been
adjusted to the Pb$+p$ convention of the ALICE data, as described in 
Ref.~\cite{ALICE_dndeta}, with
the nuclear parton density probed at high $x$ at forward (positive) rapidity
and low $x$ at backward (negative) rapidity.  The cases where the results
still appear with the assumption that the proton circulates in the direction
of positive rapidity are explicitly noted.

This paper is organized in the following fashion.  Section~\ref{sec:models}
describes the models that specifically address charged particle production.
These include saturation approaches, event generators, and perturbative 
QCD-based calculations.  Section~\ref{sec:charged_hadrons} 
compares results obtained
from models described in Sec.~\ref{sec:models} with each other and with the
available data.  The next several sections present predictions for specific
observables including identified light hadrons (Sec.~\ref{sec:part_ID}), 
quarkonium (Sec.~\ref{sec:Vogt}), direct photons (Sec.~\ref{sec:photons}), jets 
(Sec.~\ref{sec:jets}), and gauge bosons (Sec.~\ref{sec:WZ_prod}).

\section{Model descriptions}
\label{sec:models}

In this section, the models used to obtain the results for charged particle 
distributions, $dN_{\rm ch}/d\eta$, $dN_{\rm ch}/dp_T$ and the nuclear suppression
factor $R_{p{\rm Pb}}$ as a function of $p_T$ and $\eta$.  
The first calculations described
Sec.~\ref{sec:amir_ch} 
are saturation or Color Glass 
Condensate (CGC) based.  The next set of calculations are event-generator
based with results from $\mathtt{HIJING}$ in Sec.~\ref{sec:HIJING_ch}, 
$\mathtt{HIJINGB\overline{B}2.0}$ 
in Sec.~\ref{sec:Pop_ch}, and $\mathtt{AMPT}$ in 
Sec.~\ref{sec:AMPT_ch}.  Finally, calculations based on
collinear factorization in perturbative QCD are described in 
Secs.~\ref{sec:Ivan-description} and \ref{sec:Levai-description}.

In the following Section, Sec.~\ref{sec:charged_hadrons}, 
the predictions will be compared to the ALICE
Pb$+p$ test run data \cite{ALICE_dndeta,ALICE_rpa} in September 2012.


\subsection[Inclusive hadron 
production in the rcBK-CGC approach]{Inclusive hadron 
production in the rcBK-CGC approach (A. Rezaeian)}
\label{sec:amir_ch}

In the Color Glass Condensate (CGC) approach, gluon jet production in $p+A$ 
collisions can be described by $k_T$-factorization \cite{Kovchegov:2001sc},
\begin{equation} \label{M1}
\frac{d \sigma}{d y \,d^2 p_{T}}=\frac{2\alpha_s}{C_F}\frac{1}{p^2_T}\int d^2 
\vec k_{T} \phi^{G}_{p}\Lb x_1;\vec{k}_T\Rb \phi^{G}_{A}\Lb x_2;\vec{p}_T -
\vec{k}_T\Rb,
\end{equation}
where $C_F=(N_c^2-1)/2N_c$, $N_c$ is the number of colors,
$x_{1,2}=(p_T/\sqrt{s})e^{\pm y}$, $p_T$ and $y$ are the
transverse momentum and rapidity of the produced gluon
jet, and $\sqrt{s}$ is the nucleon-nucleon center-of-mass energy. 
The unintegrated gluon density, $\phi^{G}_{A}(x_i;\vec k_T)$, denotes the 
probability to find a gluon that carries fractional energy $x_i$
and transverse momentum $k_T$ in the projectile (or target) $A$. 
The unintegrated gluon
density is related to the color dipole forward scattering amplitude,  
\beq \label{M2}
\phi^G_A\Lb x_i;\vec{k}_T\Rb=\frac{1}{\alpha_s} \frac{C_F}{(2 \pi)^3}\int d^2 
\vec b_T\,d^2 \vec r_T \,
e^{i \vec{k}_T\cdot \vec{r}_T} \, \nabla^2_T \mathcal{N}_A\Lb x_i; r_T; b_T\Rb,
\eeq
with 
\beq \label{M3}
\mathcal{N}_A\Lb x_i; r_T; b_T \Rb =2 \mathcal{N}_F\Lb x_i; r_T; b_T \Rb - 
\mathcal{N}^2_F\Lb x_i; r_T; b_T \Rb,
\eeq
where $r_T$ is the transverse size of the dipole and $b_T$ is the
impact parameter of the collision. The subscript $T$ stands for the 
transverse component.  The dipole scattering amplitude $\mathcal{N}_F$ 
satisfies the nonlinear small-$x$ JIMWLK evolution equations 
\cite{JalilianMarian:1997jx,JalilianMarian:1997gr,Iancu:2000hn,Ferreiro:2001qy}, 
see below.  

In the $k_T$-factorized approach,  partons in both 
the projectile and target are assumed to be at very small $x$ so that the CGC
formalism and small-$x$ resummation is applicable to both the projectile and 
the target, assuming the projectile proton moves in the direction of forward
rapidity. This
approach is valid away from the projectile
fragmentation region. However, to treat the projectile fragmentation region in 
the forward region, an alternative approach developed
in Refs.~\cite{Dumitru:2005gt,Altinoluk:2011qy}, the so-called hybrid approach, 
is better suited. In this approach, the projectile is treated 
perturbatively within the standard collinear factorization scheme using the 
standard DGLAP picture while the target is treated employing CGC methods. 
The cross section for single inclusive hadron production at leading twist in 
asymmetric collisions such as $p+A$ in the CGC approach is given 
by~\cite{Dumitru:2005gt,Altinoluk:2011qy},
\begin{eqnarray}\label{final}
\frac{dN^{p A \rightarrow h X}}{d\eta d^2p_T}&=&\frac{K}{(2\pi)^2}\Bigg[\int_{x_F}^1 
\frac{dz}{z^2} \Big[x_1f_g(x_1,\mu_F^2)N_A(x_2,\frac{p_T}{z})D_{h/g}(z,\mu_{\rm Fr}) 
\nonumber \\
&+&\Sigma_q x_1f_q(x_1,\mu_F^2)N_F(x_2,\frac{p_T}{z})D_{h/q}(z,\mu_{\rm Fr})\Big]
\nonumber \\
&+& \frac{\alpha_s^{\rm in} }{ 2\pi^2} \int_{x_F}^1 \frac{dz}{z^2}\frac{z^4}{p_T^4}
\int_{k_T^2<\mu_F^2}d^2k_T k_T^2 N_F(k_T,x_2)\int_{x_1}^1\frac{d\xi}{\xi} \nonumber \\
&\times& \Sigma_{i,j=q,\bar q, g}w_{i/j}(\xi)P_{i/j}(\xi)x_1f_j(\frac{x_1}{\xi}, \mu_F)
D_{h/i}(z,\mu_{\rm Fr})\Bigg] \, \, .
\end{eqnarray}
A $K$-factor has been introduced to effectively incorporate 
higher-order corrections.  The parton distribution function of
the proton, $f_j(x,\mu_F^2)$, depends on the light-cone momentum fractions $x$
and the hard factorization scale $\mu_F$. The function $D_{h/i}(z,\mu_{\rm Fr})$ 
is the fragmentation function (FF) of parton $i$ to become final-state
hadron $h$ carrying a fraction $z$ of the parent parton momentum at 
fragmentation scale $\mu_{\rm Fr}$.
The inelastic weight functions, $w_{i/j}$, and the DGLAP splitting functions, 
$P_{i/j}$, are given in Ref.~\cite{Altinoluk:2011qy}. The longitudinal 
momentum fractions $x_1$ and $x_2$ are 
\begin{equation}\label{xs}
x_F\approx \frac{p_T}{\sqrt{s}}e^{\eta}; \ \ \ \ x_1=\frac{x_F}{ z}; 
\ \ \ \ \ x_2=x_1e^{-2\eta}. 
\end{equation}

The strong coupling multiplying 
the inelastic term in Eq.~(\ref{final}), is denoted $\alpha_s^{\rm in}$.  
The superscript ``in" is employed to differentiate it from the running value 
of $\alpha_s$ in the rcBK equation.  In the hybrid formulation, 
the strong coupling in the dilute regime of the projectile can differ from 
that in the rcBK description of the dense target (or dipole scattering 
amplitude). The scale at which $\alpha_s^{\rm in}$ should be evaluated cannot be 
determined in the current approximation, a full
NNLO calculation is required.  The effects of different choices of 
$\alpha_s^{\rm in}$ will be considered later.

In Eq.~(\ref{final}), the factorization scale 
$\mu_F$ is assumed to be the same in the fragmentation functions ($\mu_{\rm Fr} =
\mu_F$) and the parton 
densities. In order to investigate the uncertainties associated with choice 
of $\mu_F$, several values of $\mu_F$ are considered: $\mu_F=2p_T$; $p_T$; 
and $p_T/2$. 

In Eq.~(\ref{final}), the amplitude $N_F$ 
($N_A$) is the two-dimensional Fourier
transform of the imaginary part of the forward (proton direction) dipole-target
scattering amplitude, $\mathcal{N}_{A(F)}$, in the fundamental ($F$) or adjoint 
($A$) representations,
\begin{equation} \label{ff}
N_{A(F)}(x,k_T)=\int d^2\vec r e^{-i\vec k_T.\vec r}
\left[ 1-\mathcal{N}_{A(F)}(r,Y=\ln\Big(\frac{x_0}{x}\Big)) \right],
\end{equation}
where $r=|\vec r|$ is the dipole transverse size. The dipole 
scattering 
amplitude $\mathcal{N}_{A(F)}$ incorporates small-$x$ dynamics and can be 
calculated using the JIMWLK evolution equation 
\cite{JalilianMarian:1997jx,JalilianMarian:1997gr,Iancu:2000hn,Ferreiro:2001qy}. 
In the large $N_c$ limit, the coupled 
JIMWLK equations are simplified to the Balitsky-Kovchegov (BK) equation 
\cite{Balitsky:1995ub,Kovchegov:1999yj,Kovchegov:1999ua,Balitsky:2006wa}, 
a closed-form equation for the rapidity evolution of the dipole 
amplitude. While a numerical solution of the full next-to-leading logarithmic 
expressions is not yet 
available, the running coupling corrections to the leading log kernel, the 
so-called running-coupling BK (rcBK) equation has been very successful in 
phenomenological applications  \cite{Albacete:2010sy}. The rcBK equation has 
the following simple form  
\cite{Balitsky:1995ub,Kovchegov:1999yj,Kovchegov:1999ua,Balitsky:2006wa,Albacete:2007yr}:
\begin{eqnarray}
  \frac{\partial\mathcal{N}_{A(F)}(r,x)}{\partial\ln(x_0/x)}&=&\int d^2{\vec r_1}\
  K^{{\rm run}}({\vec r},{\vec r_1},{\vec r_2})
  \left[\mathcal{N}_{A(F)}(r_1,x)+\mathcal{N}_{A(F)}(r_2,x) \right. \nonumber \\
&-& \left. \mathcal{N}_{A(F)}(r,x)-
\mathcal{N}_{A(F)}(r_1,x)\,\mathcal{N}_{A(F)}(r_2,x)\right],
\label{bk1}
\end{eqnarray}
where $\vec r_2 \equiv \vec r-\vec r_1$.  The rcBK equation only describes the 
rapidity/energy evolution of the dipole, the initial profile and parameters of 
the dipole still need to be modeled and constrained by experimental data.  
The initial condition for the evolution generally takes a form motivated by 
the McLerran-Venugopalan model 
\cite{McLerran:1993ni,McLerran:1993ka,McLerran:1994vd},  
  \begin{equation}
\mathcal{N}(r,Y\!=\!0)=
1-\exp\left[-\frac{\left(r^2\,Q_{0s}^2\right)^{\gamma}}{4}\,
  \ln\left(\frac{1}{\Lambda\,r}+e\right)\right],
\label{mv}
\end{equation}
where the onset of small-$x$ evolution is assumed to be at
$x_0=0.01$, and the infrared scale is $\Lambda=0.241$ GeV 
\cite{Albacete:2010sy}.
The only free parameters are $\gamma$ and the initial saturation scale 
$Q_{0s}$, with $s=p$ and $A$ for proton and nuclear targets, respectively. 
Unfortunately, the current global set of small-$x$ data are very limited and 
thus cannot uniquely fix the initial dipole parameters \cite{Albacete:2010sy}. 
This problem is more severe for determining the dipole scattering amplitude on 
nuclear targets, leading to rather large unavoidable theoretical uncertainties 
on CGC predictions for $p+A$ collisions at the LHC.  In 
Ref.~\cite{Rezaeian:2012ye},
a simple scheme to test the CGC dynamics at the LHC was proposed.  This scheme
will be used to calculate the results shown later on.

\subsection[IP-Sat]{IP-Sat (P. Tribedy and R. Venugopalan)}
\label{sec:IPSat}

The impact parameter dependent dipole saturation model (IP-Sat)
\cite{Kowalski:2003hm} is a refinement of the Golec-Biernat--Wusthoff dipole 
model~\cite{GolecBiernat:1998js,GolecBiernat:1999qd} 
to give the right perturbative limit when 
the dipole radius $r_T\rightarrow0$ \cite{Bartels:2002cj}. It is equivalent to 
the expression derived in the classical effective theory of the CGC, to leading 
logarithmic accuracy~\cite{McLerran:1998nk,Venugopalan:1999wu}. The proton 
dipole cross section in this model is expressed as 
\begin{eqnarray}
\frac{d \sigma^p_{\rm dip}}{d^2b_T}(r_T,x,b_T) = 
2\left[1-\exp\left(-\frac{\pi^{2}}{2N_{c}}r_T^{2}\alpha_{s}(\mu^{2}) 
xg(x,\mu^{2})T_p(b_T)\right)\right]\, .
\label{eq:ipsat-dipole}
\end{eqnarray}
Here the scale $\mu^{2}$ is related to dipole radius $r_T$ as
\begin{equation}
\mu^{2}=\frac{4}{r_T^{2}}+\mu_{0}^{2}\, \, ,
\end{equation}
where the leading order expression for the running coupling is 
\begin{equation}
 \alpha_{s}(\mu^{2})=\frac{12\pi}{(33-2n_{f})\log(\mu^{2}/\Lambda_{\rm QCD}^{2})}
\end{equation} 
with $n_{f}=3$ and $\Lambda_{\rm QCD}=0.2$ GeV.
The model includes saturation as eikonalized power corrections to the DGLAP 
leading-twist expression and may be valid in the regime where logs in $Q^2$ 
dominate logs in $x$. 
For each value of the  dipole radius, the gluon density $xg(x,\mu^{2})$ is 
evolved from $\mu_{0}^{2}$ to $\mu^{2}$ using the LO DGLAP evolution equation 
without quarks, 
\begin{equation}
\frac{\partial xg(x,\mu^{2})}{\partial \log \mu^{2}} = \frac{\alpha_{s}(\mu^{2}
)}{2\pi}\int \limits_{x}^{1}
dz P_{gg}(z)\frac{x}{z}g\left(\frac{x}{z},\mu^{2}\right) \, \, .
\end{equation}
Here the gluon splitting function 
is 
\begin{equation}
 P_{gg}(z)=6\left[\frac{z}{(1-z)+}+\frac{1-z}{z}+z(1-z)\right]+
\left(\frac{11}{2}-\frac{n_{f}}{3}\right)\delta(1-z) \, \, .
\end{equation}
The initial gluon density at the scale $\mu^{2}_{0}$ is taken to be of the form
\begin{equation}
 xg(x,\mu^{2}_{0})=A_{g}x^{-\lambda_{g}}(1-x)^{5.6} \, \, .
\end{equation}

An important feature of the IP-Sat model is the $b$-dependence of the dipole 
cross section, introduced through the gluon density profile function $T_p(b_T)$.
This profile function, normalized to unity, is chosen to have the Gaussian 
form  
\begin{equation}
 T_{p}(b_T)=\frac{1}{2\pi B_{G}} \exp\left({-b_T^{2}\over 2B_{G}}\right)\, \, ,
\label{eq:IPsat-imp-par}
\end{equation}
where $B_G$ is a parameter fit to the HERA diffractive data. It corresponds to 
$\langle b^2\rangle = 2 B_G$, the average squared {\it gluonic} radius of the
proton.

The IP-Sat model parameters are obtained from optimal fits to HERA 
data~\cite{Kowalski:2006hc}. The parameters used in this work are listed in 
Table \ref{tab_ipsat}.
\begin{table}[pt]
\tbl{Parameters of the IP-Sat model obtained from fits to HERA 
data \protect\cite{Kowalski:2006hc}.}
{\begin{tabular}{@{}cccc@{}}\toprule
$B_G$ (GeV$^{-2}$) & $\mu_0$ (GeV$^2$) & $A_g$ & $\lambda_g$ \\
\hline
4.0 & 1.17 & 2.55 & 0.020\\ \hline
\end{tabular}}
\label{tab_ipsat}
\end{table}
The parameters of the initial gluon distribution are determined from fits to 
the HERA $F_2$ data~\cite{Chekanov:2001qu,Adloff:2000qk} with $\chi^2\sim 1$. 
The value of $B_G$ is determined primarily from the $J/\psi$ $t$-distributions 
measured by ZEUS~\cite{Chekanov:2002xi} and H1~\cite{Aktas:2005xu}. With these 
parameters, excellent agreement with the HERA exclusive vector meson and DVCS 
data is obtained. For a detailed comparison of this model to the HERA data, see
Ref.~\cite{Kowalski:2006hc}. A more recent fit to the combined ZEUS and H1 data 
has been performed in Ref.~\cite{Rezaeian:2012ji}. 

The IP-Sat model successfully describes the bulk features of the $p+p$ and 
$A+A$ data over a wide range of center-of-mass energies from RHIC to LHC
as well as the features of the RHIC d+Au data 
\cite{Tribedy:2010ab,Tribedy:2011aa}. It also provides the basis for the 
IP-Glasma model \cite{Schenke:2012wb,Schenke:2012hg} 
of initial conditions in heavy-ion collisions.

In the IP-Sat model, the dipole-nucleus cross section in a large nucleus
can be approximated as
\begin{equation}
 \frac{d\sigma^A_{\rm dip}}{d^2 s_T}  \approx 2
\left[1-\exp\left\{-\frac{AT_A(s_T)}{2} \sigma_{\rm dip}^p(r_T,
 x)\right\}\right]
\label{eq:nuc-dipole}
\end{equation}
where $AT_A(s_T)$ is the transverse density of nucleons inside a 
nucleus and $\sigma_{\rm dip}^p(r_T, x)$ is obtained by integrating the 
dipole-proton cross section in Eq.~(\ref{eq:ipsat-dipole}) over the 
impact parameter distribution in the proton. This form of the dipole-nucleus 
cross section was previously shown to give reasonable fits to the 
limited available inclusive fixed-target $e+A$ data \cite{Kowalski:2007rw}. 

In $p+A$ collisions, the LO inclusive gluon distribution can be 
expressed as \cite{Blaizot:2004wv}
\begin{equation}
\frac{{d}N_{g}^{pA}({b}_{T})}{{d}y~{d}^{2} {p}_{T}} =
\frac{ 4 \alpha_s}{\pi C_F} \frac{1}{p_{T}^2} 
\int \frac{{d}^{2} {k}_{T}}{(2\pi)^{5}} \int {d}^{2} {s}_{T} 
\frac{{d}\phi_p(x_1,{k}_{T}|{s}_{T})}{{d}^2{s}_{T}} 
\frac{{d}\phi_A(x_2,{p}_{T}-{k}_{T}|{s}_{T}-{b}_{T})}{{d}^2{s}_{T}} \, \, .
\label{eq:ktfact1}
\end{equation}
This equation is a generalization of the well known $k_T$-factorized 
expression for inclusive gluon production~\cite{Braun:2000bh} to include the 
impact parameter dependence of the unintegrated gluon distributions. 
Here $C_F = (N_c^2-1)/2 N_c$ is the Casimir for the fundamental representation. 
Using a relation between quark and gluon dipole amplitudes strictly valid in 
the large $N_c$ limit, the unintegrated gluon distribution in protons and 
nuclei can be expressed in terms of the corresponding dipole cross section 
measured in DIS as \cite{Gelis:2006tb} 
\begin{equation}
\frac{{d}\phi^{p,A}(x,{k}_{T}|{s}_{T})}{{d}^2
{s}_{T}} =\frac{{k}_T^2 N_c}{4 \alpha_s}  \int \limits_{0}^{\infty}{d}^2{r}_{T}
e^{i  \vec{k_{T}}.\vec{r_{T}}} \left[1 - \frac{1}{2}\, 
\frac{d \sigma^{p,A}_{\rm dip}}{d^2{s}_T} (r_T,x,{s}_T)\right]^{2} \, \, .
\label{eq:unint-gluon}
\end{equation}

\subsection[$\mathtt{HIJING2.1}$]{$\mathtt{HIJING2.1}$ (R. Xu, 
W.-T. Deng and X.-N. Wang)}
\label{sec:HIJING_ch}

The $\mathtt{HIJING}$ \cite{Wang:1991hta,Gyulassy:1994ew} 
Monte Carlo is based on a 
two-component model of hadron production in high-energy $p+p$, $p+A$, and 
$A+A$ collisions. The soft and hard components are separated by a cutoff 
momentum $p_0$ in the transverse momentum exchange. Hard parton scatterings 
with $p_T>p_0$ are assumed to be described by perturbative QCD (pQCD), while 
soft interactions are approximated by string excitations with an effective 
cross section $\sigma_{\mathrm{soft}}$.
In $p+A$ collisions, the single jet inclusive cross section is proportional 
to the nuclear parton densities $f_{a/A}(x_2,p_T^2,b)$,
\begin{eqnarray}
 \nonumber
\frac{{d}\sigma^{\rm jet}_{pA}}{{d}y_1{d}^2p_T}
= K \int {d}y_2 \, {d}^2b \, T_A(b) \sum_{a,b,c}x_1f_{a/p}(x_1,p^2_T) 
x_2f_{a/A}(x_2,p^2_T,b) \frac{{d}\sigma_{ab\rightarrow cd}}{{d}t} \, \, .
\end{eqnarray}
Here, $x_{1,2}=p_T (e^{\pm y_1} + e^{\pm y_2})/\sqrt{s}$ are the fractional momenta 
of the initial partons while $y_{1,2}$ are the rapidities of the final parton 
jets. Higher order corrections are absorbed into the $K$ factor. The nuclear 
thickness function is normalized to $A$, $\int {d}^2b \, T_A(b)=A$. 

There are several cold nuclear matter effects that are considered in 
$\mathtt{HIJING}$. The first is the shadowing effect.
$\mathtt{HIJING2.0}$ \cite{Deng:2010mv,Deng:2010xg} 
employs a factorized form of the
parton densities in nuclei\cite{Li:2001xa},
\begin{equation}
 f_{a/A}(x,\mu_F^2,b) = S_{a/A}(x,\mu_F^2,b)f_{a/A}(x,\mu_F^2) \label{eq:factorized}
\end{equation}
where $S_{a/A}(x,\mu_F^2,b)$ is the impact-parameter dependent nuclear 
modification 
factor. However, the shadowing employed in $\mathtt{HIJING2.0}$ does not 
include any $\mu_F^2$ dependence as in e.g. Ref.~\cite{Eskola:2009uj}. 
Therefore, shadowing effects in $\mathtt{HIJING}$ should only be valid at low 
$p_{T}$ and disappear
at larger $p_T$.  

The second cold matter effect included is the Cronin effect 
\cite{Cronin:1974zm}, the enhancement of intermediate $p_T$ hadron spectra in 
$p+A$ collisions.  Multiple scattering inside a nucleus can lead to the 
transverse momentum ($k_T$) broadening of both the initial- and final-state 
partons. A $k_T$-kick is imparted to both the initial and final-state hard 
scattered partons in each binary nucleon-nucleon scattering. The $k_T$-kick of 
each scattering follows a Gaussian distribution. Fits to the fixed-target 
$p+A$ data lead to an energy dependence of the Gaussian width,
\begin{equation}
 \langle k^2_T\rangle = [ 0.14 {\log}(\sqrt{s}/{\rm GeV})-0.43] \, 
{\rm GeV}^2/c^2 \, \, .
\end{equation}
This $k_T$-kick influences the final-state hadron rapidity distribution. 
After tuning the gluon shadowing parameter $s_g$ \cite{Li:2001xa} in 
$\mathtt{HIJING2.1}$, the charged particle rapidity distribution, 
$dN_{\rm ch}/d\eta$, in d+Au collisions at $\sqrt{s_{_{NN}}} = 200$~GeV can be 
described.  The prediction for LHC energies can be obtained by
extrapolation \cite{Xu:2012au}. 

In the default $\mathtt{HIJING}$ setting, $p+A$ and $A+A$ collision are 
decomposed into independent and sequential nucleon-nucleon collisions. Within 
each nucleon-nucleon collision, hard collisions are simulated first, followed 
by soft collisions. However, since the time scale for hard scattering is much 
shorter than soft interactions, such a sequence of hard and soft interactions 
within each binary collision might not be physical. 
In a revised scheme denoted by DHC (decoherent hard scattering), all the hard 
interactions in a $p+A$ event are simulated first. They are
subsequently followed by the soft interactions. As a consequence, the energy 
available in each hard scattering is no longer restricted by soft interactions. 

An additional cold matter effect arises from valence quark number conservation 
in the proton.  In $p+A$ collisions, the projectile proton will suffer 
multiple scatterings within the target nucleus. For each binary 
nucleon-nucleon collision, there is a finite probability for independent hard 
parton scattering involving initial partons from the projectile and target 
nucleons.  Flavor conservation limits the availability of valence quarks from 
the projectile for each of these hard interactions. This effect can change 
the relative flavor composition of produced partons per average binary 
nucleon-nucleon collision.  Since the gluon fragmentation functions are softer 
than those of the quarks, the increased fraction of produced gluon jets in 
$p+A$ collisions can lead to suppression of the final-state high-$p_T$ hadron 
spectra.

Finally, jet fragmentation can also modify the final hadron spectra in $p+A$ 
collisions. In the default $\mathtt{HIJING}$ setup, jet shower partons from 
initial- and final-state radiation are ordered in rapidity. Gluons are 
connected to the valence quark and diquark of the projectile or target 
nucleons as kinks to form systems of strings. These strings fragment into 
final-state hadrons using the Lund string fragmentation model 
\cite{Andersson:1983ia}. In $p+A$ collisions, the projectile can undergo 
multiple scatterings since its string systems have many more gluons attached 
to them than in $p+p$ collisions. Hadrons produced by the fragmentation of 
such string systems are softer than those resulting from independent 
fragmentation of individual gluons.

\subsection[$\mathtt{HIJINGB\overline{B}}$]
{$\mathtt{HIJINGB\overline{B}2.0}$ (G. G. Barnaf\"oldi, J. Barette, M. 
Gyulassy, P. Levai, M. Petrovici, and V. Topor Pop)}
\label{sec:Pop_ch}

Monte Carlo models such as 
$\mathtt{HIJING1.0}$~\cite{Wang:1991xy,Wang:1991hta},
$\mathtt{HIJING2.0}$~\cite{Deng:2010mv,Deng:2010xg} and   
$\mathtt{HIJINGB\overline{B}2.0}$
\cite{ToporPop:2011wk,ToporPop:2010qz,Barnafoldi:2011px,Pop:2012ug} have 
been developed to study hadron production in $p+p$, $p+A$ and $A+A$ 
collisions.  They are essentially two-component models which describe
the production of hard parton jets and the soft interaction between
nucleon remnants. 
Hard jet production is calculated 
employing collinearly-factorized multiple minijet production within pQCD.
A transverse momentum cut-off, $p_0$, on the final-state jet production 
is introduced so that for $p_T < p_0$ the
interaction is nonperturbative and is characterized by
a finite soft parton cross section $\sigma_{\rm soft}$. 
The jet cross sections depend on the 
parton distribution functions parameterized from  
global fits to data \cite{Deng:2010mv,Deng:2010xg}.

Nucleon remnants interact via soft gluon exchanges described by the
string models \cite{Andersson:1986gw,NilssonAlmqvist:1986rx,Bengtsson:1987kr} 
and constrained from lower energy $e^+ + e^-$, $e^\pm +p$, and $p+p$ data.  
The hard jet pairs and the two excited nucleon remnants
are connected by 
independent strings which fragment to resonances that
decay to the final-state hadrons.
Longitudinal beam-jet string fragmentation depends strongly on the 
values of the string tensions that control the
quark-antiquark ($q\bar{q}$) and 
diquark-antidiquark ($qq\overline{qq}$) creation rates
and strangeness suppression factors ($\gamma_s$). 

In $\mathtt{HIJING1.0}$ 
and $\mathtt{HIJING2.0}$, a constant (vacuum value) for the effective
value of string tension, $\kappa_0 = 1.0$ GeV/fm,  is used.
At high initial energy density, the novel nuclear physics is due to
the possibility of overlapping multiple longitudinal flux tubes 
leading to strong longitudinal color field (SCF) effects.
These effects are modeled in $\mathtt{HIJINGB\overline{B}2.0}$ 
by varying the effective string tension. 
SCFs also modify the fragmentation processes,
resulting in an increase of (strange)baryons which play an important
role in the description of the baryon to meson anomaly.
In order to describe the $p+p$ and central Pb+Pb data at
the LHC, we have shown that the energy
and mass dependence of the mean value of the string tension  
should be taken into account \cite{ToporPop:2011wk,ToporPop:2010qz}.
Moreover, to better describe the baryon to meson anomaly seen in the data,
a specific implementation of J\=J loops, has to be 
introduced. For a detailed discussion, see 
Refs.~\cite{ToporPop:2011wk,ToporPop:2010qz,Pop:2012ug}.
Similar results can be obtained by including extra diquark-antidiquark
production channels from the strong coherent fields formed in heavy-ion 
collisions \cite{Levai:2011zz}.
 
All $\mathtt{HIJING}$-type models implement nuclear effects such as 
modification of the parton distribution functions, {\em shadowing},
and {\em jet quenching} via medium-induced parton splitting. 
(Collisional energy loss is neglected \cite{Wang:1991xy,Wang:1991hta}.) 
In $\mathtt{HIJING1.0}$ and $\mathtt{HIJINGB\overline{B}2.0}$,
the Duke-Owen (DO) parameterizations of the proton parton densities 
\cite{Duke:1983gd} 
is used to calculate the jet production cross section with $p_T > p_0$.
In both codes, a constant cutoff, 
$p_0 = 2$~GeV/$c$, and a soft parton cross section,
$\sigma_{\rm soft} = 54$ mb, fit the experimental $p+p$ data.
However, for $A+A$ collisions in $\mathtt{HIJINGB\overline{B}2.0}$,
an energy and mass dependence of the cut-off parameter,
$p_0(s,A) = 0.416\sqrt(s)^{0.191} A^{0.128}$~GeV/$c$, 
was introduced \cite{ToporPop:2011wk,ToporPop:2010qz,Pop:2012ug} 
at RHIC and LHC energies in order not to violate the geometrical
limit for minijets production per unit transverse area.
The $p+p$ cutoff was kept constant at $p_0 = 2$~GeV/$c$.
In $\mathtt{HIJING2.0}$~\cite{Deng:2010mv,Deng:2010xg},
a subsequent version of $\mathtt{HIJING1.0}$~\cite{Wang:1991xy,Wang:1991hta} 
the GRV parameterization of the proton parton densities 
\cite{Gluck:1994uf} is implemented.
The GRV small $x$ gluon density is much higher
than that of the DO parameterization.
Here also an energy-dependent cutoff $p_0(s)$ and soft cross section
$\sigma_{\rm soft}(s)$ are also assumed in order to better describe the Pb+Pb 
data at the LHC.  The cutoff used in $\mathtt{HIJINGB\overline{B}2.0}$ varies 
from $p_0 = 1.5$~GeV/$c$ at the CERN SPS, $\sqrt{s} = 20$~GeV, to 4.2~GeV/$c$ at
$\sqrt{s} = 5.5$~TeV while that in $\mathtt{HIJING2.0}$, with a more complex
energy dependence, varies from 1.7 to 
3.5~GeV/$c$ in the same energy range.

One of the main uncertainties in the calculation of the
charged particle multiplicity density
in Pb+Pb collisions is the nuclear modification of parton
distribution functions, especially gluon distributions at small $x$.
In $\mathtt{HIJING}$-type models, 
the parton distributions per nucleon in a nucleus, 
$f_{a/A}(x,\mu_F^2)$, are 
factorizable into parton distributions in a nucleon, $f_{a/N}$,
and the shadowing function for parton $a$, $S_{a/A}$, as in 
Eq.~(\ref{eq:factorized}).
The shadowing parameterization in $\mathtt{HIJING1.0}$ 
\cite{Wang:1991xy,Wang:1991hta} 
is employed,
\begin{eqnarray}
        S_{a/A}(x)&\equiv&\frac{f_{a/A}(x)}{Af_{a/N}(x)} \nonumber\\
         &=&1+1.19\log^{1/6}\! A\,[x^3-1.2x^2+0.21x]\nonumber\\
 & &-s_{a}(A^{1/3}-1)\left[1 -\frac{10.8}{\log(A+1)}\sqrt{x}\right]
e^{-x^2/0.01} \, \, ,
\label{eq:shadow}
\end{eqnarray}
assuming the same dependence for quarks and gluons.  The $\mu_F^2$ evolution of
$S_{a/A}(x,\mu_F)$ is neglected.  The parameter, $s_a$, which determines the
shadowing for $x< 0.1$, the region with the strongest nuclear dependence,
is $s_a = 0.1$.  For $x > 0.1$, the $A$ dependence is rather weak.
The parameterization in Eq.~(\ref{eq:shadow}) agrees with the $x$ dependence
of the quark structure function at small and medium $x$ 
\cite{Wang:1991xy,Wang:1991hta}. 
Because the first part of Eq.~(\ref{eq:shadow}) has a weak $A$ 
dependence, impact parameter dependence is only included on part proportional
to $s_a$.  The impact parameter dependence is given as
\begin{eqnarray}
        s_a(b)=s_a\frac{5}{3}\bigg(1 - \frac{b^2}{R_A^2} \bigg) \, \, ,
                        \label{eq:rshadow}
\end{eqnarray}
where $R_A$ is the radius of the nucleus and $s_a=s_q=s_g = 0.1$.

The LHC Pb+Pb data at $\sqrt{s_{_{NN}}} = 2.76$~TeV 
\cite{ToporPop:2011wk,ToporPop:2010qz} indicate 
that impact-parameter dependent shadowing is required to understand 
the centrality dependence of the charged particle multiplicity density at
midrapidity.  These data place an indirect and model-dependent constraint
on quark and gluon shadowing.
Therefore, it is important to directly study
quark and gluon shadowing in $p+A$ collisions at the LHC.

In contrast, in $\mathtt{HIJING2.0}$~\cite{Deng:2010mv,Deng:2010xg}, 
the factor ($A^{1/3}-1$)
is raised to the power 0.6 and a stronger impact-parameter dependence, different
for quarks and gluons, $s_q = 0.1$ and $s_g=0.22-0.23$ respectively, is used 
to fit the LHC data. 
This stronger gluon shadowing requires jet quenching to be
neglected~\cite{Deng:2010mv,Deng:2010xg}.

All $\mathtt{HIJING}$-type models assume scale-independent
shadowing (independent of $Q^2$). This approximation could   
break down at sufficiently large scales due to the dominance of gluon emission
in the DGLAP \cite{Altarelli:1977zs} evolution equation.
At $Q = 2.0$ and 4.3 GeV/{\it c}, typical scales for minijet
production at RHIC and LHC respectively, low $x$ gluon
shadowing varies by $\approx 13\%$ in the EPS09 LO parameterization 
\cite{Eskola:2009uj}.

\subsection[$\mathtt{AMPT}$]{$\mathtt{AMPT}$ (Z. Lin)}
\label{sec:AMPT_ch}

The multiphase transport model $\mathtt{AMPT}$
\cite{homepage} was also used to calculate the yields and $p_T$ spectra 
of particles produced in $p+p$ and $p+$Pb collisions, 
as well as the nuclear
modification factors $R_{\rm pPb}$ in $p+$Pb collisions. The flow coefficients
have also been calculated.
Both the default, $\mathtt{AMPT-def}$ and the string melting, 
$\mathtt{AMPT-SM}$,
versions of $\mathtt{AMPT}$ \cite{Lin:2004en} have been employed.  

In the default version of $\mathtt{AMPT}$, $\mathtt{AMPT-def}$,
only minijet partons rescatter in
the parton stage. After that, Lund string fragmentation is
used for hadronization with the hadron cascade setting in 
at relatively high energy density.  The cutoff time for the hadron
cascade in these simulations is 30~fm/$c$ 
($\mathtt{NTMAX = 150}$\cite{Lin:2004en}).
On the other hand, the string-melting
version of $\mathtt{AMPT}$, $\mathtt{AMPT-SM}$, converts the usual 
initial-state hadronic strings to partonic matter when the energy density 
in the collision overlap volume is expected to be higher than that
of the QCD phase transition. $\mathtt{AMPT-SM}$ also uses a
simple quark coalescence model to describe bulk hadronization of
the resultant partonic matter.  Thus, secondary interactions are
typically dominated by hadron interactions in $\mathtt{AMPT-def}$
while dominated by parton interactions in $\mathtt{AMPT-SM}$.

Using the default $\mathtt{HIJING}$ parameters for the
Lund symmetric splitting function gives reasonable charged particle 
pseudorapidity distributions,
$dN_{\rm ch}/d\eta$ at central values of pseudorapidity for Pb+Pb collisions 
at LHC energies \cite{Xu:2011fi}. Therefore the same values ($a=0.5$ and
$b=0.9$ GeV$^{-2}$) are used for both $p+p$ and $p+$Pb collisions,
along with the same values of the strong coupling constant and parton cross
section as in Ref.~\cite{Xu:2011fi}. 

In these calculations, $p+p$ events are minimum-bias, including
diffractive events.  The MB $p+$Pb events include no
restrictions on impact parameter. 
The nuclear modification factor, $R_{p \rm Pb}$, as a function of $p_T$ is
obtained by dividing the $p+$Pb distribution by the $p+p$ distribution, both
calculated with the same version of $\mathtt{AMPT}$, normalized by the number of
binary $N+N$ collisions, $N_{\rm coll}$.  The number of collisions is assumed to
be equal to the number of participant
nucleons in the Pb nucleus ($N_{\rm coll} = N_{\rm part}^{\rm Pb}$). 
The collision centrality is defined according
to the number of charged hadrons within $|\eta|<1$. 

\subsection[Leading-order pQCD calculations]{Leading-order pQCD 
calculations (Z.-B. Kang, I. Vitev, H. Xing)}
\label{sec:Ivan-description}

The details of the calculations described here can be found in 
Ref.~\cite{Kang:2012kc}.  A summary is given here.
To leading order in the framework of factorized perturbative QCD, single 
inclusive hadron production in $p+p$ collisions, 
$p(p_1)+p(p_2)\to h(p_h)+X$, can be written as~\cite{Owens:1986mp}
\ben
\frac{d\sigma}{dy d^2p_T} &=& K\frac{\alpha_s^2}{s}\sum_{a,b,c}\int 
\frac{dx_1}{x_1}d^2k_{T_1} \, 
f_{a/N}(x_1, k_{T_1}^2)\int \frac{dx_2}{x_2}d^2k_{T_2} \, 
f_{b/N}(x_2, k_{T_2}^2) 
\nnu
&& \times
\int \frac{dz_c}{z_c^2}\,  D_{h/c}(z_c) H_{ab\to c}(\hat s,\hat t,
\hat u)\delta(\hat s+\hat t+\hat u),
\label{light}
\een
where $y$ and $p_T$ are the rapidity and transverse momentum of the produced 
hadron and $\sum_{a,b,c}$ runs 
over all parton flavors. In Eq.~(\ref{light}), $s=(p_1+p_2)^2$; 
$D_{h/c}(z_c)$ is the fragmentation function 
(FF) of parton $c$ into hadron $h$; $H_{ab\to c}(\hat s,\hat t,\hat u)$ are 
hard-scattering coefficient functions dependent on the partonic Mandelstam
invariants $\hat s,\hat t,\hat u$ \cite{Owens:1986mp}.
A phenomenological $K$ factor is included to account for higher-order QCD 
contributions. The parton distribution functions, $f_{a,b/N}(x, k_{T}^2)$, are
dependent on the longitudinal momentum fraction $x$ and the partonic transverse 
momentum $k_T$. The $k_T$-dependence is included in 
order to incorporate the Cronin effect in $p+A$ collisions. A Gaussian form 
is assumed~\cite{Owens:1986mp},
\ben
f_{a/N}(x_1, k_{T_1}^2) = f_{a/N}(x_1) \frac{1}{\pi\langle k_T^2\rangle} 
e^{-k_{T_1}^2/\langle k_T^2\rangle},
\label{gauss}
\een
where $f_{a/N}(x_1)$ are the usual collinear PDFs in a nucleon.  The 
factorization scale dependence has been suppressed in the arguments of 
$f_{a/N}$.

In $p+p$ collisions, $\langle k_T^2\rangle_{pp}=1.8$ GeV$^2/c^2$. 
The CTEQ6L1 PDFs \cite{Pumplin:2002vw} are used with the fDSS 
parameterization of the parton-to-hadron fragmentation 
functions~\cite{deFlorian:2007aj}. 
The factorization and renormalization scales are fixed 
to the transverse momentum of the produced particle, $\mu_F=\mu_R=p_T$, and
are suppressed in Eqs.~(\ref{light}) and (\ref{gauss}). 
An $\mathcal{O}(1)$ $K$-factor is found to give a good description of hadron 
production at both RHIC and LHC energies.

\subsubsection{Cold nuclear matter effects}

The $p+A$ ({\it e.g.} d+Au or $p+$Pb) nuclear modification factor, $R_{pA}$, 
is typically defined as:
\ben
R_{pA} = \left[
\frac{d\sigma_{pA}}{dy d^2p_T} \right] \left[\frac{d\sigma_{pp}}{\langle 
N_{\rm coll}\rangle dy d^2p_T}\right]^{-1} \, \, ,
\een
where $\langle N_{\rm coll}\rangle$ is the average number of binary 
nucleon-nucleon collisions.  The deviation 
of $R_{pA}$ from unity reveals the presence of cold nuclear matter (CNM) 
effects in $p+A$ collisions. 

A  variety of CNM effects can affect particle production.  This section
describes those that arise from the elastic, inelastic and coherent scattering 
of partons in large nuclei~\cite{Vitev:2006bi}. The proton and neutron 
composition of the interacting nuclei are also accounted for. In particular, 
these effects include isospin, the Cronin effect, cold nuclear matter energy 
loss and dynamical shadowing. These effects have been well 
documented in the literature.  Their implementation is briefly described
here.

\paragraph{Isospin} The isospin effect can be easily accounted for on average 
in the nuclear PDFs for a 
nucleus with  mass number $A$ and charge $Z$ by~\cite{Kang:2008wv}:
\ben
f_{a/A}(x) = \frac{Z}{A} f_{a/p}(x) + \left(1-\frac{Z}{A}\right)f_{a/n}(x) 
\, \, ,
\label{iso}
\een
assuming no modifications of the parton densities.
In Eq.~(\ref{iso}), $f_{a/p}(x)$ and $f_{a/n}(x)$ are the PDFs in a proton and a
neutron, respectively.  The neutron PDFs are related to those in the proton by 
isospin symmetry. 

\paragraph{Cronin effect} The Cronin effect has been well 
documented~\cite{Accardi:2002ik}. It can be modeled by initial-state multiple 
parton scatterings in cold nuclei and the corresponding induced parton 
transverse momentum broadening~\cite{Qiu:2003pm,Ovanesyan:2011xy}.
In particular, if the PDFs, $f_{b/A}(x_2, k_{T_2}^2)$, have a normalized Gaussian 
form, the random elastic scattering induces further $k_T$-broadening in the 
nucleus:
\ben
\langle k_{T_2}^2\rangle_{pA} = \langle k_{T_2}^2\rangle_{pp} 
+ \left\langle \frac{2\mu^2 L}{\lambda_{q,g}}\right\rangle \zeta \, \, .
\een
Here $k_{T_2}$ is the transverse momentum component of the parton prior to the 
hard scattering, $\zeta=\ln(1+\delta p_T^2)$, $\delta = 0.14$ (GeV$/c)^{-2}$, 
$\mu^2=0.12$ (GeV$/c)^2$, and $\lambda_g= (C_F/C_A) \lambda_q = 1$ fm. These 
parameters describe the RHIC data 
reasonably well.

\paragraph{Cold nuclear matter energy loss} As the parton from the proton 
undergoes multiple scattering in the nucleus before the hard collision, it can 
lose energy due to medium-induced gluon bremsstrahlung. 
This effect can be easily implemented as a shift in the momentum fraction in 
the PDFs
\ben
f_{q/p}(x_1) \to f_{q/p}\left(\frac{x_1}{1-\epsilon_{q, \, \rm eff}}\right) 
\, \, ,
\quad
f_{g/p}(x_1) \to f_{g/p}\left(\frac{x_1}{1-\epsilon_{g, \, \rm eff}}\right) 
\, \, .
\label{eloss}
\een
Ideally, Eq.~(\ref{eloss}) should include a convolution over the probability 
of cold nuclear matter energy loss~\cite{Neufeld:2010dz}. However, concurrent 
implementation of this distribution together with the Cronin effect and 
coherent power corrections is computationally very demanding. The main effect 
of fluctuations due to multiple gluon emission is an effective reduction in the
fractional energy loss $\epsilon_{q,g \, \rm  eff}$ relative to the mean value 
$\langle \epsilon_{q,g} \rangle = \langle \sum_i (\Delta E_i /E) \rangle$ where 
the sum runs over all medium-induced gluons.
Here $ \epsilon_{q,g \rm eff} = 0.7 \langle \epsilon_{q,g} \rangle$.
The average cold nuclear matter energy loss is obtained by integrating the 
initial-state medium-induced bremsstrahlung spectrum first  
derived in Ref.~\cite{Vitev:2007ve}. It also depends on the typical transverse 
momentum transfer squared per interaction between the parton and the medium 
and the gluon mean-free path $\lambda_g$. 
Therefore, the parameters are constrained to be the same as in the 
implementation of the
Cronin effect, $\mu^2=0.12$~(GeV$/c)^2$  and $\lambda_g=1$~fm. 
This calculation of initial-state cold nuclear matter energy loss 
has been shown to give a good description of the nuclear modification of 
Drell-Yan production 
in fixed-target experiments~\cite{Neufeld:2010dz}. 

\paragraph{Dynamical shadowing} 
Power-suppressed resummed coherent final-state scattering of the struck partons 
leads to shadowing effects (suppression of the cross section in the small-$x$ 
region)~\cite{Qiu:2004da}. The effect can be interpreted as the dynamical 
generation of parton mass in the background gluon field of the 
nucleus~\cite{Qiu:2004qk}.  Thus
\ben
x \to x \left(1+C_d \frac{\xi^2(A^{1/3}-1)}{-\hat t}\right)\, \, ,
\label{eq:dyn_shad}
\een
where $x$ is the parton momentum fraction in the lead ion, $C_d=C_F (C_A)$ 
if the parton $d=q (g)$ in the $2 \rightarrow 2$ parton scattering $ab\to cd$,
and $\xi^2$ represents the characteristic 
scale of the multiple scattering per nucleon.  At RHIC energies, 
$\sqrt{s_{_{NN}}}=200$ GeV, 
$\xi^2_q = C_F/C_A \xi^2_g = 0.12$~GeV$^2$~\cite{Qiu:2004da}
gives a good description of the nuclear modification in d+Au collisions for 
both single hadron and dihadron production~\cite{Kang:2011bp}. 

\subsection[Initial-state Shadowing]{Initial-state Shadowing (G. G. 
Barnaf\"oldi, J. Barette, M. Gyulassy, P. Levai, G. Papp and 
V. Topor Pop)}
\label{sec:Levai-description}

The calculations in this section use the $\mathtt{kTpQCD\_v2.0}$ code, based on 
a phenomenologically-enhanced, perturbative QCD improved parton model 
described in detail in Refs.~\cite{Zhang:2001ce,Papp:2002ub}.
The main feature of this model is the phenomenologically-generalized parton
distribution function employed to handle nonperturbative effects at
relatively low-$x$ and small $p_T$. The model includes intrinsic $k_T$ 
broadening with the average $k_T$ left as a free parameter to correct for 
nonperturbative effects.  The $k_T$ value is determined from $p+p$ data over a 
wide range of energies.  Within the framework of this model, the 
$k_T$-broadening in $p+A$ and $A+A$ collisions is related to nuclear multiple 
scattering and can generate the Cronin 
enhancement~\cite{Cronin:1974zm,Antreasyan:1978cw}
that appears within $3 \leq p_T \leq 9$~GeV/$c$ from SPS to RHIC energies.

\subsubsection{Theoretical Background}

The $\mathtt{kTpQCD\_v2.0}$ code calculates the invariant cross section for 
hadron production in $p+p$, $p+A$ and $A+A$ collisions at LO or NLO in the 
$k_T$-enhanced pQCD-improved parton model assuming collinear factorization.  
The code provides a Monte Carlo-based integration of the 
convolution~\cite{Papp:2002ub}, written here for $p+p$ collisions,
\begin{eqnarray}
\label{hadX}
 E_{h}\frac{d \sigma_h^{pp}}{d ^3p_T} &=&
        \frac{1}{s} \sum_{abc}
  \int^{1-(1-V)/z_c}_{VW/z_c} \frac{d v}{v(1-v)} \ 
  \int^{1}_{VW/vz_c} \frac{ d w}{w} 
  \int^1 {d z_c}  \\
  && \times  \int {d^2 {\vec k}_{T_1}} \ \int {d^2 {\vec k}_{T_2}}
        \, \, f_{a/p}(x_1,{\vec k}_{T_1},\mu_F^2)
        \, f_{b/p}(x_2,{\vec k}_{T_2},\mu_F^2) 
   \nonumber \\
&& \times  
 \left[
 \frac{d {\widetilde \sigma}}{d v} \delta (1-w)\, + \,
 \frac{\alpha_s(\mu_R)}{ \pi}  K_{ab,c}(\hat{s},v,w,\mu_F,\mu_R,\mu_{\rm Fr}) 
\right]  \frac{D_{c}^{h} (z_c, \mu_{\rm Fr}^2)}{\pi z_c^2}  \,\, . \nonumber
\end{eqnarray}
Here $d {\widetilde \sigma}/ d v$ represents the Born cross section 
of the partonic subprocess $ab \to cd$ while 
$K_{ab,c}(\hat{s},v,w,\mu_F,\mu_R,\mu_{\rm Fr})$ is the 
next order correction term.  The proton and parton level NLO kinematic variables
are ($s$, $V$, and $W$) and ($\hat{s}$, $v$, and $w$) respectively 
\cite{Papp:2002ub,Aversa:1988vb,Aurenche:1998gv,Aurenche:1999nz}.  
The various scales are $\mu_F$, the factorization
scale; $\mu_R$, the renormalization scale; and $\mu_{\rm Fr}$ 
the fragmentation scale.  The factorization and the 
renormalization scales are related to the momentum of the intermediate 
jet, $\mu_F=\mu_R=\kappa p_q$ where $\kappa = 2/3$, $p_q=p_T/z_c$ and 
$z_c$ is the fraction of
parton $c$ momenta transferred to the final hadron $h$.  The fragmentation 
scale is related to the final hadron momentum by $\mu_{\rm Fr}=\kappa p_T$.

The $x$-dependent proton parton distribution functions, $f_{a/p}(x,\mu_F^2)$, 
defined
in the infinite momentum frame, are generalized to three dimensions by
incorporating an initial $k_T$ dependence,
\begin{equation}
\label{gpdf}
f_{a/p}(x, \vec k_{T},\mu_F^2) = g(\vec k_{T}) f_{a/A}(x,\mu_F^2)  
\,\, .
\end{equation}
The two-dimensional initial transverse momentum distribution, $g({\vec k}_T)$, 
with intrinsic parton $k_T$ employed in these calculations is described in 
Refs.~\cite{Zhang:2001ce,Papp:2002ub,Wang:1998ww,Wong:1998pq}.
The
$k_T$ distribution is described by a Gaussian,
\begin{equation}
\label{kTgauss}
g({\vec k}_T) \ = \frac{1}{\pi \langle k^2_T \rangle} 
e^{-{k^2_T}/{\langle k^2_T \rangle}}    \,\,\, .
\end{equation}
Here $\langle k_T^2 \rangle$ is the width of the $k_T$ 
distribution, related to the magnitude of the average parton transverse 
momentum by $\langle k_T^2 \rangle = 4 \langle k_T \rangle^2 /\pi$. 
This treatment was successfully applied at LO in Ref.~\cite{Zhang:2001ce}, along
with a $K_{\rm jet}$-based NLO calculation 
\cite{Barnafoldi:2002sj,Barnafoldi:2002xp}.  In order to 
reproduce results for $NN$ collisions at relatively low $x$,
$\langle k_T^2 \rangle = 2.5$ GeV$^2/c^2$
was required.

The LO or NLO fragmentation functions, 
$D_{c}^{h}(z_c, \mu_{\rm Fr}^2)$, are the probability for parton $c$ to 
fragment into hadron $h$ with momentum fraction $z_c$ at fragmentation scale 
$\mu_{\rm Fr}$.  
The MRST(cg) \cite{Martin:2001es} parton densities are used in 
Eq.~(\ref{gpdf}), along with the KKP 
parameterization~\cite{Kniehl:2000fe} of the fragmentation
functions.  Both these sets can be applied at relatively small scales, 
$\mu_F^2 = \mu_{\rm Fr}^2 \approx 1.25$ GeV$^2$.  
Thus the results obtained in these calculations
are applicable for $p_T \geq 2$ GeV/$c$.


\subsubsection{Incorporating
Initial-State Nuclear Effects in $p+A$ and $A+A$ Collisions}

Proton-nucleus and nucleus-nucleus collisions can be described by 
incorporating the appropriate collision geometry and
and nuclear shadowing. In the Glauber framework, the cross section for
hadron production in an $A+A'$ collision can be written as an integral over 
impact parameter $b$:
\begin{equation}
\label{dAuX}
  E_{h}\frac{d \sigma_h^{AA'}}{ d ^3 p_T} =
  \int d ^2b \, d ^2r \,\, T_A(r) \,\, T_{A'}(|{\vec b} - {\vec r}|) \,
  E_{\pi} \,    \frac{d \sigma_{\pi}^{pp}(\langle k_T^2 \rangle_{pA},
\langle k_T^2 \rangle_{pA'})}{d ^3p}
\,\,\, .
\end{equation}
Here the nuclear thickness function, $T_{A}(b) = \int d z \, \rho_{A}(b,z)$,
employing the Woods-Saxon density distribution, is normalized so that 
$\int d ^2b \, T_{A}(b) = A$. 

The $p+p$ cross section from Eq.~(\ref{hadX}) includes increased $k_T$ widths 
relative to $p+p$ collisions, Eq.~(\ref{kTgauss}), as a 
consequence of multiple scattering in nuclei, see Eq.~(\ref{ktbroadpA}).  
The increased width of the $k_T$ distribution is
taken into account by adding a function, $h_{pA}(b)$, that describes the
number of effective $N+N$ collisions at impact parameter $b$, weighted by
the average transverse momentum squared imparted by each collision, to the width
in $p+p$ collisions,  $\langle k_T^2 \rangle_{pp}$,
\begin{equation}
\label{ktbroadpA}
\langle k_T^2 \rangle_{pA} = \langle k_T^2 \rangle_{pp} + C h_{pA}(b) \, \, .
\end{equation}
The function $h_{pA}(b)$ is expressed in terms of the number of collisions 
suffered by the incoming proton in the target nucleus, 
$\nu_A(b) = \sigma_{NN} T_{A}(b)$, where $\sigma_{NN}$ is the inelastic 
$N+N$ cross section:
\begin{equation}
\label{hpab}
  h_{pA}(b) = \left\{ \begin{array}{cc}
                \nu_A(b)-1 & \nu_A(b) < \nu_{m} \\
                \nu_{m}-1 & \mbox{otherwise} \\
        \end{array} \right.\ .
\end{equation}
For heavy nuclei, the maximum number of collisions is 
$3 \leq \nu_{m} \leq 4$ with $C = 0.4$~GeV$^2/c^2$.

Finally, the nuclear PDFs are modified by 
shadowing~\cite{Li:2001xa,Eskola:1998df,Eskola:2008ca,Hirai:2001np}. This 
effect, as well as the isospin 
asymmetry, are taken into account on average using the scale independent 
parameterization of $S_{a/A}(x)$ adopted from Ref.~\cite{Wang:1998ww},
\begin{equation}
f_{a/A}(x,\mu_F^2) = S_{a/A}(x) \left[\frac{Z}{A} f_{a/p}(x,\mu_F^2) + 
\left(1-\frac{Z}{A}\right) f_{a/n}(x,\mu_F^2) \right]   \,\,\,\,  ,
\label{shadow}
\end{equation}
where the neutron parton density, $f_{a/n}(x,\mu_F^2)$, is related to that of the
proton.  Results are shown with the EKS98~\cite{Eskola:1998df}, 
EPS08~\cite{Eskola:2008ca} and HKN~\cite{Hirai:2001np} parameterizations, 
as well as with the updated $\mathtt{HIJING}$ 
parameterization~\cite{Li:2001xa}. The EKS98, EPS08 and
HKN parameterizations differ for quarks, antiquarks and gluons but are 
independent of impact parameter.  The new $\mathtt{HIJING}$ parameterization
differentiates between quarks and gluons but can include impact parameter
dependence, as in Eq.~(\ref{eq:rshadow}).

\section{Charged particles}
\label{sec:charged_hadrons}

In this section, results on the charged particle multiplicity
and $p_T$ distributions and the suppression factor $R_{p{\rm Pb}}$ as a function
of $p_T$ are compiled. These results are compared with the ALICE 
test beam data where 
available.  Other, related, predictions for charged particle
observables are also shown.  
The upcoming $p+$Pb run at the LHC can place important constraints
on models of the initial state. 

Note that the 
LHC magnet design requires the magnetic rigidity of the beams in the two 
rings to be the same.  The proton beam, at 4 TeV, circulated in negative 
$z$-direction (toward negative rapidity)
in the ALICE laboratory frame while a beam of fully-stripped Pb
ions of $(82/208) \times 4$ TeV/nucleon circulated in the positive 
$z$-direction (toward positive rapidity), implying Pb$+p$ collisions if the
first-named collision partner travels in the direction of positive rapidity,
rather than $p+$Pb collisions, as the collisions are referred to throughout 
this text.  
This configuration resulted in a center of mass energy of
$\sqrt{s_{_{NN}}} = 5.02$~TeV, moving with a rapidity difference, $\Delta y_{_{NN}}
= 0.465$, in the direction of the proton beam.

It is important to note that the most of the predictions shown here were 
orginally made assuming that the proton circulated toward positive rapidity
and the nucleus toward negative rapidity, similar to the fixed-target 
configuration and also the convention for d+Au collisions at RHIC.  In such
cases, the low $x$ nuclear parton distributions are probed at positive rapidity.
Here and throughout the remainder of this paper, the results shown have been
adjusted to the Pb$+p$ convention of the ALICE data \cite{ALICE_dndeta} unless
otherwise explicitly noted.

ALICE reported the primary
charged particle pseudorapidity density in the laboratory
frame, $dN_{\rm ch}/d\eta_{\rm lab}$ in non single-diffractive (NSD) $p+$Pb
collisions. The lab frame pseudorapidity is defined as $\eta_{\rm lab} = 
- \ln \tan(\theta/2)$ where $\theta$ is the polar angle between the direction
of the produced charged particle and the beam axis.  The primary particles
are due to both prompt production in the collision and strong decays.

Calculations are typically performed in the center of mass frame but the
pseudorapidity densities shown here have been also calculated in the lab frame.
In the ALICE paper \cite{ALICE_dndeta}, calculations in the center of mass frame
were shifted by $\Delta y_{_{NN}}$ in the lab frame.  This is only approximately
correct since, at low $p_T$, the rapidity and pseudorapidity are not identical.
The uncertainty on $dN_{\rm ch}/d\eta_{\rm lab}$ due to the choice of frame
was estimated to be less than 6\% \cite{ALICE_dndeta}.  It is worth noting that
there is no ambiguity due to the calculational frame for identified particles
thus the frame dependence is not discussed in later sections.

\subsection{Multiplicity distribution}
\label{sec:dnchdeta}

The calculations of the charged particle multiplicity distributions are
described here.  Saturation model predictions are discussed first, followed
by event generator predictions and pQCD calculations with cold matter effects.

\subsubsection[Saturation Approaches]{Saturation Approaches (J. Albacete, A. Dumitru, H. Fujii, Y. 
Nara, A. Rezaeian, and R. Vogt)}
In the rcBK approach, used by Albacete and collaborators \cite{Albacete:2012xq} 
and Rezaeian for $R_{p{\rm Pb}}$ \cite{Rezaeian:2012ye,Rezaeian:2011ia},
the initial condition for the evolution of the dipole
scattering amplitude can be written as (see also Eq.~(\ref{mv})),
\begin{equation}
\mathcal{N}(r,x_0) = 1 - \exp \left[- \frac{1}{4}
\left(r^2 Q_s^2(x_0) \right)^\gamma \log \left(e+\frac{1}{r \Lambda}
  \right)\right] \,\, .  
\label{eq:N0}
\end{equation}
There are three sets of unintegrated gluon distributions that are solutions of
the rcBK small-$x$ evolution equations with the AAMQS initial condition in
Eq.~(\ref{eq:N0}).
The values of $\gamma$ and $Q_{s0}^2(x_0 = 0.01)$
for protons that provide good fits to the $e^- +p$ data are given in 
Table~\ref{tab:rcBK_IC}.  Albacete {\it et al.} use all three initial conditions
\cite{Albacete:2012xq}. Rezaeian uses g1.119 
\cite{Rezaeian:2011ia,JalilianMarian:2011dt} 
in the calculations of $R_{p{\rm Pb}}$ in Sec.~\ref{sec:RpPb} but employs the
b-CGC approach \cite{Rezaeian:2012ye,Watt:2007nr} 
to calculate $dN_{\rm ch}/d\eta$ here.

\begin{table}[pt]
\tbl{The AAMQS initial conditions used 
in the dipole evolution of the rcBK approach for protons 
\protect\cite{Albacete:2012xq,Rezaeian:2011ia}.}
{\begin{tabular}{@{}ccc@{}}\toprule
Set & $Q_{s0}^2(x_0 = 0.01)$ (GeV$^2/c^2$) & $\gamma$ \\ \hline
MV     & 0.200 & 1 \\
g1.101 & 0.157 & 1.101 \\
g1.119 & 0.168 & 1.119 \\ \botrule
\end{tabular}}
\label{tab:rcBK_IC}
\end{table}

\begin{figure}[htbp]
\begin{center}
\includegraphics[width=0.495\textwidth]{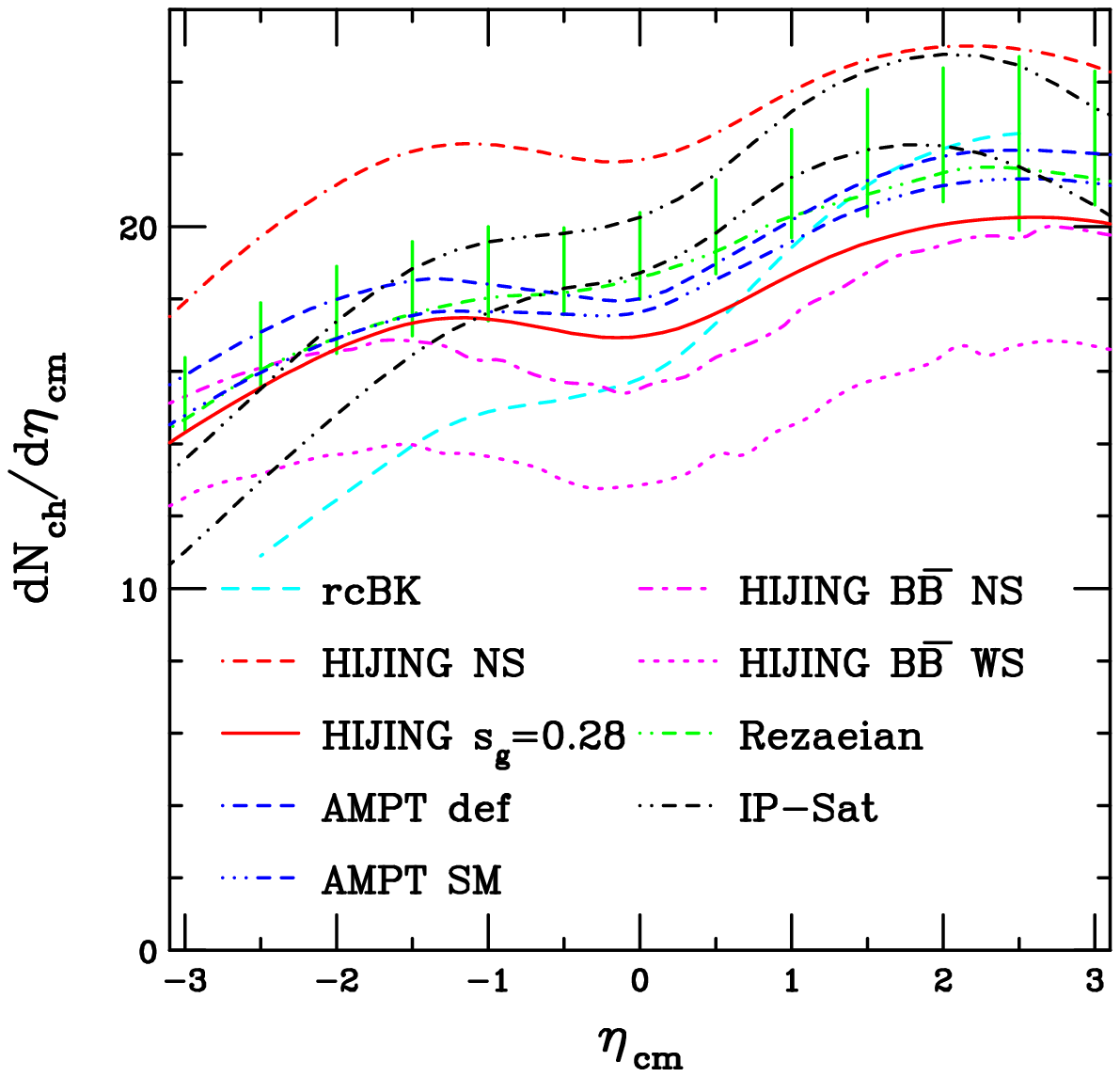}
\includegraphics[width=0.495\textwidth]{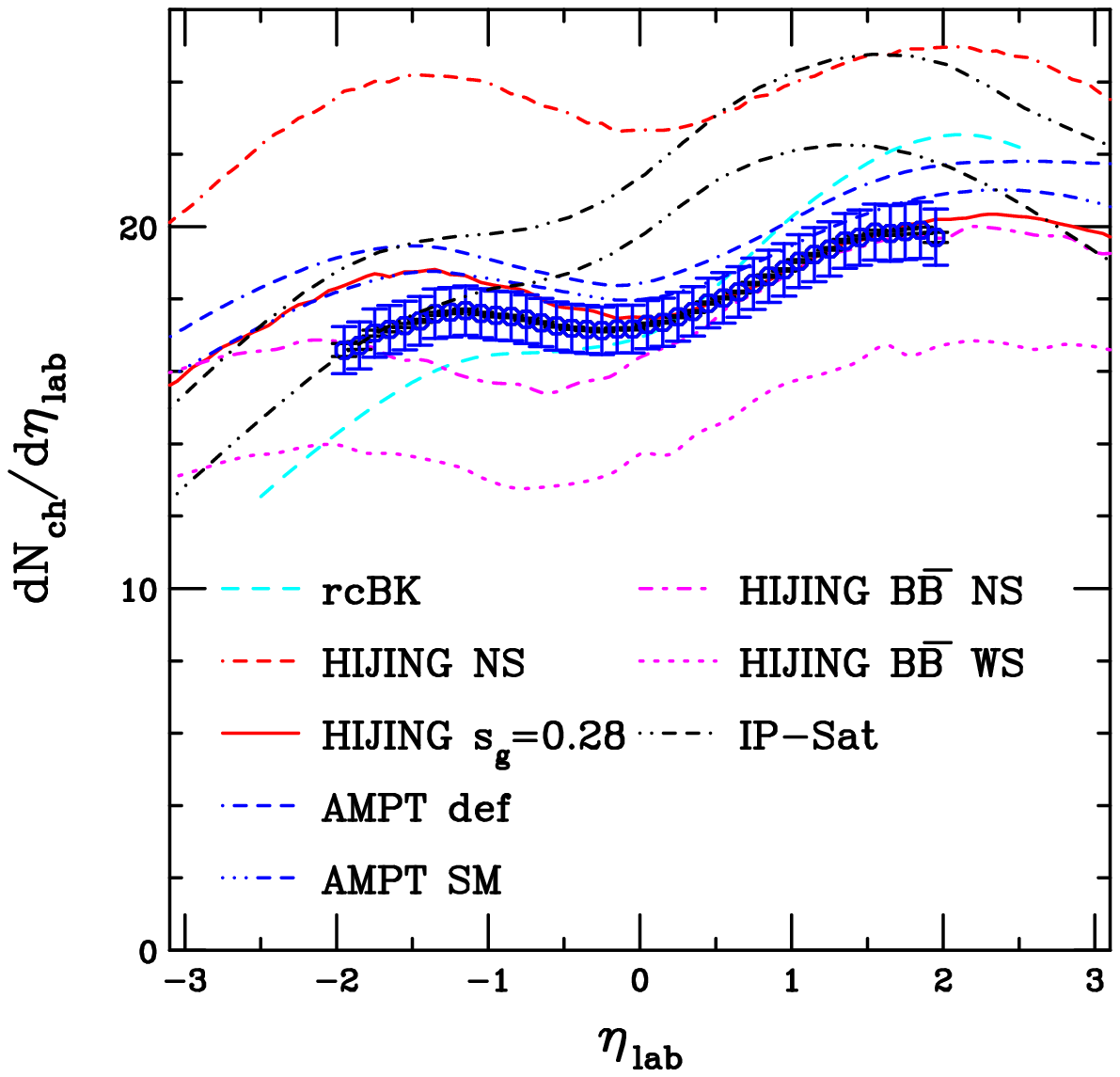}
\caption[]{Charged particle pseudorapidity distributions at $\sqrt{s_{_{NN}}}
= 5.02$ TeV in the CM (left) and lab (right) frames. The rcBK (dashed cyan)
result is from Ref.~\protect\cite{Albacete:2012xq}.  The band labeled
Rezaeian (dot-dot-dash-dashed green with vertical bars outlining the
uncertainty), described in Sec.~\protect\ref{sec:amir_ch}, is only shown in
the center-of-mass frame.  The band showing the IP-Sat result described in
Sec.~\protect\ref{sec:IPSat} is outlined by the dot-dot-dash-dashed black
curves.
The $\mathtt{HIJING2.1}$ result without (NS, 
dot-dash-dash-dashed red) and with shadowing ($s_g = 0.28$, solid red)
and the $\mathtt{HIJINGB\overline{B}2.0}$ 
result without (dot dashed magenta) and
with shadowing (dotted magenta) are also shown.  Finally, the 
$\mathtt{AMPT-def}$ (dot-dash-dash-dashed blue) and $\mathtt{AMPT-SM}$
(dot-dot-dot-dash-dash-dashed blue) are given.  The ALICE results from
Ref.~\protect\cite{ALICE_dndeta} are given on the right-hand side.  The
systematic uncertainties are shown in blue, the statistical uncertainties are
too small to be visible on the scale of the plot.
}
\label{fig:dndeta_cm_lab}
\end{center}
\end{figure}

\paragraph[rcBK]{rcBK (J. Albacete, A. Dumitru, H. Fujii and Y. 
Nara)}
The calculation of $dN_{\rm ch}/d\eta$ by Albacete {\it et al.}\ 
\cite{Albacete:2012xq}, employing 
$\gamma = 1.119$, in both the center of mass and lab frames, is shown
in the dashed cyan curves in Fig.~\ref{fig:dndeta_cm_lab}.

\paragraph[b-CGC]{b-CGC (A. Rezaeian)}
The results of calculations by Rezaeian \cite{Rezaeian:2012ye,Rezaeian:2011ia} 
in the center of mass
frame are shown by the green curves in Fig.~\ref{fig:dndeta_cm_lab}.  The
vertical lines indicate the uncertainty in the predictions.  The dipole 
forward (proton direction) 
amplitude is calculated in the b-CGC saturation model \cite{Watt:2007nr}. 
which has an explicit impact parameter dependence as well as approximately 
incorporates all known features of small-$x$ physics.  It describes the 
small-$x$ HERA data, including diffractive data \cite{Watt:2007nr}, and also 
the RHIC and LHC data at small-$x$ 
\cite{Levin:2010dw,Levin:2011hr,Rezaeian:2011ia,Levin:2010br,Rezaeian:2011ss,Levin:2010zy}. 
This framework also provides an excellent description of the 
charged hadron multiplicity in d+Au collisions at RHIC in addition to 
$p+p$ and $A+A$ collisions over a range of energies \cite{Rezaeian:2011ia}.  
For other
saturation model predictions in minimum bias $p+A$ collisions at 
$\sqrt{s_{_{NN}}} = 4.4$ TeV, see Ref.~\cite{Rezaeian:2012vc}. 

\paragraph[IP-Sat]{IP-Sat (P. Tribedy and R. Venugopalan)}
The calculation of the minimum-bias charged particle pseudorapidity distribution
was performed in the $k_T$-factorization approach using the IP-Sat model
\cite{Kowalski:2003hm} in Eq.~(\ref{eq:ktfact1}). 
For the calculation of $dN_{\rm ch}/d\eta$ in the lab frame, a constant rapidity 
shift of 0.46 has been applied. The bands shown in Fig.~\ref{fig:dndeta_cm_lab} 
arises due to uncertainties in the parameters. 

The charged particle multiplicity distributions obtained
by the event generators $\mathtt{HIJING}$, 
$\mathtt{HIJINGB\overline{B}2.0}$
and $\mathtt{AMPT}$ are now discussed.  

\subsubsection[$\mathtt{HIJING2.1}$]{$\mathtt{HIJING2.1}$ (R. Xu, W.-T. Deng 
and X.-N. Wang)}
The $\mathtt{HIJING2.1}$ predictions in Fig.~\ref{fig:dndeta_cm_lab} are 
indicated by the dot-dash-dash-dashed curves without shadowing and the 
solid curves with $s_g = 0.28$.  The difference between results with and without
shadowing are largest for this calculation.

\subsubsection[$\mathtt{HIJINGB\overline{B}}2.0$]
{$\mathtt{HIJINGB\overline{B}}2.0$ (G. G. Barnaf\"oldi, J. Barette, 
M. Gyulassy, P. Levai, M.Petrovici and V. Topor Pop)}
The $\mathtt{HIJINGB\overline{B}2.0}$ 
predictions of $dN_{\rm ch}/d\eta$ 
in minimum bias collisions
employ the values for the minijet cutoff and string-tension parameters 
of $p_0=3.1$ GeV/$c$ and $\kappa=2.0$ GeV/fm.  These values are determined 
from fits to $p+p$ and $A+A$ systematics from RHIC to the LHC, see 
Ref.~\cite{ToporPop:2011wk,ToporPop:2010qz,Pop:2012ug} for details.
Note that these calculations assume no jet quenching. 

The absolute normalization of $dN_{\rm ch}/d\eta$ is sensitive
to the low $p_T$, $p_T < 2$ GeV/$c$, nonperturbative hadronization
dynamics encapsulated in the Lund 
\cite{Andersson:1986gw,NilssonAlmqvist:1986rx} 
$\mathtt{JETSET}$ \cite{Bengtsson:1987kr} string fragmentation 
constrained by lower energy $e^+ +e^-$, $e^\pm +p$, and $p+p$ data.  The 
default $\mathtt{HIJING1.0}$ shadowing parametrization leads to substantial 
reduction of the global multiplicity at the LHC. 
The $\mathtt{HIJINGB\overline{B}2.0}$ 
results without shadowing are substantially reduced 
relative to the same predictions with $\mathtt{HIJING1.0}$ 
\cite{Wang:1991xy,Wang:1991hta}
because both the default minijet cutoff $p_0=2$~GeV/$c$
and vacuum string tension $\kappa_0=1$~GeV/fm used in $\mathtt{HIJING1.0}$ 
are generalized to vary monotonically with $\sqrt{s_{_{NN}}}$ and $A$.
As discussed in Ref.~\cite{ToporPop:2011wk,ToporPop:2010qz,Pop:2012ug}, 
systematics of 
multi-particle production in $A+A$ collisions from RHIC to the LHC 
are used to fix the $\sqrt{s_{_{NN}}}$ and $A$ dependence of $p_0$ and
$\kappa$.  The resulting dependencies, 
$p_0(s,A) = 0.416 \sqrt{s}^{0.191} A^{0.128}$~GeV/$c$ and 
$\kappa(s,A) = \kappa_0 (s/s0)^{0.04} A^{0.167}$~GeV/fm
\cite{Pop:2012ug} lead to $p_0 = 3.1$~GeV/$c$ and 
$\kappa = 2.1$~GeV/fm in $p+$Pb collisions at 5.02~TeV.

Constant values of the cutoff are employed in $p+p$ collisions, independent
of the incident energy,
$p_0^{pp} = 2$~GeV/$c$, and string tension, $\kappa_{pp} = 1.9$~GeV/fm. 
Note that, even without shadowing,
the increase of $p_0$ from 2 GeV/$c$ in $p+p$ collisions to 3.1 GeV/$c$ in 
$p+$Pb collisions causes a reduction in the minijet cross section and hence
the final pion (charged particle) multiplicity.  Such a reduction is also 
required to fit the 
slow growth (by a factor of 2.2) in the $A+A$ charged particle multiplicity 
from RHIC to LHC \cite{Harris:2012kj}. This reduction could be
interpreted as phenomenological evidence for gluon saturation beyond leading
twist shadowing.

\subsubsection[$\mathtt{AMPT}$]{$\mathtt{AMPT}$ (Z.-W. Lin)}
The $\mathtt{AMPT}$ default and string melting calculations are shown in
the blue curves in Fig.~\ref{fig:dndeta_cm_lab}.  The differences in the
two scenarios is not large.  Indeed it is much smaller than models with and
without shadowing.  These differences can arise from several sources: the
relative rescattering strengths in the parton and hadron stages; the 
hadronization models; or a combination of rescattering and hadronization.

\subsubsection[Forward/Backward difference]{Forward/Backward difference 
(R. Vogt)}
The event generator calculations produce distributions that do not show
a strong forward/backward difference between the lead and proton peaks.
The CGC-based calculations, however, show a much stronger dependence of the
results on pseudorapidity, in both frames.

The charged particle multiplicity results can be further quantified
by comparing the measured to predicted particle density at midrapidity,
near the proton peak, $\eta_{\rm lab} = -2$, and the lead peak,
$\eta_{\rm lab} =2$.  The absolute values of $dN_{\rm ch}/d\eta_{\rm lab}$ at
$\eta_{\rm lab} = -2$, 0 and 2, along with the ratio $R$ of the multiplicities
at $\eta_{\rm lab} = 2$ to $\eta_{\rm
lab} = -2)$ are given in Table~\ref{tab:etatable}.  The ALICE results as well
as the model results included both here and in the ALICE paper 
\cite{ALICE_dndeta} are given.  There are two tabulated values for both
$\mathtt{HIJINGB\overline{B}2.0}$ NS (no shadowing) and WS 
(with shadowing).  The first were shifted from the center-of-mass to the lab
frame by the ALICE collaboration and the second were direct lab frame
calculations.

\begin{table}[htbp]
\tbl{Comparison of values of $dN_{\rm ch}/d\eta_{\rm lab}$ at
$\eta_{\rm lab} = -2$, 0, 2 and the ratio $dN_{\rm ch}/d\eta_{\rm lab}
|_{\eta_{{\rm lab} = 2}}/dN_{\rm ch}/d\eta_{\rm lab}|_{\eta_{{\rm lab} = -2}}$, 
denoted by $R$ below.  The tabulated IP-Sat result is the average of the
upper and lower limits depicted on the right-hand side of 
Fig.~\ref{fig:dndeta_cm_lab}.  The $*$ on $\mathtt{HIJINGB\overline{B}2.0}$ 
indicates that the 
calculations have been shifted to the lab frame by the ALICE Collaboration
while the $\dagger$ are results provided by V.~Topor Pop {\it et al.}.  
Adapted from Ref.~\protect\cite{ALICE_dndeta}.}
{\begin{tabular}{@{}ccccc@{}}\toprule 
& \multicolumn{3}{c}{$dN_{\rm ch}/d\eta_{\rm lab}$} & $R$ \\ 
& $-2$ & 0 & 2 & \\ \hline
ALICE & $16.65 \pm 0.65$ & $17.24 \pm 0.66$ & $19.81 \pm 0.78$ & $1.19 \pm 0.05$
\\ \hline
Saturation Models & & & & \\
IP-Sat & 17.55 & 20.55 & 23.11 & 1.32 \\
KLN    & 15.96 & 17.51 & 22.02 & 1.38 \\
rcBK   & 14.27 & 16.94 & 22.51 & 1.58 \\ \hline
$\mathtt{HIJING}$-based & & & & \\
$\mathtt{HIJING2.1}$ NS (no shad) & 23.58 & 22.67 & 24.96 & 1.06 \\
$\mathtt{HIJING2.1}$ WS ($s_g = 0.28$) & 18.30 & 17.49 & 20.21 & 1.10 \\
$\mathtt{HIJINGB\overline{B}2.0}$ NS$^*$ & 20.03 & 19.68 & 23.24 & 1.16 \\
$\mathtt{HIJINGB\overline{B}2.0}$ NS$^\dagger$ & 16.84 & 16.39 & 19.68 & 1.16 \\
$\mathtt{HIJINGB\overline{B}2.0}$ WS$^*$ & 12.97 & 12.09 & 15.16 & 1.17 \\ 
$\mathtt{HIJINGB\overline{B}2.0}$ WS$^\dagger$ & 13.98 & 13.71 & 16.73 & 1.20 \\ 
\hline
$\mathtt{AMPT}$ & & & & \\
$\mathtt{AMPT-Def}$ & 19.07 & 18.56 & 21.65 & 1.14 \\
$\mathtt{AMPT-SM}$  & 18.14 & 18.10 & 20.84 & 1.15 \\ \hline
$\mathtt{DPMJET}$ & 17.50 & 17.61 & 20.67 & 1.18 \\ \hline
\end{tabular}}
\label{tab:etatable}
\end{table}

\subsubsection{Centrality Dependence of $dN_{\rm ch}/d\eta$}

Here we present two calculations of the centrality dependence of the charged
particle multiplicity distributions.

\begin{figure}[htbp]
\vspace*{1in}
\begin{center}
\includegraphics[width=0.55\textwidth]{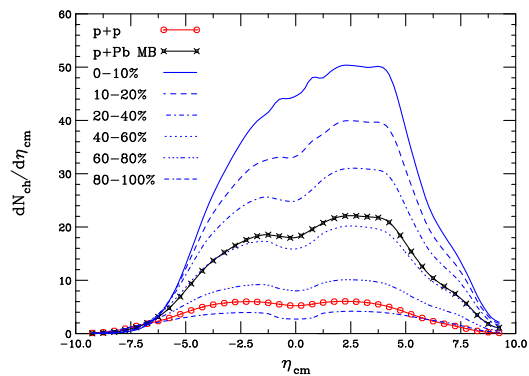}
\end{center}
\caption[]{Charged particle pseudorapidity distributions in the center of mass
frame for $p+p$, $p+$Pb minimum bias and six of the centrality bins defined
in Table~\protect\ref{table1} as calculated in $\mathtt{AMPT}$.}
\label{fig:dnchdeta_cent}
\end{figure}

\paragraph[$\mathtt{AMPT}$]{$\mathtt{AMPT}$ (Z. Lin)}
\label{sec:AMPT_cent}
Figure~\ref{fig:dnchdeta_cent} shows the pseudorapidity rapidity 
distributions calculated in the center-of-mass
frame by $\mathtt{AMPT-def}$. Results are shown for $p+p$ and minimum bias 
$p+$Pb collisions as well as for six different
centrality classes.  In $\mathtt{AMPT}$, the $p+p$ events are minimum bias
events, including diffractive events. 

\begin{table}[htpb]
\tbl{Centrality classes of $p+$Pb events from $\mathtt{AMPT-SM}$
The centrality is determined from the number of
 charged hadrons within $|\eta|<1$ in the center-of-mass frame.}
{\begin{tabular}{@{}cccccccc@{}}\toprule
Bin & $\langle b \rangle$ (fm) & $b_{\rm min}$ (fm) & $b_{\rm
  max}$ (fm) & $N^{\rm Pb}_{\rm part}$   & $N^{\rm Pb}_{\rm part, \, in}$ &
$\langle N_{\rm ch}(|\eta|<1) \rangle$ \\
\hline
MB  & 5.84 & 0.0 & 13.2 & 7.51 & 5.37 & 36.1 \\
0-5\% & 3.51 & 0.0 & 8.8 & 15.70 & 12.12 & 99.9 \\
5-10\% & 3.76 & 0.0 & 9.4 & 14.19 & 10.71 & 79.6 \\
10-20\% & 4.00 & 0.0 & 9.9 & 12.93 & 9.59 & 66.5 \\
20-40\% & 4.56 & 0.0 & 12.0 & 10.70 & 7.69 & 49.6 \\
40-60\% & 5.65 & 0.1 & 13.2 & 7.30 & 5.03 & 31.4 \\
60-80\% & 7.08 & 0.1 & 13.2 & 3.77 & 2.50 & 15.9 \\
80-100\% & 8.08 & 0.2 & 13.2 & 1.85 & 1.11 & 5.5 \\
\hline
\end{tabular}}
\label{table1} 
\end{table}

Since the ratio $R_{p{\rm Pb}}$ is normalized by a factor proportional
to the number of binary 
nucleon-nucleon collisions, $N_{\rm coll}$, the number of collisions in the
same centrality classes as included in Fig.~\ref{fig:dnchdeta_cent} is
shown in Table~\ref{table1}.  In $\mathtt{AMPT}$, $N_{\rm coll}$ is assumed to
be equivalent to the total 
number of participant nucleons in the Pb nucleus, $N^{\rm Pb}_{\rm part}$. 
Because there is some model dependence in the definition of centrality
bins, it is worthwhile noting that other calculations of this same quantity
may give somewhat different results.  Here, the centrality of $p+$Pb collisions 
is defined according to the number of charged hadrons within $|\eta|<1$. 
Table~\ref{table1} shows various information for each 
centrality class in the center-of-mass frame, 
including the average, minimum and maximum impact parameters, 
$N^{\rm Pb}_{\rm part}$, the number of participant Pb nucleons involved in
inelastic collisions, $N^{\rm Pb}_{\rm part, \, in}$, 
and the average number of charged particles within $|\eta|<1$ 
from $\mathtt{AMPT-SM}$.  Note that the $\mathtt{AMPT-SM}$ values in 
Table~\ref{table1} are essentially the same as those for $\mathtt{AMPT-def}$
except for the average number of charged particles (last column).

In Fig.~\ref{fig:dnchdeta_cent}, the $40-60$\% centrality bin
gives a distribution that is very close to the min-bias result, in particular
on the proton side.  The most central bin is the most asymmetric distribution,
as well as the largest in magnitude.
On the other hand, the distribution for the $80-100$\% centrality bin is
symmetric around $\eta_{\rm cm} = 0$ and lower in magnitude than the $p+p$
distribution.

\begin{figure}[htbp]
\begin{center}
\includegraphics[width=0.55\textwidth]{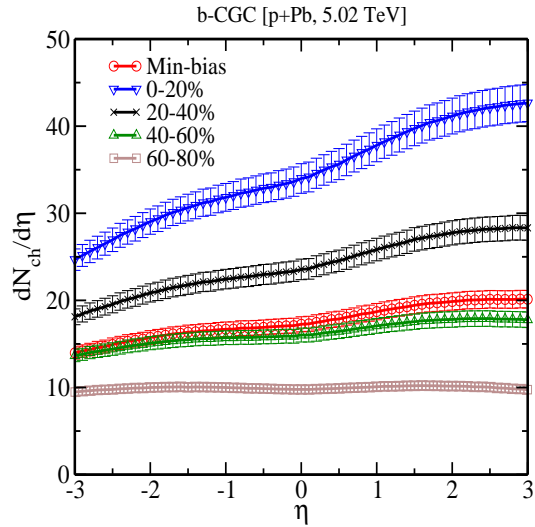}
\caption[]{Charged particle pseudorapidity distributions in the center of mass
frame of $p+$Pb collisions at various centralities within the b-CGC saturation
 model. The $\sim 5$\% theoretical uncertainties arising from fixing the 
overall normalization at RHIC are also shown.  From 
Ref.~\protect\cite{Rezaeian:2012ye}.}
\label{fig-m}
\end{center}
\end{figure}

\paragraph[b-CGC]{b-CGC (A. Rezaeian)}
\label{sec:Amir_cent}
In \fig{fig-m}, the charged hadron multiplicity
obtained from $k_T$ factorization in the b-CGC approach is
shown the $0-20\%$, $20-40\%$, $40-60\%$, $60-80\%$ centrality bins as well
as minimum-bias collisions \cite{Rezaeian:2012ye}.  The impact parameter 
dependence of the saturation 
model is crucial for defining the collision centrality.  The largest asymmetry 
in the multiplicity distribution is observed in more central collisions while, 
for peripheral collisions such as the $60-80\%$ most central, the system
becomes more similar to that produced in $p+p$ collisions.  This is reflected 
in the total charged hadron multiplicity distribution. 

Note that for all 
results shown in \fig{fig-m}, a fixed minijet mass equal to current-quark mass 
is assumed for all energies/rapidities and centralities. Since the minijet mass 
is related to pre-hadronization/hadronization stage and cannot be obtained from 
saturation physics, it was fixed by fitting lower energy minimum-bias data. 
In very peripheral collisions where the system becomes more similar to 
symmetric $p+p$ collisions, this assumption is less reliable. More importantly, 
one should also note that the $k_T$ factorization employed here is only proven 
in asymmetric $p+A$ collisions at small $x$. Therefore, for more peripheral 
collisions, the current CGC prescription may be less reliable.

\subsection{Transverse Momentum distribution}
\label{sec:dnchdpt}

\subsubsection[Compilation of midrapidity results]{Compilation of midrapidity 
results (J. Albacete, A. Dumitru, H. Fujii, Y. Nara, R. Xu, W.-T. Deng, X.-N. Wang, 
G. G. Barnaf\"oldi, J. Barette, M. Gyulassy, P. Levai, M. Petrovici, V. Topor Pop, 
Z. Lin and R. Vogt)}
\begin{figure}[htbp]
\begin{center}
\includegraphics[width=0.495\textwidth]{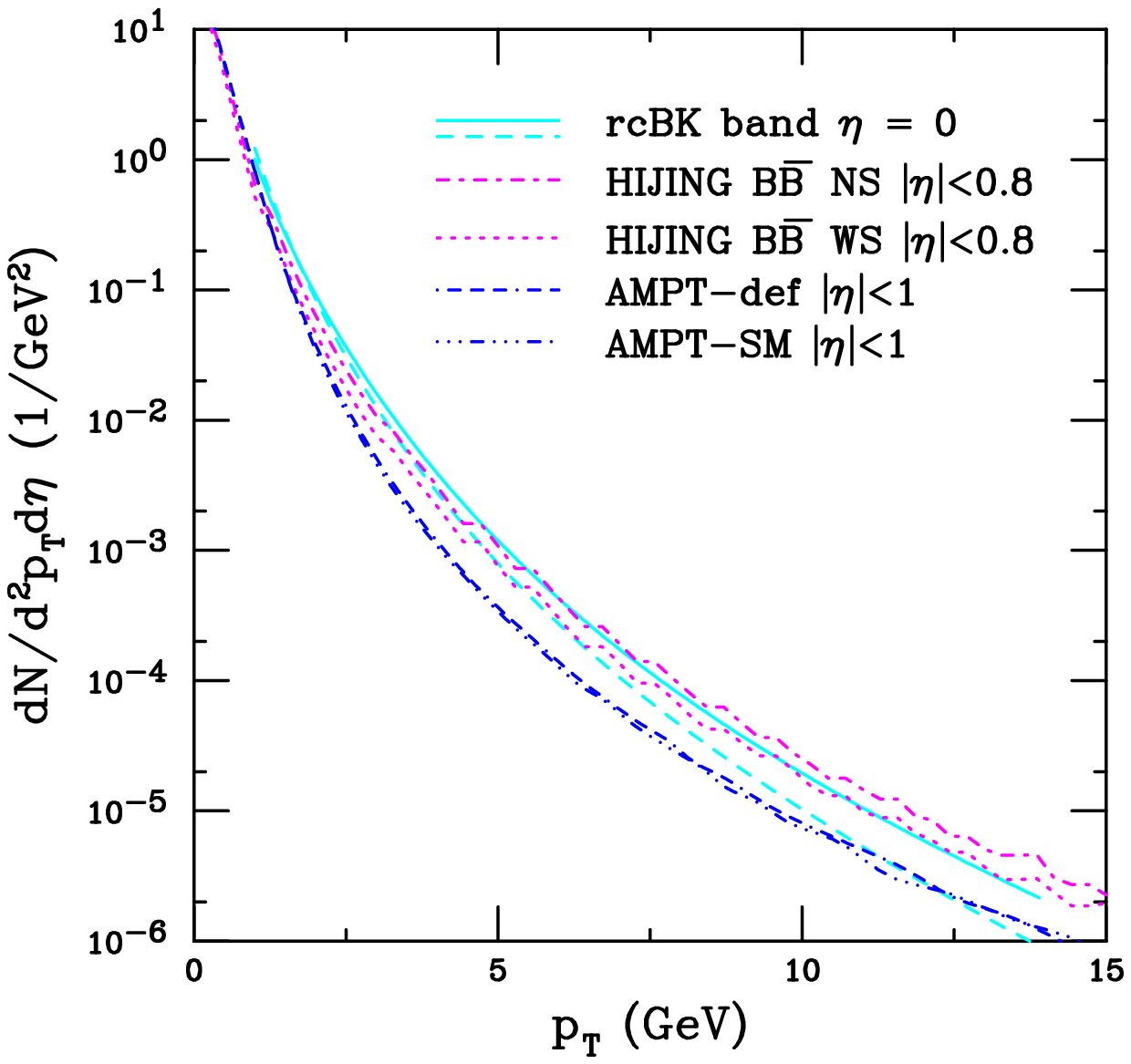}
\includegraphics[width=0.495\textwidth]{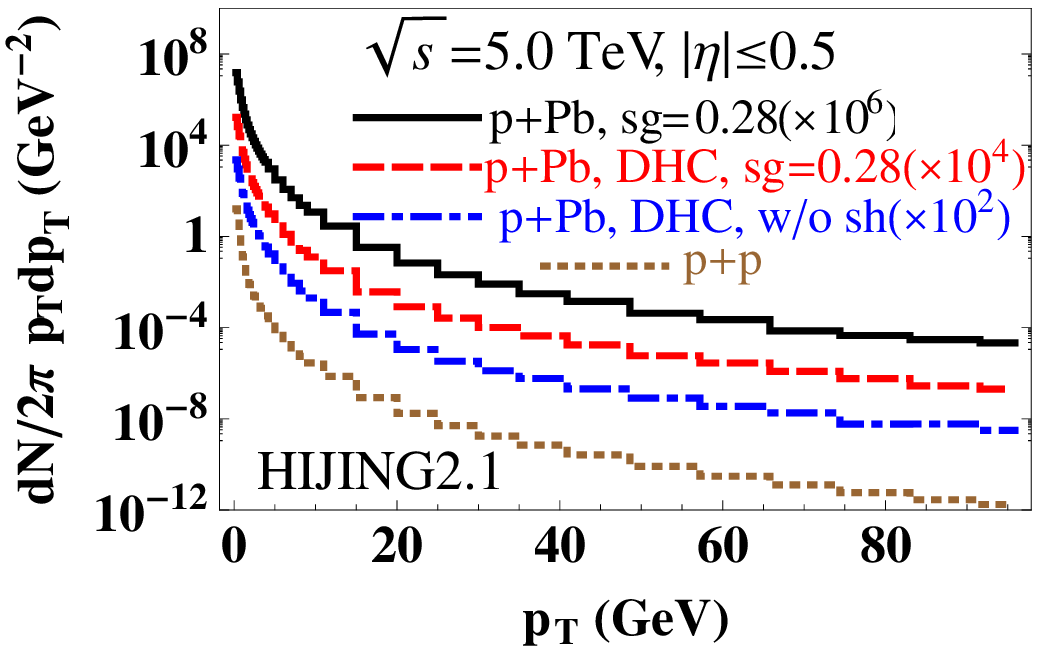}
\caption[]{(Left) Charged particle $p_T$ distributions at $\sqrt{s_{_{NN}}}
= 5.02$ TeV.  The solid and dashed cyan curves outline the rcBK band
calculated by Albacete {\it et al.} \protect\cite{Albacete:2012xq}. 
The magenta curves, calculated with $\mathtt{HIJINGB\overline{B}2.0}$ 
as described in Sec.~\protect\ref{sec:Pop_ch} are presented
without (dot-dashed) and with (dotted) shadowing. The $\mathtt{AMPT}$ results,
given by the dot-dash-dash-dashed (default) and dot-dot-dot-dashed (SM) blue
curves, are described in Sec.~\protect\ref{sec:AMPT_ch}. (Right) The charged 
hadron $p_T$ distribution in $p+$Pb collisions 
with different $\mathtt{HIJING2.1}$ options, scaled by the indicated factors to
separate the curves.  The $p+p$ distribution is
shown for comparison.
}
\label{fig:dnchdpt_comp}
\end{center}
\end{figure}

The midrapidity, $|\eta|<0.8$, charged hadron $p_T$ distributions are shown on
the left-hand side of 
Fig.~\ref{fig:dnchdpt_comp} for rcBK \cite{Albacete:2012xq}, 
$\mathtt{HIJINGB\overline{B}2.0}$, and $\mathtt{AMPT}$.
The $\mathtt{HIJINGB\overline{B}2.0}$ distributions are similar to
the rcBK results, albeit somewhat higher for $p_T>10$ GeV/$c$. 
The $\mathtt{AMPT}$ distributions, on the other hand, drop faster at low
$p_T$ than the other results but then become harder at high $p_T$.
The $\mathtt{AMPT}$
results are essentially independent of whether
string melting is included or not while the  $\mathtt{HIJINGB\overline{B}2.0}$
results without shadowing lie above those with shadowing.

The right-hand side of Fig.~\ref{fig:dnchdpt_comp} shows several options for
cold matter effects in $\mathtt{HIJING2.1}$ relative to $p+p$ collisions.
The $p+p$ result is unscaled while the $p+$Pb curves with decoherent hard
scatterings (DHC) without shadowing, DHC with shadowing, and shadowing only
are separated from each other, starting from the $p+p$ result, by a factor
of 100.

\subsubsection[reBK at $y=0$, 2]{reBK at $y=0$, 2 (J. Albacete, A. Dumitru, 
H. Fujii and Y. Nara)}
\begin{figure}[htpb]
\begin{center}
\includegraphics[width=0.495\textwidth]{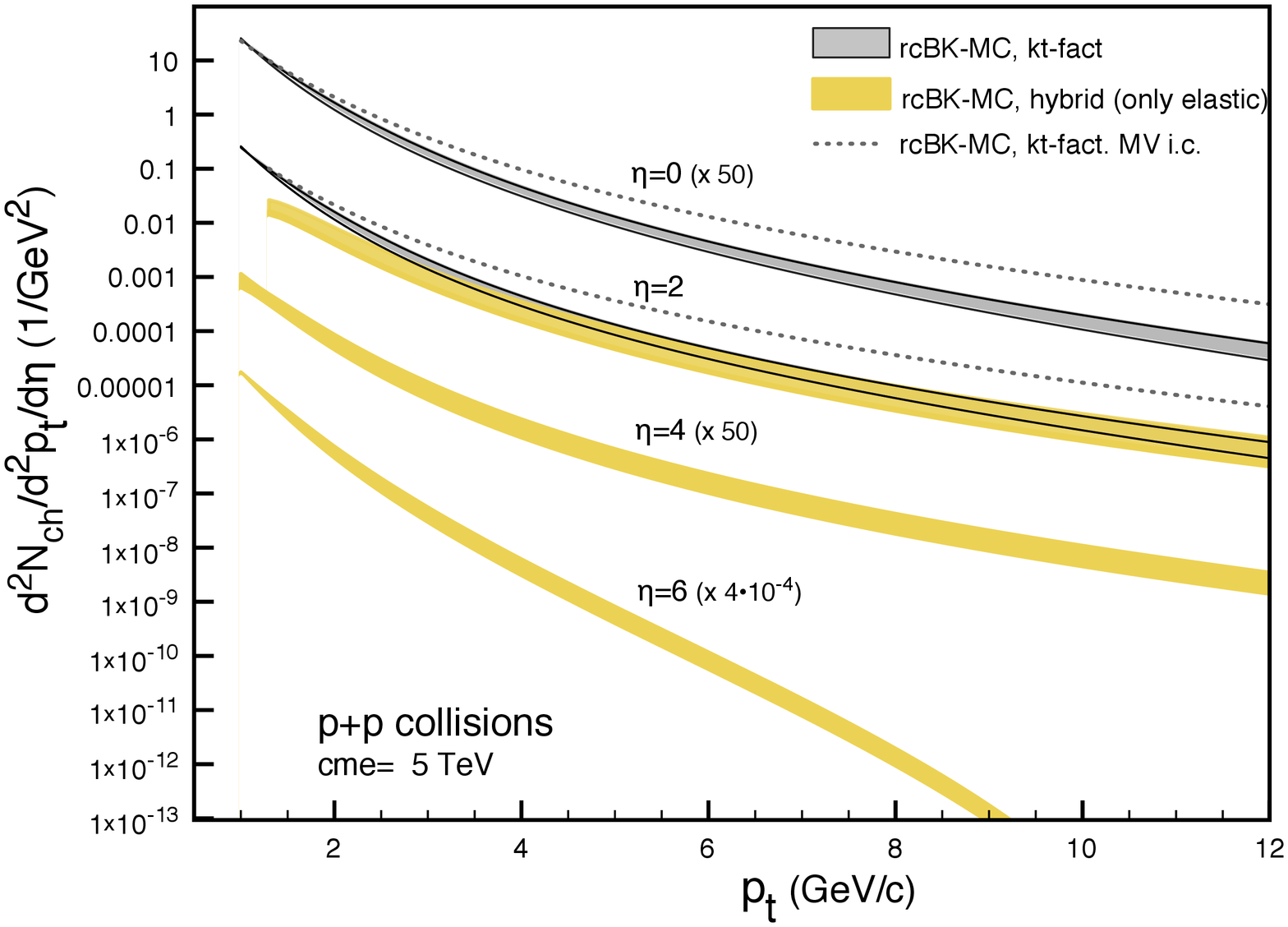}
\includegraphics[width=0.495\textwidth]{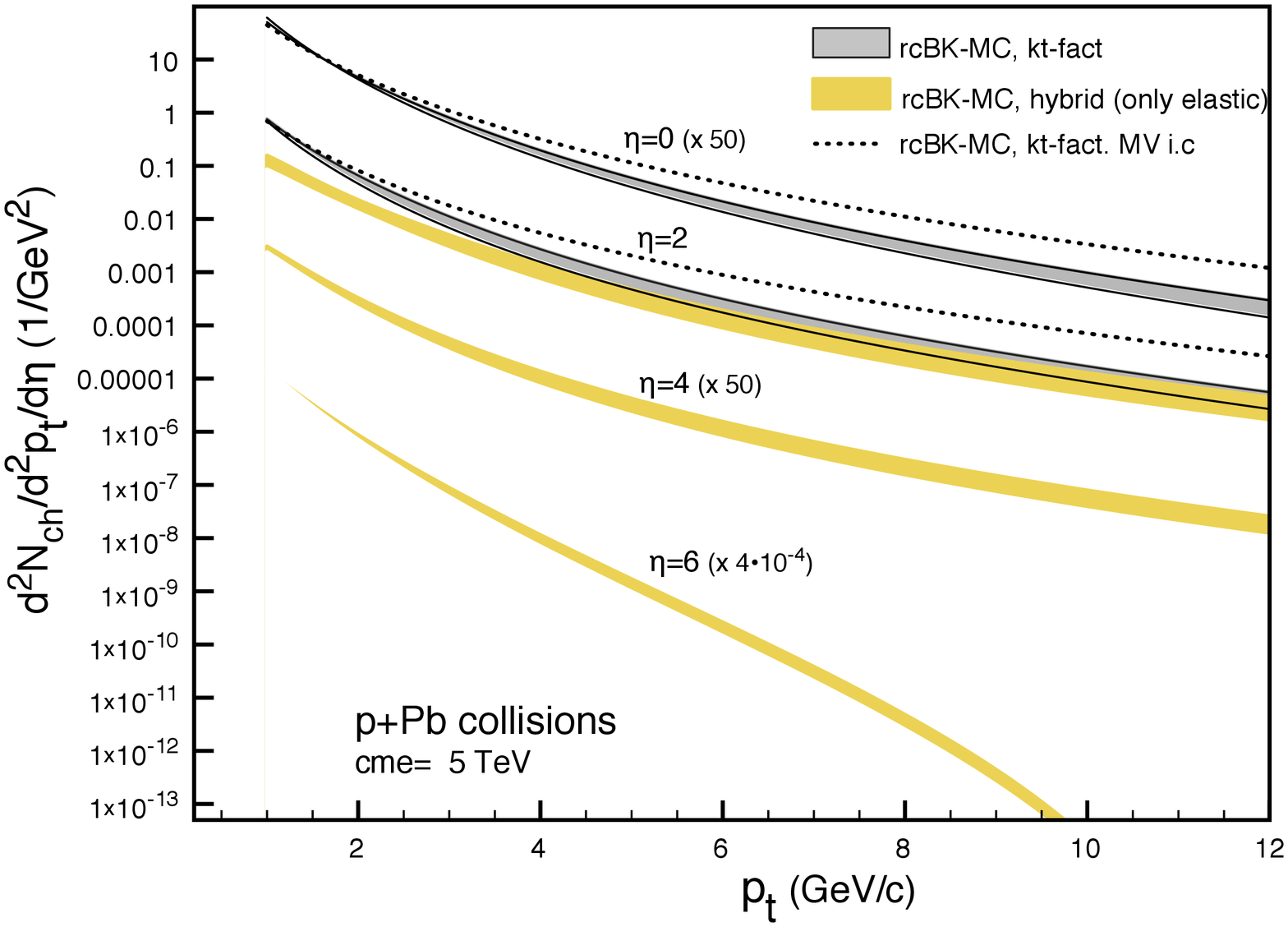}
\end{center}
\vspace*{-.4cm}
\caption[]{The predicted transverse momentum spectrum in $p+p$ (left) and
  minimum-bias $p+$Pb (right) collisions at $\surd s=5$~TeV at different
  rapidities (with the convention that the proton beam moves toward
forward rapidity).  From Ref.~\protect\cite{Albacete:2012xq}.}
\label{fig:pPb_spectrum}
\end{figure}

Fig.~\ref{fig:pPb_spectrum} shows the $p_T$ spectrum in $p+p$ (left) and
$p+$Pb (right)
collisions at different rapidities for the AAMQS initial conditions,
see Table~\ref{tab:rcBK_IC}. Near
central rapidity, $k_T$-factorization is employed while at forward rapidities
(in the proton direction)
the hybrid formalism is applied. The bands correspond
to uncertainty estimates due to small variations in the scale entering
the coupling and fragmentation functions.

\subsubsection[flow coefficients with AMPT]{Flow Coefficients obtained with
$\mathtt{AMPT}$ (Z. Lin)}

The flow coefficients $v_n\{2\}$ $(n=2$, $3$, $4)$ in this study are 
calculated employing the two-particle cumulant method with 
$v_n=\sqrt {\left < \cos \left [ n (\phi_i - \phi_j ) \right ] \right >}$,
where $\cos \left [ n (\phi_i - \phi_j ) \right ]$ is averaged over all particle
pairs in the specified phase space with both particles of each pair coming 
from the same event.

\begin{figure}[htbp]
\begin{center}
\includegraphics[width=0.495\textwidth]{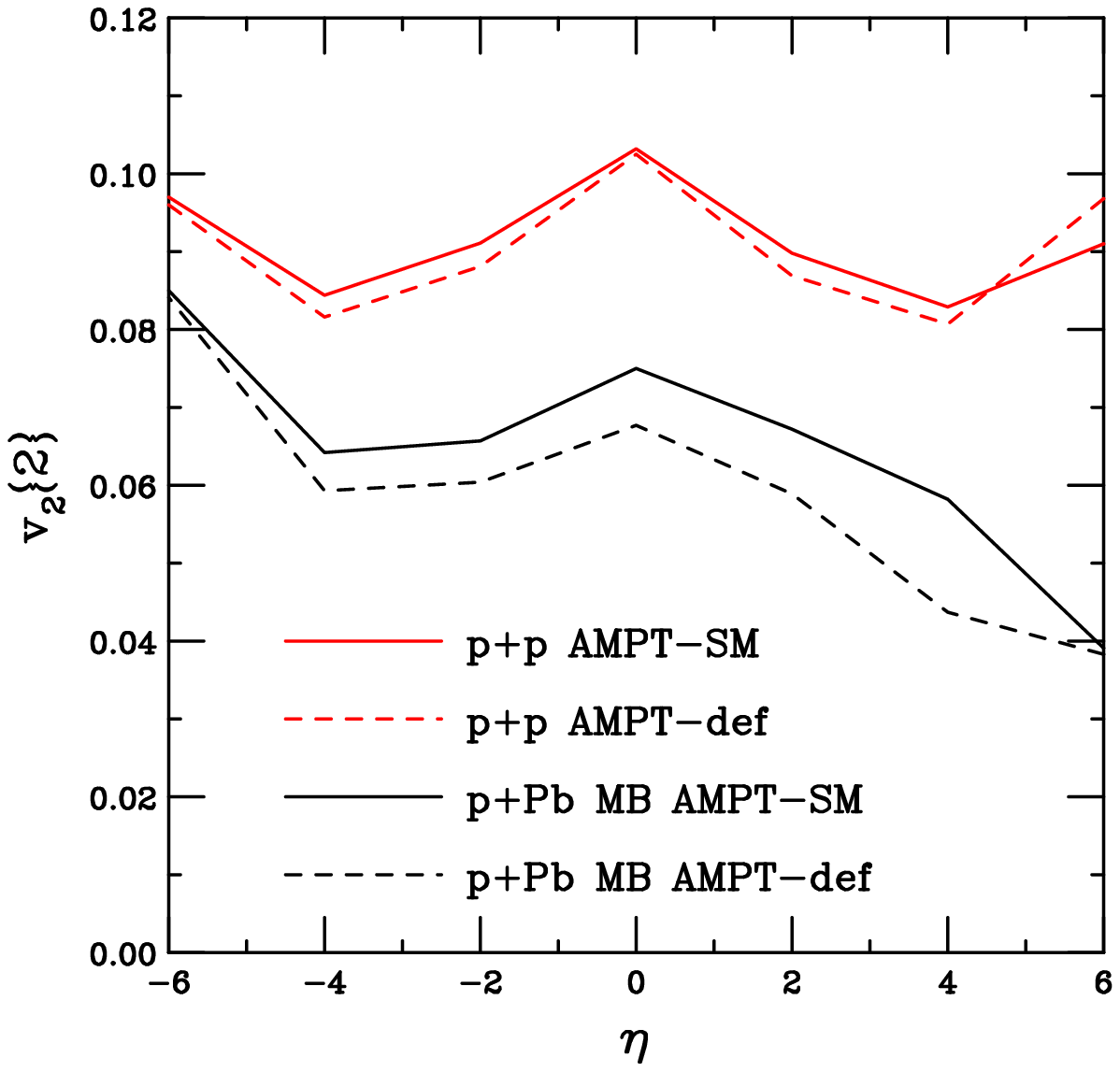}
\includegraphics[width=0.495\textwidth]{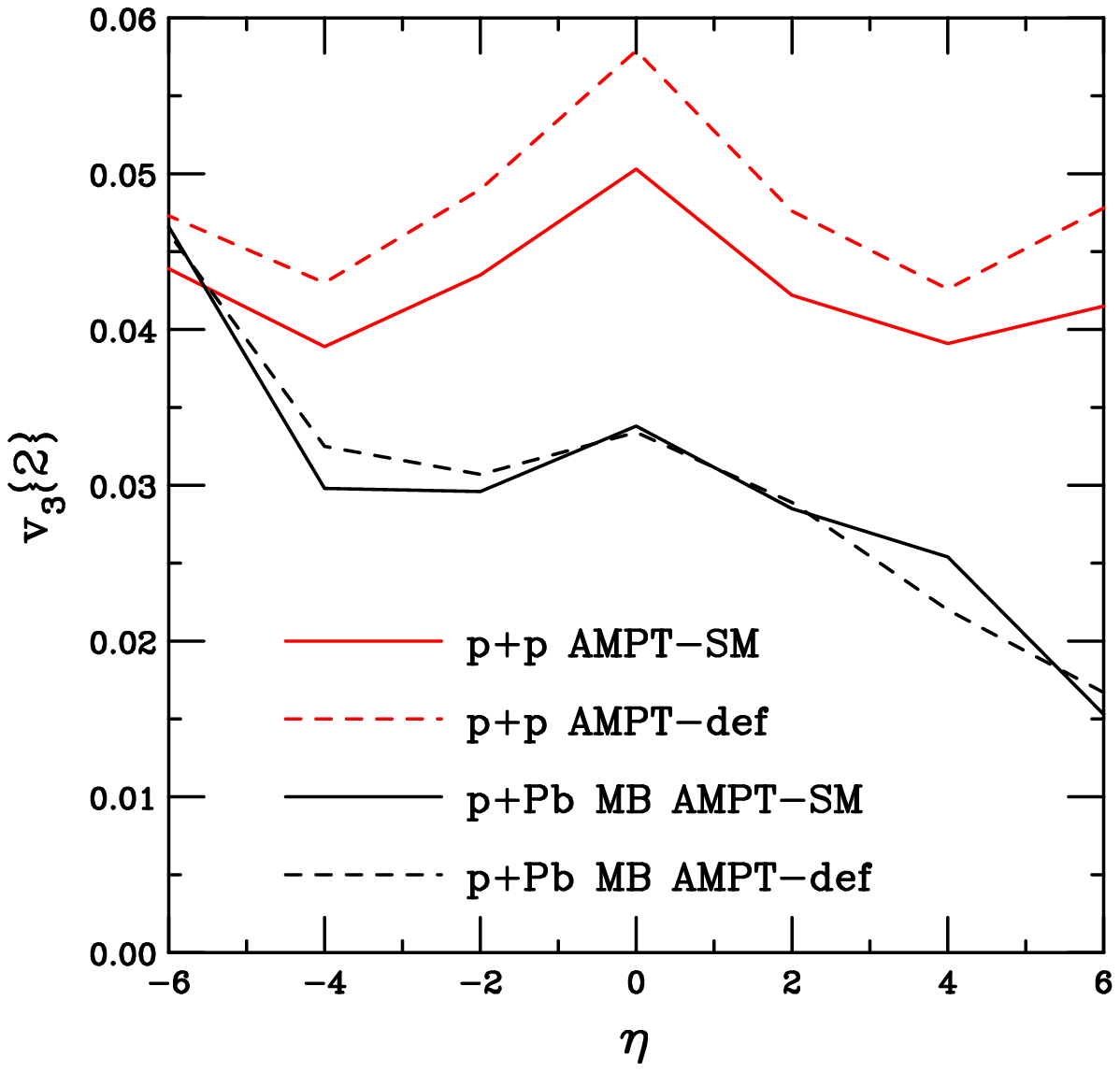} \\
\includegraphics[width=0.495\textwidth]{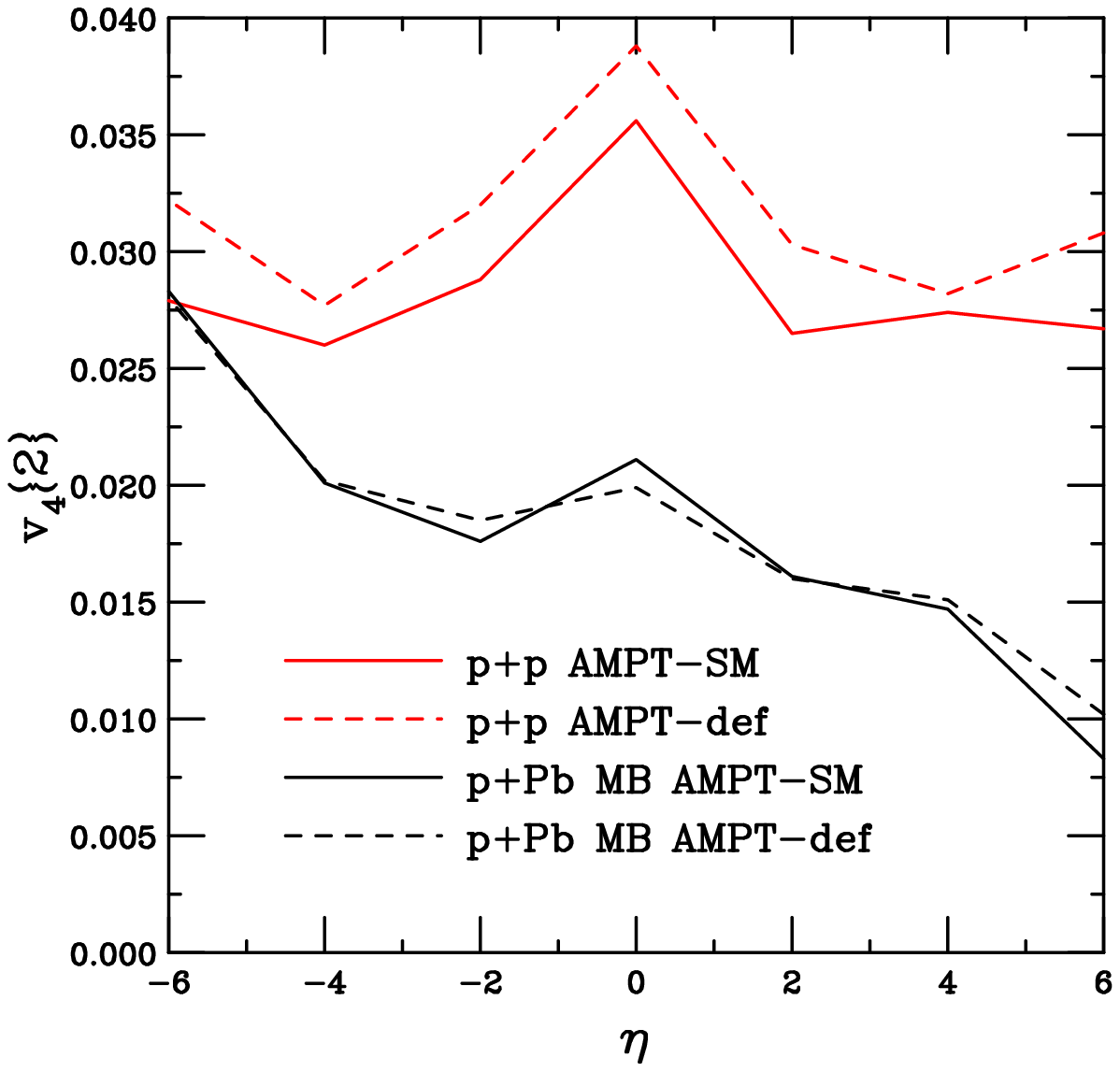}
\end{center}
\caption[]{The flow coefficients $v_2\{2\}$ (top left), $v_3\{2\}$ (top right),
and $v_4\{2\}$ (bottom) of charged particles as a function of $\eta$ in $p+p$ 
and minimum-bias $p+$Pb collisions. The results are shown for both
$\mathtt{AMPT-def}$ and $\mathtt{AMPT-SM}$.}
\label{fig:vn_eta}
\end{figure}

Figure~\ref{fig:vn_eta} shows $v_2\{2\}$, $v_3\{2\}$ and $v_4\{2\}$ for 
unidentified charged particles as a function of $\eta$ in $p+p$ and 
minimum-bias $p+$Pb collisions from $\mathtt{AMPT-def}$ and $\mathtt{AMPT-SM}$. 
The magnitude of the flow coefficients generally decreases with increasing
$n$. There is a local maximum at $\eta \sim 0$. 
While the $p+p$ results are approximately symmetric, as expected, the $p+$Pb 
results exhibit a strong asymmetry. In the direction of the proton beam,
the $p+$Pb coefficients increase to approach the magnitudes of the 
$p+p$ coefficients. 

\begin{figure}[htbp]
\begin{center}
\includegraphics[width=0.495\textwidth]{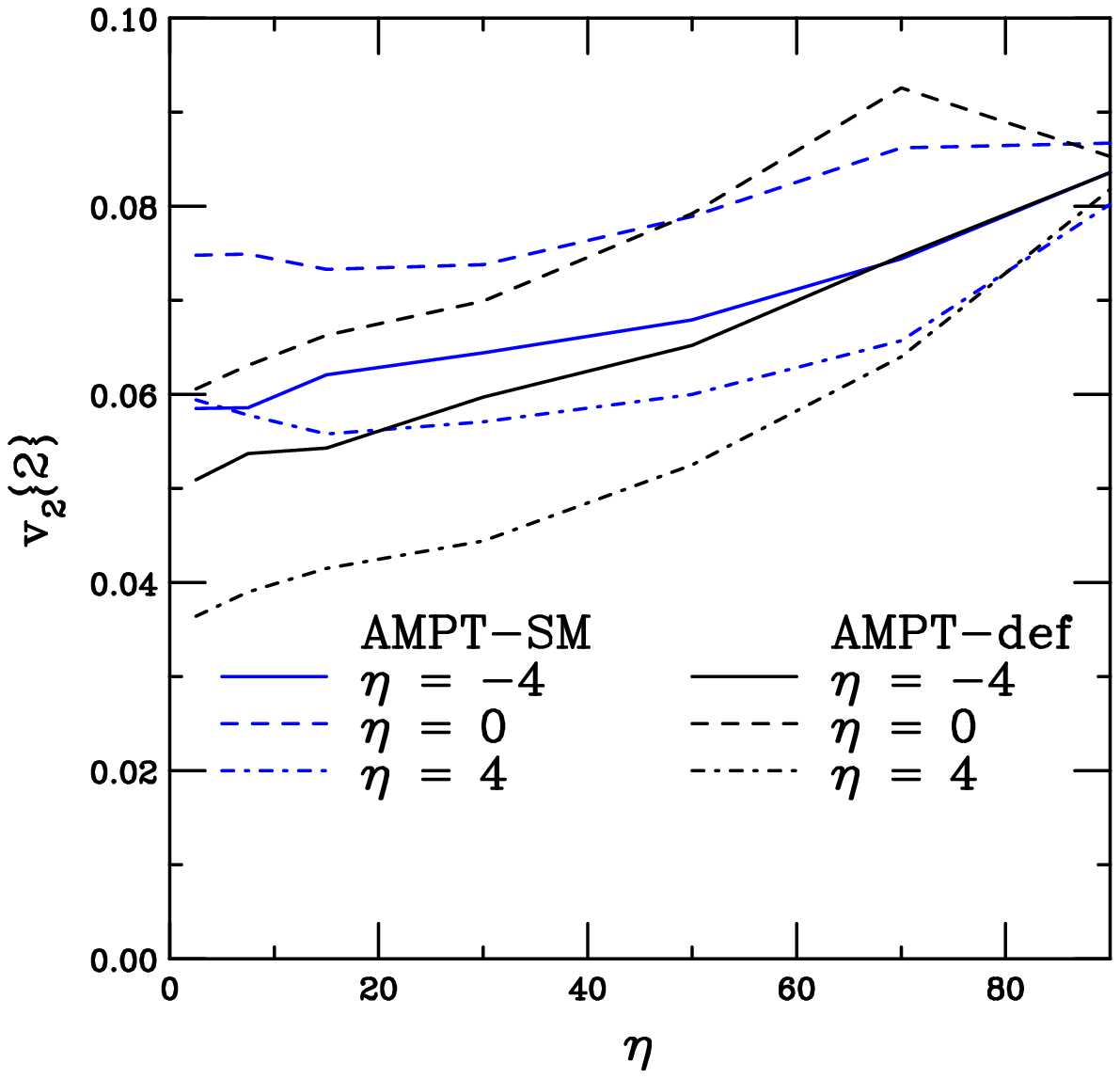}
\includegraphics[width=0.495\textwidth]{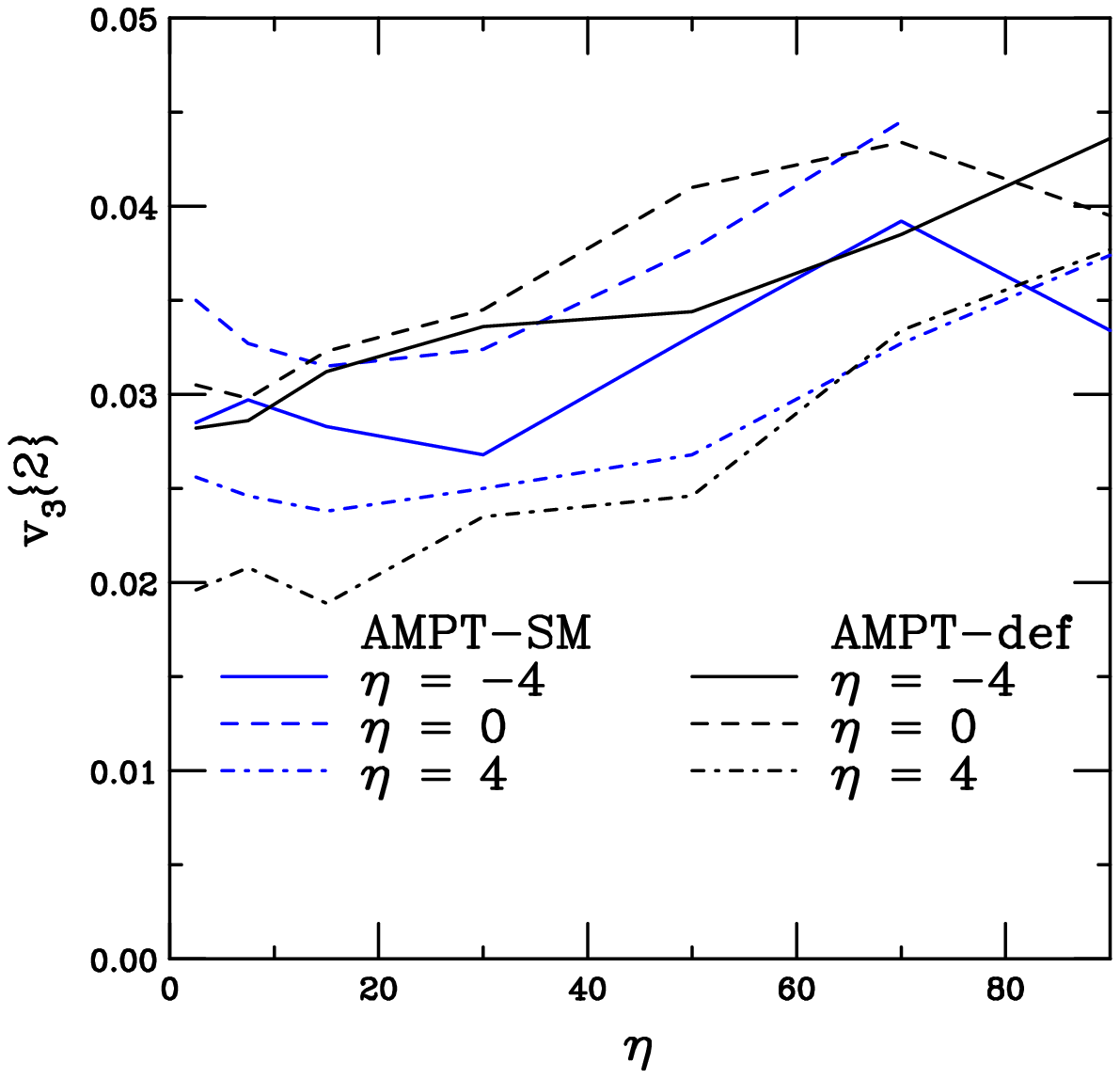} 
\end{center}
\caption[]{The flow coefficients $v_2\{2\}$ (left) and $v_3\{2\}$ (right)
as a function of centrality in $p+$Pb collisions at $\langle \eta \rangle
= 4$, $0$, and $-4$. 
The results are shown for both
$\mathtt{AMPT-def}$ and $\mathtt{AMPT-SM}$.}
\label{fig:vn_cent}
\end{figure}

Figure~\ref{fig:vn_cent} shows the centrality dependence of $v_2\{2\}$ and 
$v_3\{2\}$ in $p+$Pb collisions.  Three different $\eta$ ranges are
shown: $-5\leq \eta \leq -3$, $|\eta| \leq 1$, and $3 \leq \eta \leq 5$
($\langle \eta \rangle = -4$, 0 and 4 respectively.  The coefficients increase 
from central to peripheral events for both $\mathtt{AMPT-SM}$ and 
$\mathtt{AMPT-def}$. The coefficients are largest at $\langle \eta \rangle =
0$ while those at $\langle \eta \rangle = 4$, in the direction of the Pb
beam, are smallest, consistent with the results in Fig.~\ref{fig:vn_eta}.

\subsection{Medium Modification Factor $R_{p{\rm Pb}}$}
\label{sec:RpPb}

\subsubsection{$R_{p{\rm Pb}}(p_T)$ at $\eta \sim 0$}

\begin{figure}[htbp]
\begin{center}
\includegraphics[width=0.495\textwidth]{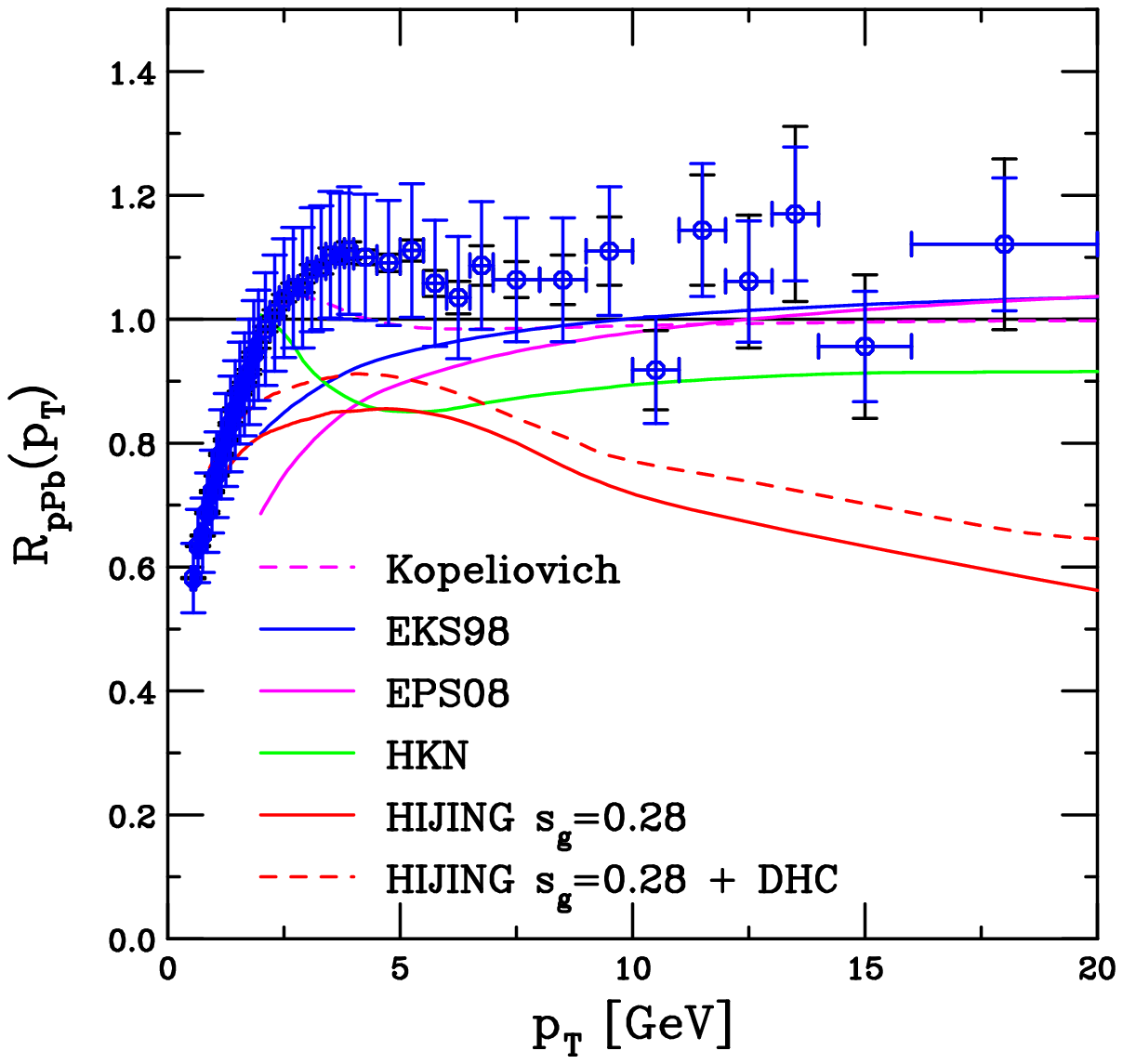}
\includegraphics[width=0.495\textwidth]{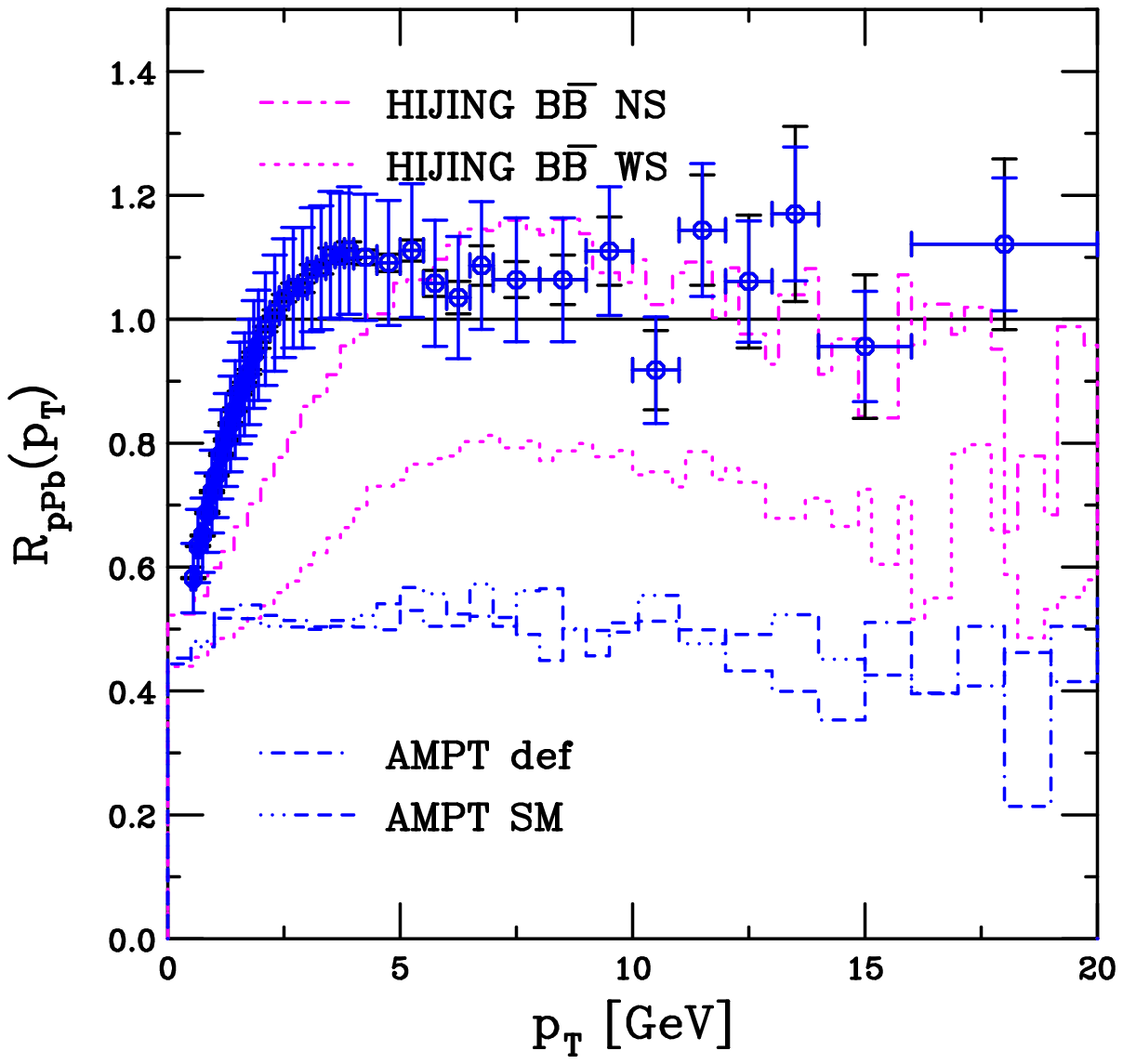} \\
\includegraphics[width=0.495\textwidth]{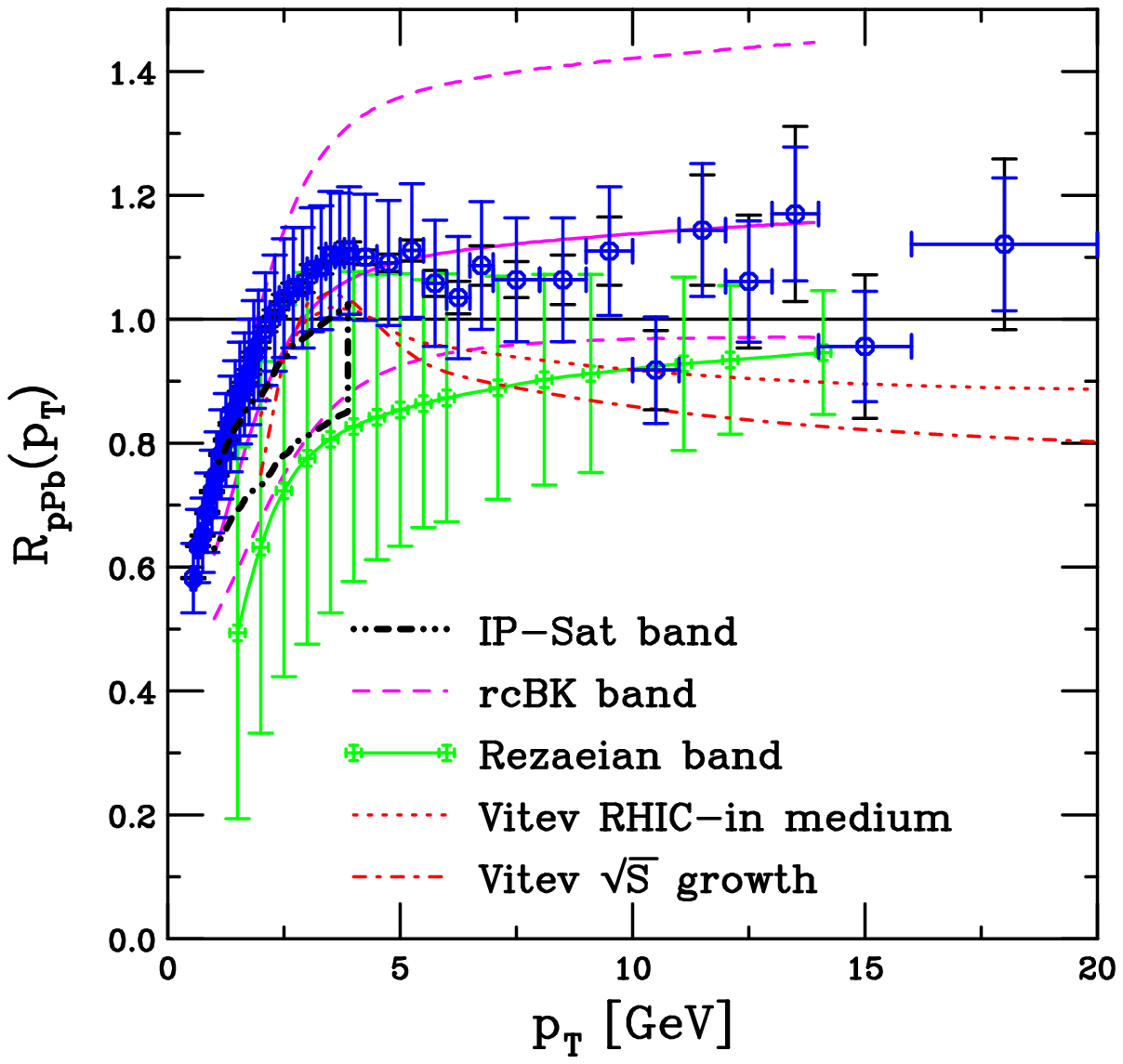}
\caption[]{Charged particle $R_{p{\rm Pb}}p_T$ at $\sqrt{s_{_{NN}}}
= 5.02$ TeV at $\eta \sim 0$. (Top left) Results with more `standard' shadowing
(labeled EKS98, EPS08 and HKN) described in 
Sec.~\protect\ref{sec:Levai-description}, 
Refs.~\protect\cite{Kopeliovich:2002yh,Kopeliovich:1999am} (labeled
Kopeliovich) and the $\mathtt{HIJING2.1}$ shadowing parameterization in 
Eq.~(\protect\ref{eq:shadow}) are compared.  
The difference in the $\mathtt{HIJING2.1}$ curves depends on whether the
hard scatterings are coherent or not, see Sec.~\protect\ref{sec:HIJING_ch}. 
(Top Right) $\mathtt{HIJINGB\overline{B}2.0}$ with and without shadowing 
(Sec.~\protect\ref{sec:Pop_ch}) compared to $\mathtt{AMPT}$ 
default and with string melting (Sec.~\protect\ref{sec:AMPT_ch}).
(Bottom) The band from rcBK saturation model calculations by Albacete 
{\it et al.} and Rezaeian with $N=5$ and varying $\alpha_s^{\rm in}$ 
described in Sec.~\protect\ref{sec:amir_ch} are compared to IP-Sat calculations
by Tribedy and Venugopalan in Sec.~\protect\ref{sec:IPSat} and
calculations by Vitev {\it et al.} discussed in 
Sec.~\protect\ref{sec:Ivan-description}. (More detailed results for the 
uncertainties in Rezaeian's calculation at other rapidities
can be found in Fig.~\protect\ref{rp-h}.) The ALICE results from
Ref.~\protect\cite{ALICE_rpa} are also shown.  The
systematic uncertainties are shown in blue, the statistical uncertainties are
in black.}
\label{fig:comp_RpPb}
\end{center}
\end{figure}

We now turn to the medium modification or suppression factors, defined as
the ratio of the $p+A$ to $p+p$ cross section,
\begin{eqnarray}
R_{p {\rm Pb}}(p_T, \eta;b) = 
\frac{1}{N_{\rm coll}(b)}\frac{d\sigma_{p {\rm Pb}}/d\eta dp_T}
{d\sigma_{pp}/d\eta dp_T} \, \, .
\end{eqnarray}
Results are shown first for $R_{p{\rm Pb}}(p_T)$ for charged hadrons
(pions) in Fig.~\ref{fig:comp_RpPb}.
Standard shadowing calculation are shown on the left-hand side, results with
event generators in the center, while CGC calculations and results with energy 
loss in cold matter are shown on the right-hand side.

\paragraph[Cronin effect expected at the LHC: dipole formulation]{Cronin effect expected at the LHC: dipole formulation (B. Z. Kopeliovich and J. Nemchik)}
\label{Boris}
At LHC energies, hadron production is dominated by gluon fragmentation.  
The dipole formalism, on the light cone, is employed in
the target rest frame, leading to the factorized
expression for hadron production: a convolution of the projectile gluon PDF, 
the gluon fragmentation function, and the gluon splitting, $g\to gg$, 
cross section \cite{Kopeliovich:2002yh}.
The gluon splitting cross section can be written in the dipole representation 
as
\begin{eqnarray*}
\lefteqn{\frac{d\sigma(gA\to g_1gX)}
{d^2p_T\,dy_{1}} =
\int d^2b\int d^2r_1d^2r_2\,
e^{i\vec p_T \cdot (\vec r_1-\vec r_2)}\,
\overline{\Psi_{gg}^*(\vec r_1,\alpha)
\Psi_{gg}(\vec r_2,\alpha)}}
\label{10}
\\ && \times
\left[1 - e^{-{1\over2}\sigma^N_{3g}(r_1,x)T_A(b)}
-e^{-{1\over2}\sigma^N_{3g}(r_2,x)T_A(b)} +
e^{-{1\over2}\sigma^N_{3g}(\vec r_1-\vec r_2,x)T_A(b)} 
\right]\ \, \, , \nonumber
 \end{eqnarray*}
where the subscript 1 indicates the hadronizing gluon involved in inclusive
high $p_T$ hadron production by gluon fragmentation.  The momentum fraction 
of radiation gluon $g_1$ is defined as $\alpha = p^+_{g_1}/p^+_g$, the transverse
coordinates of the emitted gluons are $r_1$ and $r_2$, $x$ is the momentum
fraction of the gluon in the target nucleus and $T_A(b)$ is the nuclear 
thickness function.  The distribution function, $\Psi_{gg}(\vec r,\alpha)$,
derived in \cite{Kopeliovich:1999am}, describes the $|gg\rangle$ Fock component 
of the projectile gluon light-cone wavefunction, including the nonperturbative
gluon interaction.  It is characterized by the localization of gluons at a 
short relative transverse separation, $r_0\approx 0.3$~fm. 
The $3g$ dipole cross section, $\sigma^N_{3g}$, is expressed as a combination
of the $\bar qq$ dipole cross sections, extracted from phenomenology and DIS 
data,
$\sigma^N_{3g}(r) = {2\over9}\Bigl\{\sigma_{\bar qq}(r) + \sigma_{\bar qq}(\alpha r) 
+ \sigma_{\bar qq}[(1-\alpha)r]\Bigr\}$\footnote{The $x$ dependence of
$\sigma^N_{3g}$ has been suppressed.}.

Multiple interactions of higher Fock components (containing more than one gluon)
of the incident proton in the nucleus leads to a suppression of the dipole
cross section, known as gluon shadowing. The magnitude of the suppression 
factor $R_g$ was evaluated in Ref.~\cite{Kopeliovich:1999am} and, as a 
consequence of
the small gluon separation, $r_0$, the modification was found to be rather 
small.  Thus, at first order, gluon shadowing can be implemented by the simple 
replacement $\sigma^N_{3g} \rightarrow R_g(x,Q^2,b) \sigma_{3g}$ in
Eq.~(\ref{10}).  The results Ref.~\cite{Kopeliovich:2002yh} at 
$\sqrt{s}=5$~TeV are
given by the solid cyan curve on the left-hand side of Fig.~\ref{fig:comp_RpPb}.

The  Cronin enhancement is rather weak with a
maximum of $R_{pPb}\sim 1.05$ at $p_T\sim 3$~GeV$/c$. The height
of the maximum is extremely sensitive to the strength of gluon shadowing.
The enhancement could completely disappear, or even change sign, if the 
strength of gluon shadowing was underestimated in 
Ref.~\cite{Kopeliovich:1999am}.

\paragraph[Results with standard shadowing parameterizations]{Results with standard shadowing parameterizations (G. G. Barnaf\"oldi, J. Barette, M. Gyulassy, P. Levai, G. Papp and V. Topor Pop)}

The left-hand side of Fig.~\ref{fig:comp_RpPb} also shows $R_{p{\rm Pb}}(p_T)$ for
several standard shadowing functions: EKS98~\cite{Eskola:1998df} (blue), 
EPS08~\cite{Eskola:2008ca} (magenta), and HKN~\cite{Hirai:2001np} (green).  
All three were
calculated in the rapidity interval $|\eta| < 0.3$, in the center of mass
frame.  Since none of these parameterizations include impact parameter 
dependence, the results are independent of collision centrality.  For details, 
see Ref.~\cite{Barnafoldi:2011px}. For other results employing
$\mathtt{kTpQCD\_v2.0}$, see 
Refs.~\cite{Zhang:2001ce,Cole:2007ru,Barnafoldi:2002xp,Barnafoldi:2008ec,Adeluyi:2008gj}. 

\paragraph[$\mathtt{HIJING2.1}$]{$\mathtt{HIJING2.1}$ (R. Xu, W.-T. Deng and X.-N.
Wang)}
Two $\mathtt{HIJING2.1}$ calculations including shadowing with parameter
$s_g = 0.28$ are shown with the standard shadowing parameterizations
on the left-hand side of Fig.~\ref{fig:comp_RpPb}.  They are lower than
the other calculations at high $p_T$.  Including the decoherent hard scatterings
moves $R_{p {\rm Pb}}$ closer to unity over all $p_T$.

\paragraph[$\mathtt{HIJINGB\overline{B}2.0}$]{$\mathtt{HIJINGB\overline{B}2.0}$ (G. G. Barnaf\"oldi, J. Barette, M. Gyulassy, P. Levai, M. Petrovici and V. Topor Pop)}
The $\mathtt{HIJINGB\overline{B}2.0}$ results without (NS) and with
(WS) shadowing in the central rapidity region, $|\eta|< 0.8$, are shown in 
the central panel of Fig.~\ref{fig:comp_RpPb}. 
Including shadowing and strong color field effects reduces $R_{p{\rm Pb}}$ 
from unity to $\sim 0.7$ for $5<p_T< 10$ GeV/$c$.  This is similar to CGC 
predictions in the KKT04 model~\cite{Kharzeev:2003wz}. 
Further, similar, results for $R_{p {\rm Pb}}(p_T)$ 
are found in Ref.~\cite{Levai:2011qm,Barnafoldi:2008ec} using LO pQCD 
collinear factorization with the
$\mathtt{HIJING2.0}$ shadowing parameterization \cite{Li:2001xa}, 
the GRV proton PDFs \cite{Gluck:1994uf}, and hadron fragmentation functions 
from Ref.~\cite{Kniehl:2000fe}. 

\paragraph[$\mathtt{AMPT}$]{$\mathtt{AMPT}$ (Z. Lin)}
The $\mathtt{AMPT}$ results, evaluated in the center of mass frame
pseudorapidity interval $|\eta|<1$ are also shown in the central panel of
Fig.~\ref{fig:comp_RpPb}.  In minimum bias collisions, $N_{\rm coll} = 7.51$ in
$\mathtt{AMPT}$.  There is an $\approx 50$\% suppression of charged hadron
production for the entire $p_T$ range shown, considerably lower than that 
obtained in the other calculations shown.  The statistical uncertainty, not
shown in Fig.~\ref{fig:comp_RpPb},  
becomes large for $p_T > 10$ GeV/$c$.  There is little difference in $R_{p {\rm
Pb}}$ between the default and string melting options of $\mathtt{AMPT}$.

The right-hand side of Fig.~\ref{fig:comp_RpPb} compares rcBK results from
Albacete {\it et al.} and Rezaeian to IP-Sat calculations by Tribedy and
Venugopalan and pQCD calculations by Vitev {\it et al.}
including energy loss in cold matter.

\paragraph[rcBK predictions]{rcBK predictions (J. Albacete, A. Dumitru, H. Fujii, Y. Nara, A. Rezaeian and R. Vogt)}
\label{Amir}

Rezaeian's calculation assumes $N_{\rm coll}=6.9$ in minimum bias $p+$Pb 
collisions \cite{d'Enterria:2003qs}.  To compare these calculations to data, 
it is necessary to rescale the results to match the normalization $N_{\rm coll}$
to the experimental value.  The NLO MSTW proton PDFs 
\cite{Martin:2007bv,Martin:2009iq} and the NLO 
KKP fragmentation functions \cite{Kniehl:2000fe} are employed. 
The rcBK equation, Eq.~(\ref{bk1}), accounts for the rapidity/energy evolution
of the dipole but does not include impact parameter dependence explicitly. 
In the minimum-bias analysis here the initial saturation scale, 
$Q_{0s}(x_0=0.01)$, is considered as an impact-parameter averaged value and 
extracted from minimum-bias data.  Thus, fluctuations in particle production 
and any other possible nonperturbative effects are incorporated into this 
average value of $Q_{0s}$. It was found that $Q_{0p}^2(x_0=0.01) \approx 
0.168$~GeV$^2/c^2$ with $\gamma \approx 1.119$ provides a good description of 
small-$x$ proton data from the LHC, HERA and RHIC 
\cite{Albacete:2010sy,Albacete:2010ad}, see also 
Ref.\,\cite{JalilianMarian:2011dt}.  These values are employed in the rcBK 
description of the projectile proton.  

Here the difference between protons and nuclei originates from the different 
initial saturation scales, $Q_{0s}$, in the rcBK equation, see \eq{mv}. The 
initial nuclear saturation scale $Q_{0A}$ at a given centrality is generally 
less constrained than that in the proton because the small-$x$ data from d+Au 
collisions at RHIC and DIS data on heavy target are limited with rather large
uncertainties, leading to correspondingly larger uncertainties on 
$Q_{0A}$ \cite{Rezaeian:2012ye},  
\begin{equation} \label{qa}
 (3 - 4)~Q_{0p}^2 \le Q_{0A}^2 (x_0=0.01) \le (6 - 7)\,Q_{0p}^2 \, \, .
\end{equation}
The role of fluctuations on particle production and other nonperturbative
effect, including the impact-parameter dependece of the initial nuclear 
saturation scale are effectively incorporated into the average value of
$Q_{0A}^2$ assuming $Q_{0A}^2 = N Q_{0p}^2$ is constrained by experimental data.
Figure~\ref{fig:comp_RpPb} shows results at $\eta=0$
with $N=5$ and varying $\alpha_s^{\rm in}$. For results over the full range of
$N$ in Eq.~(\ref{qa}), see Fig.~\ref{rp-h}.

On the other hand, Albacete {\it et al} let the initial nuclear saturation 
scale be proportional to the nuclear density at each point in the transverse 
plane.  They compare two different methods \cite{Albacete:2012xq},
\begin{eqnarray}
Q_{0A}^2(b,x_0) & = & N_{\rm part}(b) Q_{0p}^2(x_0)  \,\, \, {\rm and} \\
              & = & N_{\rm part}^{1/\gamma}(b) Q_{0p}^2(x_0) \, \, \, .
\end{eqnarray}
The second option is an ad hoc way of correcting for the violation in the 
additivity of the nucleon number in the dipole scattering amplitude at small 
dipole sizes resulting from the fact that $\gamma \ne 1$.  The position of the 
nucleons in the transverse plane is simulated by Monte Carlo methods makes it
possible to account for geometry fluctuations which can have large numerical 
impact \cite{Albacete:2012xq}.

\paragraph[IP-Sat]{IP-Sat (P. Tribedy and R. Venugopalan)}
The nuclear modification factors for inclusive charged particles in 
minimum-bias collisions employing the IP-Sat approach is done using 
$k_T$-factorization approach at $y=0$ in the center-of-mass frame. 
The LO KKP fragmentation functions~\cite{Kniehl:2000fe} are used to compute 
the inclusive charged hadron spectrum from the gluon distribution. The details 
of the parameters used can be found in 
Refs.~\cite{Tribedy:2010ab,Tribedy:2011aa}. 
The band shows the uncertainty in the calculation.  Note that $R_{p {\rm Pb}}$ 
approachs unity with increasing $p_T$.

\paragraph[Cold matter energy loss]{Cold matter energy loss (Z.-B. Kang, I. Vitev and H. Xing)}

The red curves on the right-hand side of Fig.~\ref{fig:comp_RpPb} are results
including cold matter energy loss at $y=0$, see Sec.~\ref{sec:Ivan-description}.
The upper curve corresponds to parameters determined for RHIC while the lower 
curve includes a potential enhancement in these parameters for LHC energies.
There is a very small ``Cronin peak'' in the low  $p_T$ region. The peak is 
very close to unity and not as pronounced as that seen in lower energy 
fixed-target experiments because dynamical shadowing strongly suppresses
particle production in this region. Initial-state energy loss is also larger 
because diagrams with incident gluons make a bigger contribution to the cross 
section. At high $p_T$, a small, $\sim 15$ \%, suppression remains, also due to 
cold nuclear matter energy loss.

The preliminary data are most compatible with the saturation model calculations,
albeit the uncertainties of these calculations are still rather large. The
calculation by Kopeliovich is also in relative agreement with the data.  
However, we note that none of the calculations available for both 
$dN_{\rm ch}/d\eta$ and $R_{p{\rm Pb}}$ at $\eta \sim 0$ agree with both observables
simultaneously.

\subsubsection[$\mathtt{HIJING2.1}$ parton vs. hadron $R_{pA}$]{$\mathtt{HIJING2.1}$ parton vs. hadron $R_{pA}$ (R. Xu, W.-T. Deng and X.-N. Wang)}

\begin{figure}[htpb]
  \centering
\includegraphics[width=0.495\textwidth]{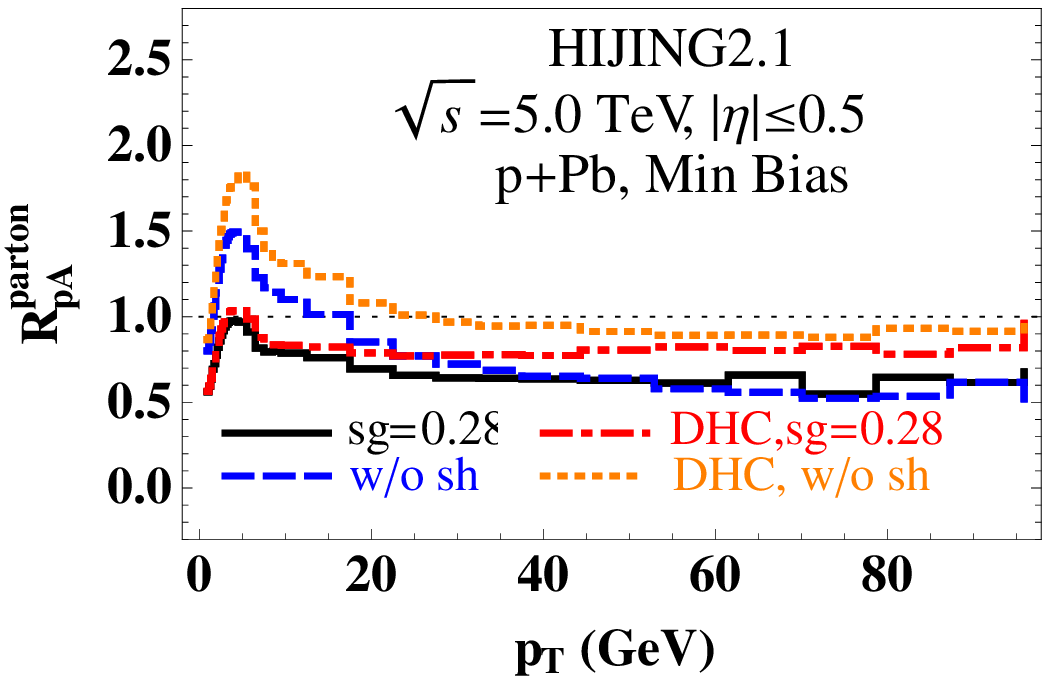}
\includegraphics[width=0.495\textwidth]{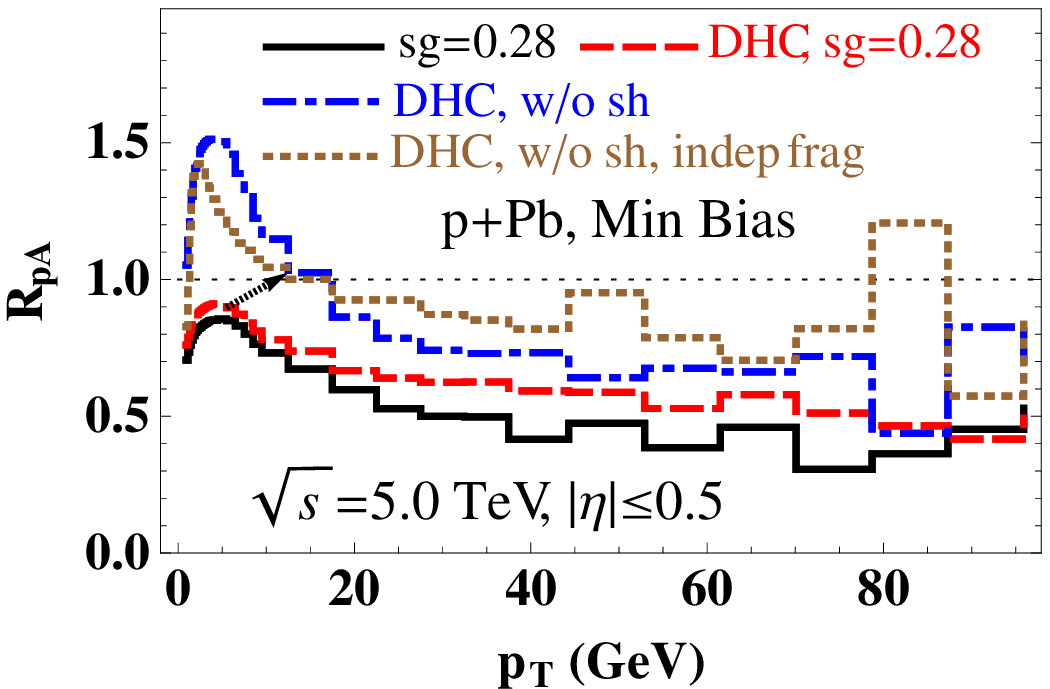}
\caption[]{(Left) 
The nuclear modification factor of the parton $p_T$ spectra in 
$p+$Pb collisions. (Right) The charged hadron nuclear modification 
factor with different $\mathtt{HIJING2.1}$ options. 
The arrow indicates the most probable trend of the nuclear modification 
factor to transition from the low to the high $p_{T}$ regions.
}
\label{fig:RpA_HIJING}
\end{figure}

On the left-hand side of Fig.~\ref{fig:RpA_HIJING}, 
the nuclear modification factor of the 
final-state parton  $p_T$ spectra, $R_{pA}(p_T)$,
is shown for $p+$Pb collisions at the LHC.  Here $\langle N_{\rm coll}\rangle$ 
is the average number of binary nucleon-nucleon interactions in $p+A$ 
collisions.  It is clear that both nuclear shadowing and the hard-soft coupling 
can suppress the $p_T$ spectra of the produced partons. These features will 
transferred to the final hadron spectra after hadronization, as shown on
the right-hand side of the Figure.

The predicted charged hadron nuclear modification factors are shown on
the right-hand side of
Fig.~\ref{fig:RpA_HIJING}. The enhancement at intermediate $p_T$ is due to 
the $k_T$ broadening arising from multiple scattering.  Parton shadowing, 
the modified soft-hard coupling and enhanced density of gluon jets due to 
valence quark conservation all contribution to suppression of the charged 
hadron spectra. In addition, string fragmentation can further suppress the 
high $p_T$ hadron spectra compared to independent fragmentation.

Since the effects of parton shadowing will disappear at large $p_{T}$ 
\cite{Eskola:2009uj} due to the QCD evolution not yet included here while hard 
and soft scatterings will decohere, the
nuclear modification factor will likely follow the default result, including 
DHC, at low $p_{T}$. At large $p_T$, the result should approach that with 
DHC effects but no shadowing. There may be further possible modifications due 
to the hadronization of multiple jets.  In Fig.~\ref{fig:RpA_HIJING}, this
probable trend is indicated by the arrow joining these two results at 
intermediate $p_{T}$.

\subsubsection{$R_{p {\rm Pb}}(p_T)$ at $|\eta| \neq 0$}
\label{forward_rpPb}

Here we show several results for the charged particle $R_{p {\rm Pb}}$ away from
central rapidity.  Two employ the rcBK approach but with different initial
conditions.  The last is a pQCD calculation including cold matter energy loss.

\paragraph[Rezaeian rcBK]{Rezaeian rcBK (A. Rezaeian)}

\begin{figure}[htpb]       
\begin{tabular}{cc}
\includegraphics[width=0.45\textwidth,height=0.4\textheight] {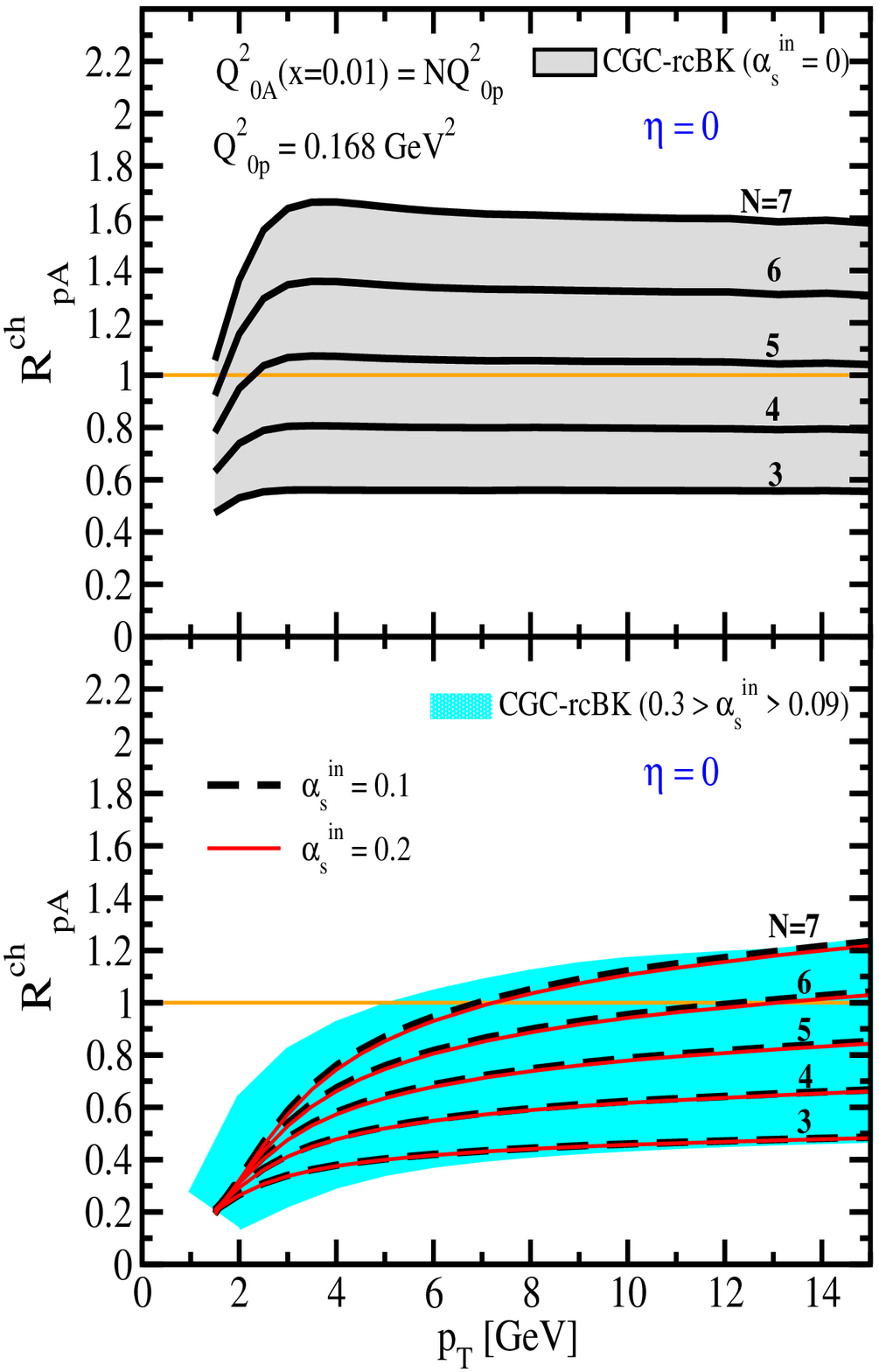} &
\includegraphics[width=0.45\textwidth,height=0.4\textheight] {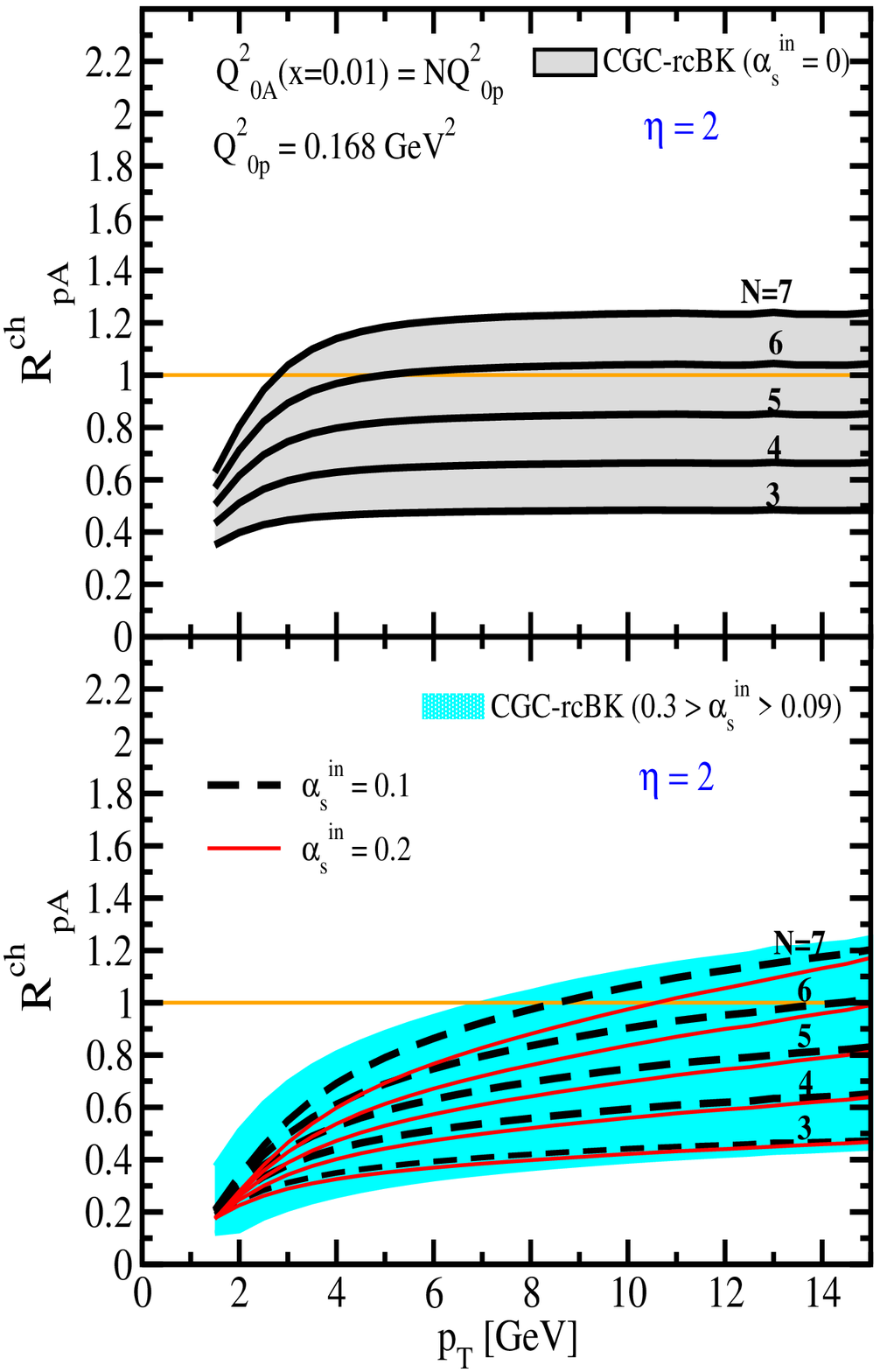} \\     
\includegraphics[width=0.45\textwidth,height=0.4\textheight] {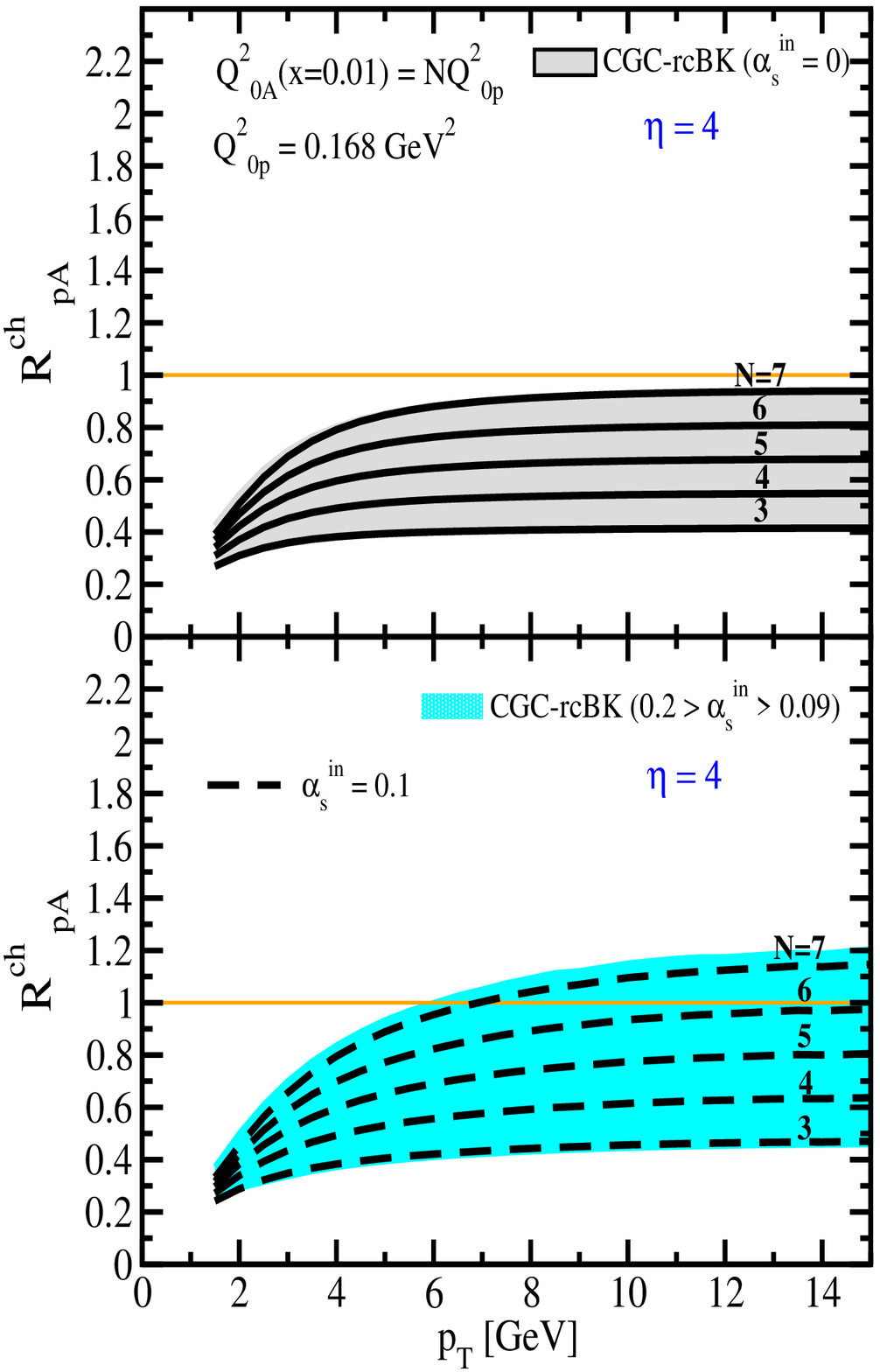} &      
\includegraphics[width=0.45\textwidth,height=0.4\textheight] {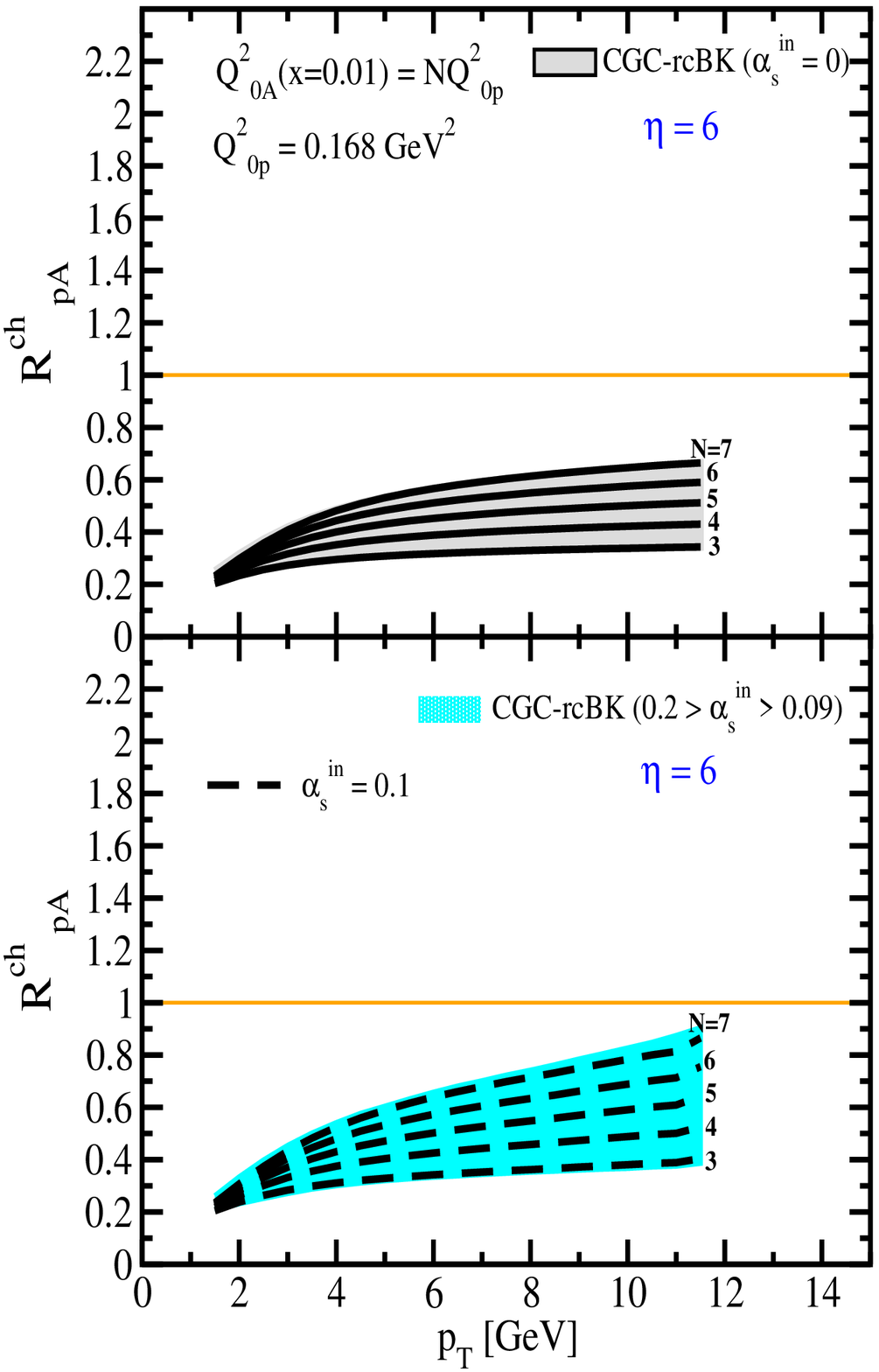} 
\end{tabular}
\caption[]{The nuclear modification factor $R_{pA}^{\rm ch}$ for charged hadron
production in minimum bias $p+$Pb collisions at $\eta=0$, $2$, $4$, and $6$
(with the convention that the proton beam moves toward
forward rapidity)
obtained from the hybrid factorization in \eq{qa}  assuming different values
of the saturation scale in the nucleus, $Q_{0A}^2$.  The lines
labeled by a given value of $N$, for $3<N<7$, are results with fixed 
factorization scale $\mu_F=p_T$ 
and fixed saturation scale $Q_{0A}^2=N Q_{0p}^2$ and 
$Q_{0p}^2=0.168$~GeV$^2/c^2$.  The bands shown the variation in the results with 
the choice of factorization scale.
Two panels are shown for each rapidity.
The upper panel shows results obtained by taking $\alpha_s^{in}=0$ (assuming
only elastic contribution) while the bottom panel shows the variation
of $\alpha_s^{\rm in}$ in the range $0.09 \ge \alpha_s^{in} \ge 0.3$.  In the
bottom panels for $\eta = 0$ and 2, results with both $\alpha_s^{\rm in} = 0.1$
and 0.2 are shown, while for $\eta = 4$ and 6, only $\alpha_s^{\rm in} = 0.1$
is shown.  The plots are taken from Ref.~\protect\cite{Rezaeian:2012ye}.}
\label{rp-h}           
\end{figure}

In \fig{rp-h}, $R_{pA}^{\rm ch}(p_T)$ is shown for minimum bias collisions at 
$\eta=0$, 2, 4, and 6 (with the convention that the proton beam moves toward
forward rapidity) obtained from the hybrid factorization \eq{final} 
supplemented with rcBK evolution. The band labeled CGC-rcBK includes 
uncertainties associated with the variation of initial nuclear saturation scale,
$Q_{0A}^2$, constrained in \eq{qa}, as well as uncertainties due to the choice 
of factorization scale,  $\mu_F=2p_T$, $p_T$, and $p_T/2$, in \eq{final}.  The
values of $N$ scale the nuclear saturation scale relative to that in the proton,
$Q_{0A}^2 = NQ_{0p}^2$, with $3<N<7$.  The variation of the results with $N$
represents an effective variation with centrality and incorporating possible
effects of fluctuations.  Going forward in proton rapidity 
reduces the range of the results and lessens the dependence on $N$.

As shown in Ref.~\cite{JalilianMarian:2011dt},
 the RHIC data unfortunately cannot fix the value of  
$\alpha_s^{\rm in}$ in \eq{final}: with a reasonable $K$-factor, 
$0 \le \alpha_s^{\rm in} \le 0.3$ is consistent with hadron production at RHIC 
both in $p+p$ and d$+A$ collisions.  Figure~\ref{rp-h} shows the effect of 
varying $\alpha_s^{\rm in}$ in \eq{final}.  For each value of $\eta$, results
are shown both for $\alpha_s^{\rm in}=0$ and variation within the range 
$0.09 \ge \alpha_s^{\rm in} \ge 0.3$.  For $\eta = 0$ and 2, both $\alpha_s^{\rm
in} = 0.1$ and 0.2 are shown.  There is only a small difference at $\eta = 0$
while a larger difference can be seen at forward proton rapidity, especially for
$N>5$.  With $\eta = 4$ and 6, only $\alpha_s^{\rm in} = 0.1$ is shown.
Note that taking $\alpha_s^{\rm in} > 0$ changes the $p_T$ slope considerably,
as may be expected.

In order to quantify the uncertainties due to different initial nuclear 
saturation scales, in Fig.~\ref{rp-h} $R_{pA}^{\rm ch}$ is shown for different 
initial saturation scales $Q_{0A}^2=N Q_{0p}^2$ for $3<N<7$ with factorization 
scale $Q=p_T$. Unfortunately, $R^{\rm ch}_{pA}$ is 
very sensitive to $Q_{0A}^2$. However, by comparing to measurements of 
$R^{\rm ch}_{pA}$ at one rapidity, the predictions in Fig.~\ref{rp-h} can be used 
to determine $Q_{0A}^2$ at that rapidity.  Then, if the measured $R_{pA}^{\rm ch}$ 
at one rapidity lies between two lines labeled $N_1$ and $N_2$, the results at 
other rapidities should correspond to the same values of $N$ and thus the same 
$Q_{0A}^2$.  Only these results should be considered to be true CGC predictions. 
In this way, knowing $R_{pA}^{\rm ch}$ at one rapidity 
significantly reduces theoretical uncertainties associated with $Q_{0A}^2$ at 
other rapidities.  Therefore, despite rather large theoretical uncertainties, 
it is still possible to test the main CGC/saturation dynamics at the LHC.

It is clear that CGC approaches predict more suppression at forward proton
rapidities compared to collinear factorization 
\cite{QuirogaArias:2010wh,Arleo:2011gc}. Moreover, the small-$x$ 
evolution washes away the Cronin-type peak at low $p_T$ at all rapidities. 
Therefore, observation of a Cronin peak in $p+$Pb collisions at the LHC, 
regardless of whether it is accompanied by overall enhancement or suppression, 
can potentially be a signal of non-CGC physics since it is difficult to 
accommodate this feature in the current accuracy of the CGC approach 
\cite{Rezaeian:2012ye}. 

\paragraph[Albacete {\it et al} rcBK]{Albacete {\it et al} rcBK (J. Albacete, A. Dumitru, H. Fujii and Y. Nara)}

\begin{figure}[htpb]
\begin{center}
\includegraphics[width=0.495\textwidth]{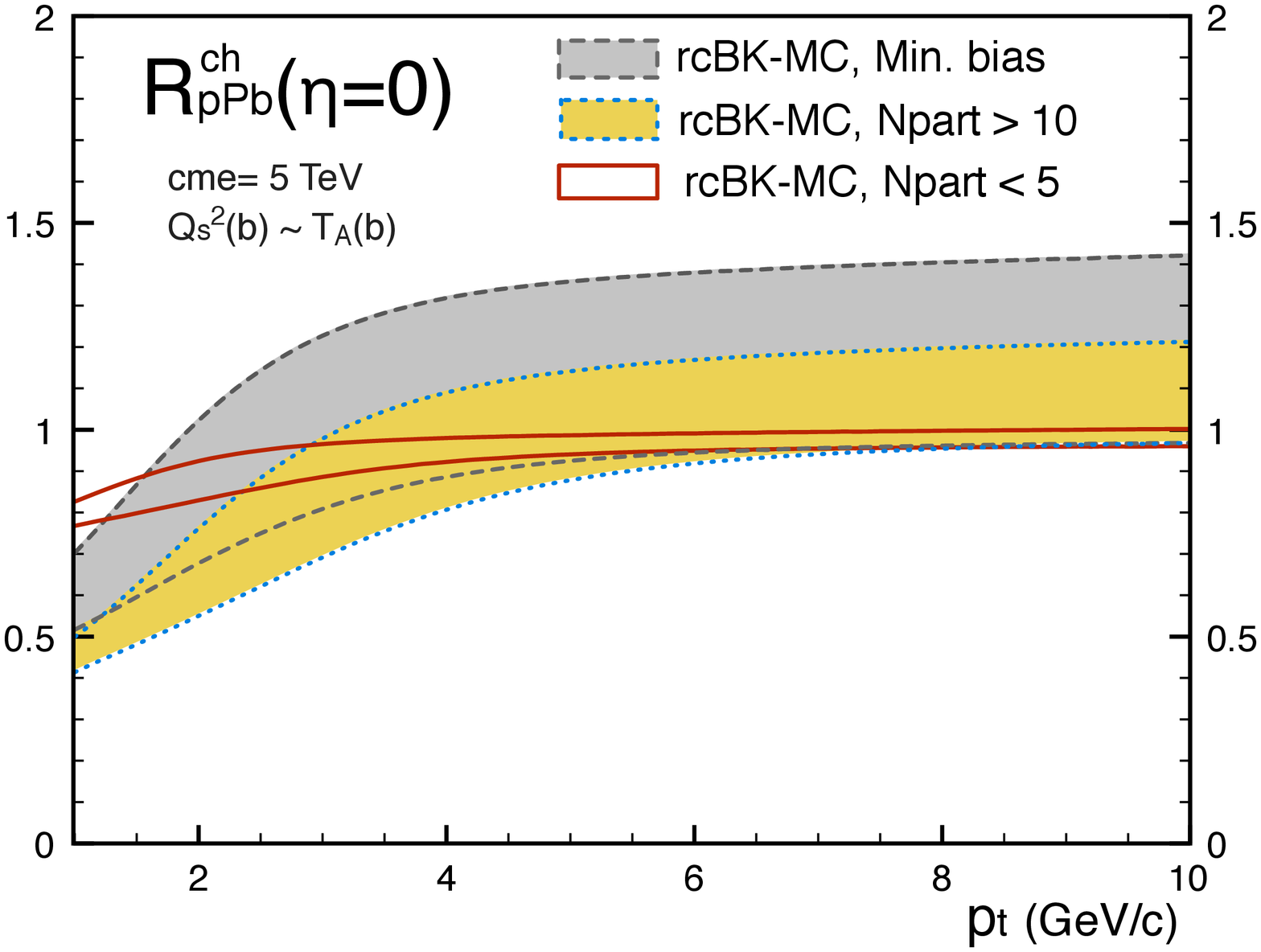}
\includegraphics[width=0.495\textwidth]{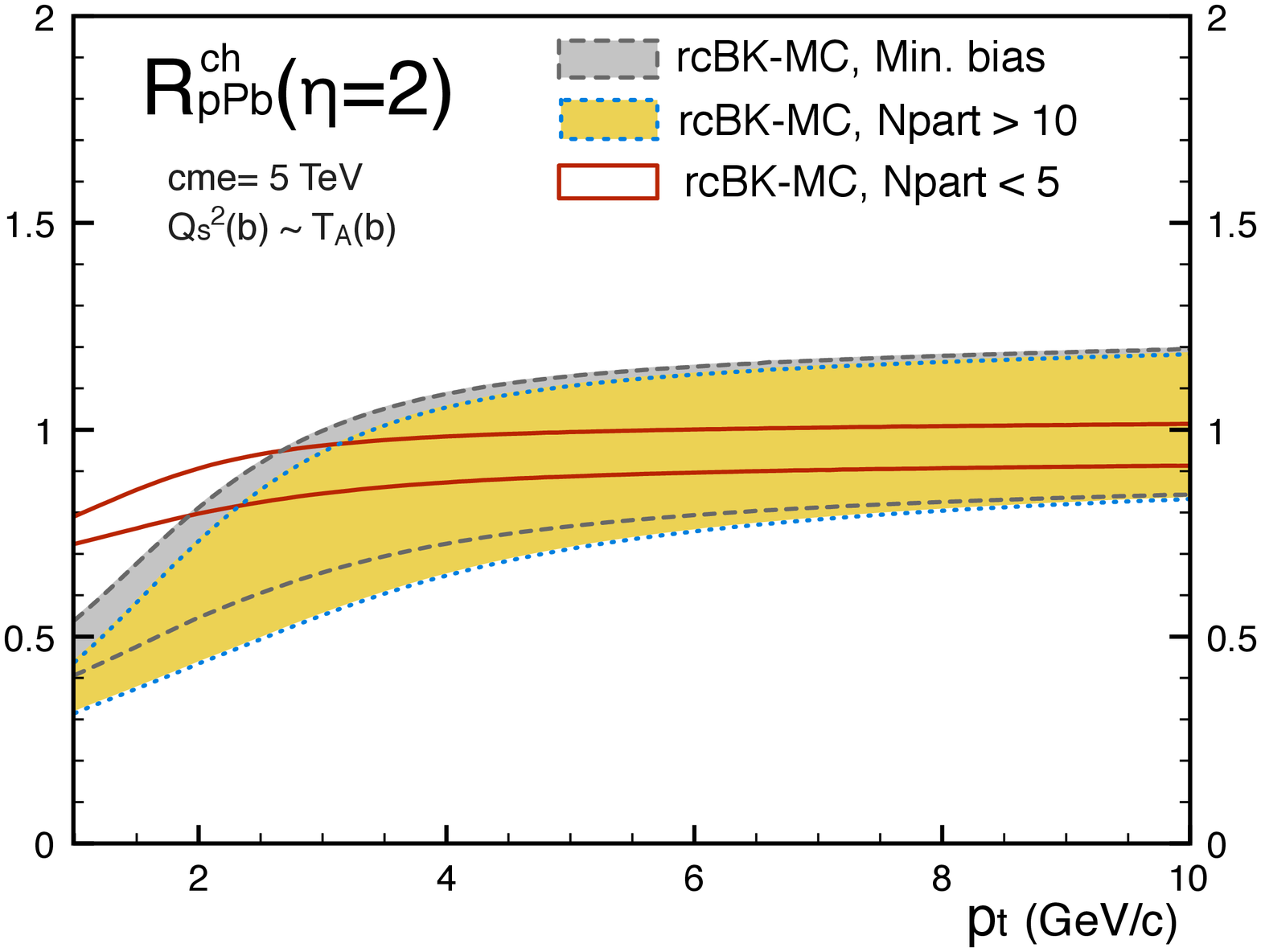}
\end{center}
\vspace*{-0.4cm}
\caption[]{The nuclear modification factor for three different
  centrality classes assuming $k_{T}$-factorization.  The $\eta=2$
result is obtained with the convention that the proton beam moves toward
forward rapidity.
}
\label{fig:RpPb_centr}
\end{figure}

Figure~\ref{fig:RpPb_centr} shows bands for $R_{pA}(p_T)$ at $\eta = 0$ and 2
with rcBK-MC.  The upper edges of the bands are
calculated with $\gamma=1.119$ and KKP-LO
fragmentation functions. the lower limit of the bands corresponds to
the McLerran-Venugopalan initial condition ($\gamma=1$) with DSS-NLO
fragmentation functions, taken from Ref.~\cite{Albacete:2012xq}.

There is suppression of $R_{p {\rm Pb}}(p_T \sim 1 \, {\rm GeV}/c) = 
0.6 \pm 0.1$ at midrapidity.  The ratio increases monotonically to
$p_T \sim 3$~GeV/$c$ where it flattens.  There is no clear Cronin
peak. For all unintegrated gluon densities, $R_{p {\rm Pb}}$ decreases with 
increasing rapidity.  In addition to the minimum bias results, two separate
centrality classes are also shown, a central bin, $N_{\rm part}>10$, and the
most peripheral bin, $N_{\rm part}<5$.  There is a stronger effect for the most
central collisions, as expected, for low $p_T$ at both $\eta = 0$ and 2.  (Note
that there is a smaller difference between minimum bias and the most central
collisions at $\eta = 2$.  The overall effect is very
weak for the peripheral bin with suppression persisting only to $p_T \sim 
2-3$~GeV/$c$.  
For more details, see Ref.~\cite{Albacete:2012xq}.

\paragraph[Cold matter energy loss]{Cold matter energy loss (Z.-B. Kang, I. Vitev 
and H. Xing)}

\begin{figure}[htpb]
\begin{center}
\includegraphics[width=0.5\textwidth]{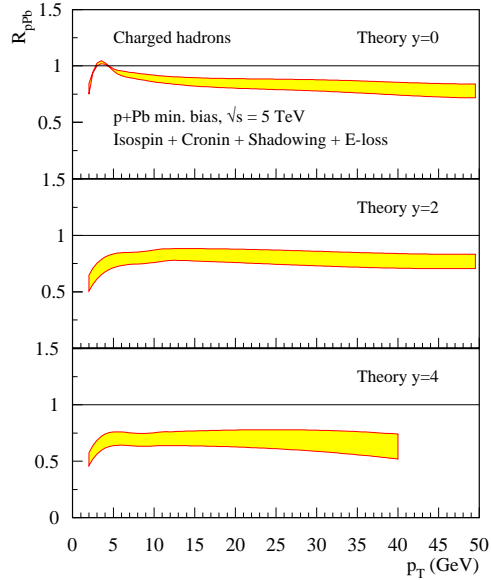}
\end{center}
\caption[]{Predictions for the nuclear modification factor $R_{pPb}$ as a 
function
of $p_T$ for charged hadron production in minimum 
bias $p+$Pb collisions.  Results are shown for three rapidities: $y=0$ (top), 
$y=2$ (center), and $y=4$ (bottom) with the convention that the proton beam 
moves toward forward rapidity, see Ref.~\protect\cite{Kang:2012kc}.
}
\label{lhc_h}
\end{figure}

Figure~\ref{lhc_h} presents model predictions for $R_{p {\rm Pb}}(p_T)$ in 
minimum bias $p+$Pb collisions at $y=0$ (top), $y=2$ (middle), and $y=4$ 
(bottom). 
The upper edge of the bands corresponds to the RHIC scattering parameters while
The lower edge allows for a potential high-energy enhancement of the parameters.

The $y=0$ results were already presented in Fig.~\ref{fig:comp_RpPb}.
At large proton rapidity the CNM effects are all
amplified due to the larger values of the proton momentum fraction $x_2$ 
(relevent for cold nuclear matter energy loss with the proton moving in the 
direction of forward rapidity), the smaller values of the 
nuclear momentum fraction $x_1$ (relevent for dynamical shadowing with the
nucleus moving in the direction of backward rapidity) and more 
steeply-falling $p_T$ spectra (relevent for the Cronin effect).  
As a result, at low $p_T$ and large ion rapidity, dynamical 
shadowing can dominate 
and lead to stronger suppression of inclusive particle production (note the
disappearance of the small Cronin enhancement at $y>0$).  At high $p_T$ the 
suppression is a combined effect of cold nuclear matter energy loss and the 
Cronin effect.  As will be seen later, the 
behavior of $R_{p{\rm Pb}}$ for $\pi^0$ and direct photon production is 
qualitatively
similar to the $R_{p {\rm Pb}}$ dependence for charged hadrons shown here.

Away from midrapidity, these predictions suggest a stronger effect than that
found with EPS09 parton shadowing alone.  On the other hand, the effect shown
here is weaker than the results using the rcBK-MC approach also shown in this
section.\\[2ex]

\begin{figure}[htpb]
\begin{center}
\includegraphics[width=0.5\textwidth]{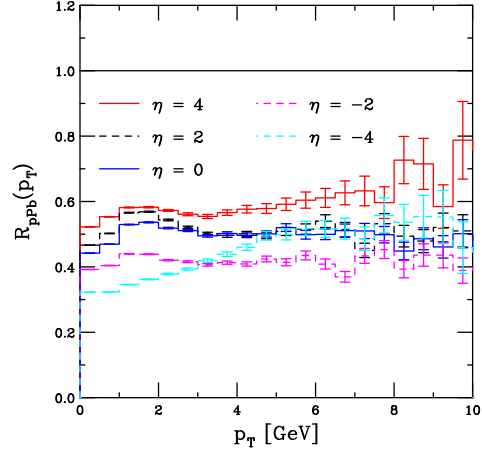}
\end{center}
\caption[]{Predictions for the nuclear modification factor $R_{pPb}$ as a 
function
of $p_T$ for charged hadron production in minimum 
bias $p+$Pb collisions for $\eta = -4$, $-2$, 0, 2 and 4 calculated using
$\mathtt{AMPT-def}$.}
\label{AMPT-etanot0}
\end{figure}

\paragraph[AMPT at $\eta \neq 0$]{$\mathtt{AMPT}$ at $\eta \neq 0$ (Z. Lin)}
Figure~\ref{AMPT-etanot0} shows results with the $\mathtt{AMPT-def}$ 
calculation in 
minimum bias $p+$Pb collisions at five different $\eta$ values. At $\eta = 4$,
the ratio is larger than at lower rapidities.  The ratios are very 
similar for $\eta = 0$ and 2.  The backward rapidity results have lower values
of $R_{p {\rm Pb}}$.  While the ratios are indepedent of $p_T$ within the
uncertainties for $\eta = -2$, 0 and 2, the ratios increase with $p_T$ for
$\eta = 4$ and $-4$ with the largest increase for $\eta = -4$.

\subsubsection{Forward-backward asymmetry(G. G. Barnaf\"oldi, J.Barette, M. Gyulassy, P. Levai, G. Papp, M. Petrovici, V. Topor Pop, Z. Lin and R. Vogt)}
 

One method of determining the difference between the proton and lead rapidity
regions is studying the forward-backward asymmetry in charged hadron
production, given by
\begin{eqnarray}
Y_{\rm asym}^{h}(p_T) & = & 
\frac{E_h d^3\sigma^h_{p\rm Pb}/d^2p_T d\eta |_{\eta>0}}{E_h d^3
\sigma^h_{p {\rm Pb}}/d^2p_T d\eta|_{\eta<0}} \, \, , \nonumber \\
& = & \frac{R^h_{p {\rm Pb}}(p_T,\eta>0)}{R^h_{p{\rm Pb}}(p_T,\eta<0)} \ .
\label{yasym}
\end{eqnarray}
where the `forward' direction, $\eta > 0$, is that of the lead beam, toward 
positive rapidity while the `backward' direction, $\eta < 0$ is that of the
proton beam, toward negative rapidity.

\begin{figure}[htpb]
\begin{center}
\includegraphics[width=0.55\textwidth]{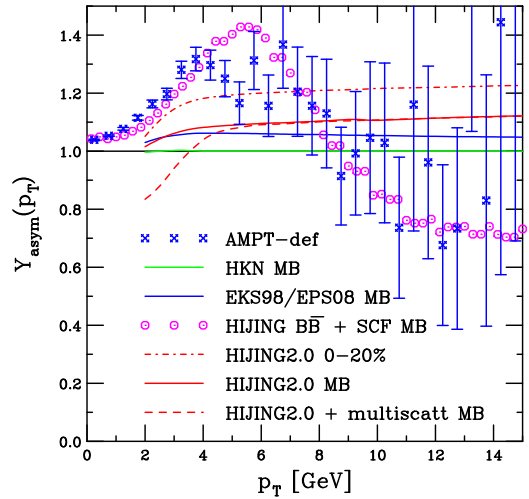}
\end{center}
\caption[]{Predictions for the forward-backward asymmetry, $Y_{\rm asym}^{h}(p_T)$,
from Refs.~\protect\cite{Levai:2011qm,Barnafoldi:2008ec}.  
Centrality independent results are shown
for the HKN, EKS98 and EPS08 parameterizations (labeled MB).  Minimum bias
results are also shown for $\mathtt{HIJINGB\overline{B}2.0}$ and 
$\mathtt{HIJING2.0}$ with multiple scattering.  In addition, 
$\mathtt{HIJING2.0}$ results in MB collisions and for the 20\% most central
collisions (using the parameter $s_a$ in Eq.~(\protect\ref{eq:rshadow}))
are also shown.  The blue points are the $\mathtt{AMPT-def}$ results.
}
\label{fig:ggb:2}
\end{figure}

The asymmetries, calculated in the center of mass frame
in the range $0.3<|\eta|<0.8$ \cite{Adeluyi:2008gj},
both for standard shadowing and event generators, 
are shown in Fig.~\ref{fig:ggb:2}.   
Two calculations of the forward-backward asymmetry in $p+$Pb collisions 
are shown. The first result is in the $\mathtt{kTpQCD\_v2.0}$ approach with
several standard shadowing parameterizations.  The second is obtained with 
$\mathtt{AMPT}$.

Results with $\mathtt{kTpQCD\_v2.0}$  
including the $\mathtt{HIJING2.0}$~\cite{Li:2001xa}, 
EKS98~\cite{Eskola:1998df}, EPS08~\cite{Eskola:2008ca}, and 
HKN~\cite{Hirai:2001np} shadowing parameterizations are presented.
Since the EKS98, EPS08 and HKN parameterizations are independent of impact
parameter, there is no difference between minimum bias results and results
with a centrality cut.  Thus, these are all labeled MB for minimum bias.
With the HKN parameterization, $Y_{\rm asym} \approx 1$ independent of $p_T$.
The EKS98 and EPS08 parameterizations, producing identical ratios, 
predict a small enhancement, $Y_{\rm asym}>1$.  

Results with the
$\mathtt{HIJING2.0}$ shadowing parameterization with and without multiple
scattering differ for $p_T<5$ GeV/$c$ but converge at higher $p_T$.  
When the collision centrality is taken into account, the forward-backward
asymmetry is enhanced by the centrality cut alone.  Chaning the strength of
low $x$ shadowing, as in Eq.~(\ref{eq:rshadow}), does not 
change the shape of $Y_{\rm asym}(p_T)$ for $p_T>5$ GeV/$c$.

Finally, results for the minimum bias asymmetry is also shown for
$\mathtt{HIJINGB\overline{B}2.0}$ \cite{ToporPop:2011wk}. These calculations 
include both shadowing and strong color field effects (indicated ``+ SCF''). 
Here the ratio rises from near unity at $p_T \sim 0$ to a peak at $p_T \sim
5.5$~GeV/$c$.  It then decreases to $Y_{\rm asym} \sim 0.75$ for 
$p_T \geq 10$~GeV/$c$.  


The forward-backward asymmetry of $R_{p{\rm Pb}}$, calculated in the center of 
mass frame with $\mathtt{AMPT-def}$
in the range $0.3 < |\eta| < 0.8$, is shown by the blue points
in Fig.~\ref{fig:ggb:2}
for charged particles in minimum-bias events with simulations employing 
$\mathtt{AMPT-def}$.  The initial enhancement first increases with $p_T$ and 
then decreases.  Although the statistical errors become very large above 
$p_T \sim 10$~GeV/$c$, it is clear that the result is similar to the general
behavior of the $\mathtt{HIJINGB\overline{B}2.0}$ result. 

The low $p_T$ enhancement may arise because the asymmetry in $dN/d\eta$ in
$p+$Pb relative to $p+p$ in the lead direction (positive rapidity) introduces 
a natural enhancement in $R_{p {\rm Pb}}$ in the forward direction.  The enhanced
particle production in $p+$Pb is most likely at low $p_T$, see e.g. 
Fig.~\ref{AMPT-etanot0}.


The $\mathtt{AMPT-def}$ results include the impact-parameter dependent nuclear 
shadowing parameterization implemented in $\mathtt{HIJING1.0}$.  
Therefore the asymmetry at other centralities would differ from 
that in minimum-bias events.

\subsubsection[$R_{p {\rm Pb}}(\eta)$]{$R_{p {\rm Pb}}(\eta)$ (G. G. Barnaf\"oldi, J. Barette, M. Gyulassy, P. Levai, M. Petrovici and V. Topor Pop)}

Finally, we end this section with the predicted ratio $R_{p{\rm Pb}}$ as a
function of pseudorapidity in $\mathtt{HIJINGB\overline{B}2.0}$.  As discussed
previously, increasing the minijet cutoff parameter $p_0$ from 2 GeV/$c$ in
$p+p$ collisions to 3.1 GeV$/c$ in $p+$Pb causes a slow growth in the per 
nucleon multiplicity which could be interpreted as evidence for gluon 
saturation.  It is difficult to directly relate $p_0$ to the 
saturation scale $Q_s$ because
in $\mathtt{HIJING}$ hadronization proceeds through longitudinal
string fragmentation.  The low $p_T$ part of the minijet spectrum is 
particularly
sensitive to the $\sqrt{s_{_{NN}}}$ and $A$ dependence of minijet suppression
while the $p_T>5$ GeV/$c$ minijet tails
are unaffected.   The energy and $A$ dependence 
of the string tension arises from strong color field (color rope) effects 
not included in CGC phenomenology that assumes hadronization by 
$k_T$-factorized gluon fusion.

\begin{figure}[htpb]
\centerline{\includegraphics[width=0.55\textwidth]{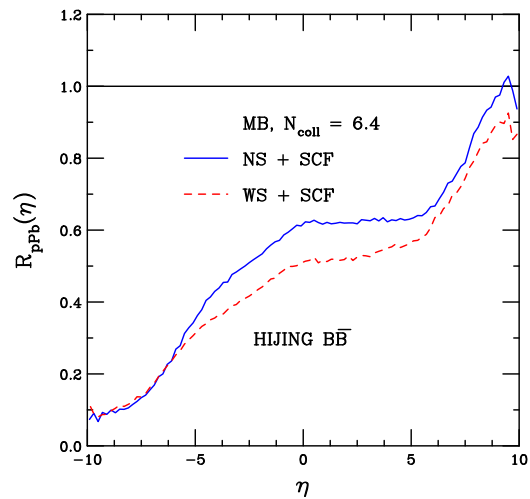}}
\caption[]{The suppression factor $R_{p {\rm Pb}}$ as a function of pseudorapidity,
$\eta$, for $\mathtt{HIJINGB\overline{B}2.0}$ with strong color field effects,
without (solid) and with (dashed) shadowing.}
\end{figure}

$\mathtt{HIJING}$ minijet hadronization does not proceed by 
independent fragmentation,
as in $\mathtt{PYTHIA}$ \cite{Bengtsson:1987kr}, 
but through string fragmentation 
with gluon minijets represented as kinks in the strings. The interplay
between longitudinal string fragmentation dynamics and minijets
is a nonperturbative feature of $\mathtt{HIJING}$-type models.  The 
effect of string fragmentation on the multiplicity is manifested in the 
behavior of $R_{p {\rm Pb}}(\eta)$ in the fragmentation regions $|\eta|>5$.
In the nuclear beam fragmentation region, the ratio is approximately linearly
increasing from $\eta = 5$ to the kinematic limit.  There is an approximate
plateau in the ratio over $0< \eta < 5$, followed by a decrease toward the 
value $1/N_{\rm coll}$ in the proton fragmentation region, a Glauber geometric
effect first explained in Refs.~\cite{Brodsky:1977de,Adil:2005qn}
and a feature of string fragmentation in $\mathtt{HIJING}$. 

\section{Identified Particles}
\label{sec:part_ID}

\subsection[$\mathtt{AMPT}$]{$\mathtt{AMPT}$ (Z. Lin)}
\label{sec:AMPT_pid}

\begin{figure}[htpb]
\begin{center}
\includegraphics[width=0.495\textwidth]{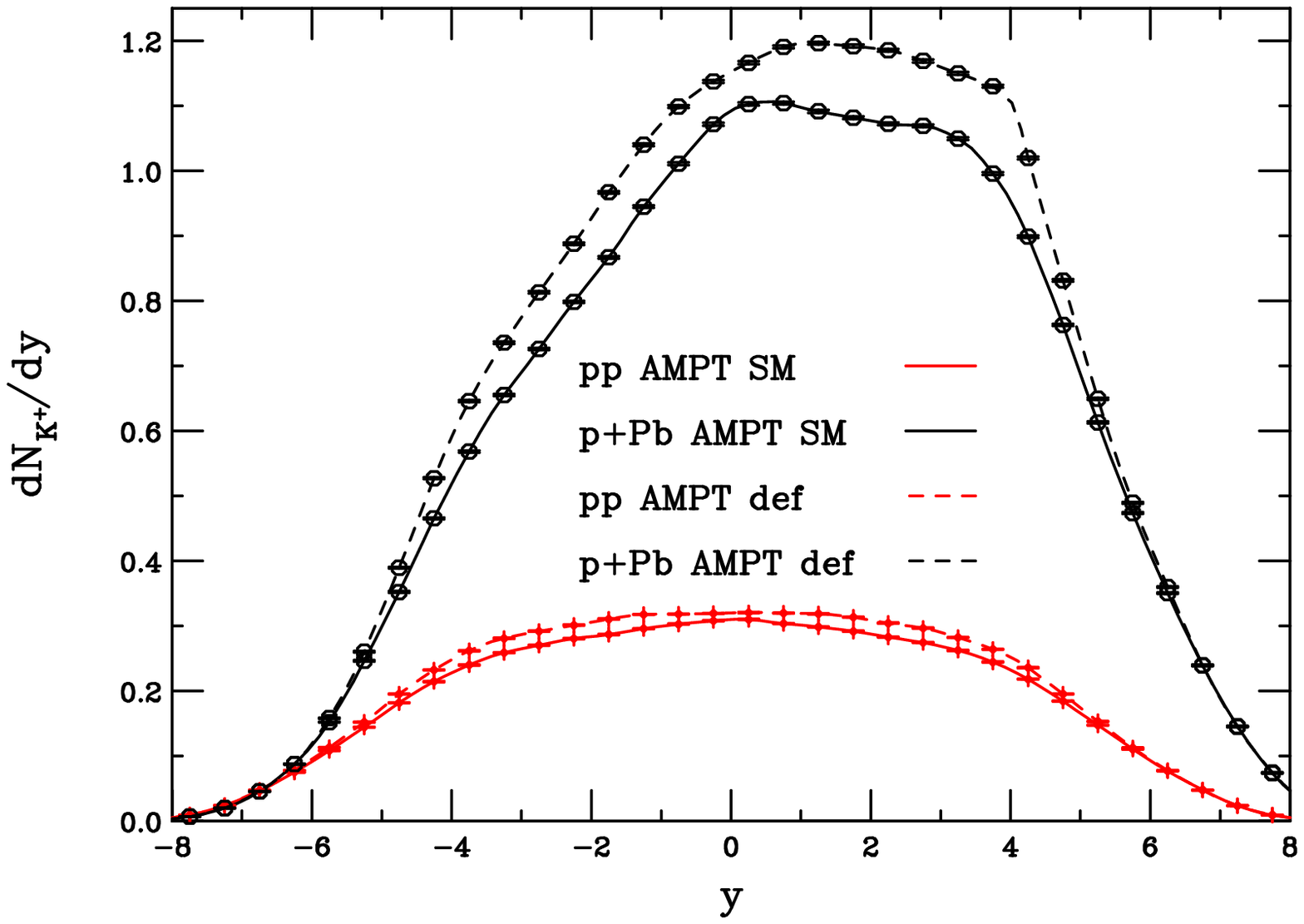}
\includegraphics[width=0.495\textwidth]{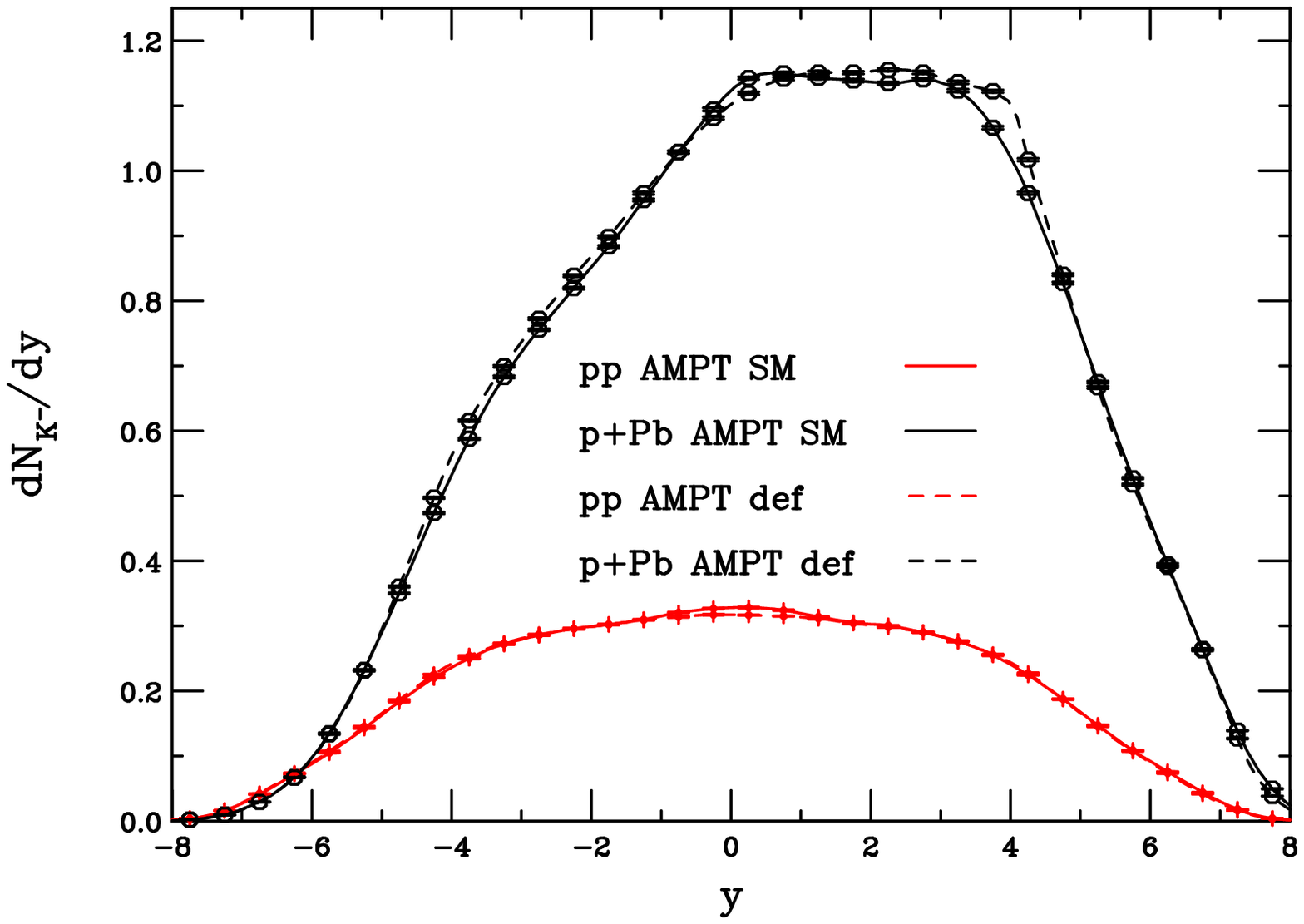}
\end{center}
\caption[]{Rapidity distribution, $dN/dy$, of $K^+$ (left) and $K^-$ (right)
mesons.}
\label{fig:AMPT-Kpm}
\end{figure}

\begin{figure}[htpb]
\begin{center}
\includegraphics[width=0.495\textwidth]{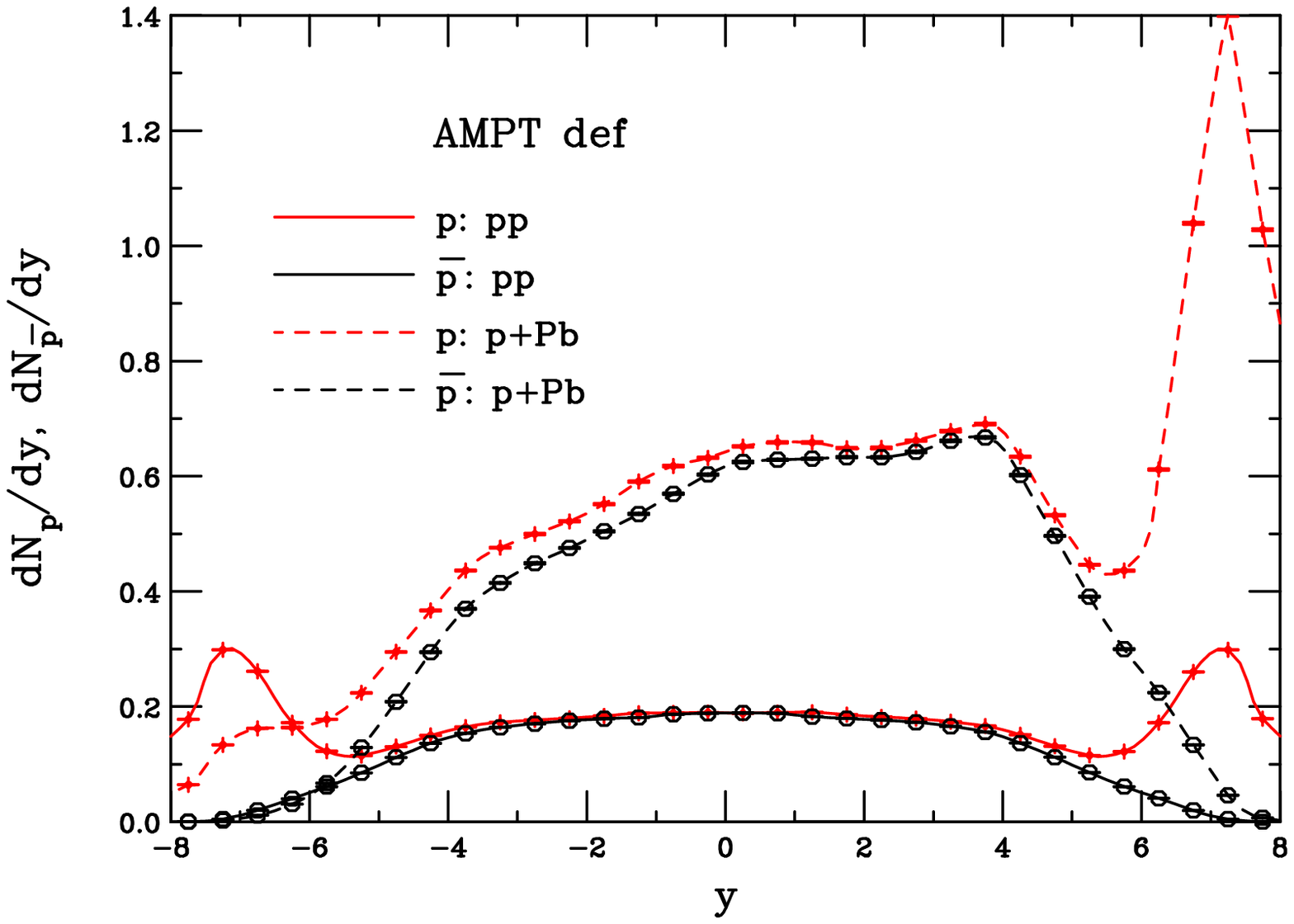}
\includegraphics[width=0.495\textwidth]{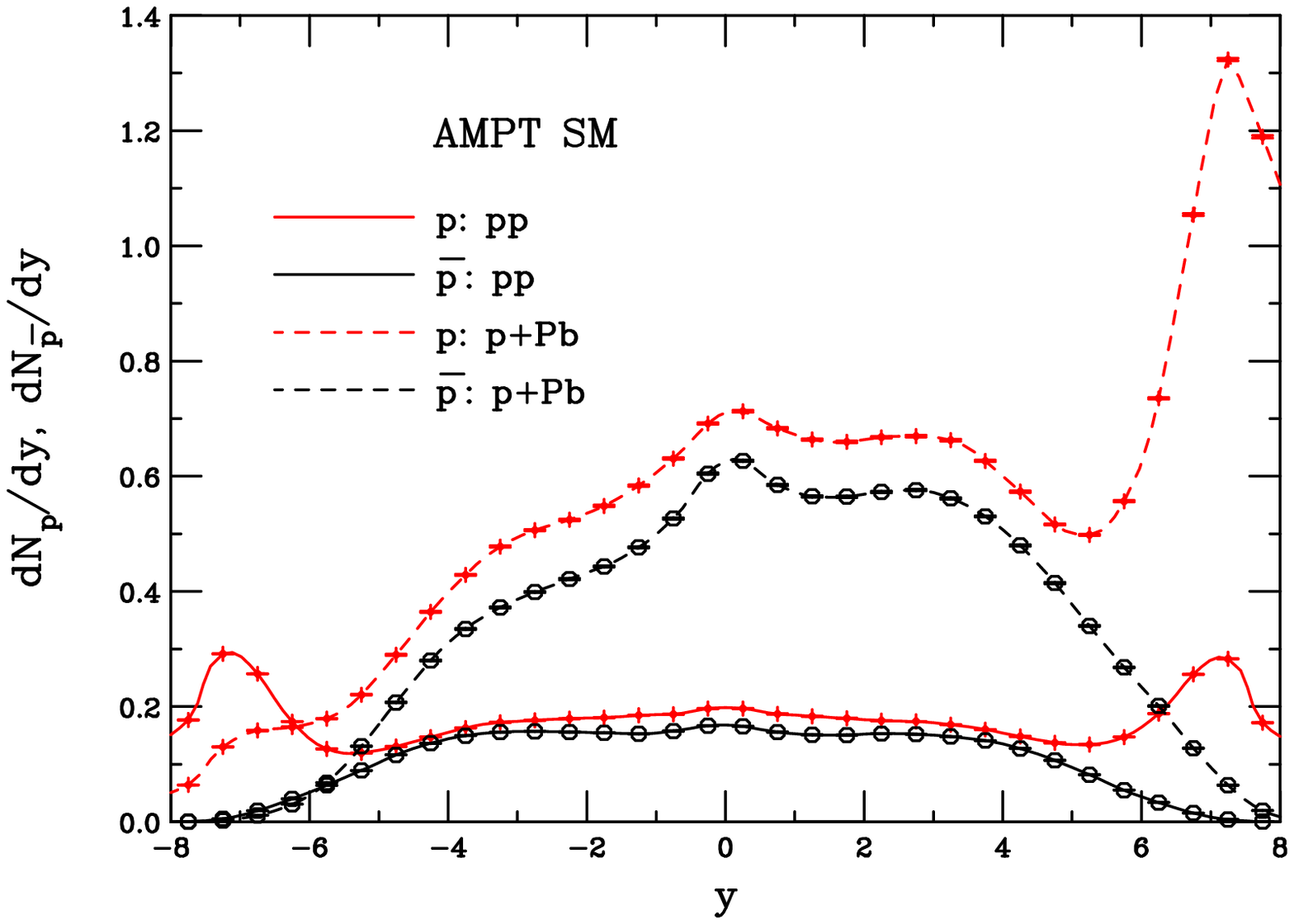}
\end{center}
\caption[]{Rapidity distribution, $dN/dy$, of and $p$ and
$\overline p$ in the default (left) and SM (right) scenarios.}
\label{fig:AMPT-p-pbar}
\end{figure}

Some representative $\mathtt{AMPT}$ results for identified particles
are shown here.  The rapidity distributions for $K^+$ and $K^-$ production
in the default and string melting versions are shown in Fig.~\ref{fig:AMPT-Kpm}.
The default and string melting versions of the rapidity distributions of
protons and antiprotons are shown in Fig.~\ref{fig:AMPT-p-pbar}.

There are
differences between the default and string melting versions, even in 
$p+p$ collisions.  However, these differences are typically small.  Note
that the bump at midrapidity in the proton and antiproton rapidity
distributions in the string-melting version is likely a result of the simple
quark coalescence model implemented in $\mathtt{AMPT}$ (see e.g.\ Fig.~38
of Ref.~\cite{Lin:2004en}).

\begin{figure}[htpb]
\begin{center}
\includegraphics[width=0.495\textwidth]{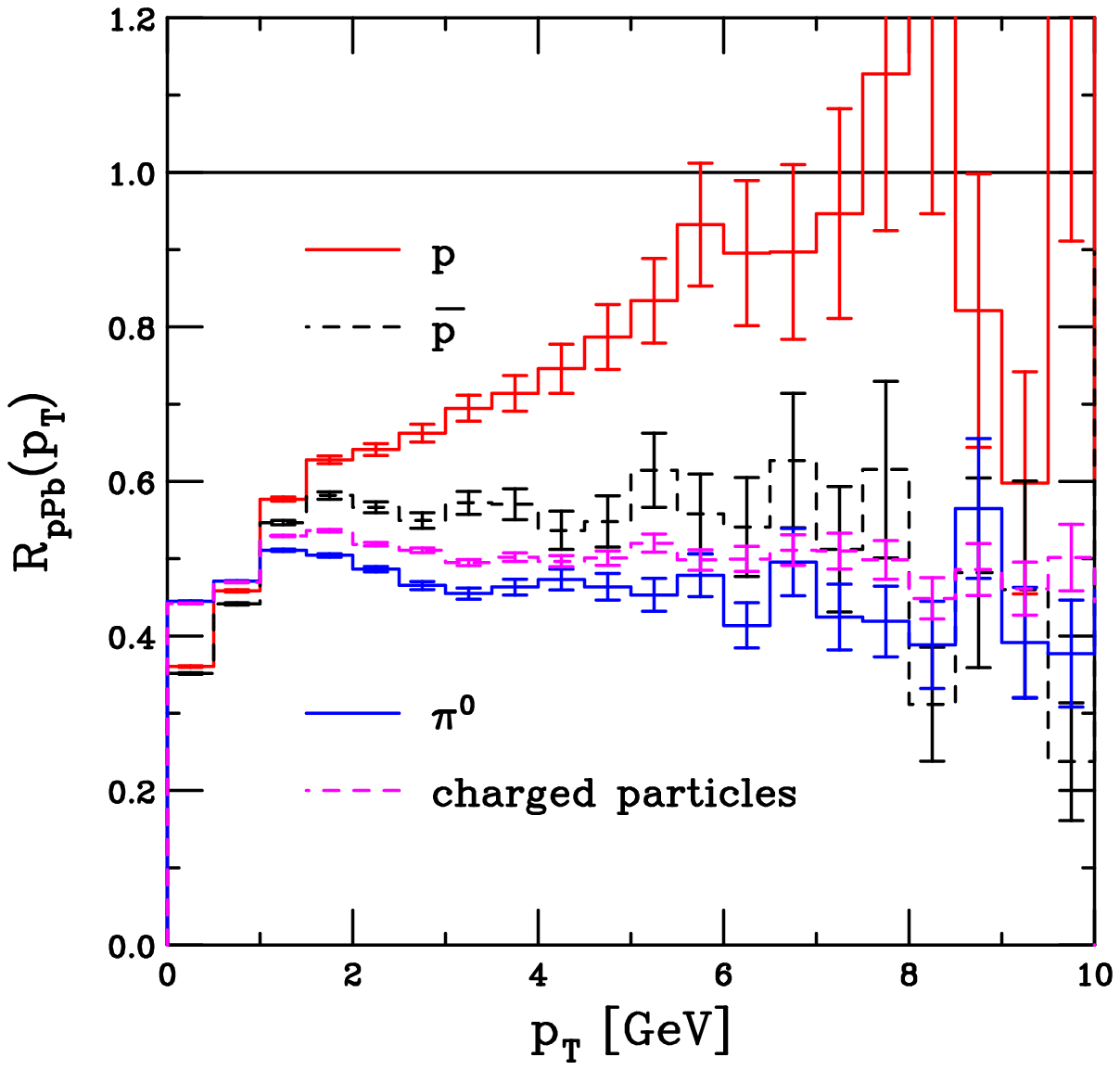}
\includegraphics[width=0.495\textwidth]{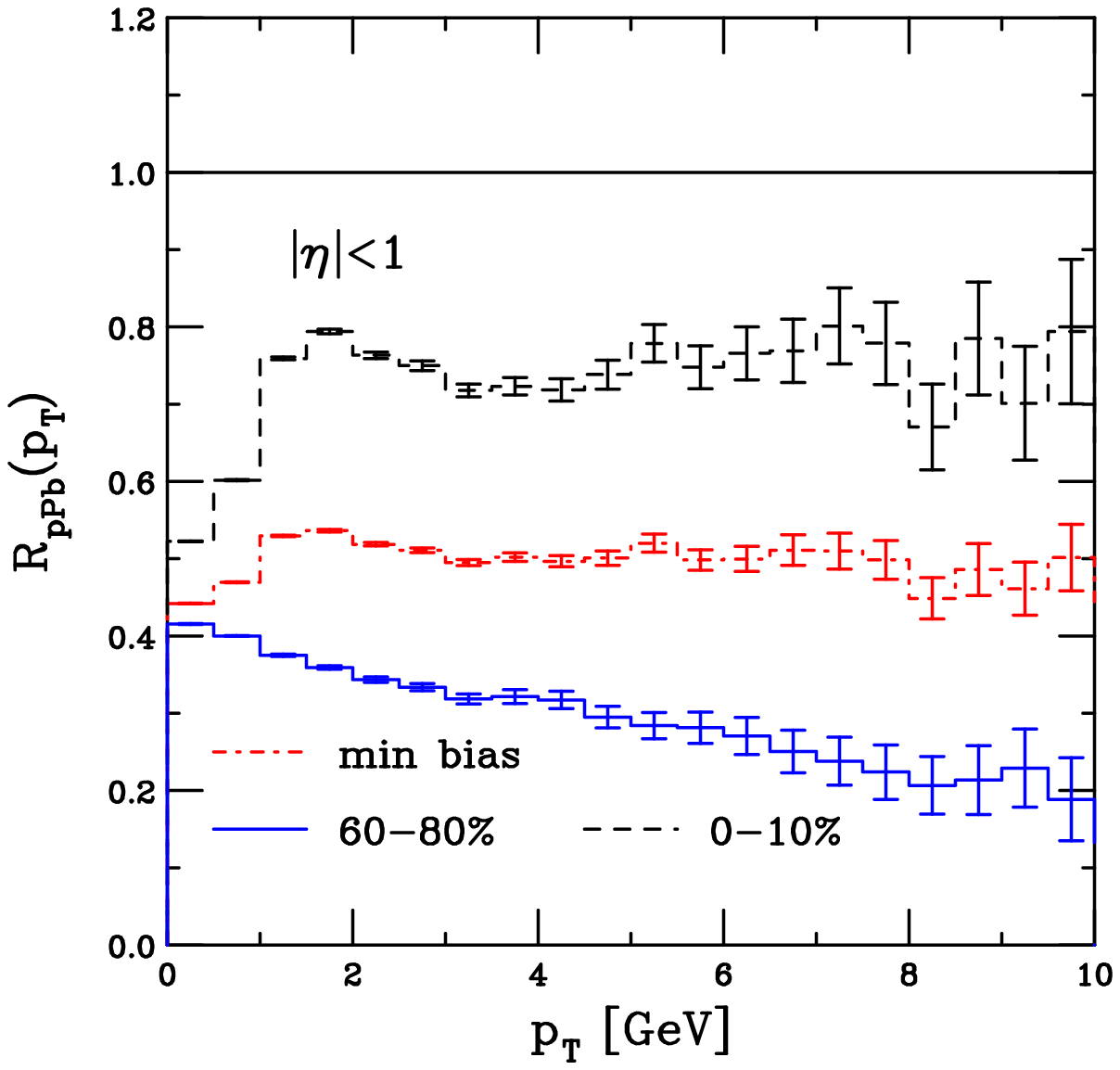}
\end{center}
\caption[]{(Left) The minimum bias 
results at $y=0$ for $\pi^0$, $p$ and $\overline p$ are shown with the
charged hadron ratio for $\eta = 0$.  All results are calculated with
$\mathtt{AMPT-def}$.
(Right) The charged hadron results at midrapidity, $|\eta|<1$, are shown for 
minimum bias collisions as well as in two centrality bins:
the 10\% most central and the 60-80\% most central.  All results are 
calculated employing $\mathtt{AMPT-def}$.}
\label{fig:AMPT-rpA}
\end{figure}

The left-hand side of Fig.~\ref{fig:AMPT-rpA} shows $R_{p {\rm Pb}}$ for protons,
antiprotons, neutral pions and charged particles.  The identified particles are
shown for $y=0$ while the charged particle ratio is for $\eta = 0$.
The ratios are all very similar except for protons which grows almost linearly
with $p_T$.

The right-hand side of Fig.~\ref{fig:AMPT-rpA} presents results for  the 
suppression factor for charged particles within $|\eta|<1$ in three centrality
bins: minimum bias; the 10\% most central collisions; and a peripheral bin,
the $60-80$\% most central collisions.  Interestingly, the ratio is smallest
and decreasing with $p_T$ for the more peripheral bin while the largest value
of $R_{p{\rm Pb}}$ is the most central bin.

\subsection[Centrality dependent $R_{\rm pPb}^{\pi^0}(p_T,y=0)$ with 
EPS09s nPDFs]{Centrality dependent $R_{p{\rm Pb}}^{\pi^0}(p_T,y=0)$ with 
EPS09s nPDFs (I. Helenius and K. J. Eskola)}
\label{sec:Eskola}

Predictions of the centrality dependence of $\pi^0$ production at midrapidty
are presented here.   Two spatially-dependent nuclear PDF (nPDF) sets, EPS09s 
and EKS98s were determined in Ref.~\cite{Helenius:2012wd}. The key component
is a power series ansatz of the nuclear thickness function ($T_A(\vec{s})$) 
for the spatial dependence of the nPDF modifications,
\begin{equation}
r_i^A(x,Q^2,\vec{s}) = 1 + \sum\limits_{j=1}^{n} 
c^i_j(x,Q^2)\left[T_A(\vec{s})\right]^j \, \, .
\label{eq:ta_series}
\end{equation}
The $A$ dependence of the earlier global nPDF fits EPS09 \cite{Eskola:2009uj} 
and EKS98 \cite{Eskola:1998df} was exploited to obtain the values of the 
$A$-independent coefficients $c^i_j(x,Q^2)$.  It was found that $n=4$ is 
sufficient for reproducing the $A$ systematics of the globally-fitted nPDFs.

Predictions are presented for the nuclear modification factor 
$R_{p {\rm Pb}}^{\pi^0}$ 
of inclusive neutral pion production at the LHC using the new EPS09s NLO nPDF 
set. As in Ref.~\cite{Helenius:2012wd}, the centrality classes are defined
in terms of impact parameter intervals.  For a given centrality class, 
$R_{p{\rm Pb}}^{\pi^0}$ is defined as 
\begin{equation}
R_{p{\rm Pb}}^{\pi^0}(p_T,y; b_1,b_2) \equiv \dfrac{\left\langle\dfrac{d^2 
N_{p {\rm Pb}}^{\pi^0}}{dp_T dy}\right\rangle_{b_1,b_2}}{ 
\dfrac{\langle N_{\rm coll}^{p {\rm Pb}} \rangle_{b_1,b_2}} {\sigma^{NN}_{\rm in}}
\dfrac{d^2\sigma_{pp}^{\pi^0}}{dp_T dy}} 
= \dfrac{\int_{b_1}^{b_2} d^2 \vec{b} \dfrac{d^2 N_{p {\rm Pb}}^{\pi^0}(\vec{b})}{dp_T 
dy} }{ \int_{b_1}^{b_2} d^2 \vec{b} \,T_{p {\rm Pb}}(\vec{b})
\dfrac{d^2\sigma_{pp}^{\pi^0}}{dp_T dy}},
\label{eq: R_pPb}
\end{equation}
where the impact parameter limits $b_1$ and $b_2$ are calculated from the 
optical Glauber model. The proton is assumed to be point-like, i.e. 
$T_{p {\rm Pb}}(\vec{b}) = T_{\rm Pb}(\vec{b})$. The minimum bias ratio
$R_{p {\rm Pb}}^{\pi^0}$ is obtained with $b_1 = 0$ and $b_2 \rightarrow \infty$. 
The cross sections are calculated in NLO using the $\mathtt{INCNLO}$ 
package\footnote{http://lapth.in2p3.fr/PHOX\_FAMILY/readme\_inc.html} 
\cite{Aversa:1988vb} with the CTEQ6M proton PDFs \cite{Pumplin:2002vw} along
with three different fragmentation functions: KKP \cite{Kniehl:2000fe}; 
AKK \cite{Albino:2008fy}; and fDSS \cite{deFlorian:2007aj}. The uncertainty 
band reflecting the nPDF uncertainties is calculated using the error sets of 
EPS09s with the fDSS fragmentation functions according to the prescription 
described in the original EPS09 paper \cite{Eskola:2009uj}.

\begin{figure}[htbp]
\centering
\hspace{-0.5cm}\includegraphics[width=0.495\textwidth]{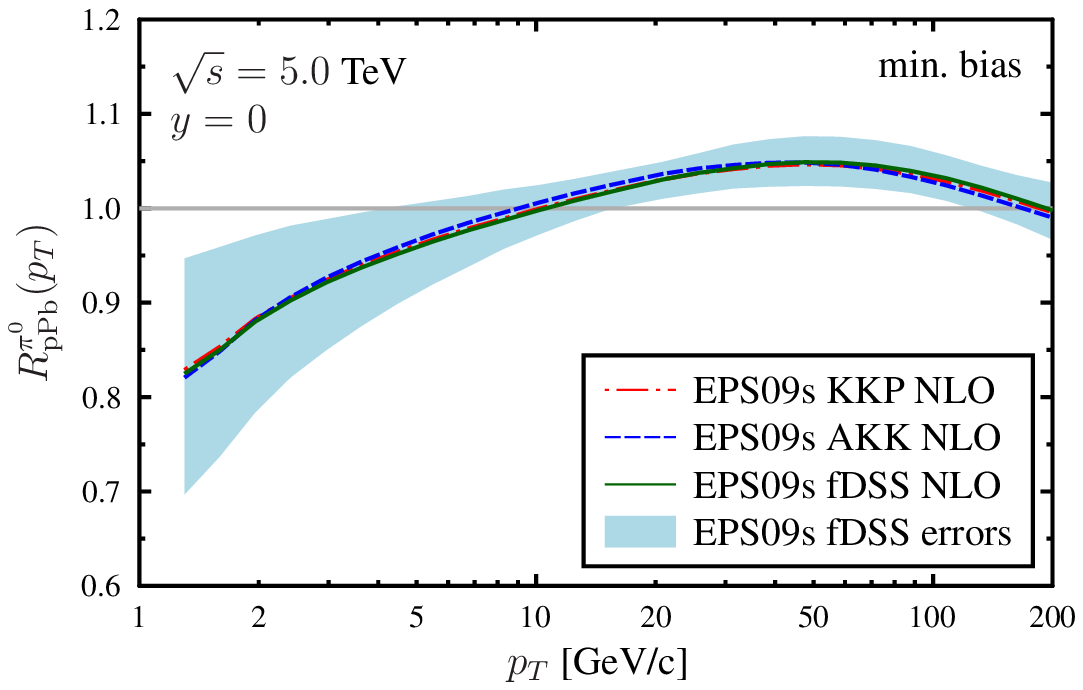}
\includegraphics[width=0.475\textwidth]{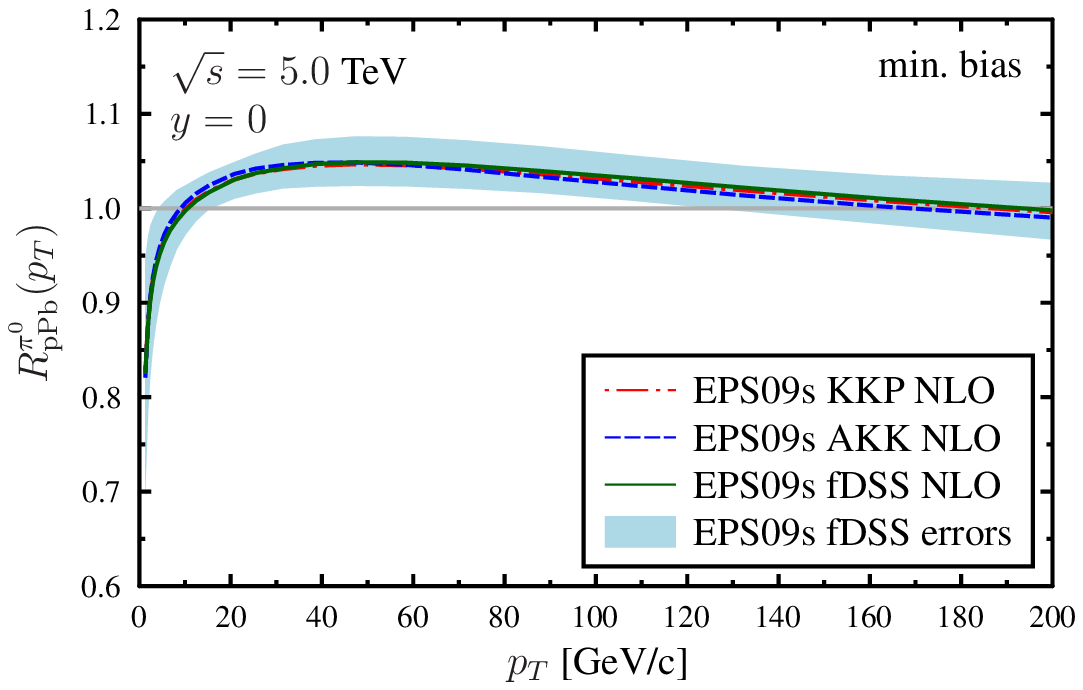}
\caption[]{The nuclear modification factor for inclusive $\pi^0$ production in 
$p+$Pb collisions at $y=0$ for minimum bias collisions calculated with the 
EPS09s NLO nPDFs, plotted on logarithmic (left) and linear (right) $p_T$ 
scales.  The blue error band is calculated employing the EPS09s error sets. All 
scales are fixed to the pion $p_T$. Based on 
Ref.~\protect\cite{Helenius:2012wd}.}
\label{fig:R_pPb_mb}
\end{figure}

In Fig.~\ref{fig:R_pPb_mb} the minimum bias $R_{p{\rm Pb}}^{\pi^0}$ is shown as a 
function of $p_T$ at midrapidity ($y = 0$) in the nucleon-nucleon center-of-mass
frame. Both logarithmic and linear $p_T$ scales are shown. In the region 
$p_T < 10 \text{ GeV}/c$ 
a suppression due to small-$x$ shadowing in the nPDFs is
observed while at $10 < p_T < 200 \text{ GeV}/c$ there is a small enchancement 
due to the antishadowing. Even though the different fragmentation functions 
may yield different absolute cross sections, these differences cancel in 
the ratio $R_{p{\rm Pb}}^{\pi^0}$.

\begin{figure}[htbp]
\centering
\includegraphics[width=0.995\textwidth]{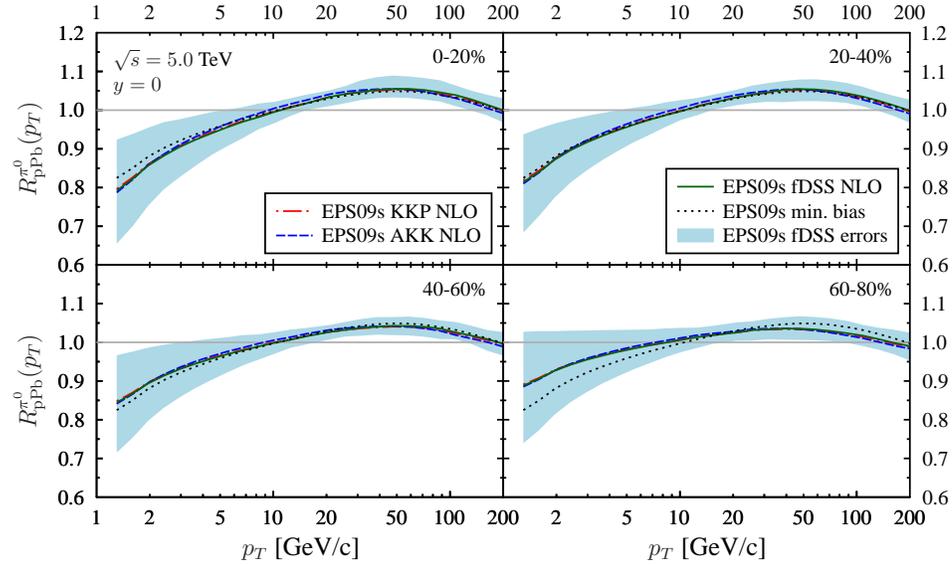}
\caption[]{The nuclear modification factor for inclusive $\pi^0$ production in 
$p+$Pb collisions at $y=0$ in four centrality classes calculated with the 
EPS09s NLO nPDFs. The blue error band is calculated employing the EPS09s error 
sets. All scales are fixed to the pion $p_T$. For comparison, the minimum bias 
result is also shown. Based on Ref.~\protect\cite{Helenius:2012wd}.}
\label{fig:R_pPb}
\end{figure}

In Fig.~\ref{fig:R_pPb} $R_{p {\rm Pb}}^{\pi^0}$ is presented in four different 
centrality classes, $(0-20)~\%$, $(20-40)~\%$, $(40-60)~\%$ and $(60-80)~\%$. 
The impact parameter limits for these centrality classes can be found in 
Ref.~\cite{Helenius:2012wd}. For comparison, the minimum bias results are
also shown.
Slightly larger nuclear effects are observed in the most central collisions 
than in the minimum bias collisions.  Nuclear effects in the most peripheral 
collisions are about a factor of two smaller than those in the most central 
collisions. 

\subsection[Cold matter effects with energy loss]{Cold matter effects with 
energy loss (Z.-B.Kang, I. Vitev and H. Xing)}
\label{sec:Vitev_pi0}

The predictions for $R_{p {\rm Pb}}^{\pi^0}$, like those for charged hadrons shown
in Fig.~\ref{lhc_h} are based on perturbative QCD factorization.  Cold nuclear
matter effects are implemented separately. 
The advantage of this approach is that all CNM effects have a clear physical 
origin, generally centered around multiple parton scattering. 
The implementation of these calculable CNM effects is well documented, see
Sec.~\ref{sec:Ivan-description}.  
Isospin effects, the Cronin effect, cold nuclear 
matter energy loss, and dynamical shadowing are all included. 

\begin{figure}[htpb]
\begin{center}
\includegraphics[width=0.5\textwidth]{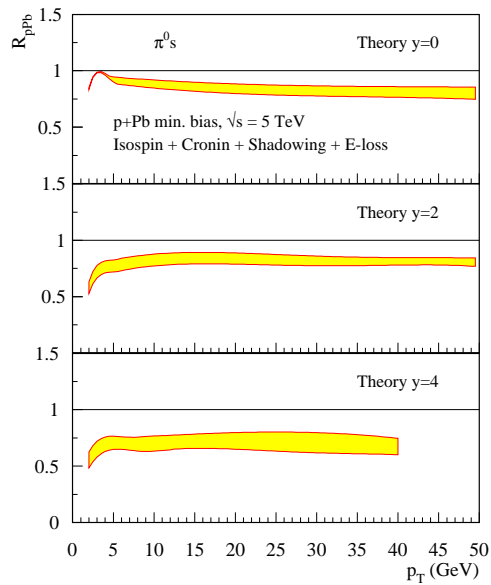}
\end{center}
\caption[]{Predictions for the nuclear modification factor 
$R_{p{\rm Pb}}^{\pi^0}(p_T)$ as a function in minimum bias collisions at
$\sqrt{s_{_{NN}}}=5$~TeV. Results for three rapidities: $y=0$ (top), $y=2$
(middle), and $y=4$ (bottom) are shown. The calculations have  been made
with the convention that the proton beam moves toward
forward rapidity, see Ref.~\protect\cite{Kang:2012kc}.}
\label{lhc_pi0}
\end{figure}

The neutral pion results in Fig.~\ref{lhc_pi0} are rather similar to those
shown in Fig.~\ref{lhc_h}.  The upper edges of the bands at $y=0$ (top), 
$y=2$ (middle), and $y=4$ (bottom) correspond to the RHIC scattering parameters 
while the lower edges correspond to potential enhancement of these parameters.

As seen by comparison with the EPS09 minimum bias results for $y=0$ in
Fig.~\ref{fig:R_pPb_mb}, 
the combined effects included here are stronger than with
shadowing alone.  Antishadowing in EPS09 produces a ratio larger than unity
for $10 < p_T < 200$ GeV/$c$ while the ratio is smaller than unity over the
entire range calculated here ($p_T < 50$ GeV/$c$).

\section[Quarkonium]{Quarkonium (R. Vogt)}
\label{sec:Vogt}

The predictions for the $J/\psi$ suppression factor, considering only shadowing
effects on the parton densities are described in this section.  
There are a number of possible cold matter effects on $J/\psi$ production,
including modifications of the parton densities in nuclei (shadowing); breakup
of the quarkonium state due to inelastic interactions with nucleons 
(absorption); and energy loss in cold matter.   Since the quarkonium
absorption cross section decreases with center-of-mass energy, we can expect
that shadowing is the most important cold matter effect at midrapidity,
see Refs.~\cite{Lourenco:2008sk,McGlinchey:2012bp}.  
Here we show results for the
rapidity and $p_T$ dependence of shadowing at $\sqrt{s_{_{NN}}} = 200$ GeV for
d+Au collisions at RHIC and the rapidity dependence at $\sqrt{s_{_{NN}}} = 5$ TeV
$p+$Pb collisions, neglecting absorption.

The results are
obtained in the color evaporation model (CEM) at next-to-leading order in the
total cross section.
In the CEM, the quarkonium 
production cross section is some fraction, $F_C$, of 
all $Q \overline Q$ pairs below the $H \overline H$ threshold where $H$ is
the lowest mass heavy-flavor hadron,
\begin{eqnarray}
\sigma_C^{\rm CEM}(s)  =  F_C \sum_{i,j} 
\int_{4m^2}^{4m_H^2} ds
\int dx_1 \, dx_2~ f_i^p(x_1,\mu_F^2)~ f_j^p(x_2,\mu_F^2)~ 
\hat\sigma_{ij}(\hat{s},\mu_F^2, \mu_R^2) \, 
\, , \label{sigtil}
\end{eqnarray} 
where $ij = q \overline q$ or $gg$ and $\hat\sigma_{ij}(\hat s)$ is the
$ij\rightarrow Q\overline Q$ subprocess cross section.    
The normalization factor $F_C$ is fit 
to the forward (integrated over $x_F > 0$) 
$J/\psi$ cross section data on only $p$, Be, Li,
C, and Si targets.  In this way, uncertainties due to 
ignoring any cold nuclear matter effects which are on the order of a few percent
in light targets are avoided.  The fits are restricted to the forward cross 
sections only.

The same values of the central charm quark
mass and scale parameters are employed as those 
found for open charm, $m = 1.27 \pm 0.09$~GeV/$c^2$,
$\mu_F/m = 2.10 ^{+2.55}_{-0.85}$, and $\mu_R/m = 1.60 ^{+0.11}_{-0.12}$ 
\cite{Nelson:2012bc}.
The normalization $F_C$ is obtained for the central set,
$(m,\mu_F/m, \mu_R/m) = (1.27 \, {\rm GeV}, 2.1,1.6)$.  
The calculations for the extent of the mass and scale uncertainties are
multiplied by the same value of $F_C$ to
obtain the extent of the $J/\psi$ uncertainty band \cite{Nelson:2012bc}.
The results shown here are not the same as those calculated at leading order 
\cite{Vogt:2010aa} because the LO and NLO gluon shadowing parameterizations 
differ significantly at low $x$ \cite{Eskola:2009uj}.

\begin{figure}[htpb]
\begin{center}
\begin{tabular}{cc}
\includegraphics[width=0.495\textwidth]{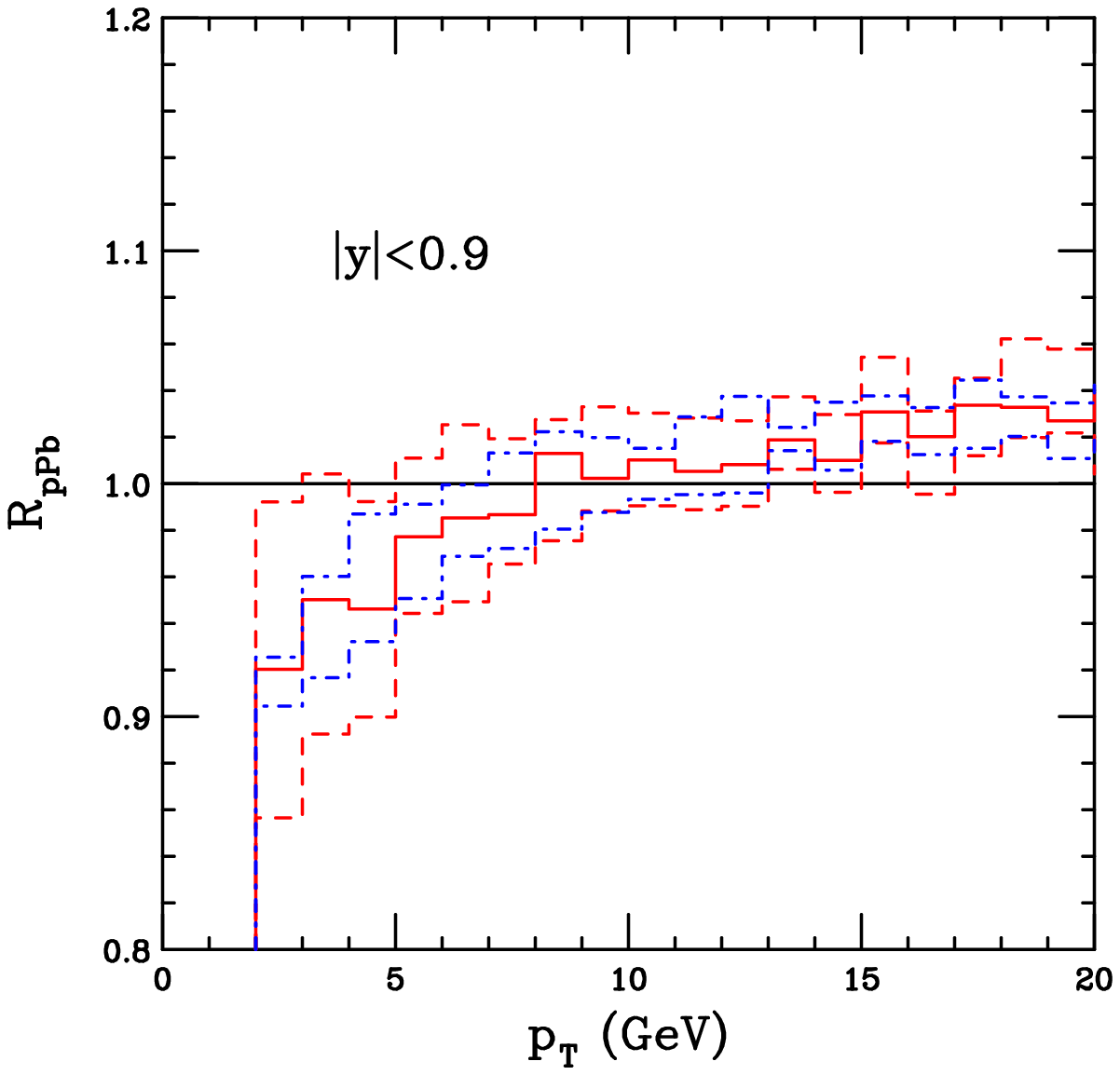} &
\includegraphics[width=0.495\textwidth]{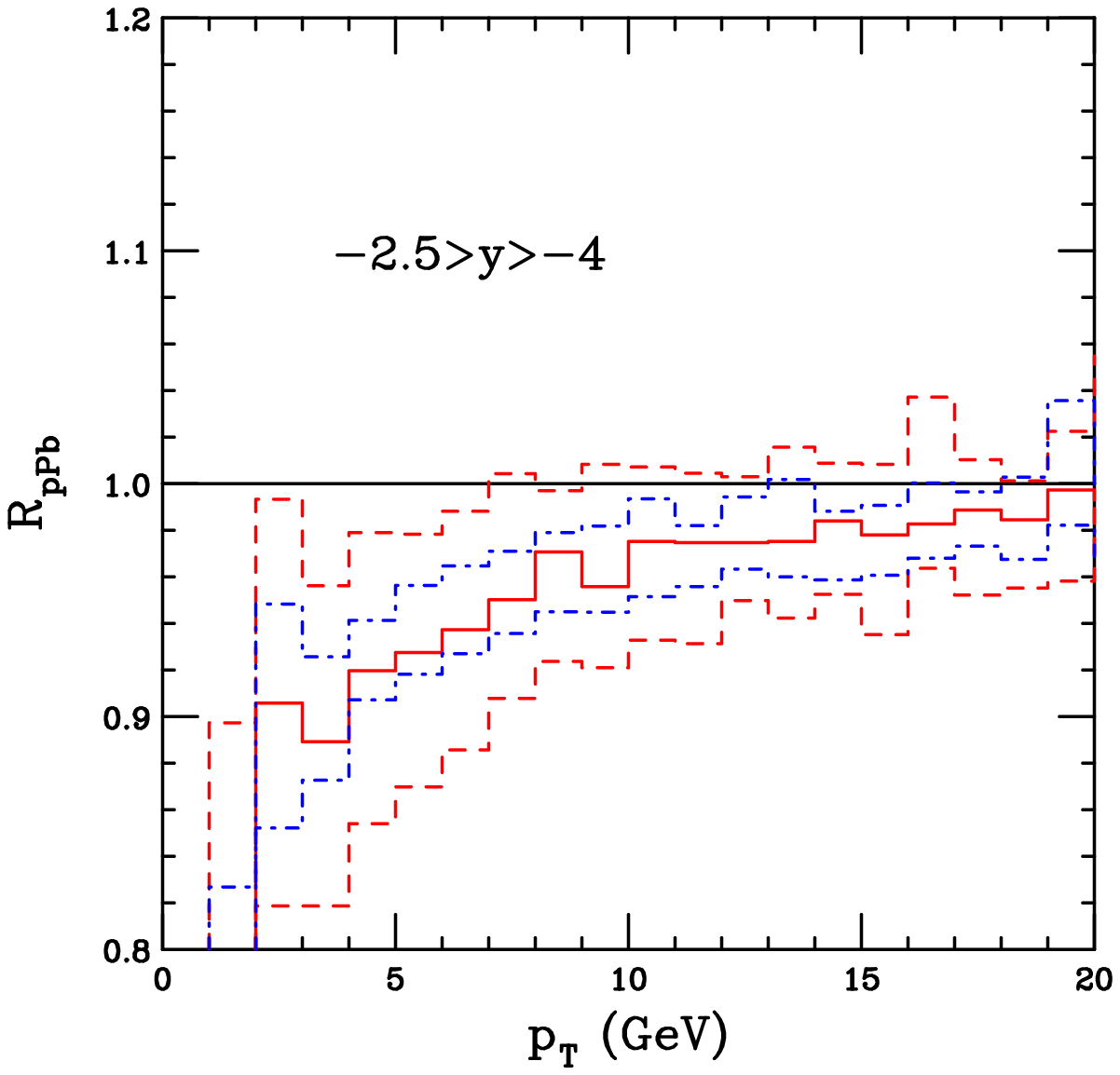} \\
\includegraphics[width=0.495\textwidth]{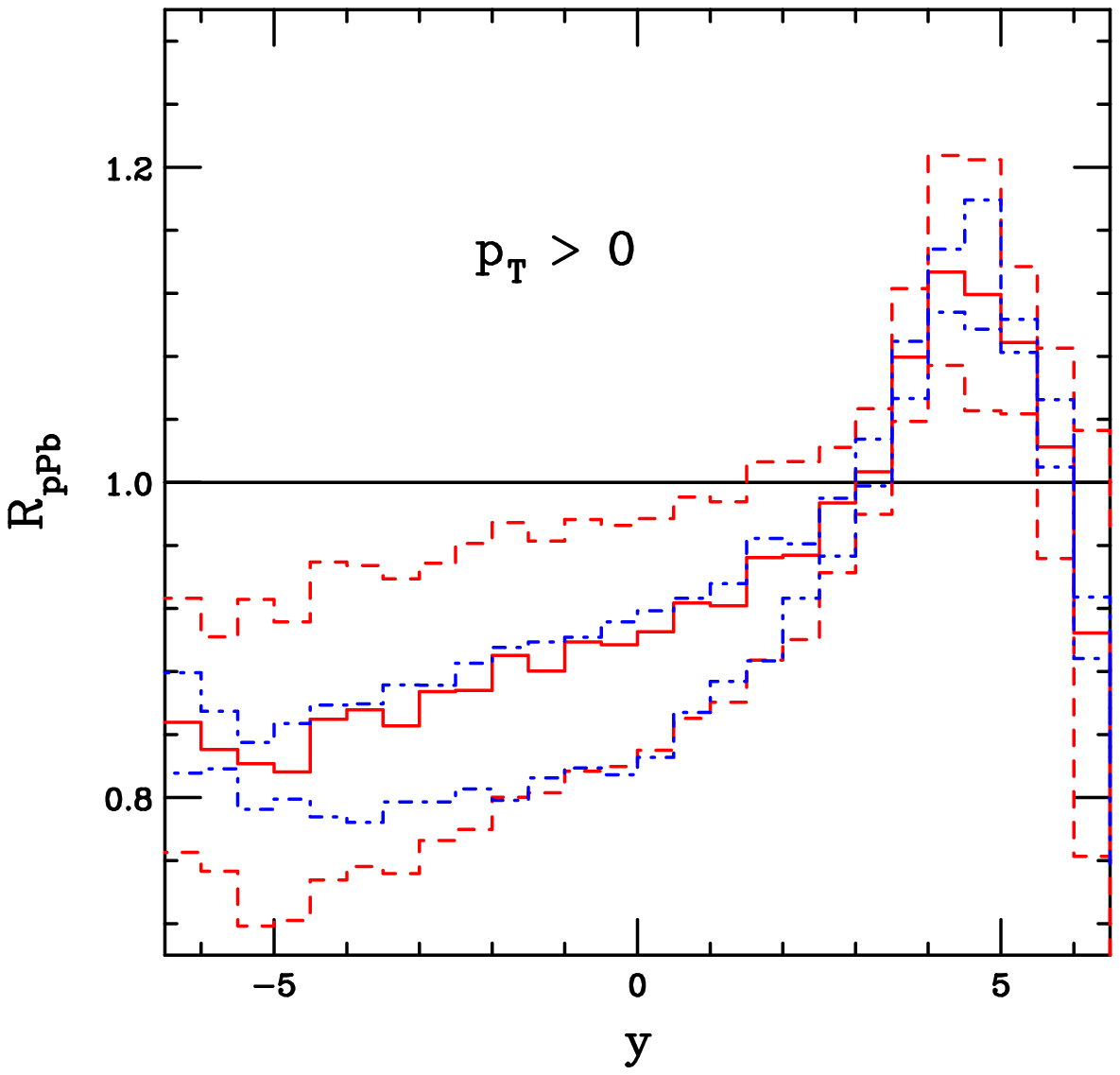} &
\includegraphics[width=0.495\textwidth]{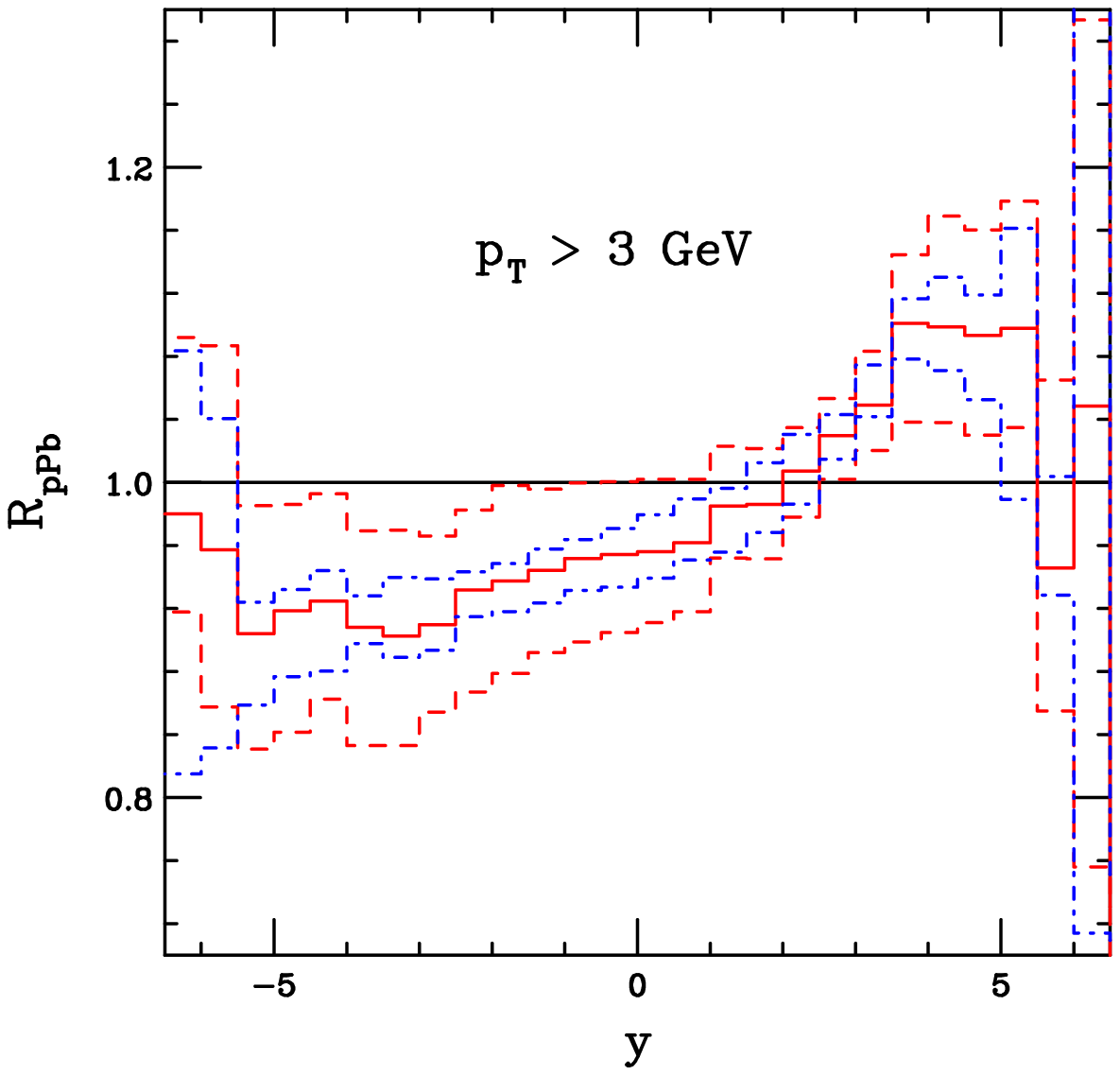} 
\end{tabular}
\end{center}
\caption[]{
The ratio $R_{\rm dAu}$ (left) and $R_{\rm AuAu}$ (right) at $\sqrt{s} = 5$ TeV.
The dashed red histogram shows the EPS09 uncertainties while the dot-dashed
blue histogram shows the dependence on mass and scale.  The $p+p$ denominator
is also calculated at 5 TeV and does not include any rapidity shift in $p+$Pb
collisions.  For a discussion about normalizing the results to $p+p$
collisions at different energies, see Ref.~\protect\cite{Vogt:2010aa}.}
\label{fig:Jpsi}
\end{figure}

Figure~\ref{fig:Jpsi} shows the uncertainty
in the shadowing effect due to uncertainties in the EPS09 shadowing 
parameterization \cite{Eskola:2009uj}
(red) as well as those due to the mass and scale uncertainties
obtained in the fit to the total charm cross section (blue) calculated with
the EPS09 central set.  All the 
calculations are NLO in the total cross section and assume that the intrinsic
$k_T$ broadening is the same in $p+p$ as in $p+$Pb.
Note that the rapidity-dependent
ratios are adjusted so that the lead nucleus moves toward negative rapidity and
the small $x$ region of the nucleus is at large negative rapidity, as is the
case for pseudorapidity distributions and the 
rapidity-dependent ratios shown previously.

The mass and scale uncertainties are calculated 
based on results using the one standard deviation uncertainties on
the quark mass and scale parameters.  If the central, upper and lower limits
of $\mu_{R,F}/m$ are denoted as $C$, $H$, and $L$ respectively, then the seven
sets corresponding to the scale uncertainty are  $\{(\mu_F/m,\mu_F/m)\}$ =
\{$(C,C)$, $(H,H)$, $(L,L)$, $(C,L)$, $(L,C)$, $(C,H)$, $(H,C)$\}.    
The uncertainty band can be obtained for the best fit sets by
adding the uncertainties from the mass and scale variations in 
quadrature. The envelope containing the resulting curves,
\begin{eqnarray}
\sigma_{\rm max} & = & \sigma_{\rm cent}
+ \sqrt{(\sigma_{\mu ,{\rm max}} - \sigma_{\rm cent})^2
+ (\sigma_{m, {\rm max}} - \sigma_{\rm cent})^2} \, \, , \label{sigmax} \\
\sigma_{\rm min} & = & \sigma_{\rm cent}
- \sqrt{(\sigma_{\mu ,{\rm min}} - \sigma_{\rm cent})^2
+ (\sigma_{m, {\rm min}} - \sigma_{\rm cent})^2} \, \, , \label{sigmin}
\end{eqnarray}
defines the uncertainty. 
The EPS09 band is obtained by calculating the deviations from the central value
for the 15 parameter variations on either side of the central set and adding
them in quadrature.  With the new uncertainties on the charm cross section,
the band obtained with the mass and scale variation is narrower than that with
the EPS09 variations.

\section{Photons}
\label{sec:photons}

Photons are ideal probes in heavy ion collisions due to their lack of 
final-state interactions with either a quark-gluon plasma or a hot hadron gas. 
Baseline measurements in $p+p$ and $p+$Pb collisions are very important 
to subtract photons from hard initial parton scatterings which are unrelated
to QGP and, in particular, to determine cold nuclear matter effects on
photon production. Measured
nuclear modification factors for photons at high transverse momenta ($p_T >5$ 
GeV/$c$) measured in Pb+Pb collisions at the LHC and Au+Au collisions at RHIC, 
$R_{AA} \approx 1$, have demonstrated that hard processes scale with the
number of binary nucleon-nucleon collisions in nucleus-nucleus collisions. 
An especially important question for heavy-ion physics at the LHC is the 
magnitude of low-$x$ effects on the parton densities, particularly for
gluons. The presence of
shadowing, {\it i.e.}\ suppression of the number of low-$x$ partons, has a
significant effect on benchmarking hard processes at intermediate $p_T$. 
Direct photons and other electromagnetic probes are in particular needed for
a quantitatively precise determination of the nuclear parton distributions.

This section includes an NLO calculation of direct photon production in $p+p$
and $p+$Pb collisions; LO calculations of cold matter effects on photon
production at several rapidities; saturation effects on direct and inclusive
prompt photon production and photon-hadron correlations; and a discussion
on gluon saturation and shadowing in dilepton and photon production.

\subsection[Direct photon cross sections]{Direct photon cross sections (R. J. Fries and S. De)}
\label{sec:Fries}

Predictions of the $p_T$ and $y$ 
dependence of direct photon production are given here.
The impact-parameter averaged EPS09 shadowing parameterizations 
\cite{Eskola:2009uj} are used with the CTEQ6.6 parton densities
\cite{Nadolsky:2008zw} to calculate results in minimum bias $p+$Pb
collisions.  These calculations can thus
be employed to check the validity of the EPS09 modifications. The per
nucleon cross sections given here can be directly compared to the minimum
bias $p+$Pb data by converting the cross sections to differential
yields scaling the results by $\sigma_{\rm in} = 67$ mb and the average number
of $NN$ collisions in a given centrality bin.

These NLO in $\alpha_s$ calculations are performed with $\mathtt{JETPHOX 1.3.1}$
\cite{Catani:2002ny,Aurenche:2006vj}.  Hard prompt photons and fragmentation 
photons (obtained with the BFG-II fragmentation functions \cite{Bourhis:1997yu})
are both included.  An isolation cut of
$E_{T,\mathrm{hadron}} < 5$ GeV on hadronic energy is imposed within a $R=0.4$ 
isolation cone around the photon, similar to the CMS analysis method 
for $p+p$ collisions at 7 TeV. 

\begin{figure}[htpb]
\begin{center}
\includegraphics[width=0.5\textwidth]{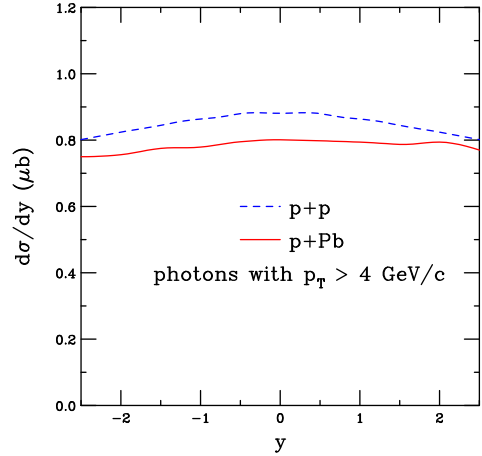}
\end{center}
\caption[]{The direct photon cross section, $d\sigma/dy$, as a function of 
rapidity in the laboratory frame for photons with $p_T >4$~GeV/$c$ and 
an isolation cut $E_{T,\mathrm{hadron}} < 5$ GeV. Both $p+p$ and
$p+$Pb results are shown. The $p+$Pb results are normalized to the per
nucleon cross section.
} 
\label{fig:gammay}
\end{figure}

Figure~\ref{fig:gammay} shows the rapidity dependence, $d\sigma/dy$, of
direct photon production in $p+p$ and $p+$Pb collisions, normalized per nucleon.
A $p_T$ cut, $p_T > 4$~GeV/$c$ is imposed.  The results are presented in
the lab frame.  Recall that there is a difference, $\Delta y = 0.465$, between
the center-of-mass and lab frames at the current energy. 

\begin{figure}[htpb]
\includegraphics[width=0.495\textwidth]{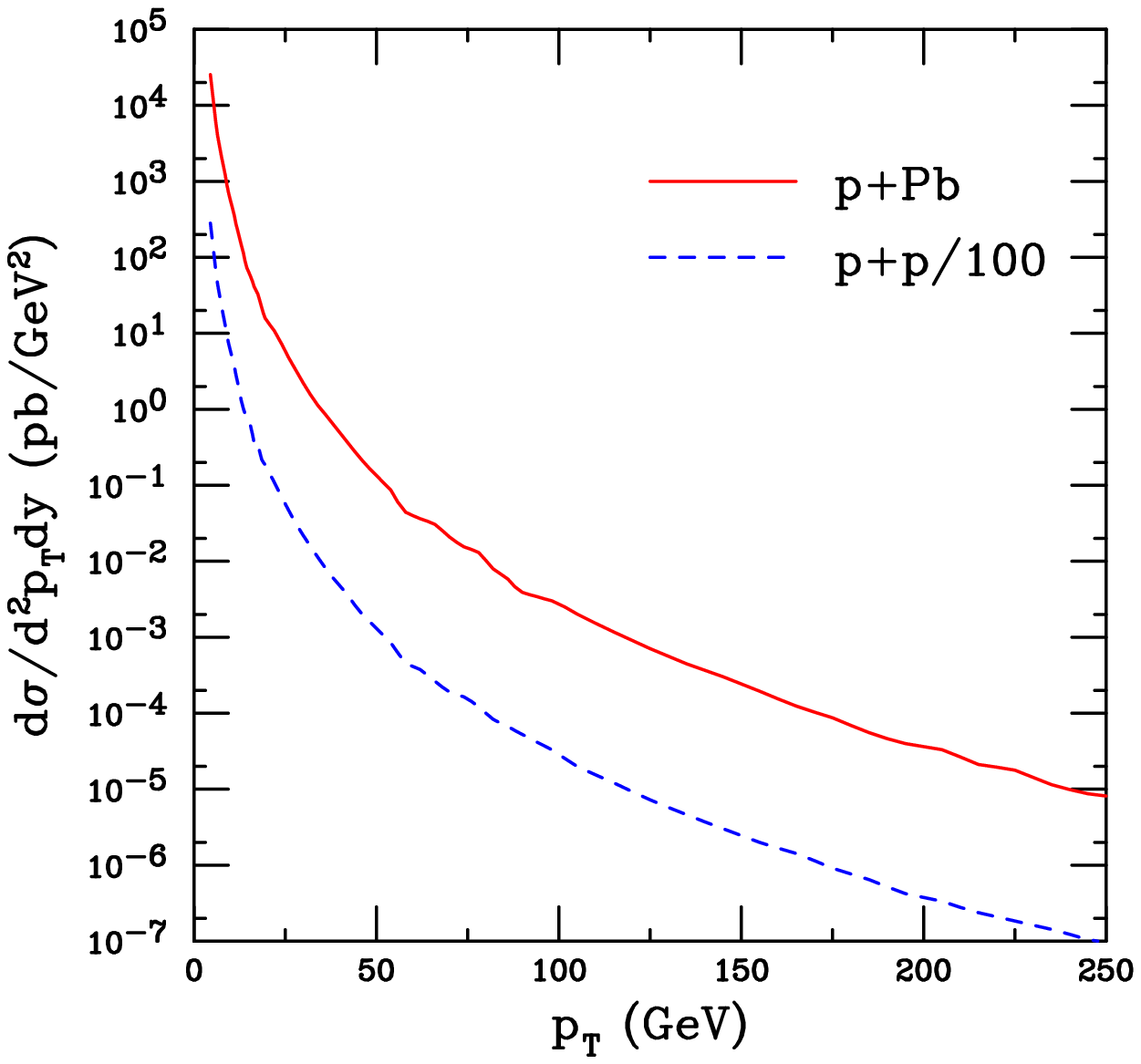}
\includegraphics[width=0.495\textwidth]{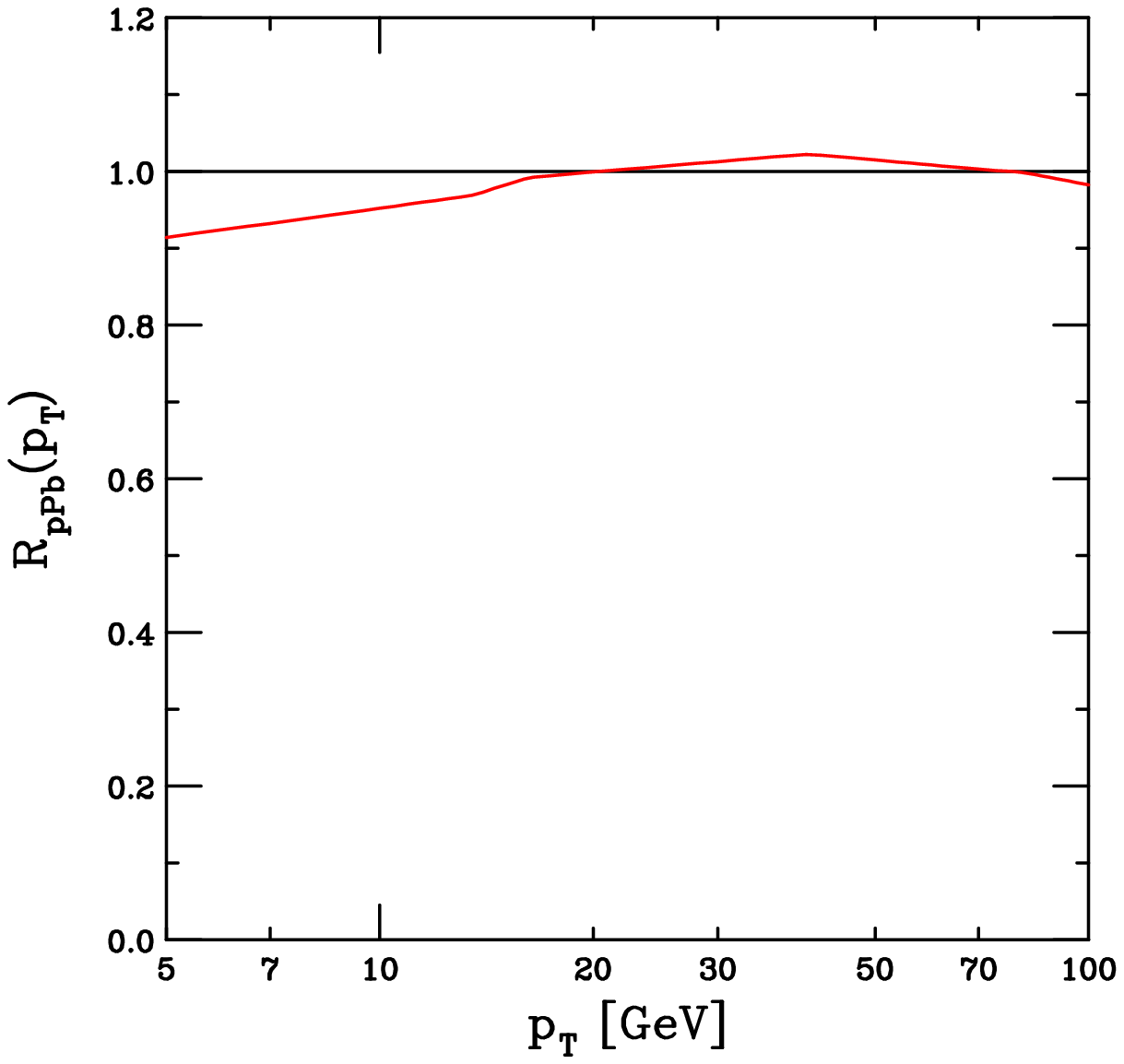}
\caption[]{(Left) The direct photon $p_T$ distribution at $y=0$ in the lab frame. 
The $p+p$ distribution is scaled down by two orders of magnitude.
(Right)  The corresponding modification factor $R_{p {\rm Pb}}(p_T)$.
Note the logarithmic $p_T$ scale. 
  \label{fig:gammapt}}
\end{figure}

On the left-hand side of Fig.~\ref{fig:gammapt}, the $p_T$ spectrum at $y=0$, 
$d\sigma/d^2p_Tdy$, is shown for $p+p$ and $p+$Pb collisions. 
The $p+p$ result is scaled down by a factor of 100 to separate the two
curves.  The corresponding nuclear modification factor is shown on the
right-hand side.  Note
that a logarithmic scale is used on the $x$-axis here to highlight the 
modification of the low $p_T$ part of the photon spectrum.

\subsection[Cold matter effects on photon production]{Cold matter effects on photon production (Z.-B. Kang, I. Vitev and H. Xing)}
\label{sec:Vitevphotons}

Prompt photon production in $p+p$ collisions has two 
components, the direct and fragmentation contributions~\cite{Owens:1986mp}:
\ben
\frac{d\sigma}{dy d^2p_T} = \frac{d\sigma_{\rm dir}}{dy d^2p_T} +
\frac{d\sigma_{\rm frag}}{dy d^2p_T}.
\een
The fragmentation contribution is given by:
\ben
\frac{d\sigma_{\rm frag}}{dy d^2p_T}&=&K \frac{\alpha_s^2}{s}\sum_{a,b,c}
\int \frac{dx_1}{x_1}d^2k_{T_1} \, 
f_{a/N}(x_1, k_{T_1}^2)\int \frac{dx_2}{x_2}d^2k_{T_2} \, 
f_{b/N}(x_2, k_{T_2}^2) 
\nnu
&&\times
\int \frac{dz_c}{z_c^2} \, D_{\gamma/c}(z_c)H_{ab\to c}(\hat s,\hat t,
\hat u)\delta(\hat s+\hat t+\hat u).
\label{frag}
\een
The expression is exactly the same as Eq.~(\ref{light}) if the 
parton-to-hadron fragmentation function,
$D_{h/c}(z_c)$, is replaced by the parton-to-photon fragmentation function, 
$D_{\gamma/c}(z_c)$. The direct 
contribution can be written as:
\ben
\frac{d\sigma_{\rm dir}}{dy d^2p_T} &=&
K\frac{\alpha_{\rm em} \alpha_s}{s}\sum_{a,b} 
\int \frac{dx_1}{x_1} d^2k_{T_1} f_{a/N}(x_1, k_{T_1}^2) \, 
\int \frac{dx_2}{x_2}d^2k_{T_2} f_{b/N}(x_2, k_{T_2}^2) \nnu &&\times
H_{ab\to \gamma}(\hat s,\hat t,\hat u)\delta\left(\hat s+\hat t+\hat u\right),
\label{dir}
\een
where $H_{ab\to \gamma}$ are the partonic hard-scattering functions for 
direct photon production~\cite{Owens:1986mp,Xing:2012ii}.  

In $p+p$ collisions, $\langle k_T^2\rangle_{pp}=1.8$ GeV$^2/c^2$, along with a $K$
factor of $\mathcal{O}(1)$, which gives a good description of production at
RHIC and LHC energies. 
The CTEQ6L1 PDFs \cite{Pumplin:2002vw} are employed with the GRV 
parametrization for parton-to-photon fragmentation 
functions~\cite{Gluck:1992zx}. The factorization and renormalization 
scales are fixed 
to the transverse momentum of the produced photon, $\mu_F = \mu_R = p_T$. 

\begin{figure}[htpb]
\begin{center}
\includegraphics[width=0.5\textwidth]{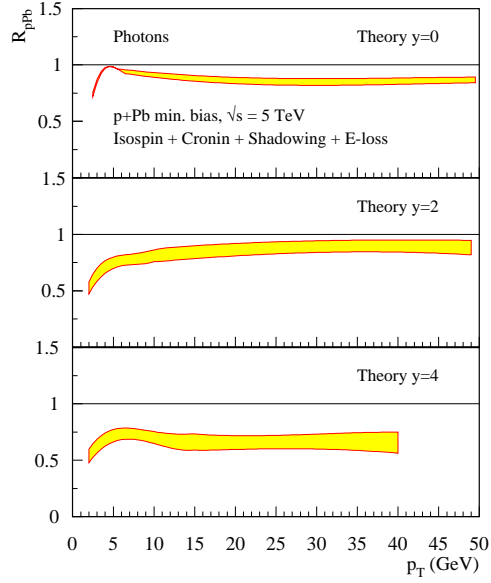}
\end{center}
\caption[]{Predictions of the nuclear modification factor $R_{p {\rm Pb}}$ as a 
function of $p_T$ for prompt photon production in minimum 
bias $p+$Pb collisions for $y=0$ (top), $y=2$ (middle), 
and $y=4$ (bottom), see Ref.~\protect\cite{Kang:2012kc}.}
\label{lhc_phot}
\end{figure}

The results shown here are calculated employing the same cold matter
effects described
in Sec.~\ref{sec:Ivan-description}.
Figure~\ref{lhc_phot} presents predictions for the nuclear
modification factor in prompt photon production as a function of $p_T$ for 
$y=0$ (top), $y=2$ (middle), and $y=4$ (bottom). 
The upper edge of the bands corresponds to the RHIC scattering parameters 
The lower edge represents a high-energy enhancement of the parameters.
The 
behavior of $R_{pPb}$ for direct photons is qualitatively the same
as the $\pi^0$s and charged hadron results shown earlier.

\subsection[rcBK calculation of photon production]{rcBK calculation of 
photon production (A. Rezaeian)}
\label{sec:Amirphotons}

The cross section for semi-inclusive prompt photon-quark production in $p+A$ 
collisions at leading twist in the CGC formalism is 
\cite{Gelis:2002ki,Baier:2004ti},   
\bea
&&{d\sigma^{q\, A \rightarrow q(l)\,\gamma(p^\gamma)\, X}
\over d^2\vec{b_T}\, d^2\vec{p_T}^{\gamma}\, d^2\vec{l_T}\, d\eta_{\gamma}\, 
d\eta_h} = {Ke_q^2\, \alpha_{\rm em} \over \sqrt{2}(4\pi^4)} \, 
{p^-\over  (p_T^\gamma)^ 2 \sqrt{s}} \, {1 + (l^-/ k^-)^2 \over
[p^- \, \vec{l_T} - l^- \vec{p_T}^\gamma]^2}
\nonumber \\
&& \times \delta \bigg[x_q - {l_T \over \sqrt{s}} e^{\eta_h} - 
{p_T^\gamma \over \sqrt{s}} e^{\eta_\gamma} \bigg] 
\, [ 2 l^- p^-\, \vec{l_T} \cdot \vec{p_T}^\gamma + 
p^- (k^- -p^-)\, l_T^2 + l^- (k^- -l^-)\,(p_T^\gamma)^2 ]  \nonumber \\ && \times
N_F (|\vec{l_T} + \vec{p_T}^\gamma|,  x_g) \, ,
\label{cs}
\eea
where $p^\gamma$, $l$, and $k$ are the $4$-momenta of the produced prompt 
photon, the outgoing $q$ ($\overline q$) and the incident $q$ ($\overline q$), 
respectively. 
A $K$ factor was introduced to absorb higher-order corrections. 
The light-cone fraction $x_q$ is the ratio of the incoming quark and proton 
energies, $x_q =k^-/\sqrt{s/2}$. The pseudorapidities of the
outgoing prompt photon, $\eta_\gamma$, and quark, $\eta_h$, are defined as 
$p^-= (p_T^\gamma /\sqrt{2}) e^{\eta_{\gamma}}$
and $l^-= (l_T/\sqrt{2}) e^{\eta_h}$. The angle between the final-state 
quark and the prompt photon, $\Delta \phi$, is defined as $\cos(\Delta \phi) 
\equiv (\vec{l}_T \cdot \vec{p}_T^\gamma)/(l_T p_T^\gamma)$.  Only high 
$p_T$ light hadron production is considered here.  Therefore rapidity and 
pseudorapidity are equivalent.  The semi-inclusive photon-hadron production 
cross section in 
proton-nucleus collisions can be obtained by convoluting the partonic cross 
section, \eq{cs}, with the quark distribution functions of the proton 
and the quark-hadron fragmentation function, 
\begin{eqnarray*}\label{qh-f}
\lefteqn{\frac{d\sigma^{p\, A \rightarrow h (p^h)\, \gamma (p^\gamma)\, X}}{d^2\vec{b_T} \, 
d^2\vec{p_T}^\gamma\, d^2\vec{p_T}^h \, d\eta_{\gamma}\, d\eta_{h}}
= } \\ && \int^1_{z_{f}^{\rm min}} \frac{dz_f}{z_f^2} \, \int\, dx_q\,
f _q(x_q,Q^2)  \frac{d\sigma^{q\, A \rightarrow q(l)\,\gamma(p^\gamma)\, X}}
{ d^2\vec{b_T}\, d^2\vec{p_T}^{\gamma}\, d^2\vec{l_T}\, d\eta_{\gamma}\, d\eta_h}  
D_{h/q}(z_f,Q^2) \, ,
\end{eqnarray*}
where $p^h_T$ is the transverse momentum of the produced hadron. 
The sum over quark and antiquark flavors in Eq.~(\ref{qh-f}) is implied.
The light-cone momentum fractions $x_q$, $x_{\bar q}$, and $x_g$ in 
Eqs.~(\ref{cs}) and (\ref{qh-f}) are related to the transverse momenta and
rapidities of the produced hadron and prompt photon \cite{JalilianMarian:2012bd},
\[ \begin{array}{ll}
\normalsize
x_q = x_{\bar{q}} = \frac{1}{\sqrt{s}}\left(p_T^\gamma\, e^{\eta_{\gamma}}+
\frac{p_T^h}{z_f}\, e^{\eta_h}\right) \, \, , &
x_g = \frac{1}{\sqrt{s}}\left(p_T^\gamma\, e^{-\eta_{\gamma}} + 
\frac{p_T^h}{z_f}\, e^{-\eta_{h}}\right) \, \, , \\ 
z_f = \frac{p_T^h}{l_T} \, \, , &
z_{f}^{\rm min}=\frac{p_T^h}{\sqrt{s}}
\left(\frac{e^{\eta_h}}
{1 - (p_T^\gamma/ \sqrt{s}) \, e^{\eta_\gamma}}\, 
\right) \, \, . 
\end{array} \]

The single inclusive prompt photon cross section in the CGC framework can be 
obtained from \eq{cs} by integrating over the momenta of the final state quark
or antiquark. 
The corresponding cross section can be then divided into two contributions: 
fragmentation (first term) and direct (second term) photons \cite{JalilianMarian:2012bd},
\begin{eqnarray*}\label{pho4}
\lefteqn{\frac{d\sigma^{p\, A \rightarrow \gamma (p^\gamma) \, X}}{d^2\vec{b_T}
d^2\vec{p_T}^\gamma d\eta_{\gamma}}
= \frac{K}{(2\pi)^2} \Big[\int_{x_q^{\rm min}}^1 d x_q f_q(x_q, Q^2)\frac{1}{z}\, 
N_F(x_g,p_T^\gamma/z)D_{\gamma/q}(z, Q^2)  } \\
&& + \frac{e_q^2 \alpha_{\rm em}}{2\pi^2(p_T^\gamma)^4}\int_{x_q^{\rm min}}^1 
d x_q f_q(x_q, Q^2)z^2[1+(1 - z)^2] \int_{l_T^2<Q^2}d^2\vec{l_T}\,l_T^2\, 
N_F(\bar{x}_g,l_T)\Big], \
\end{eqnarray*}
where 
$D_{\gamma/q}(z,Q^2)$ is the leading order quark-photon fragmentation function 
\cite{Owens:1986mp,Bourhis:1997yu,Gluck:1992zx}. Similar to the hybrid formalism for 
inclusive hadron production, 
\eq{final}, $Q$ is a hard scale. Although the cross sections given in 
\eq{final} and \eq{pho4} describe different final-state particle production, 
they are strikingly similar. The light-cone momentum fractions 
$x_g$, $\bar{x}_g$, and $z$ above are related to the transverse momentum and
rapidity of the prompt photon \cite{JalilianMarian:2012bd}, 
\[ \begin{array}{ll}
\normalsize
x_g =  x_q \, e^{-2\, \eta_\gamma} \, \, , &
\bar{x}_g = \frac{1}{x_q\, s} \left[{(p_T^\gamma)^2\over z} + 
\frac{(l_T-p_T^\gamma)^2}{1-z}\right] \, \, , \\ 
x_q^{\rm min}=\frac{p_T^\gamma}{\sqrt{s}}e^{\eta_{\gamma}} \, \, , &
z = \frac{p_T^\gamma}{x_q\, \sqrt{s}}e^{\eta_{\gamma}} \, \, .
\end{array} \]
The expression in \eq{pho4}  was obtained using a hard cutoff to subtract the 
collinear singularity \cite{JalilianMarian:2012bd}. The use of a cutoff may 
result in a mismatch 
between the finite corrections to \eq{pho4} and those included in 
parameterizations of the photon fragmentation function. However, this mismatch 
is higher-order in the coupling constant and its proper treatment requires 
a full NLO calculation.
 
In Eqs.~(\ref{qh-f}) and (\ref{pho4}), the factorization scale 
$\mu_F$ is assumed to be the same in the fragmentation functions and the parton 
densities. In order to investigate the uncertainties associated with choice 
of $\mu_F$, several values of $\mu_F$ are considered: $\mu_F=2p^\gamma_T$; 
$p^\gamma_T$; and $p^\gamma_T/2$.
The amplitude $N_F$ in Eqs.~(\ref{cs}) and (\ref{pho4}) is defined in 
Eq.~(\ref{ff}).

\begin{figure}[htpb]       
\begin{center}
\includegraphics[width=0.49\textwidth] {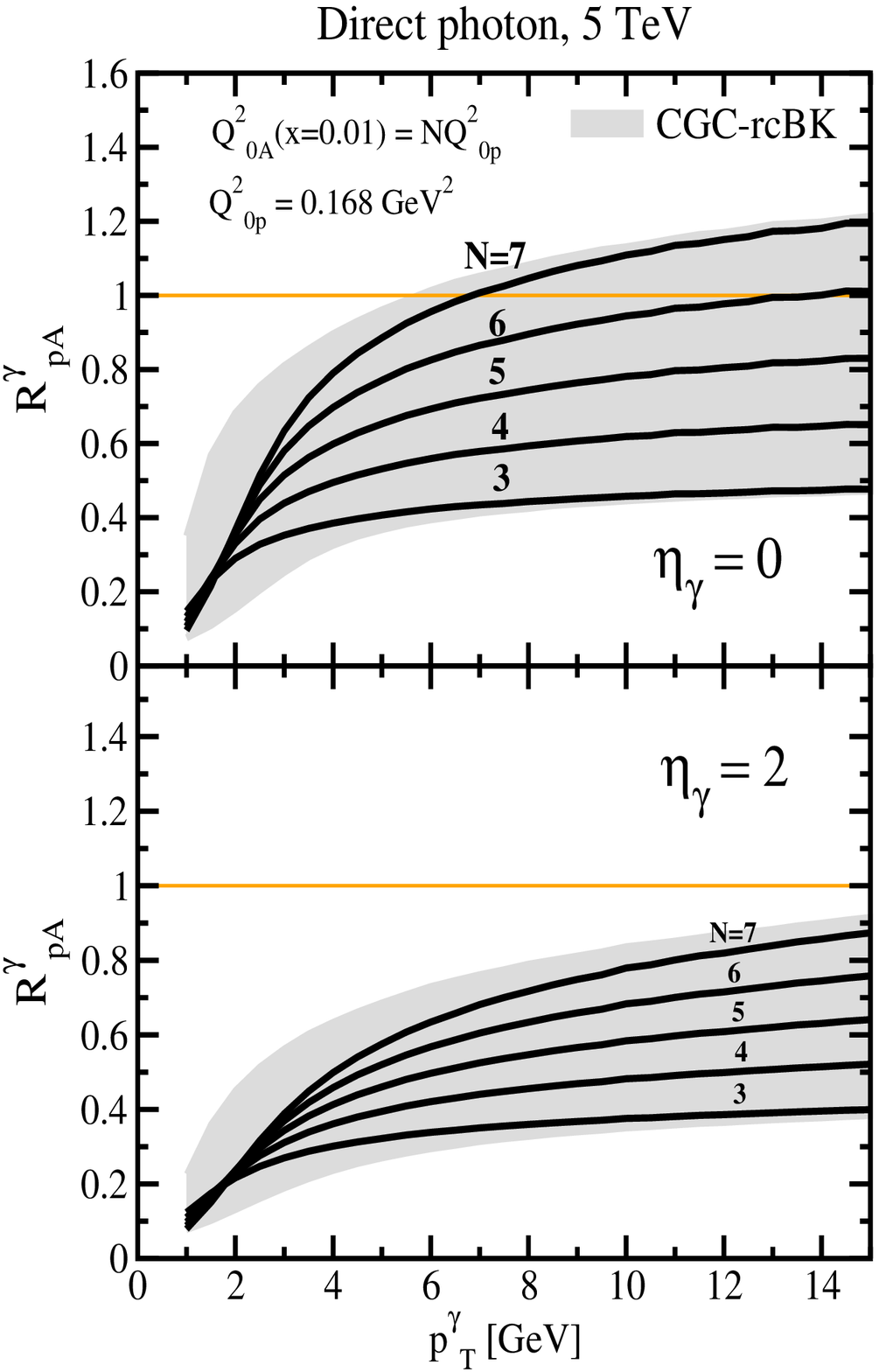}       
\includegraphics[width=0.49\textwidth] {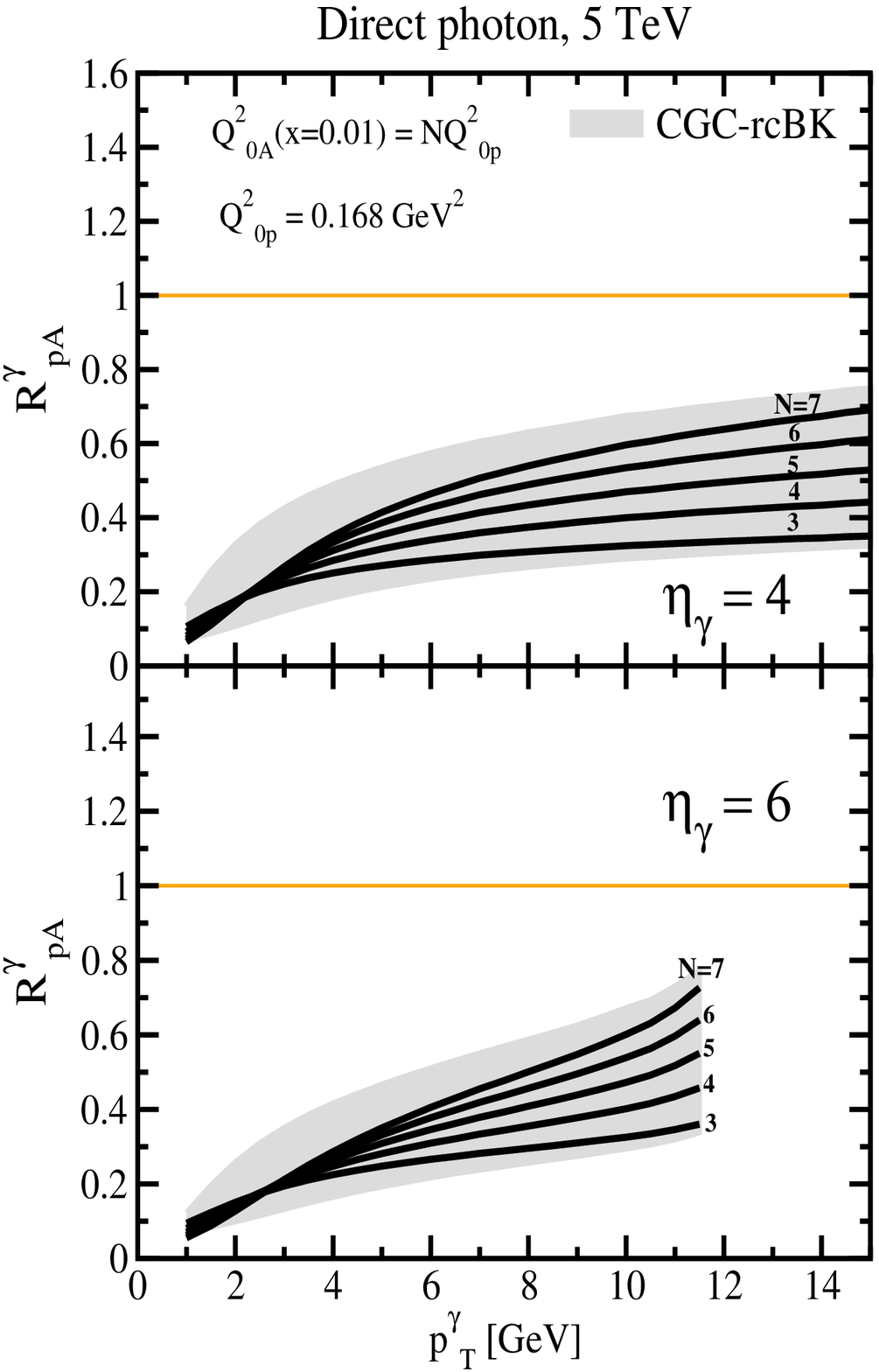}      
\end{center}
\caption[]{The nuclear modification factor $R_{pA}^{\gamma}$ for direct photon 
production in minimum bias $p+$Pb collisions at rapidities 
$\eta_{\gamma} =0$, $2$, $4$, and $6$ 
(with the convention that the proton beam moves toward
forward rapidity) obtained from \protect\eq{pho4} with 
solutions of the rcBK equation with different initial nuclear saturation 
scales.  The band labeled CGC-rcBK includes uncertainties due to the variation 
of the nuclear saturation scale and the factorization scale $\mu_F$.  
Similar to \fig{rp-h}, the lines labeled $N$ are results with a fixed 
factorization scale, $\mu_F=p^{\gamma}_T$, and a fixed saturation scale 
$Q_{0A}^2=N Q_{0p}^2$ with $Q_{0p}^2=0.168\,\text{GeV}^2/c^2$.
The range $3<N< 7$ is constrained in \eq{qa}. 
See Ref.~\protect\cite{Rezaeian:2012ye}.}
\label{rp-p}
\end{figure}

\begin{figure}[htpb]       
\begin{center}
\includegraphics[width=0.49\textwidth] {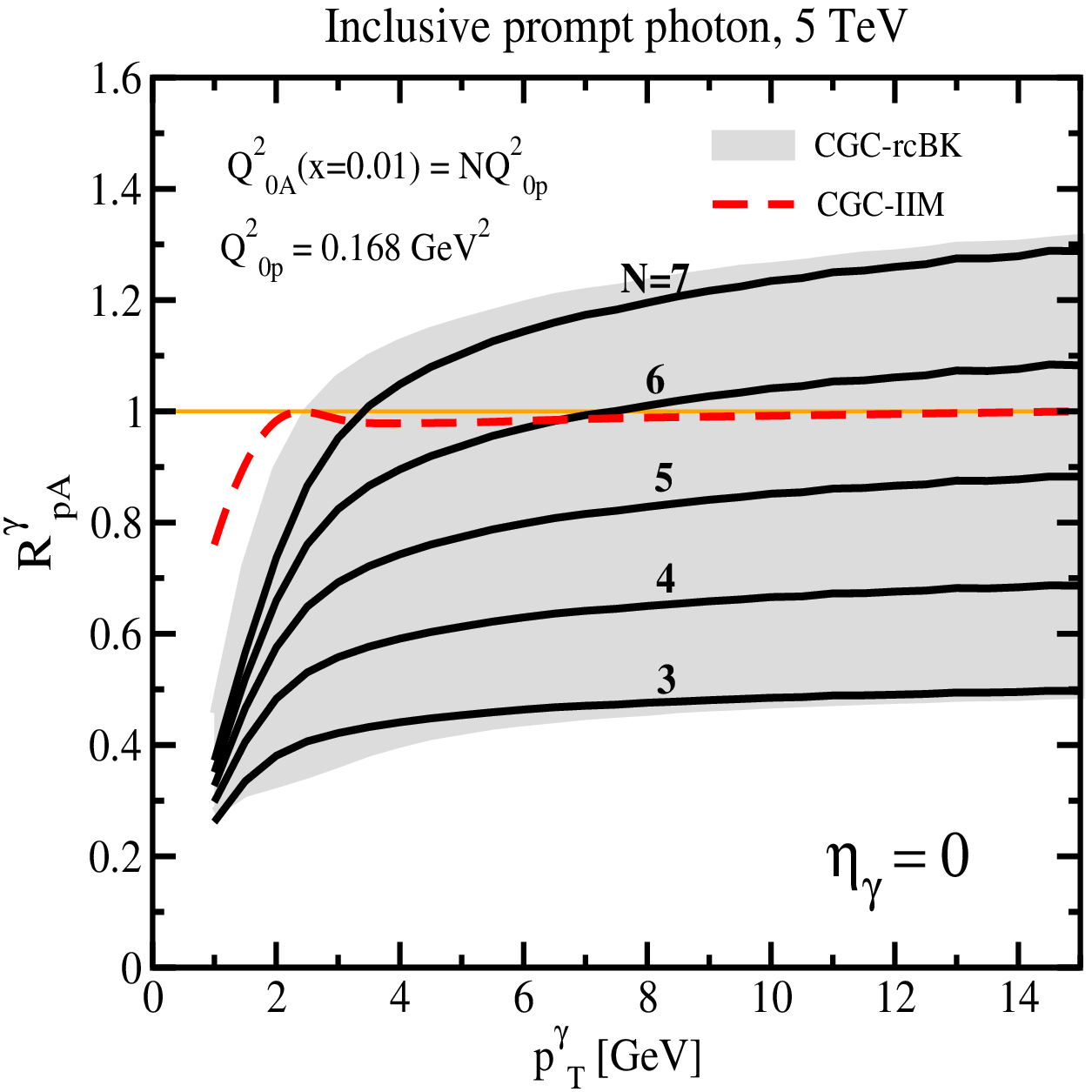}      
\includegraphics[width=0.49\textwidth] {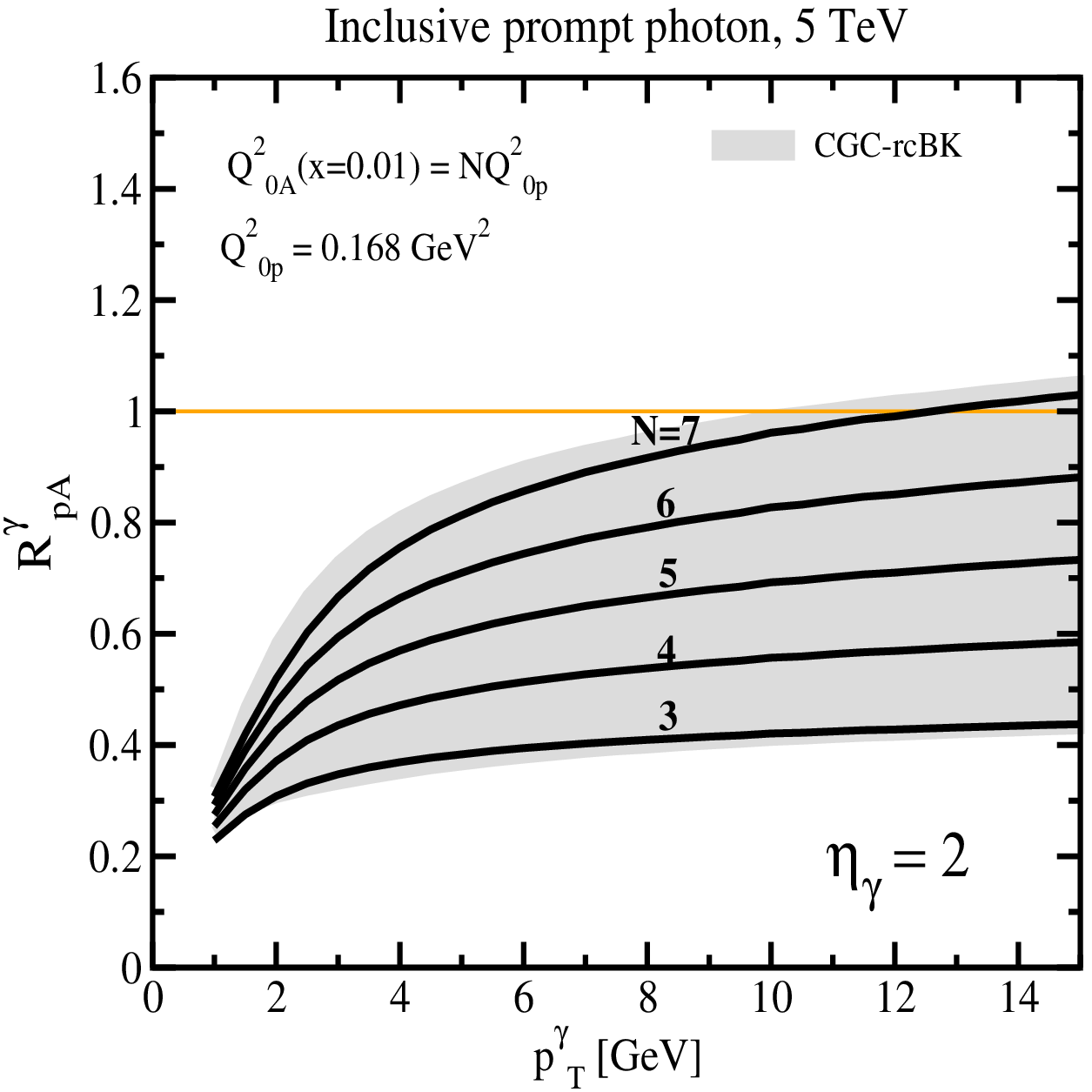}      
\end{center}
\caption[]{The nuclear modification factor $R_{pA}^{\gamma}$ for single inclusive 
prompt photon production in minimum bias $p+$Pb collisions at $\eta_{\gamma}=0$
and 2 (with the convention that the proton beam moves toward
forward rapidity).  The curves are described in \protect\fig{rp-p}.  
The dashed red line 
labeled CGC-IIM was calculated according to the IIM dipole saturation model 
\protect\cite{Rezaeian:2009it}.  See Ref.~\protect\cite{Rezaeian:2012ye}.}
\label{rp-p2}
\end{figure}

In Figs.~\ref{rp-p} and \ref{rp-p2}, predictions of the direct photon and 
single inclusive prompt photon production nuclear 
modification factors $R_{pA}^{\gamma}(p_T)$ in minimum bias $p+$Pb collisions 
at several rapidities.  The solutions of the rcBK equation were obtained using 
Eq.\,(\ref{pho4}) with different initial nuclear saturation scales.  The band 
labeled CGC-rcBK includes uncertainties due to the variation of $Q_{0A}^2$ 
within the range given in \eq{qa} with factorization scales 
$\mu_F=2p_T^\gamma$, $p_T^\gamma$, and $p_T^\gamma/2$. In \fig{rp-p2}, the 
$\eta_\gamma=0$ results include inclusive prompt photon production calculated 
in Ref.~\cite{Rezaeian:2009it} employing the Iancu-Itakura-Munier (IIM) 
saturation model which also provides a good description of the HERA data 
\cite{Iancu:2003ge}.  In the IIM method, saturation is approached from the BFKL 
region.  Therefore the IIM small-$x$ evolution is different from that obtained
with the rcBK equation. 
 
As discussed in Sec.~\ref{forward_rpPb}, the results in Figs.~\ref{rp-p}
and \ref{rp-p2} on $R_{pA}^{\gamma}$, together with those on $R_{pA}^{\rm ch}$ in
Fig.~\ref{rp-h}, can be used to fix the nuclear saturation scale.  Once 
$Q_{0A}^2$ has been established at one rapidity, the predictions at other
rapidities are fixed and the CGC/saturation dynamics can be tested.

\subsubsection{Photon-hadron correlations}

Photon-hadron azimuthal correlations in $p+A$ and $p+p$ collisions could be 
an excellent probe of small-$x$ dynamics \cite{JalilianMarian:2012bd,Rezaeian:2012wa}. This correlation 
can be defined as \cite{JalilianMarian:2012bd,Rezaeian:2012wa},
\begin{equation}\label{az}
P(\Delta \phi)=\bigg[ {d\sigma^{p\, A \rightarrow h(p_T^h)\,\gamma(p_T^\gamma)\, X} 
[\Delta \phi]\over d^2\vec{b_T}\,p_T^h dp_T^h\, p_T^\gamma dp_T^\gamma\, 
d\eta_{\gamma}\, d\eta_h\, d\phi} \bigg] \bigg[{d\sigma^{p\, A \rightarrow 
h(p_T^h)\,\gamma(p_T^\gamma)\, X} [\Delta \phi= \Delta \phi_c] 
\over d^2\vec{b_T}\,p_T^h dp_T^h\, p_T^\gamma dp_T^\gamma\, d\eta_{\gamma}\, 
d\eta_h\, d\phi} \bigg]^{-1} \, \, .
\end{equation}
The correlation function $P(\Delta \phi)$ is the probability of semi-inclusive 
photon-hadron pair production in a certain kinematic region at angle  
$\Delta \phi$ relative to production in the same kinematics at fixed reference 
angle, $\Delta \phi_c=\pi/2$ \cite{JalilianMarian:2012bd,Rezaeian:2012wa}. 

\begin{figure}[htpb]       
\includegraphics[width=0.495\textwidth] {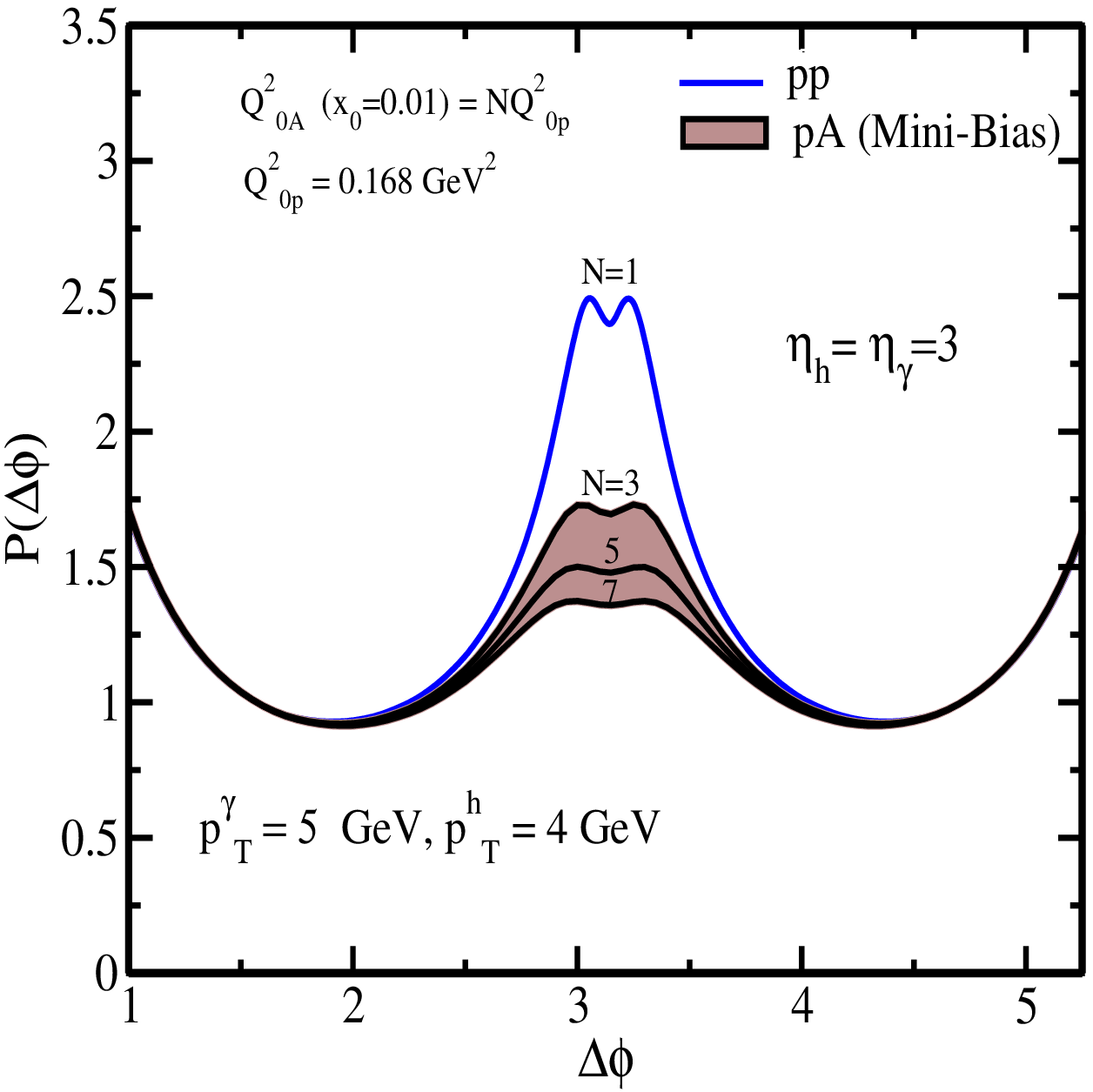}        
\includegraphics[width=0.495\textwidth] {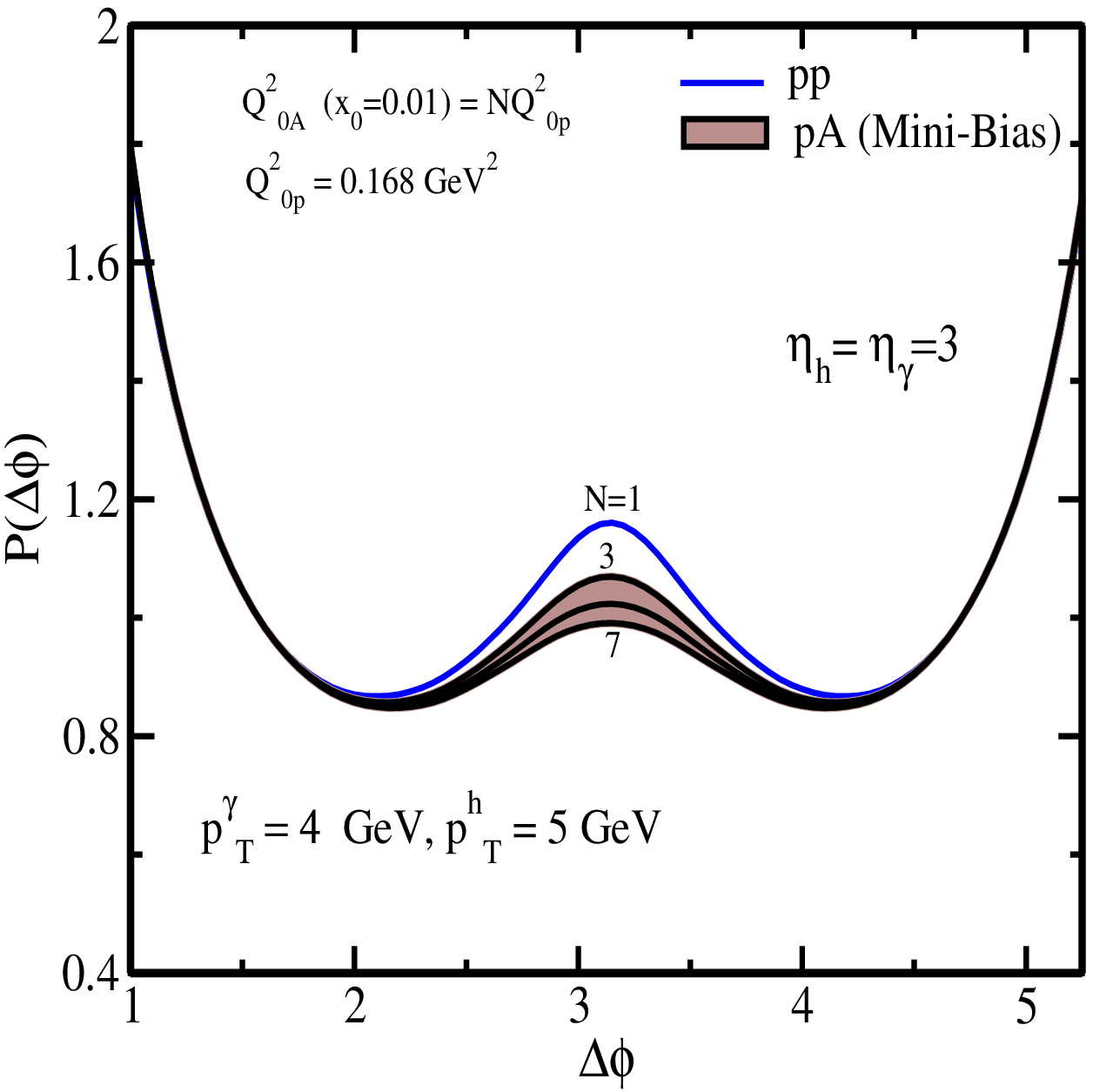}    
\caption[]{The $\gamma-\pi^0$ correlation $P(\Delta \phi)$ defined 
in \protect\eq{az} in minimum bias $p+p$ and $p+$Pb collisions at $\sqrt{S}=5$ 
TeV and $\eta_{h}=\eta_{\gamma}=3$ 
(with the convention that the proton beam moves toward
forward rapidity) obtained via the rcBK evolution equation with 
several initial saturation scales, $Q^2_{0A}=NQ^2_{0p}$ with $N=3$, 5, and 7, in 
two different transverse momentum regions: $p^h_T< p^{\gamma}_T$ (left) and  
$p^h_T>p^{\gamma}_T$ (right). Taken from 
Ref.~\protect\cite{Rezaeian:2012ye,Rezaeian:2012wa}.}
\label{rp-c}
\end{figure}

Fig.~\ref{rp-c} shows the predicted $P(\Delta \phi)$ in minimum bias $p+p$ and 
$p+$Pb collisions at 5 TeV for  $p^h_T< p^{\gamma}_T$ (left) and 
$p^h_T>p^{\gamma}_T$ (right).  Given the $p_T$ of the produced photon and hadron, 
the corresponding correlation can have either a double or single peak 
structure. In Ref.~\cite{Rezaeian:2012wa} it was shown that this feature is 
related to saturation physics and is governed by the ratio $p_T^h/p_T^\gamma$. 
The change from a double to a single peak correlation, depending on the relative
$p_T$, is unique to semi-inclusive photon-hadron production: since the trigger
particle in dihadron correlations is always a hadron, it consistently exhibits
a single-peak structure.

Photon-hadron correlations can also be quantified by the coincidence 
probability.  In contrast to production of a more symmetric final state such as
dihadron production, in photon-hadron production the trigger particle can
be either the prompt photon or the hadron \cite{Rezaeian:2012wa}.  When the 
photon is used as the trigger, the coincidence probability is defined as 
$CP_h(\Delta \phi)=N^{\text{pair}}_h (\Delta \phi)/N_{\gamma}$ where 
$N^{\text{pair}}_h (\Delta \phi)$ is the photon-hadron yield.  The momentum of
the photon trigger, the leading ($L$) particle, is denoted $p^\gamma_{T,L}$
while the momentum of the associated ($S$) hadron (typically a $\pi^0$) is
denoted $p^h_{T,S}$.  
The azimuthal angle between the photon and the $\pi^0$ is 
$\Delta \phi$ \cite{Rezaeian:2012wa},
\begin{eqnarray} 
CP_{h}(\Delta \phi;  p^h_{T,S}, p^\gamma_{T,L}; \eta_\gamma,\eta_h)&=&\frac{2\pi 
\int_{p^\gamma_{T,L}} dp_T^\gamma p_T^\gamma \int_{p^h_{T,S}} dp_T^h p_T^h
\frac{dN^{p\, A \rightarrow h(p_T^h)\,\gamma(p_T^\gamma)\, X} }
{d^2\vec{p_T}^\gamma\, d^2\vec{p_T}^h\, d\eta^{\gamma}\, d\eta^h }}
{\int_{p^\gamma_{T,L}} d^2\vec{p_T}^\gamma\, 
\frac{dN^{p\, A \rightarrow \gamma(p_T^\gamma)\, X}}
{ d^2\vec{p_T}^\gamma\,d\eta_{\gamma} }} \,\, . \label{cp1}
\end{eqnarray}
The integrals are carried out within momentum intervals defined by 
$p^\gamma_{T,L}$ and $p^h_{T,S}$.  The yields in the 
above expression are defined in Eqs.~(\ref{cs}) and (\ref{pho4}).  The 
correlation defined in  \eq{az}  can be considered as a snapshot of the 
integrand in the coincidence probability defined in Eq.\,(\ref{cp1}). 
In the same fashion, one can define a hadron-triggered coincidence probability 
\cite{Rezaeian:2012wa}.  If the $\pi^0$ is the trigger particle, the momenta are
instead denoted by $p^\gamma_{T,S}$ and $p^h_{T,L}$ which then become the lower 
limits on the integrals in \eq{cp1}. 
The away-side coincidence probability for the azimuthal 
correlation of photon-hadron pairs can also have a double-peak or single-peak 
structure, depending on the trigger particle selection and kinematics 
\cite{Rezaeian:2012wa}.  

\begin{figure}[htpb]       
\begin{center}
\includegraphics[width=0.5\textwidth] {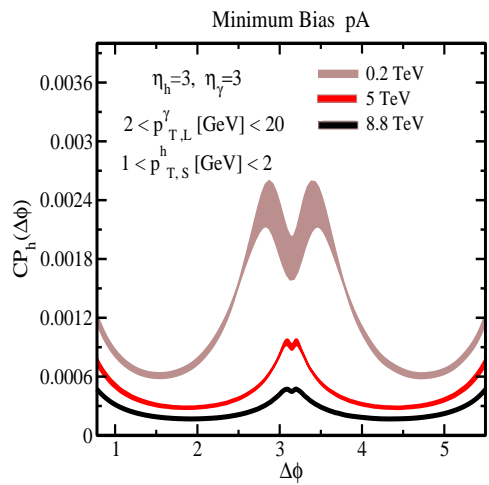}
\end{center}
\caption[]{The $\gamma-\pi^0$ coincident 
probability, $CP_{h}(\Delta \phi)$, defined in \protect\eq{cp1} in 
minimum-bias $p+A$ collisions at $\eta_h=\eta_\gamma=3$ (with the convention 
that the proton beam moves toward
forward rapidity) for $\sqrt{s} = 0.2$, 5
and 8.8 TeV. Taken from 
Ref.~\protect\cite{Rezaeian:2012ye,Rezaeian:2012wa}.}
\label{rp-c2}
\end{figure}

In \fig{rp-c2}, predictions of the 
azimuthal correlations between the produced prompt photon and hadron, 
calculated employing the coincidence probability $CP_h$ in $p+$Pb collisions 
are shown for $\sqrt{s}=0.2$, $5$, and $8.8$ TeV. Equation~(\ref{pho4}) is
used in the denominator of \eq{cp1}.  The collinear divergence was removed 
from \eq{cp1} by introducing the photon fragmentation function.  The
numerator is calculated using \eq{cs}.  An overall normalization problem may
result given that the away-side correlation at $\Delta \phi=\pi/2$ is not 
sensitive to the collinear singularity.  Proper treatment of this problem 
requires a full NLO calculation which is currently unavailable. However, 
choosing a different photon fragmentation function 
\cite{Owens:1986mp,Bourhis:1997yu,Gluck:1992zx}
will only slightly change the results given the freedom to choose the 
fragmentation scale and the rather large uncertainties due to $Q_{0A}^2$. 
This possible normalization problem is not present in the correlation defined 
via \eq{az}. 

Although there are theoretical uncertainties associated with the strength of 
the photon-hadron correlations, due to both higher-order corrections and the
less constrained saturation scale $Q_{0A}^2$.  Nevertheless, the decorrelation
of away-side photon-hadron production increases with energy, rapidity and 
density.  This decorrelation, together with the appearance of a double or 
single-peak structure, are robust predictions of CGC/saturation effects in
the leading-log approximation.

\subsection[Proton-nucleus dilepton and photon production at the LHC:
gluon saturation and shadowing]{Proton-nucleus dilepton and photon production at the LHC:
gluon saturation and shadowing (R. Baier, F. Gelis, A. H. Mueller and D. Schiff)}
\label{sec:Baier}

Here, a qualitative discussion
on saturation and shadowing effects on photon and dilepton production at very
large rapidity in high-energy proton-nucleus collisions is given.
The aim is to describe the various steps of assumptions leading to the
present understanding.

In $\sqrt{s_{_{NN}}} = 200$ GeV d+Au data on high-$p_T$ hadron production at 
large rapidity (on the deuteron side) there is a significant suppression of
hadron production compared to the expectation from $p+p$
collisions. This result suggests that there may be
a significant amount of leading-twist gluon shadowing in the nuclear 
wavefunction in the region probed by forward hard scattering at RHIC, 
with even more expected at the LHC.

In many ways hard photons (or dileptons arising from virtual
photons) are a better probe than high-$p_T$ hadrons.  Photons
are less sensitive to fragmentation effects while final-state effects are 
almost absent. Thus, at $p_T \sim 2 - 3$ GeV/$c$, leading-twist 
factorization is expected to be accurate. Hence the nuclear gluon distribution
probes $x$ values somewhat smaller than $10^{-4}$ in $p+$Pb collisions at the 
LHC.

Increasing the photon rapidity into the forward (proton rapidity)
region, $y > 0$, the gluon
$x$ probed decreases rapidly, {\it e.g.}\ for virtual photons
of mass $M$ and rapidity $y$ produced in the process $q + g \rightarrow 
\gamma^*$, $x \simeq (M/\sqrt{s})\exp{(-y)}$.  With
$\sqrt{s} = 5500$~GeV, $M = 5.0$~GeV/$c^2$, and $y = 3.5$, $x = 2.7 \times
10^{-5}$, so that indeed a fast quark in the direction of proton rapidity
produces virtual photons that probe 
the small $x$ gluon distribution in the nucleus.

It is extremely interesting to explore, at least qualitatively,
the size of the suppression one might expect in such reactions.
The treatment here follows Ref.~\cite{Baier:2004ti} which can be examined for
further details.  This discussion
is based on a picture derived from the Color Glass Condensate (CGC) effective
theory for the gluon distribution\footnote{For recent reviews, see
Refs.~\cite{Gelis:2010nm,Triantafyllopoulos:2012uv} and references therein.}
together with BFKL evolution (see e.g. Ref.~\cite{Mueller:2002zm}) 
to reach higher values of $y$.

\subsubsection{Factorized formula for the inclusive $\gamma^* $ cross section}

\begin{figure}[htpb]
\begin{center}
\epsfig{bbllx=0,bblly=0,bburx=340,bbury=100,
file=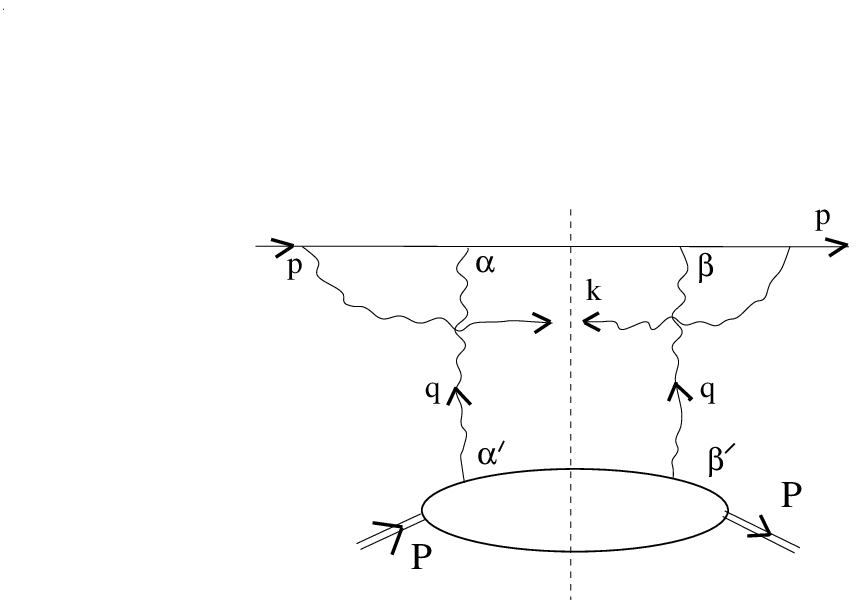,width=11cm}
\caption[]{Two gluon exchange graph for quark ($p$) + nucleus ($P$) production 
of a real or virtual photon,
$\rightarrow ~ \gamma^{(*)} (k) ~X$. }
\label{fig:Baier}
\end{center}
\end{figure}

A dimensionless observable, $\sigma(\vec Q,Y)$, can be written in the 
$k_\bot$-factorized form 
\begin{equation}
\sigma (\vec \mu, Y) = \int \frac{d^2 q_T}{\pi q^2_T} \,
H (\vec q_T , \vec \mu ) \, 
\phi_G (\vec  q_T , Y) \, ,
\label{(4.5)}
\end{equation}
where $\mu$ is the hard scale of the reaction, equal to 
the transverse momentum, $k_T$, of the (massive) photon, i.e.
\begin{equation}
\sigma ( \vec k_{T} , Y) \equiv
\sigma (\vec b_{T} , \vec k_{T} , Y) = 
\frac{k^{~2}_T d \sigma}{d^2 b_T d^2 k_T d  \ln z}  \, . 
\label{(4.7)}
\end{equation}
For simplicity incident quarks are considered instead of protons, 
$q+A\rightarrow \gamma^* X$, see Fig.~\ref{fig:Baier}.
Here $k_T$ is the transverse momentum of the $\gamma^*$, and $z$ is
the longitudinal momentum fraction of the $\gamma^*$
with respect to the incident quark momentum, $z=k_+ / p_+$, where 
$p_+ \rightarrow \infty$.  The impact parameter of the $q+A$ collisions
is denoted $\vec b$ and $Y=\ln{1/x}$.

The leading-twist contribution involves only two exchanged gluons, as
shown in Fig.~\ref{fig:Baier}. $H$ is the hard part of the
interaction in the $k_T$-factorized 
form while $\phi_G (\vec q_T , Y)/q^2_T$ is considered to be proportional 
to the differential high energy $q+A \rightarrow q+A$ cross section.

\subsubsection{Anomalous scaling and shadowing - specific predictions}

While there are theoretical uncertainties, there are also robust characteristic
qualitative results.  The function $\phi_G ( k_T , Y)$ can be approximated
by the scaling function, 
\begin{equation}
\phi_G ( k_T , Y) = \phi_G ( k_T / Q_s (\vec b, Y)) \approx 
\left( k^2_T / Q^2_s (\vec b, Y) \right)^{\lambda_0 - 1} , 
\label{(5.7)}
\end{equation}
with anomalous dimension $\lambda_0 \simeq 0.37$ and
saturation scale $Q_s$.  In $\vec b = 0$ collisions at fixed $k_T$ and $Y$,
an anomalous $A$ dependence is predicted, together with 
shadowing/suppression of the ratio $R_{pA}$,
\begin{equation}
R_{pA} \approx 
A^{-\lambda_0 / 3} ~.
\label{(5.5a)}
\end{equation}

In order to obtain results at large photon rapidities based
on the BFKL evolution in the presence of saturation, the rapidity 
dependence of the saturation scale is chosen to be compatible with 
phenomenology,  
 \begin{equation}
Q^2_s (\vec b = 0 , Y =  y) =
 Q^2_s (Y =0) \exp(\lambda \,  y) , \, \,
 ~~ \lambda \simeq 0.3  \, \, .
\label{(5.9)}
\end{equation}
There is significant suppression expected,
$R_{pA} \simeq 0.5$, 
especially when the $\gamma^{(*)}$ 
is produced at forward (in the proton direction)
rapidities, e.g.\  $y >  3$ and  $k_T > 2$~GeV/$c$.
More detailed predictions of forward inclusive prompt photon production and 
semi-inclusive photon-hadron
correlations in high energy proton-nucleus collisions at the LHC 
using the CGC formalism are presented in 
Refs.~\cite{JalilianMarian:2012bd,Rezaeian:2012wa}.

Measurements of photons and dileptons, in addition to hadrons with
moderate transverse
momenta but at large rapidities in $p+$Pb collisions at the LHC will
provide important information supporting the saturation picture of high density
gluon dynamics at high energies and small values of $x$.

\section{Jets}
\label{sec:jets}

In this section, several results for jet production and modification in media
are presented.  Multi-jet production is discussed and predictions for the jet
yields are shown in Sec.~\ref{sec:Nestor}.  Cold nuclear matter effects on
jet and dijet production are presented in Sec.~\ref{sec:Zhang_jets}.  The
angular decorrelation of dijets due to saturation effects is discussed in
Sec.~\ref{sec:kutak}.  Predictions of long-range near-side azimuthal
collimation event with high multiplicity are given in Sec.~\ref{sec:Dusling}.
Finally, enhanced transverse momentum broadening in
medium is described in Sec.~\ref{sec:Ivan_jet}.

\subsection[Multi-jet Production]{Multi-jet Production (N. Armesto)}
\label{sec:Nestor}

The study of jet production in Pb+Pb collisions at the LHC 
\cite{Aad:2010bu,:2012is,Chatrchyan:2011sx,Chatrchyan:2012nia,Chatrchyan:2012gt,Chatrchyan:2012gw} and their 
medium modifications, are important for probing the properties of the hot and 
dense matter formed in heavy-ion collisions and is thus
a very hot topic in heavy-ion physics. 
Therefore, studies of (multi-)jet production in $p+A$ collisions are of
great importance as a cold QCD matter benchmark.  Here the jet rates in
minimum bias $p+$Pb 
collisions at the LHC (4+1.58 TeV per nucleon) are computed
using the Monte Carlo code in
Refs.~\cite{Frixione:1995ms,Frixione:1997np,Frixione:1997ks} that implements 
fixed-order perturbative QCD up to next-lo-leading order. 
This code produces at most three jets and contains neither parton cascades nor 
hadronization corrections. 
 
In the computation, the renormalization and factorization scales have been set 
equal and fixed to $\mu=\mu_F=\mu_R=E_T$, where $E_T$ is the total
transverse energy in the event.  The central set of the
NLO MSTW2008 parton densities (MSTW2008nlo68cl) \cite{Martin:2009iq} have been 
used for the unmodified nucleons. The nuclear modification of parton densities 
was examined using the EKS98 \cite{Eskola:1998iy,Eskola:1998df} and 
EPS09NLO \cite{Eskola:2009uj} sets.  The standard Hessian method was employed 
to estimate the uncertainties coming from nuclear parton densities. The
precision of the computation, limited by CPU time, gives statistical 
uncertainties smaller than 10\% for the bin with the highest $E_T$ in the 
results shown here.  The anti-$k_T$ jet finding algorithm with $R=0.5$ 
\cite{Cacciari:2008gp} was used. Only jets with $E_T>20$ GeV are considered.
The uncertainties due to the choice of nucleon parton
densities, isospin corrections, and scale fixing, together with
the influence of the jet-finding algorithm and the choice of nuclear
targets and collision energies, were discussed
elsewhere~\cite{Accardi:2003be,Accardi:2003gp} and are not considered here.

\begin{figure}[htbp]
\begin{center}
\setlength{\epsfxsize=1.0\textwidth}
\centerline{\epsffile{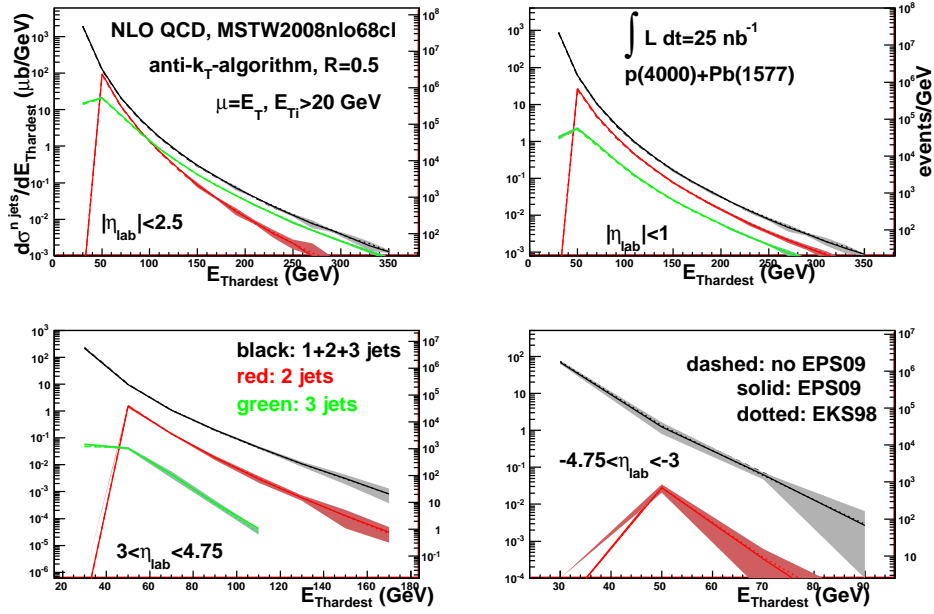}}
\end{center}
\caption[]{The 1+2+3- (black), 2- (red) and 3-jet (green) cross sections as a 
function of the $E_T$ of the hardest jet within the acceptance. Results in
different pseudorapidity windows in the laboratory frame computed for 
minimum bias $p+$Pb collisions are shown.  The dashed lines are the results 
without nuclear modifications of the PDFs.  Results with EPS09NLO 
\cite{Eskola:2009uj} (solid) and EKS98 (dotted) 
\cite{Eskola:1998iy,Eskola:1998df} are also shown. The bands correspond to 
uncertainties computed using the Hessian method for EPS09 \cite{Eskola:2009uj}.
The right vertical axes gives the scale for the corresponding
yields with an integrated luminosity of 25 nb$^{-1}$. See the text for further 
details.}
\label{fig:jetppb}
\end{figure}

Figure \ref{fig:jetppb} shows the sum of the 1-, 2- and 3-jet yields, as well
as the individual 2- and 3-jet yields within four pseudorapidity windows
(two central, one backward and one forward)
in the lab frame
as a function of the $E_T$ of the hardest jet within the acceptance, 
$E_{T \, _{\rm hardest}}$. The yields, computed 
for luminosity ${\cal L}=0.25\times 10^{29}$ cm$^{-2}$s$^{-1}$ integrated over a
one month ($10^6$ s) run (corresponding to an integrated luminosity of 
25 nb$^{-1}$) \cite{pPblumi}, are quite large. The yields are given on the 
right-hand vertical axes in 
Fig.~\ref{fig:jetppb}.  High yields are required to study cold matter effects
in high $p_T$ multi-jet production.  Figure~\ref{fig:jetppb} demonstrates that
such studies are feasible.  For example, in the backward region, $-4.75<\eta
<-3$, sufficiently high rates for the sum per GeV of 1+2+3-jet events can be 
achieved for $E_{T_{\rm hardest}}< 50$ GeV. The only regions where the yields from
certain channels are too low for statistically significant results are those
for 3-jet events in the forward region and 2- and 3-jets in the backward
region (only the yields for 2-jet events are shown). 
Note that the turnover in the 2-jet yields when $E_{T_{\rm hardest}} \rightarrow 
20$ GeV comes from singularities that appear in the NLO calculation due to
large logarithms because $E_{T_1}\sim E_{T_2} (\sim E_{T_3})$.  
Resummation techniques are required to obtain reliable 
results in this kinematic region.

Nuclear modifications of the PDFs are very small, maximum ${\cal O}$(20\%),
hardly visible in the yields in Fig.~\ref{fig:jetppb}. Note that the wide 
uncertainty bands in the largest $E_{T_{\rm hardest}}$ bins in the forward and 
backward regions are due to statistical fluctuations in the Monte Carlo and 
do not have a physical origin. The corresponding hot nuclear matter
effects in Pb+Pb collisions are expected to be much larger.

Further work to compare these yields with experimental data would require 
consideration of hadronization corrections and the effects of background 
subtraction such as in the well-established jet area method 
\cite{Cacciari:2007fd}. These should be the subject of future studies.

\subsection[Cold matter effects on jet and dijet production]{Cold matter effects on jet and dijet production (Y. He, B.-W. Zhang and E. Wang)}
\label{sec:Zhang_jets}

Inclusive NLO 
\cite{Kunszt:1992tn,Ellis:1990ek,Ellis:1992en,Vitev:2009rd,He:2011pd,He:2011sg}
jet and dijet production are studied in $p+$Pb collisions. 
Cold nuclear matter effects are included by employing 
the EPS09~\cite{Eskola:2009uj}, DS11~\cite{deFlorian:2012qw,deFlorian:2011fp} 
and HKN07~\cite{Hirai:2007sx} parametrizations of nuclear parton densities. 
The numerical results for the inclusive jet 
spectrum, the dijet transverse energy and mass spectra, and the 
dijet triply-differential cross sections, all using a jet cone size $R=0.4$ in 
minimum bias $p+$Pb collisions are shown.

\begin{figure}[htpb]
\begin{center}
\includegraphics[trim=0 150 40 10,width=0.48\textwidth]{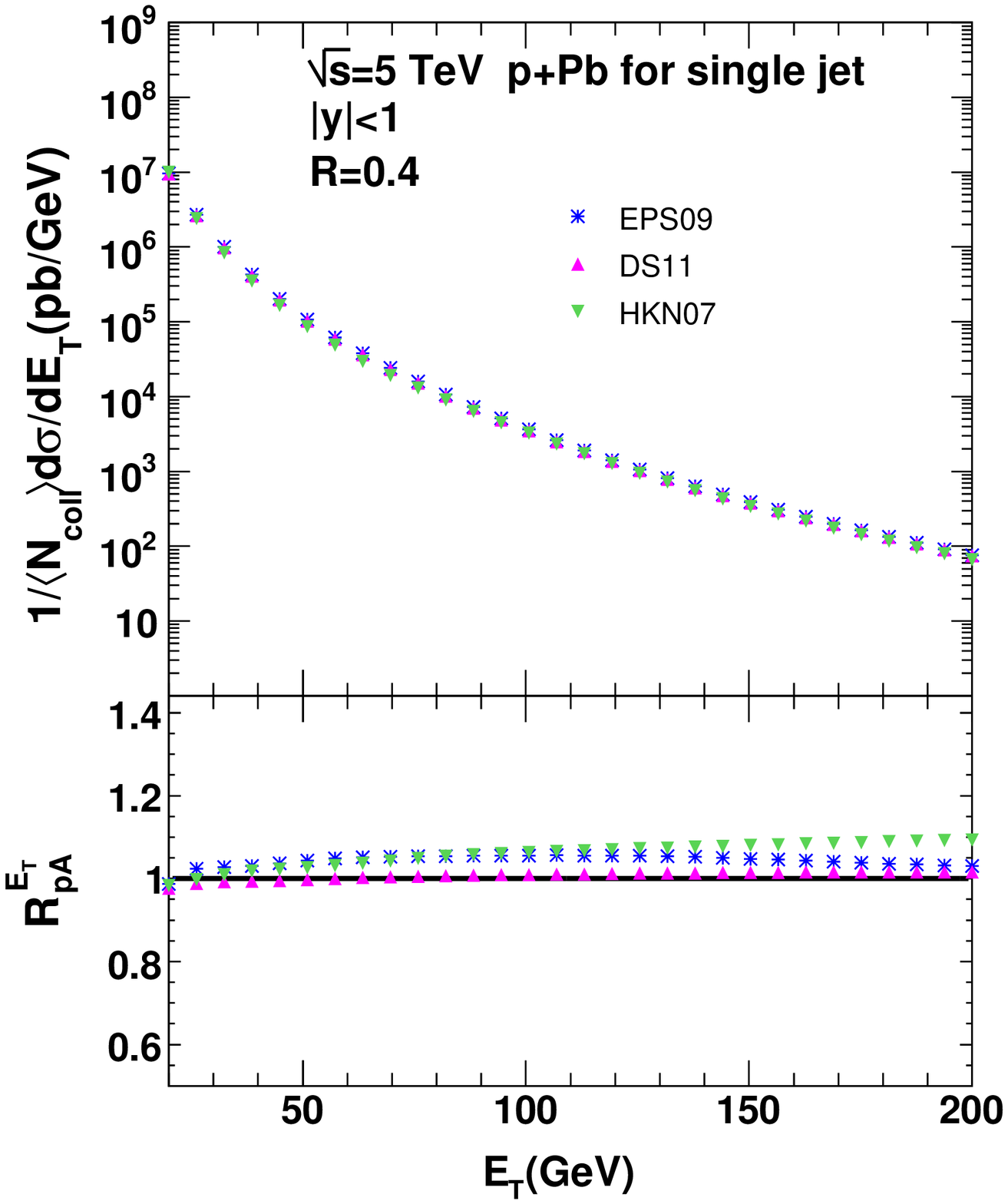}
\includegraphics[trim=0 150 40 10,width=0.48\textwidth]{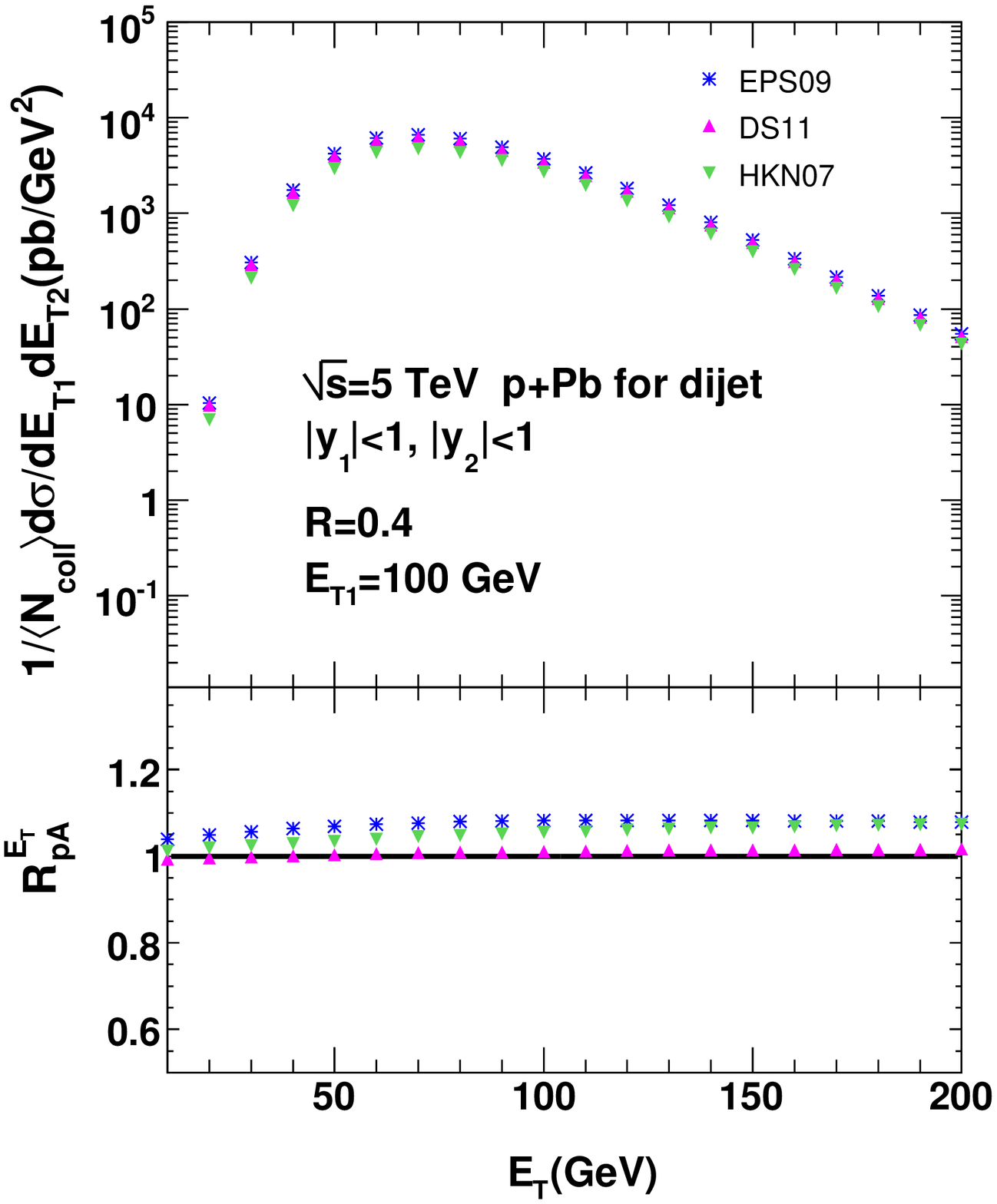}
\end{center}
\caption[]{The inclusive jet (left) and dijet (right) results in $p$+Pb 
collisions at $\sqrt{s}=5$~TeV.  The dijet results, presented as a function of
$E_{T2}$ are given for fixed
energy $E_{T1} = 100$ GeV for jet 1.  The rapidity acceptance for all jets is
$|y|<1$.  The $E_T$ spectra are shown in the upper part of the
plots while the nuclear modification factors are shown in the lower panels of
the figures.}
\label{1+2}
\end{figure}

The left-hand side of Fig.~\ref{1+2} 
illustrates the inclusive jet spectra scaled by $N_{\rm coll}$ including the
nuclear modifications in the central rapidity region, $|y|<1$. Results for
$R_{p{\rm Pb}}$ are also shown.
The spectra with EPS09 and HKN07 show
an enhancement in the transverse energy range $30<E_T< 200$~GeV. However, the 
CNM effects are negligible over the entire region of jet $E_T$ with DS11.

The right-hand side of Fig.~\ref{1+2} displays the rescaled dijet $E_T$ 
spectra as a function of $E_{T2}$ with jet 1 at a fixed transverse 
energy of $E_{T1}=100$~GeV employing the same modifications of the parton
densities as before. Both jets are within the central rapidity region, 
$|y_1|<1$ and $|y_2|<1$. The cross sections in the range $10<E_T< 200$~GeV
exhibit a peak near $E_T=100$ GeV. Again the spectra with EPS09 and HKN07 are 
enhanced.  On the other hand, the DS11 set shows very small suppression and 
enhancement as shown in Fig~\ref{1+2}.

\begin{figure}[htpb]
\begin{center}
\includegraphics[trim=0 150 40 10,width=0.48\textwidth]{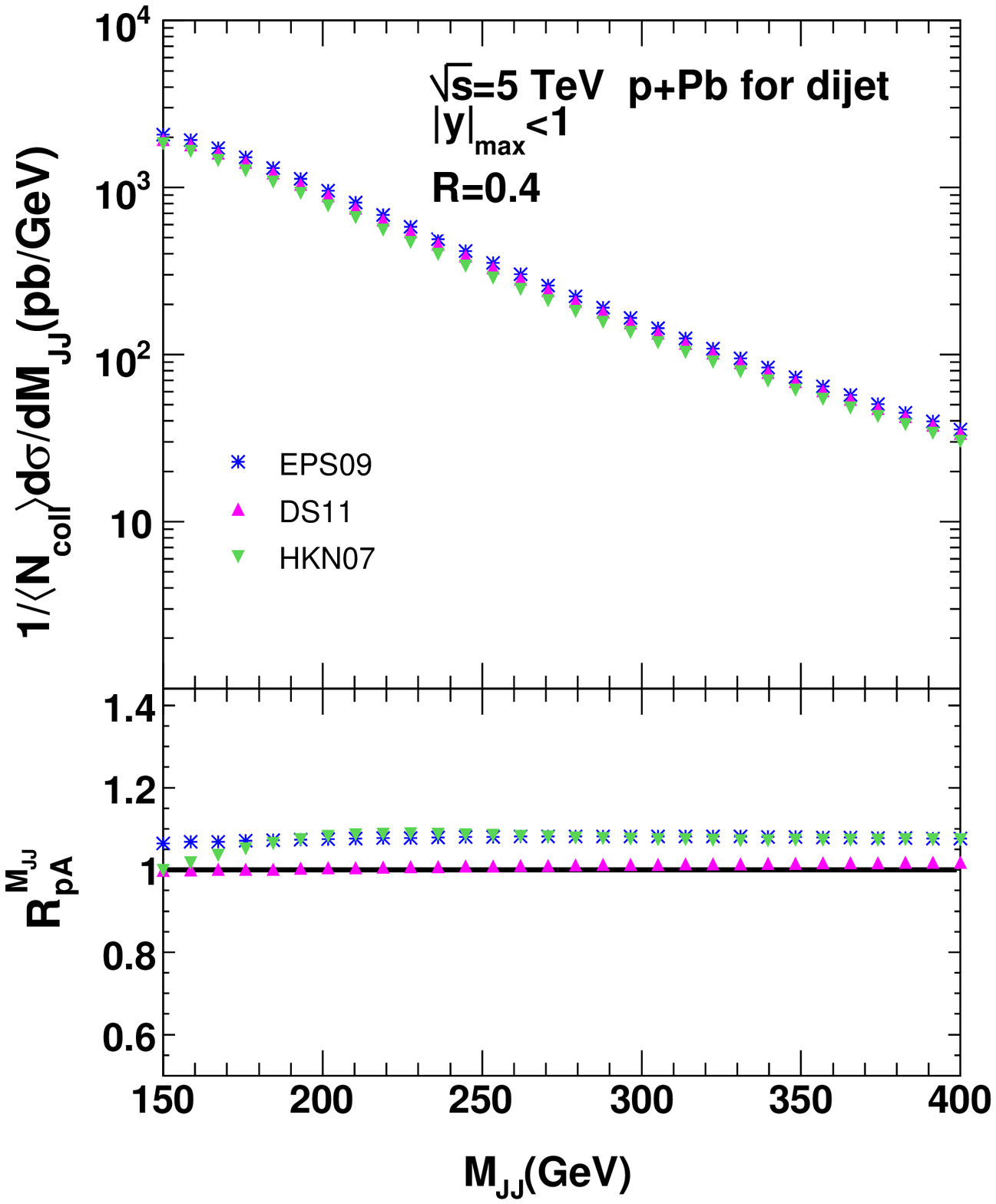}
\includegraphics[trim=0 150 40 10,width=0.48\textwidth]{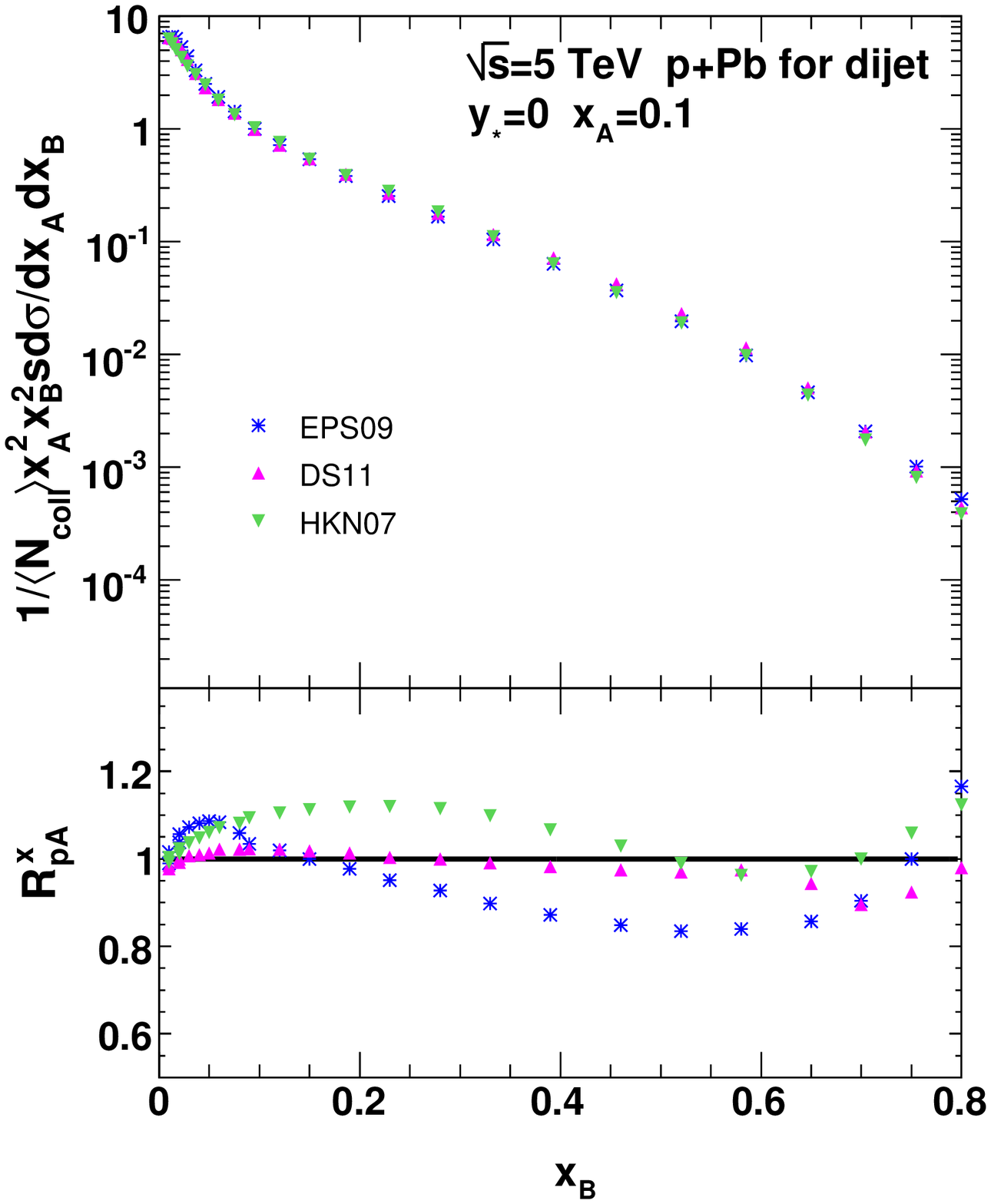}
\end{center}
    \caption[]{(Left) The dijet invariant mass spectra in $p+$Pb collisions.
(Right) The dijet triply differential cross sections in $p+$Pb collisions.
The nuclear modification factors, $R_{p{\rm Pb}}$, are also shown.
}
    \label{3-4}
\end{figure}

Figure~\ref{3-4} gives the rescaled dijet invariant mass $M_{JJ}$ spectra and 
the corresponding nuclear modification factors. Here $M_{JJ}$ is defined as  
$[(\sum p^{\mu}_n)^2]^{1/2}$ where the sum is over all particles in the two 
jets~\cite{Aad:2010ad}. The maximum rapidity of the two leading jets, 
$|y|_{\max}=\max(|y_1|,|y_2|)$ is defined so that $|y|_{\max}<1$.  Jets with 
energies greater than 40~GeV in the rapidity range $|y|<2.8$ are selected. 
The CNM effects with EPS09 and HKN07 also enhance the $M_{JJ}$ spectra over 
a wide $M_{JJ}$ region while the effect with DS11 is modest.

Figure~\ref{3-4} also shows the rescaled dijet triply differential cross 
sections~\cite{Ellis:1994dg} 
and their nuclear modifications. The momentum fractions 
$x_1$ and $x_2$ are defined as
\begin{equation}
x_1=\sum\limits_{i\in {\rm jet}} \frac{E_{Ti}}{\sqrt{s}}e^{y_i} \, \, ,~~
x_2=\sum\limits_{i\in {\rm jet}} \frac{E_{Ti}}{\sqrt{s}}e^{-y_i} \, \, .
\label{V}
\end{equation}
where the sums run over all particles in the jets. The results in Fig.~\ref{3-4}
are shown for the rapidity difference between the jets, 
$y_* = 0.5(|y_1 - y_2|)$
where $y_1$ and $y_2$ are the rapidities of the two leading jets. The
results as a function of $x$ show obvious deviations between nuclear 
modification factors which implies that the dijet triply-differential cross 
sections could be a good observable to distinguish between different 
shadowing parameterizations~\cite{HZW2012}.

\subsection[Angular decorrelation of dijets as a signature of gluon 
saturation]{Angular decorrelation of dijets as a signature of gluon 
saturation (K. Kutak and S. Sapeta)}
\label{sec:kutak}

\begin{figure}[htpb] \centering
  \includegraphics[width=0.45\textwidth]{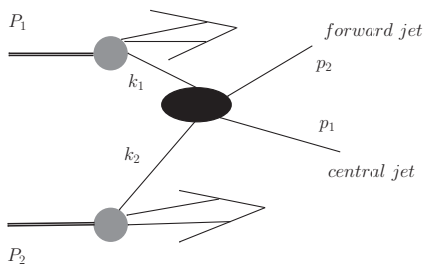}
  \caption[]{Jet production in the forward (assuming the proton moves toward
positive rapidity) region in hadron-hadron collisions.
  } 
  \label{fig:jet_production} 
\end{figure}

Here predictions for the emergence of saturation \cite{Gribov:1984tu} effects 
on dijet production in $p+$Pb scattering at the LHC. The results are 
based 
on a study performed in Refs.~\cite{Kutak:2012rf,Deak:2010gk}. 
The saturation scale characterizes formation of a dense system of partons. 
There is growing evidence that the phenomenon of gluon saturation indeed occurs
\cite{Albacete:2010pg,Dumitru:2010iy}. 
This calculation employs a high-energy factorization formalism which accounts 
for both the high energy scale of the scattering and the hard
momentum scale $p_T$ provided by produced hard system
\cite{Catani:1990eg}.  
Dijets separated by large rapidity gaps \cite{Marquet:2003dm} are considered
here.  More specifically, the focus is on a case where one jet is measured in 
the central rapidity region of the detectors while the other is at large 
rapidity, see Fig.~\ref{fig:jet_production}. 
Such final states probe the parton density in one of the hadrons at low
momentum fraction $x$ while the other is at relatively large $x$. 
Predictions of azimuthal
decorrelations of dijets production in $p+$Pb relative to $p+p$ collisions
are presented.

Assuming, without loss of generality, that $x_1\simeq 1$ and
$x_2 \ll 1$ (if the proton moves in the direction of forward rapidity), 
the cross section takes the form
\begin{eqnarray}
  \frac{d\sigma}{dy_1dy_2dp_{T1}dp_{T2}d\Delta\phi} 
&  = &
  \sum_{a,c,d} 
  \frac{p_{T1}p_{T2}}{8\pi^2 (x_1x_2 s)^2} \nonumber \\
& & \mbox{} \times  {\cal M}_{ag\to cd}
  x_1 f_{a/A}(x_1,\mu^2)\,
  \phi_{g/B}(x_2,k^2_T)\frac{1}{1+\delta_{cd}}\, \, ,
  \label{eq:cs-fac}
\end{eqnarray}
where
\beq
  k_T^2 = p_{T1}^2 + p_{T2}^2 + 2p_{T1}p_{T2} \cos\Delta\phi\, \, ,
\eeq
and $\Delta\phi$ is the azimuthal distance between the outgoing
partons. The squared matrix element ${\cal M}_{ag\to cd}$ includes  $2\to 2$
process with one off-shell initial-state gluon and three on-shell partons
$a$, $c$, and $d$.
The following partonic subprocesses contribute to the production of the dijet
system: $qg  \to  qg$;  $gg  \to q\bar q$; and $gg\to gg$ \cite{Deak:2009xt}.
The off-shell gluon in Eq.~(\ref{eq:cs-fac}) is described by the
unintegrated gluon density $\phi_{g/B}(x_2,k^2_T)$, a solution of a nonlinear 
evolution equation \cite{Kutak:2003bd,Kutak:2004ym} that depends on the
longitudinal momentum fraction $x_2$,
and the transverse momentum of the off-shell gluon, $k_T$. 
On the side of the on-shell parton, probed at high proton $x_1$, the collinear
parton density $f_{a/A}(x_1,\mu^2)$ is appropriate. 
\begin{figure}[htpb]
  \centering
  \includegraphics[width=0.475\textwidth,angle=-90]{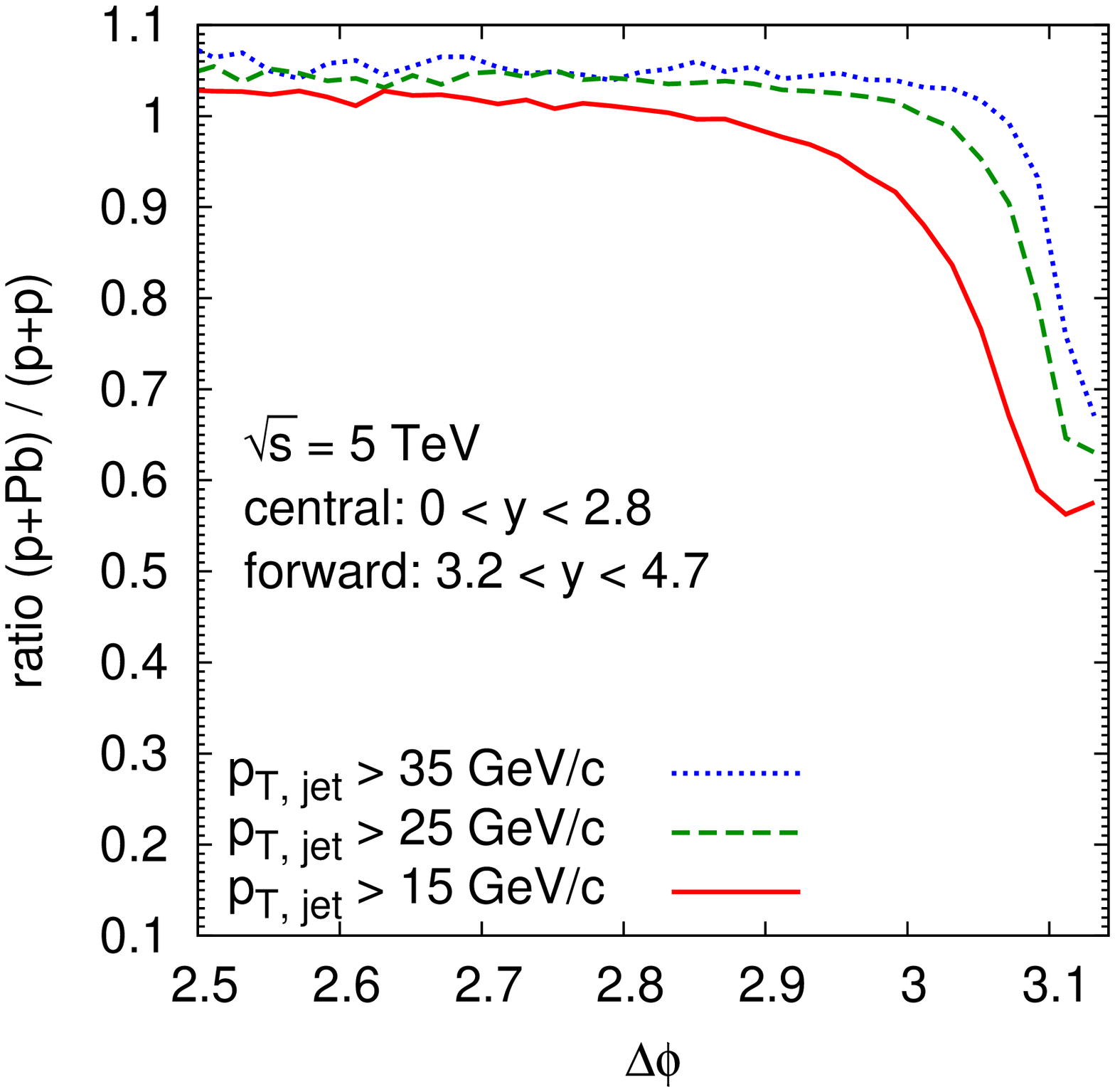}
  \includegraphics[width=0.475\textwidth,angle=-90]{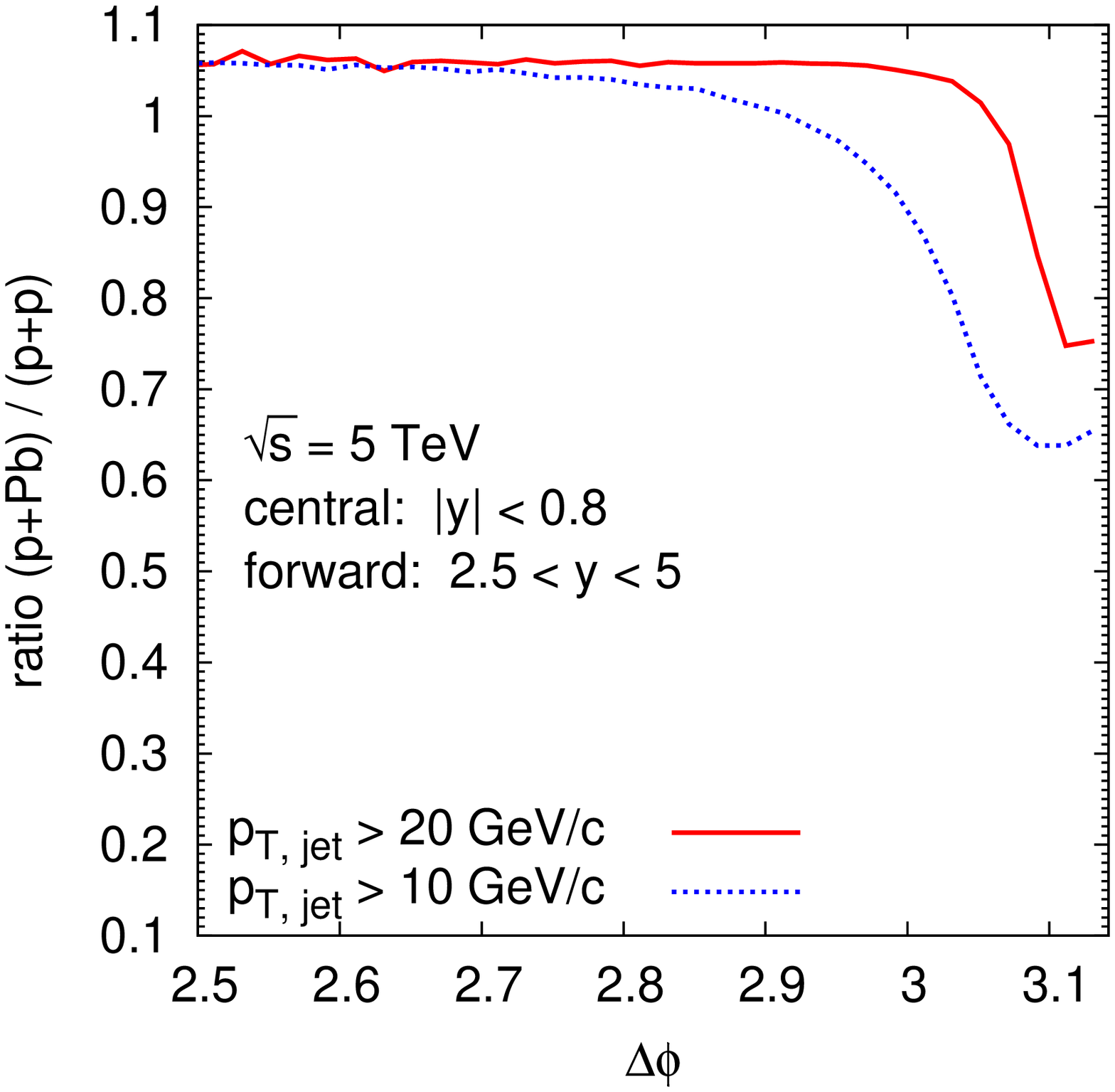}
  \caption[]{Ratio of differential cross sections for central-forward dijet
  production at $\sqrt{s} = 5 \text{ TeV}$ as functions of azimuthal distance
  between the jets $\Delta\phi$ for three different jet $p_T$ cuts. 
  On the left, the selected rapidities correspond to the CMS detector while,
  on the right, the rapidity range appropriate for ALICE is shown.
The calculations have  been made
with the convention that the proton beam moves toward
forward rapidity.}
  \label{fig:decor-p-Pb}
\end{figure}

Figure~\ref{fig:decor-p-Pb} shows the ratios of the differential cross
sections for central-forward dijet production in $p+$Pb relative to $p+p$ 
as a function of the azimuthal distance between the jets, $\Delta \phi$.
In the region $\Delta\phi \sim \pi$, the gluon density is probed at small 
$k_T$, where it is strongly
suppressed due to nonlinear effects.  The ratio is a signal of saturation that
is sensitive to the enhancement of saturation effects afforded by the larger
$A$ of the Pb nucleus.

The left-hand side of Fig.~\ref{fig:decor-p-Pb} shows the results for various  
jet $p_T$ cuts: $p_{T \,{\rm jet}} > 15$, 25, and 35 GeV/$c$ 
while the jet rapidities are restricted to
positive values within the coverage of the central and forward CMS detectors.  
On the right-hand side of Fig.~\ref{fig:decor-p-Pb}, lower $p_T$ cuts,
$p_T > 10$ and 20 GeV/$c$, are used and the rapidity is restricted to positive 
values within the central and larger rapidity range of the ALICE detector.  
The results show that, when the two jets are back-to-back, the cross section
ratio is significantly smaller than one.
This is a consequence of stronger gluon saturation in the
Pb nucleus than in a proton.  Thus the unintegrated gluon distribution in
the region of small and medium $k_T$ is more suppressed in Pb than in protons,
as shown in Fig.~\ref{fig:decor-p-Pb}.  In addition to the effects shown here, 
the region near $\Delta\phi \simeq \pi$ 
is sensitive to Sudakov effects which further suppress the cross
section.  Therefore, refinement along those lines could be envisaged in the
future. However, the ratio shown here is less sensitive to
these corrections.  It is worth emphasizing that the
suppression due to saturation predicted in Fig.~\ref{fig:decor-p-Pb} is 
strong and extends over a large enough $\Delta\phi$ range for
experimental observation, even if the region near $\Delta\phi=\pi$ could be 
further refined.

\subsection[Predictions for long-range near-side azimuthal collimation in high
multiplicity p+Pb collisions]{Predictions for long-range near-side 
azimuthal collimation in high
multiplicity p+Pb collisions (K. Dusling and R. Venugopalan)}
\label{sec:Dusling}

Rapidity-separated dihadron correlation measurements can provide valuable
insight into the gluon dynamics of the nuclear wavefunction.  For
example, it was recently realized~\cite{Dumitru:2010iy} that a novel 
``near-side" azimuthal collimation in high multiplicity $p+p$ 
events~\cite{Khachatryan:2010gv} is a consequence of nonlinear gluon 
interactions and was found to be in excellent agreement with computations in 
the Color Glass Condensate (CGC) effective field theory 
\cite{Dusling:2012iga,Dusling:2012cg}.  

The situation in $p+$Pb collisions is more attractive.  The degree of 
near-side collimation is highly sensitive to the fine structure of the 
unintegrated gluon distribution.  It can be shown that the collimated yield 
is enhanced by $\left(Q_s^{\rm Pb}/Q_s^{p}\right)^2$ in asymmetric collisions.  
Here $Q_s^{\rm Pb}$ is the saturation scale probed in the lead nucleus and can 
be estimated to scale with the number of participants, 
$Q_s^{2,\rm Pb}\approx N_{\rm part}Q_s^{2, p}$, where $Q_s^{p}$ is the saturation 
scale for the proton.  Therefore, for high multiplicity events where many 
nucleons in the lead nucleus participate, a significant near-side collimation 
is observed \cite{CMS:2012qk}.  Similar to the $p+p$ data, calculations within 
the CGC effective theory are able to explain the data \cite{Dusling:2012wy}.  

\begin{figure}[htpb]
\vspace{50pt}
\includegraphics[width=\textwidth]{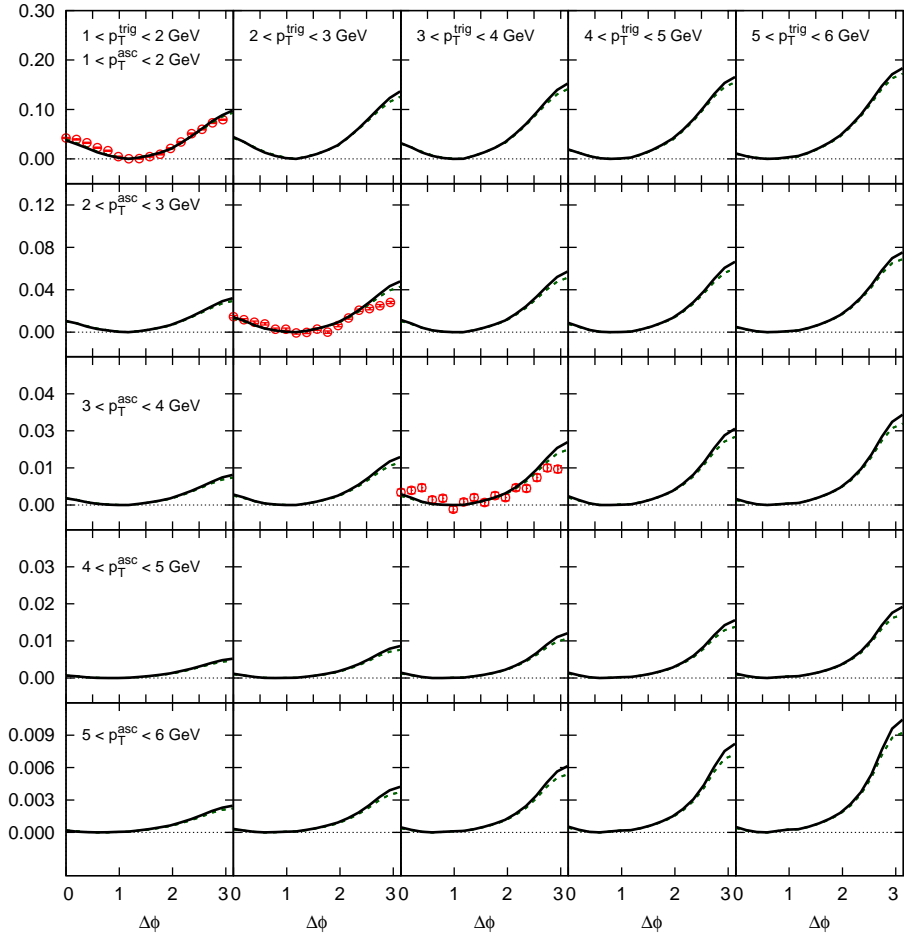}
\caption[]{Correlated yield $1/N_{\rm trig} d^2N/d\Delta\phi$ after ZYAM (zero 
yield at minimum, used to remove the $\Delta \Phi$-independent pedestal) 
as a function of $\vert \Delta \phi\vert$ integrated over $2\leq 
\vert\Delta\eta\vert \leq 4$ for the most central multiplicity bin 
$N_{\rm trk}^{\rm offline} \geq 110$.  The CMS data \protect\cite{CMS:2012qk} 
have currently only been provided for 
the diagonal components, $p_T^{\rm trig} \sim p_T^{\rm assoc}$, of the correlation 
matrix.  The theory curves are the result of adding the BFKL contribution 
responsible for the away-side jet and the ``dipole"-like Glasma contribution.
The solid black curve is the result for $Q_{0p}^2 = 0.504$ GeV$^2$ on 
$N_{\rm part}^{\rm Pb} =14$ and the dashed green is for $Q_{0p}^2 = 0.336$ 
GeV$^2$ on $N_{\rm part}^{\rm Pb} =16$.}
\label{fig:pPb_matrix_central}
\end{figure}

This last study combined the physics of saturation and BFKL dynamics in order 
to understand  the near-side collimation and the recoiling away-side 
jet.  Both are required for a full understanding of the azimuthal structure of 
dihadron correlations.  The per-trigger yields shown in 
Fig.~\ref{fig:pPb_matrix_central} are well described by this framework for the 
symmetric $p_T^{\rm trig} \sim p_T^{\rm assoc}$ windows where data are currently 
available.  Predictions for the asymmetric  $p_T^{\rm trig} \neq p_T^{\rm assoc}$
windows are also shown.

\subsection[Enhancement of transverse momentum broadening]{Enhancement of transverse momentum broadening (H. Xing, Z.-B. Kang, I. Vitev and E. Wang)}
\label{sec:Ivan_jet}

Both  initial-state  and final-state multiple scattering lead to 
acoplanarity, or momentum imbalance of the two  leading final-state 
particles. To 
quantify this effect, the transverse momentum imbalance $\vec{q}_{T}$  
is defined as:
\ben
\vec{q}_{T}=\vec{p}_{T 1}+\vec{p}_{T 2},
\een
with the average squared transverse momentum imbalance 
\ben
\langle q_T^2\rangle = \left(  \int d^2 \vec{q}_T q_T^2
\frac{d\sigma}{d\mathcal{PS} \, d^2\vec{q}_T}  \right) \left( 
\frac{d\sigma}{d\mathcal{PS}} \right)^{-1} \, \, .
\een
Here, $d\sigma/d\mathcal{PS}$ is the differential cross section with 
the appropriate phase space factor, determined separately for each process. 
For example, in $p+A$ collisions, $d\mathcal{PS}=dy_1dy_2 dp_T^2$ 
for  photon+jet production and $d\mathcal{PS}=dy_1dy_2 dp_{T 1} dp_{T 2}$ for 
photon+hadron production.

The enhancement  of the transverse momentum imbalance (or nuclear broadening)  
in $h+A$ ($h=p$, $\gamma^*$) collisions relative to $h+p$ collisions can be 
quantified by the difference:
\ben
\Delta \langle q_T^2\rangle = \langle q_T^2\rangle_{hA} 
- \langle q_T^2\rangle_{hp} \, \, .
\label{master}
\een
The broadening $\Delta \langle q_T^2\rangle$ is a result of multiple quark and 
gluon scattering and is a direct probe of the properties of the nuclear medium.
Both initial-state and final-state multiple parton interactions are taken into
account to calculate the nuclear broadening 
$\Delta \langle q_T^2\rangle$ for photon+jet (photon+hadron) and heavy 
quark (heavy meson) pair production in $p+A$ collisions.

The nuclear broadening $\Delta \langle q_T^2\rangle$ 
can be calculated in perturbative QCD. A specific method based on double 
parton scattering has been discussed in detail in 
Refs.~\cite{Kang:2008us,Kang:2011bp}. The derivation discussed here closely 
follows Ref.~\cite{Kang:2011bp}. The leading contribution to the nuclear 
broadening comes from double scattering: either in the initial-state
or the final-state.
The contributions from these diagrams in the covariant gauge can be
calculated to obtain the following 
expression for the nuclear broadening of photon+jet production in $p+A$ 
collisions:
\ben
\Delta\langle q_T^2\rangle
& = & \left(\frac{8\pi^2\alpha_s}{N_c^2-1}\right) \nonumber \\ 
&  & \mbox{} \times
\frac{ \sum_{a, b}f_{a/p}(x') 
\left[T_{b/A}^{(I)}(x) H^I_{ab\to\gamma d} (\hat s, \hat t, \hat u)+
T_{b/A}^{(F)}(x) H^F_{ab\to\gamma d} (\hat s, \hat t, \hat u)\right]}
{\sum_{a,b} f_{a/p}(x') f_{b/A}(x) 
H^{U}_{ab\to\gamma d}(\hat s, \hat t, \hat u)} \, \, .
\label{imbalance}
\een
Here $T_{b/A}^{(I)}(x) = T_{q/A}^{(I)}(x)$ (or $T_{g/A}^{(I)}(x)$) are twist-4 
quark-gluon (or gluon-gluon) correlation functions associated with 
initial-state multiple scattering, defined as~\cite{Kang:2008us,Kang:2011bp}
\ben
T_{q/A}^{(I)}(x) &=&
 \int \frac{dy^{-}}{2\pi}\, e^{ixp^{+}y^{-}}
 \int \frac{dy_1^{-}dy_{2}^{-}}{2\pi} \,
      \theta(y^{-}-y_1^{-})\,\theta(-y_{2}^{-}) \nonumber \\
 & & \mbox{} \times     \frac{1}{2}\,
     \langle p_{A}|F_{\alpha}^{\ +}(y_{2}^{-})\bar{\psi}_{q}(0)
                  \gamma^{+}\psi_{q}(y^{-})F^{+\alpha}(y_1^{-})
     |p_{A} \rangle \, \, ,
\label{TqA}
\\
T_{g/A}^{(I)}(x) &=&
 \int \frac{dy^{-}}{2\pi}\, e^{ixp^{+}y^{-}}
 \int \frac{dy_1^{-}dy_{2}^{-}}{2\pi} \,
      \theta(y^{-}-y_1^{-})\,\theta(-y_{2}^{-}) \nonumber \\
& & \mbox{} \times \frac{1}{xp^+}\,
\langle p_A| F_\alpha^{~+}(y_2^-)
F^{\sigma+}(0)F^+_{~\sigma}(y^-)F^{+\alpha}(y_1^-)|p_A\rangle\, \, .
\label{TgA}
\een
The corresponding twist-4 correlation functions associated with final-state
multiple scatter are $T_{q/A}^{(F)}(x)$ and $T_{g/A}^{(F)}(x)$.
They are defined as in Eqs.~(\ref{TqA}) and (\ref{TgA}), except  
the $\theta$-functions are replaced such that \cite{Kang:2008us,Kang:2011bp},
\ben
\theta(y^{-}-y_1^{-})\,\theta(-y_{2}^{-})
\to
\theta(y_1^{-}-y^{-})\,\theta(y_{2}^{-}) \, \, .
\label{theta}
\een
The broadening in photon + hadron production is calculated similarly.

Nuclear broadening is evaluated in a formalism where multiple scattering 
contributes to the cross section via higher-twist
matrix elements in the nuclear state. This framework follows a well-established 
QCD factorization formalism 
for particle production in $p+A$ collisions and
has previously been used to describe 
cold nuclear matter effects such as energy loss,
dynamical shadowing and broadening.
This work differs 
from more generic parton broadening phenomenology in because the color and 
kinematic structures of the hard part are evaluated exactly.  In particular, 
the nuclear enhancement of the transverse momentum imbalance is studied in 
dijet and photon+jet production. 

\begin{figure}[htpb]
\begin{center}
\includegraphics[width=0.495\textwidth]{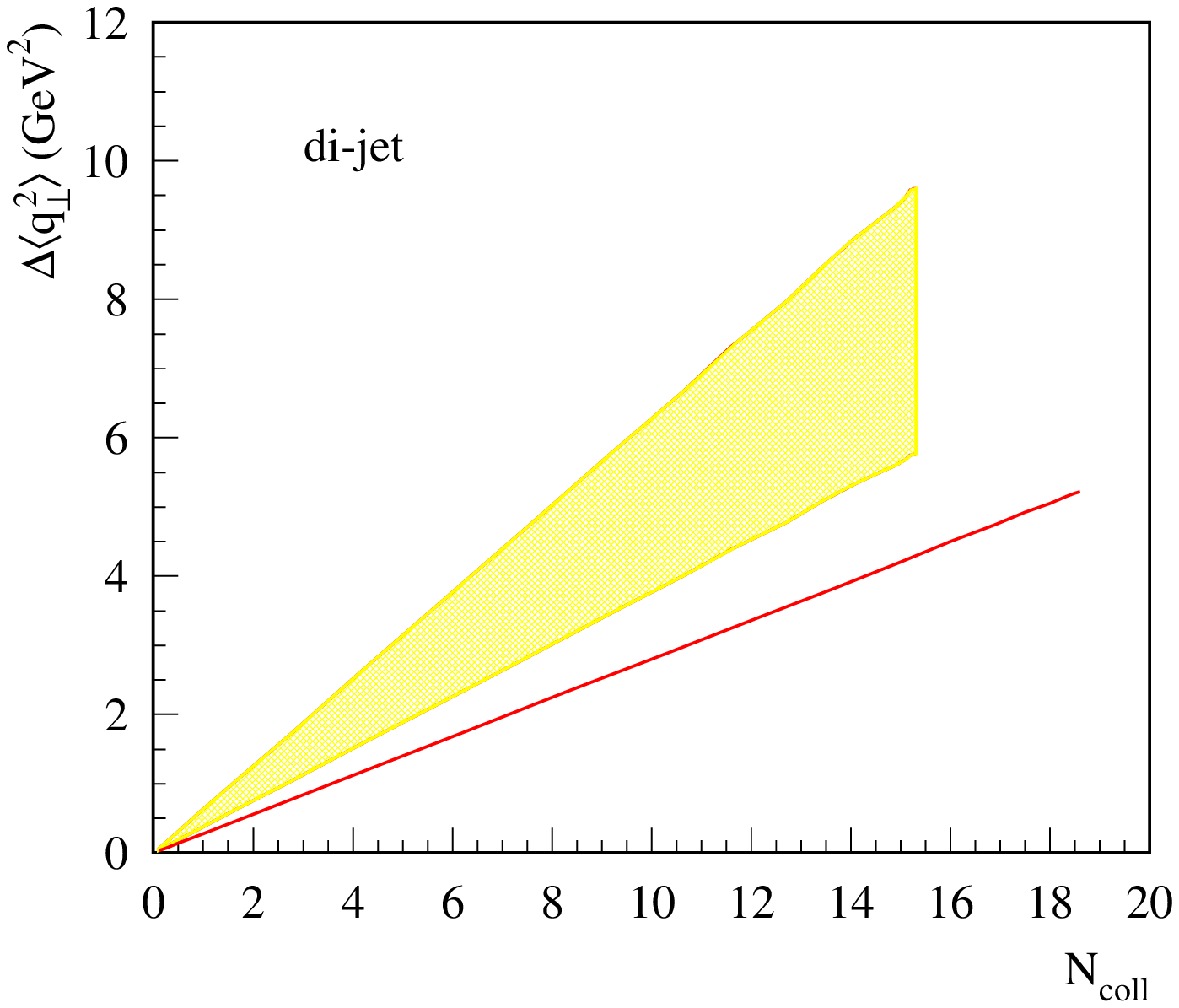}
\includegraphics[width=0.495\textwidth]{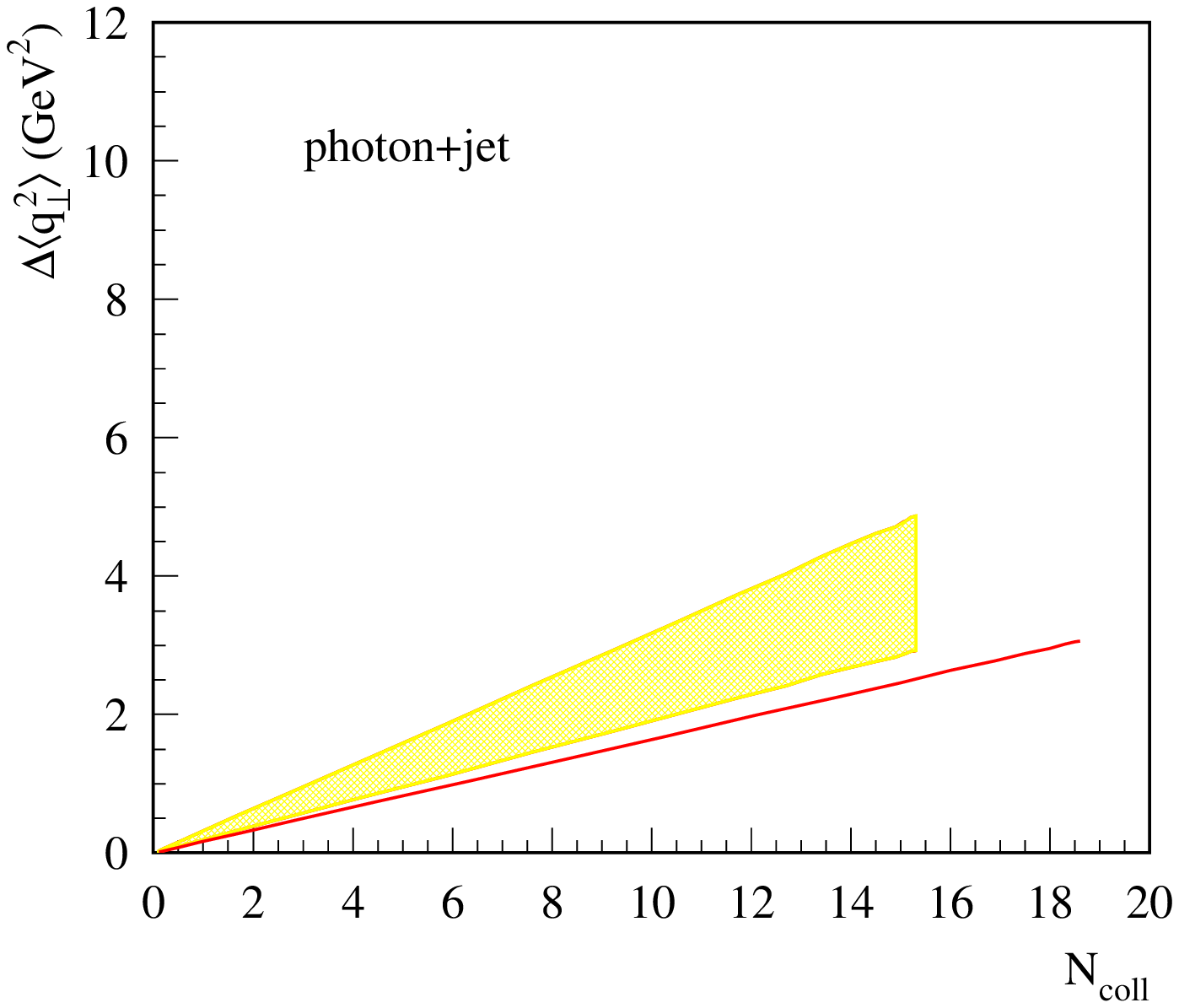}
\end{center}
\caption[]{Nuclear broadening, $\Delta\langle q_{T}^2\rangle$, is shown for 
dijet (left) and photon+jet (right) production at specified rapidities
in $p+A$ collisions as a function of $N_{\rm 
coll}$. Results are shown for $y_1=y_2=2$ in $\sqrt{s}=5$ TeV $p+$Pb collisions
and $y_1=y_2=1$ at $\sqrt{s}=200$ GeV d+Au collisions. (The calculations have  
been made with the convention that the proton beam moves toward
forward rapidity.) The jet transverse
momentum is integrated over $30<p_T<40$~GeV/$c$ at $\sqrt{s} = 5$ TeV and
$15<p_T<25$~GeV/$c$ at $\sqrt{s} = 200$ GeV. The red line shows the result
for RHIC kinematics with scattering parameter $\xi^2 = 0.12$ GeV$^2$ while the 
yellow band represents the variation of $\xi^2$ in the LHC kinematics. 
}
\label{fig:heavy}
\end{figure}

The results are shown in Fig.~\ref{fig:heavy}.  The line gives the baseline
$\Delta \langle q_T^2 \rangle$ determined from RHIC as a function of 
$N_{\rm coll}$.  The band is the result in the LHC kinematics for a plausible
range of the scattering parameter $\xi^2$, defined in Eq.~(\ref{eq:dyn_shad}).  
The band is broader for dijet
production and $\Delta \langle q_T^2 \rangle$ is also larger.  There is also a
somewhat larger deviation of the bottom edge of the band from the RHIC result
for dijet production.

\section{Gauge Bosons}
\label{sec:WZ_prod}

Gauge boson production is discussed in this section.  Section~\ref{sec:Zhang_WZ}
makes predictions for $W$ and $Z^0$ $p_T$ and rapidity distributions in $p+p$
and $p+$Pb collisions.  The $W^{\pm}$ charge asymmetry as a function of
the decay lepton rapidity is also shown. Section~\ref{sec:Zsum} describes a
calculation of the resummed $Z^0$ $p_T$ distribution while transverse broadening
of vector boson production is discussed in Sec.~\ref{sec:Zbroad}.

\subsection[$W$ and $Z$ production and $W^\pm$ charge asymmetry]{$W$ and $Z$ production and $W^\pm$ charge asymmetry (P. Ru, E. Wang, B.-W. 
Zhang and W.-N. Zhang)}
\label{sec:Zhang_WZ}

Production of the gauge bosons, $W^+, W^-$ and $Z^0$, is discussed.  The 
transverse momentum and the rapidity distributions in min-bias
$p+$Pb collisions at $\sqrt{s}=5$~TeV are calculated
at next-to-leading order and next-to-next-to-leading order.
The fiducial cross sections $\sigma_{\rm fid}$ are calculated in the fiducial 
phase space for vector boson production used by ATLAS Collaboration
~\cite{Aad:2011gj,Aad:2011fp} in both $p+$Pb and $p+p$ collisions.

The fiducial $Z^0$ cross section is the inclusive cross section $p+ \, {\rm Pb}
\to Z^0/\gamma^*+X$ multiplied by the branching ratio for $Z^0/\gamma^* \to
l^+l^-$ within the fiducial acceptance. Here $X$ denotes the
underlying event and the recoil system. The fiducial acceptance is
assumed to be the same as that defined by ATLAS in $p+p$ 
collisions~\cite{Aad:2011gj} with the following cuts on the lepton transverse
 momentum and pseudorapidity, and the dilepton invariant mass: 
$p_T^l>20$~GeV/$c$; $|\eta^l|<2.4$;
and $66 < m_{ll} < 116$~GeV/$c^2$.

The fiducial $W^\pm$ cross sections are the inclusive cross sections $p+ \,
{\rm Pb} \to W^\pm +X$ multiplied by the branching ratios for $W^\pm \to
l\nu$ within the fiducial acceptance.  Following the ATLAS 
definition~\cite{Aad:2011fp}, the acceptance cuts on lepton and neutrino 
transverse momentum and pseudorapidity as well as the $W$ transverse mass are: 
$p_T^l>20$~GeV/$c$; $|\eta_{l, \nu}|<2.4$; $p_T^{\nu}>25$~GeV/$c$; and
$m_T=\sqrt{2p_T^lp_T^{\nu}(1-\cos(\phi^l-\phi^{\nu}))}>40$~GeV/$c^2$.

Results are simulated to NLO ($\mathcal {O}(\alpha_s)$) and NNLO 
($\mathcal {O}(\alpha_s^2)$) in the total cross section employing 
$\mathtt{DYNNLO}$ for Drell-Yan-like production in hadron-hadron
collisions~\cite{Catani:2009sm}. 
The MSTWNLO and MSTWNNLO proton parton densities
are used, along with the EPS09~\cite{Eskola:2009uj} and DSSZ~\cite{deFlorian:2011fp}
shadowing parameterizations.

Tables~\ref{wpwmz_sigpptable} and \ref{wpwmz_sigpAtable}
shows the fiducial cross sections, normalized to
their per nucleon values for better comparison. The NLO and NNLO values are
shown for comparison, both for $p+$Pb and $p+p$ collisions at the same energy.
The ratio between the NNLO and NLO cross section, giving some indication of
the theoretical uncertainty and the convergence of the perturbative expansion
for gauge boson production, is $\sim 1.02$ showing that the higher order
corrections are small.  When the NLO and NNLO results are compared in $p+$Pb
collisions with the EPS09 shadowing parameterization, a similar correction
is found.  The difference between the EPS09 and DSSZ parameterizations, both
calculated at NLO, is also quite small, on the order of 1\%.  There is a
slight decrease in the total fiducial $Z^0$ cross section.  There is a larger
decrease for $W^+$ and an enhancement in the fiducial cross section in $p+$Pb
collisions.  This is less an effect of shadowing than it is of isospin since
$u \overline d \to W^+$ and $\overline u d \to W^-$ and there are more $d$
quarks in the lead nucleus, causing the enhanced cross section.

\begin{table}[htbp]
\tbl{The total vector boson production
cross sections in the fiducial phase space, $\sigma_{\rm fid}$, in units of nb
in $p+p$ collisions. The results at NLO and NNLO are compared.}
{\begin{tabular}{@{}ccc@{}}\toprule 
\multicolumn{3}{c}{$p+p$ $\sigma_{\rm fid}$ (nb)} \\
Decay channel  & MSTWNNLO & MSTWNLO \\ \hline
$Z\to e^+e^-$   & 0.339 & 0.332 \\
$W^+\to e^+\nu$ & 2.35 & 2.30 \\
$W^-\to e^-\nu$ & 1.47 & 1.44 \\
 \hline
\end{tabular}}
\label{wpwmz_sigpptable}
\end{table}

\begin{table}[htbp]
\tbl{The total vector boson production
cross sections per nucleon in the fiducial phase space, 
$\sigma_{\rm fid}/\langle N_{\rm bin}\rangle$
in $p+$Pb collisions.  Columns 2 and 3 compare the results
at NLO and NNLO calculated with EPS09 while columns 3 and 4 compare
the EPS09 and DSSZ shadowing parameterizations at NLO.}
{\begin{tabular}{@{}cccc@{}}\toprule 
\multicolumn{4}{c}{$p+$Pb $\sigma_{\rm fid}/\langle N_{\rm coll}\rangle$ (nb)} \\
 & MSTWNNLO & MSTWNLO & MSTWNLO \\
Decay channel & EPS09 & EPS09 & DSSZ \\ \hline
$Z\to e^+e^-$ & 0.338 & 0.328 & 0.329 \\
$W^+\to e^+\nu$ & 2.06 & 2.02 & 2.05  \\
$W^-\to e^-\nu$ & 1.54 & 1.52 & 1.53  \\
 \hline
\end{tabular}}
\label{wpwmz_sigpAtable}
\end{table}

\begin{figure}[htpb]
\begin{center}
\includegraphics[width=0.495\textwidth]{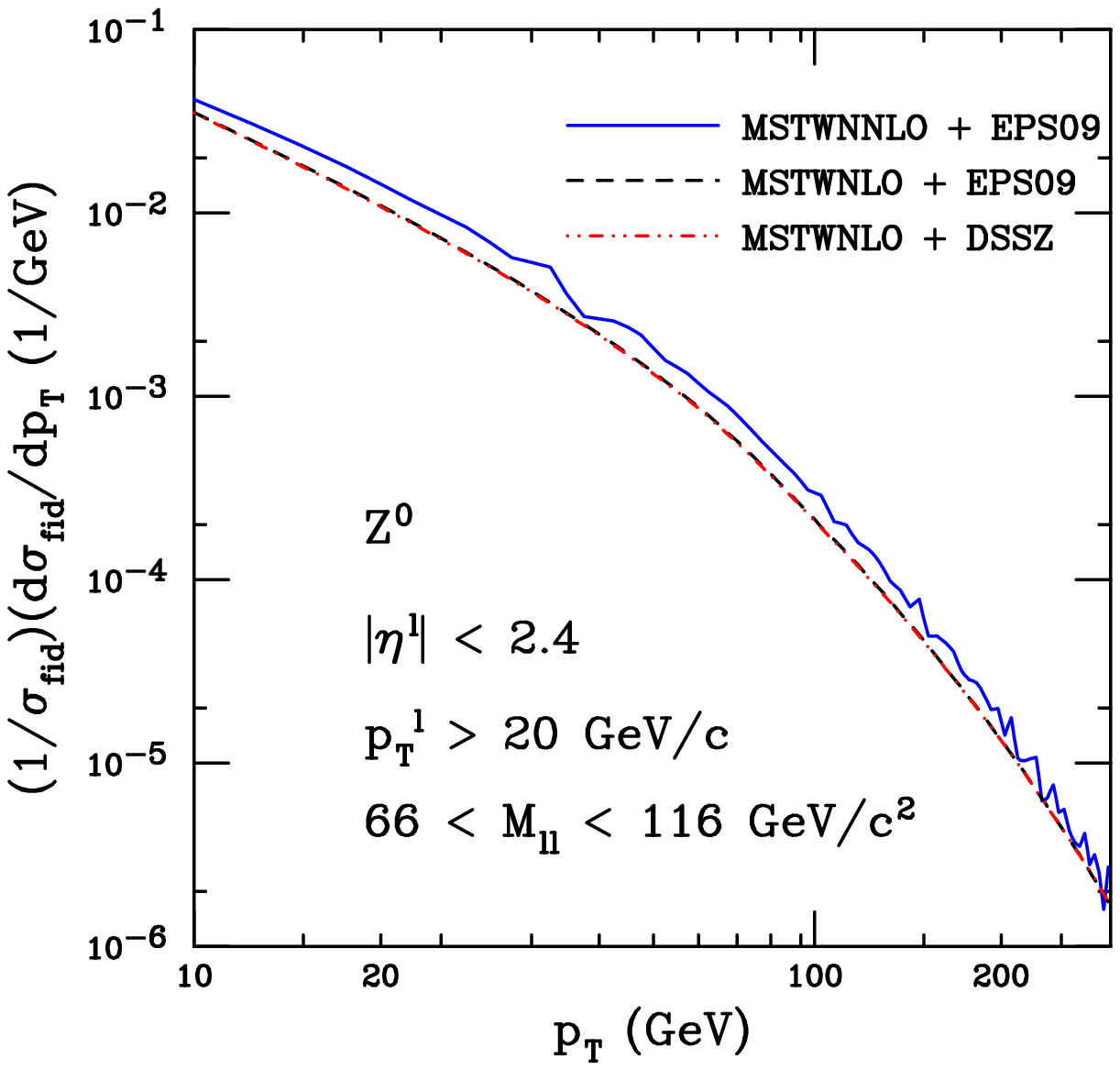}
\includegraphics[width=0.495\textwidth]{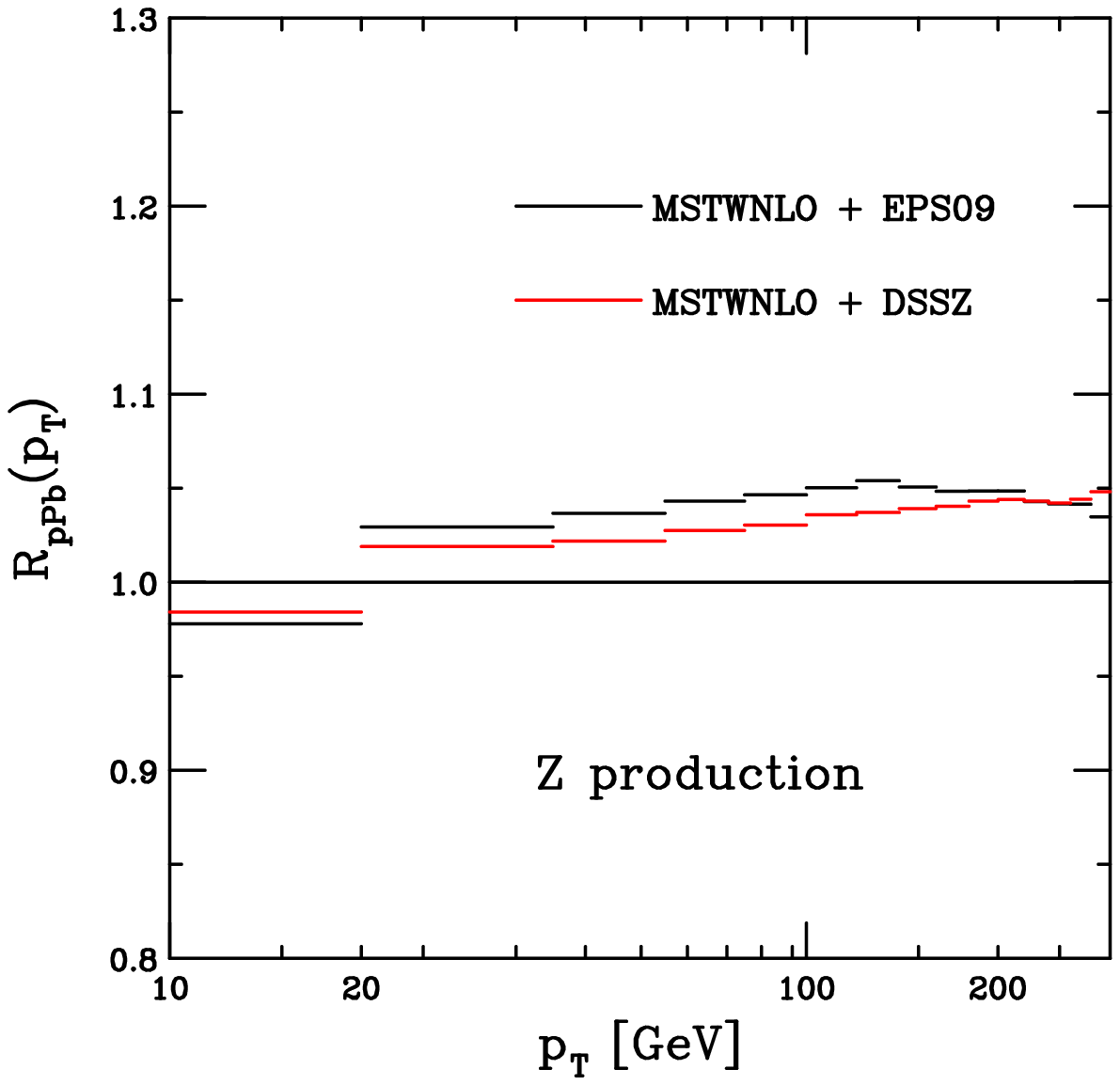}
\end{center}
\caption[]{(Left) Normalized $Z^0$ differential cross section
$(1/\sigma_{\rm fid})(d\sigma_{\rm fid}/dp_T)$. (Right) The suppression factor
$R_{pA}(p_T)$ in 20~GeV/$c$ $p_T$ bins.}
\label{fig:zdist_zrpa}
\end{figure}

Figure~\ref{fig:zdist_zrpa} shows the $Z^0$ $p_T$ distributions in $p+$Pb
collisions at both NLO and NNLO.  The NNLO result is somewhat higher and not
as smooth as the NLO calculations which appear independent of the choice of
shadowing parameterization.  Differences between the results with EPS09 and
DSSZ, which can be attributed to shadowing effects rather than isospin, are 
only apparent when the ratio $R_{p \, {\rm Pb}}(p_T)$ is formed.  The slight
decrease in per nucleon yield in $p+$Pb relative to $p+p$ seen in the total
cross sections in Tables~\ref{wpwmz_sigpptable} and \ref{wpwmz_sigpAtable}
are due to the lowest $p_T$ bin,
$p_T < 20$~GeV/$c$.  At higher $p_T$, the ratio increases above unity.  
However, the effect is not significantly larger than 5\% over the entire
$p_T$ range.

\begin{figure}[htpb]
\begin{center}
\includegraphics[width=0.495\textwidth]{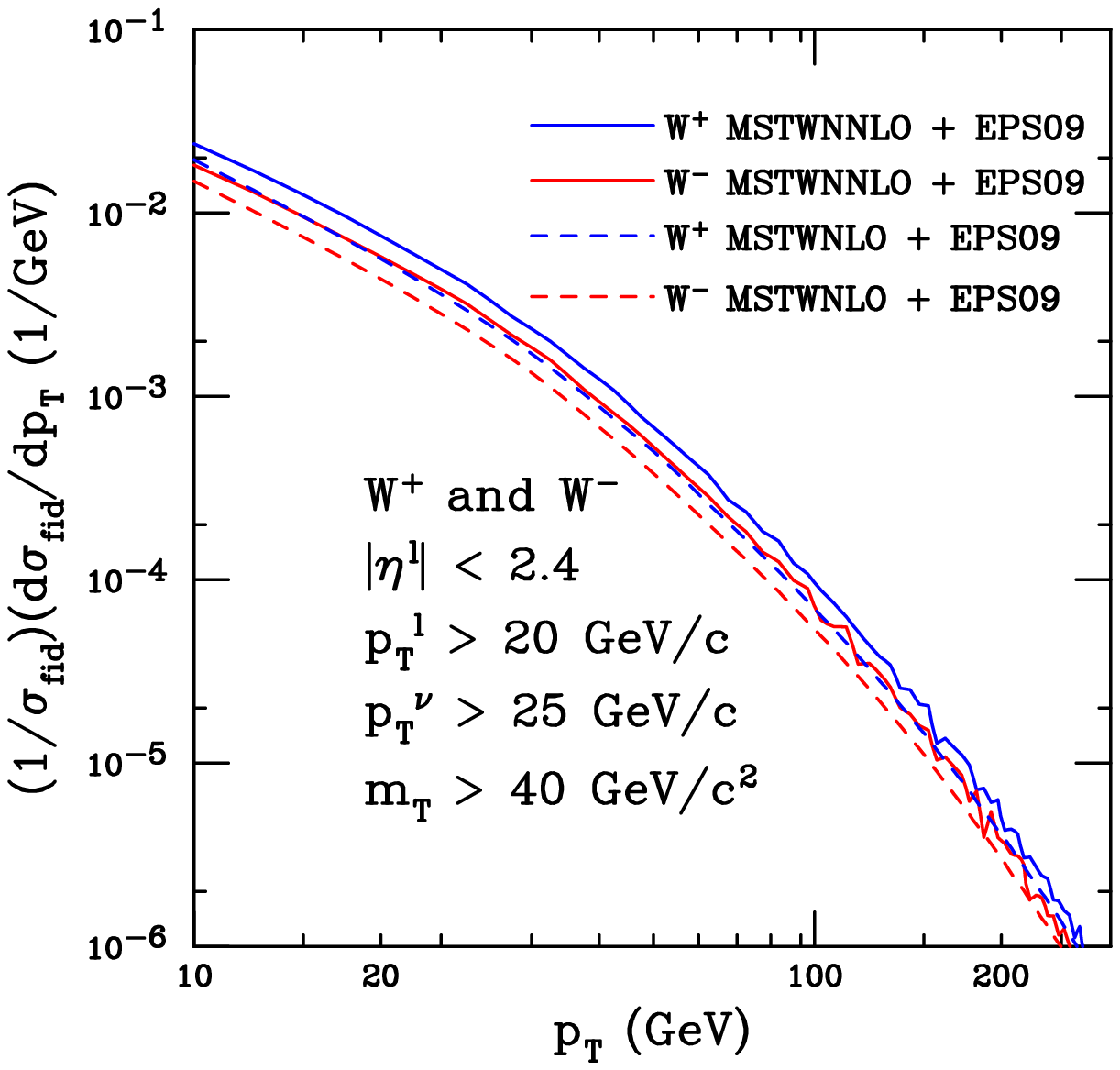}
\includegraphics[width=0.495\textwidth]{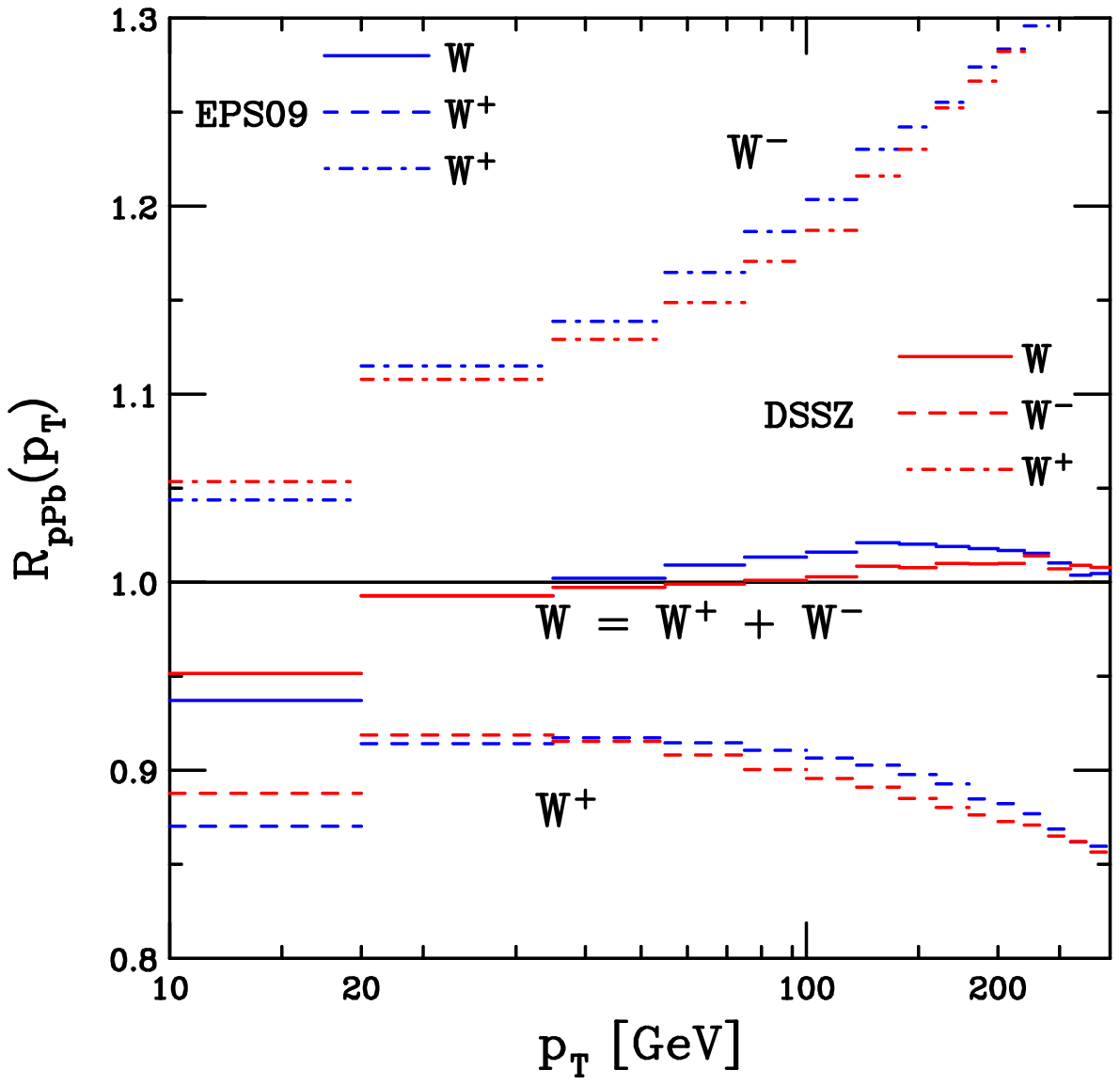}
\end{center}
\caption[]{(Left)
The normalized $W = W^+ + W^-$ differential cross section 
$(1/\sigma_{\rm fid})(d\sigma_{\rm fid}/dp_T)$.
(Right) The suppression factor $R_{pA}(p_T)$ for $W$ production given in
20~GeV/$c$ $p_T$ bins.}
\label{fig:wdist_wrpa}
\end{figure}

The individual $W^+$ and $W^-$ $p_T$ distributions are shown at NLO and NNLO
on the left-hand side of Fig.~\ref{fig:wdist_wrpa}.  Again, the NLO results are
smoother than the NNLO calculations.  The difference in the overall cross
sections are clearly observable: in the fiducial range of the calculations,
the $W^+$ cross section is $\approx 35$\% greater than the $W^-$ 
cross section.  The ratio of $p+$Pb to $p+p$ is shown on the right-hand side
of Fig.~\ref{fig:wdist_wrpa}.  The $W^-$ ratio is larger than unity and increasing
strongly with $p_T$.  The effect is due to the greater abundance of $d$
quarks in the Pb nucleus (126 neutrons vs.\ 82 protons).  The valence $d$ quark
distributions in the neutrons, equivalent to the valuence $u$ quark 
distributions in the protons, have a larger density at relatively high $x$,
causing the observed increase.  Conversely, the lower density of valence $u$
quarks in the Pb nucleus causes $R_{p {\rm Pb}} < 1$ over the entire $p_T$
range.  The sum of the two charged gauge bosons shows a trend very similar to
that of the $Z^0$ in Fig.~\ref{fig:zdist_zrpa}, revealing a result closer to
the true shadowing effect.  Even though the $W^+$ cross
section is greater than that of the $W^-$, the isospin effect on the $W^-$ is
large enough to make the ratio larger than unity at high $p_T$.

\begin{figure}[htpb]
\begin{center}
\includegraphics[width=0.495\textwidth]{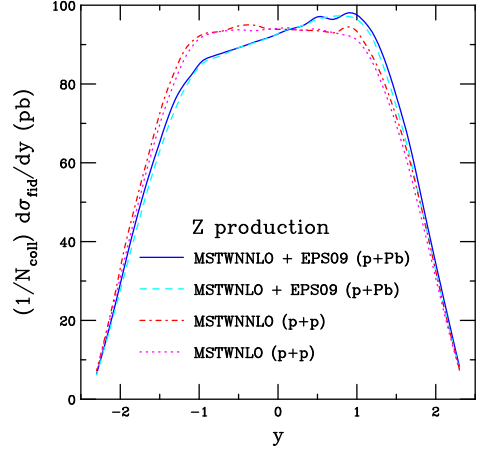}
\end{center}
\caption[]{The $Z^0$ rapidity distribution 
$(1/\langle N_{\rm coll}\rangle)(d\sigma_{\rm fid}/dy)$.  Results are shown for
both $p+$Pb and $p+p$ collisions in the center-of-mass frame for both systems.}
\label{fig:zydist}
\end{figure}

\begin{figure}[htpb]
\begin{center}
\includegraphics[width=0.495\textwidth]{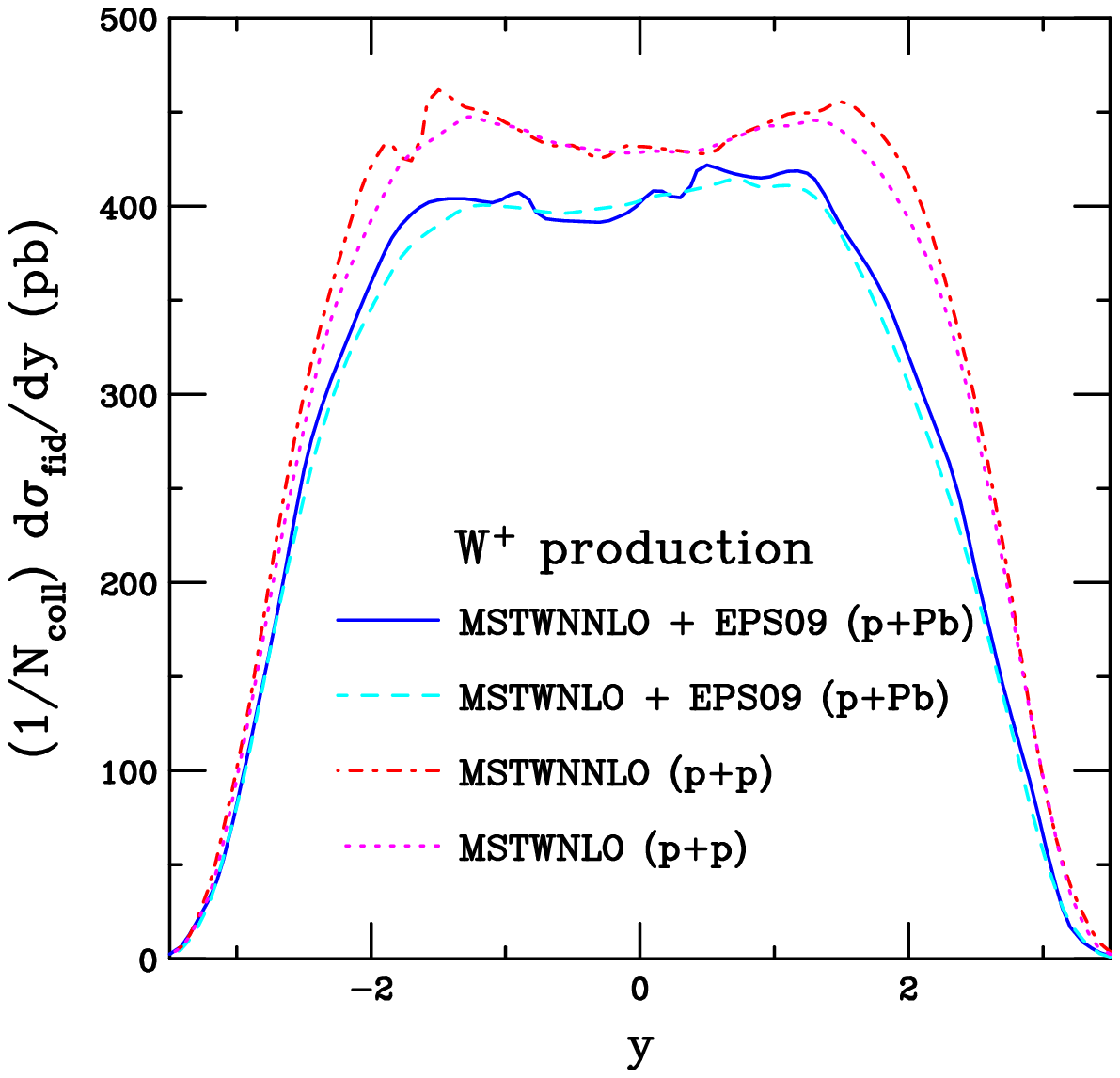}
\includegraphics[width=0.495\textwidth]{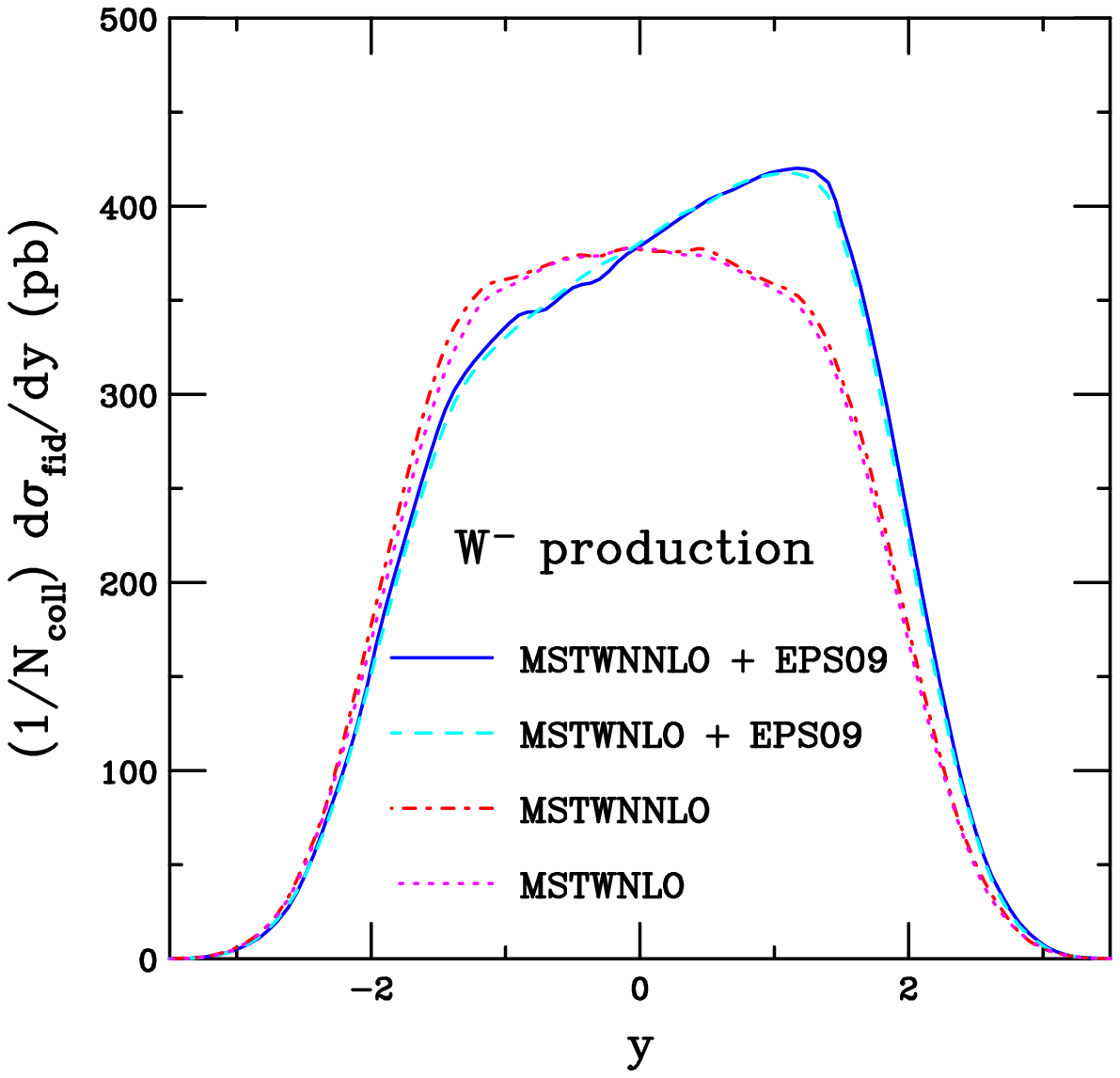}
\end{center}
\caption[]{The $W^+$ (left)
and $W^-$ (right) rapidity distributions 
$(1/\langle N_{\rm coll}\rangle)(d\sigma_{\rm fid}/dy)$.  Results are shown for
both $p+$Pb and $p+p$ collisions in the center-of-mass frame for both systems.}
\label{fig:wpwm_ydist}
\end{figure}

Figures~\ref{fig:zydist} and \ref{fig:wpwm_ydist} show the $Z^0$, $W^+$ and
$W^-$ rapidity distributions, normalized per nucleon.  For comparison the 
$p+p$ and $p+$Pb distributions are shown both at NLO and NNLO.  Aside from
numerical fluctuations at NNLO, the order of the calculation makes little
difference in either the shape or the magnitude of the rapidity distributions.
The $p+p$ distributions are all symmetric around $y=0$ while the $p+$Pb
distributions are peaked in the direction of the Pb nucleus.

There are significant differences between the $W^+$ and $W^-$ distributions
even for $p+p$ collisions.  The $W^+$ distribution is considerably broader
with peaks away from $y=0$, at $|y| \sim 1.7$ due to the larger average
momentum fraction $x$ of the
valence $u$ quarks in the proton.
The greater density of valence $u$ quarks in the proton leads to the larger
overall cross section.  The isospin effect tends to make the $W^+$ rapidity
distribution more symmetric since the valence $u$ quark distribution in the
neutron, equivalent to the valence $d$ distribution in the proton, has a smaller
average $x$ and lower density which reduces the cross section while removing
the peaks away from midrapidity.  The $W^-$ distribution in $p+p$ collisions
is both smaller and narrower than the $W^+$.  This distribution shows the
strongest isospin effect with rapidity.

\begin{figure}[htpb]
\begin{center}
\includegraphics[width=0.55\textwidth]{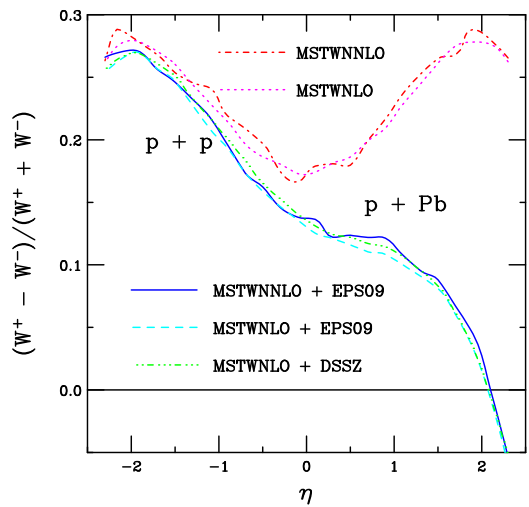}
\end{center}
\caption[]{The $W^\pm$ charge asymmetry, $(N_{W^+}-N_{W^-})/(N_{W^+}+N_{W^-})$, as a 
function of the charged lepton pseudorapidity
in both $p+p$ and $p+$Pb collisions in the center-of-mass frame for both 
systems.}
\label{fig:wpwm_asym}
\end{figure}

These differences are reflected in the $W^\pm$ charge asymmetry, defined
as $(N_{W^+}-N_{W^-})/(N_{W^+}+N_{W^-})$ and shown as a function of the decay
lepton pseudorapidity.  The results are given in Fig.~\ref{fig:wpwm_asym}.
The $p+p$ asymmetry is symmetric around $y=0$ with a strong dip at midrapidity.
The origin is clear from the individual $W^+$ and $W^-$ rapidity distributions.
There is, however, a strong forward/backward asymmetry in $p+$Pb collisions.
In the direction of the proton beam, the asymmetry follows that of the $p+p$
result.  It falls off and becomes negative in the direction of the lead beam.
This trend is independent of the order of the calculation and the shadowing
parameterization used.

\subsection[Nuclear modification of the transverse momentum 
spectrum of $Z^0$ production]{Nuclear modification of the transverse 
momentum 
spectrum of $Z^0$ production (Z.-B. Kang and J.-W. Qiu)}
\label{sec:Zsum}
 
The Collins-Soper-Sterman formalism \cite{Collins:1984kg} is used to calculate 
$Z^0$ production over the full $p_T$ range at the LHC,
\begin{eqnarray}
\frac{d\sigma_{A+B\rightarrow Z^0+X}}{dy\, dp_T^2} &=&
\frac{1}{(2\pi)^2}\int d^2b\, e^{i\vec{p}_T\cdot \vec{b}}\,
\tilde{W}(b,M_Z,x_1,x_2) 
+ Y(p_T,M_Z,x_1,x_2)\, .
\label{css-gen}
\end{eqnarray}
The $\tilde{W}$ term gives the dominant contribution when
$p_T\ll M_Z$ while the $Y$ term is perturbatively calculable, see 
Ref.~\cite{Qiu:2000hf}, allowing a smooth transition from the resummed low 
$p_T$ region to $p_T\sim M_Z$ where the fixed-order perturbative QCD 
calculations work well.  
In Eq.~(\ref{css-gen}), $x_1= e^y\, M_Z/\sqrt{s}$ and 
$x_2= e^{-y}\, M_Z/\sqrt{s}$ while $\tilde{W}$
is given by \cite{Qiu:2000hf}
\begin{equation}
\tilde{W}(b,M_Z,x_1,x_2) = \left\{
\begin{array}{ll}
 \tilde{W}^{\rm P}(b,M_Z,x_1,x_2) & \quad \mbox{$b\leq b_{\rm max}$} \\
 \tilde{W}^{\rm P}(b_{\rm max},M_Z,x_1,x_2)\,
 \tilde{F}^{\rm NP}(b,M_Z,x_1,x_2;b_{\rm max})
                        & \quad \mbox{$b > b_{\rm max}$}
\end{array} \right.
\label{qz-W-sol}
\end{equation}
where $b_{\rm max} \sim 1/$(few GeV) is a parameter that specifies the region in 
which $\tilde{W}^{\rm P}$ is perturbatively valid, and $\tilde{F}^{\rm NP}$ is a 
nonperturbative function determining the large $b$ behavior of $\tilde{W}$ 
and is defined below.  
In Eq.~(\ref{qz-W-sol}), $\tilde{W}^{\rm P}(b,M_Z,x_1,x_2)$ 
includes all powers of large perturbative logarithms resummed  
from $\ln(1/b^2)$ to $\ln(M_Z^2)$
\cite{Collins:1984kg} 
\begin{equation}
\tilde{W}^{\rm P}(b,M_Z,x_1,x_2) = 
{\rm e}^{-S(b,M_Z)}\, \tilde{W}^{\rm P}(b,c/b,x_1,x_2)\, ,
\label{css-W-sol}
\end{equation}
where $c$ is a constant of order one \cite{Collins:1984kg,Qiu:2000hf}, and
\begin{equation}
S(b,M_Z) = \int_{c^2/b^2}^{M_Z^2}\, 
  \frac{d{\mu}^2}{{\mu}^2} \left[
  \ln\left(\frac{M_Z^2}{{\mu}^2}\right) 
  A(\alpha_s({\mu})) + B(\alpha_s({\mu})) \right],
\end{equation}
with perturbatively-calculated coefficients $A(\alpha_s)$ and $B(\alpha_s)$ 
given in Ref.~\cite{Qiu:2000hf} and references therein.  
The perturbative factor in Eq.~(\ref{css-W-sol}), 
$\tilde{W}^{\rm P}(b,c/b,x_1,x_2)$, has no large logarithms.  It is expressed
as
\begin{equation}
\tilde{W}^{\rm P}(b,c/b,x_1,x_2) =\sigma_0 \sum_{i=q,\bar{q}}
f_{i/A}(x_1,\mu=c/b)\, f_{\bar{i}/B}(x_2,\mu=c/b)\, 
\label{css-W-pert}
\end{equation}
where $\sigma_0$ is the leading order $q\overline q \rightarrow Z^0$ partonic 
cross section \cite{Qiu:2000hf}.  The functions $f_{i/A}$ and $f_{\bar{i}/B}$ 
are the modified parton distributions given by  \cite{Collins:1984kg} 
\begin{equation}
f_{i/A}(x_1,\mu) = \sum_a 
  \int_{x_1}^1\frac{d\xi}{\xi}\, 
  C_{i/a}(x_1/\xi,\mu) \, \phi_{a/A}(\xi,\mu)
\label{mod-pdf}
\end{equation}
where $\sum_a$ is over $a=q, \bar{q}, g$, 
$\phi_{a/A}(\xi,\mu)$ are the normal proton or effective nuclear parton 
distribution functions (PDFs) 
and $C_{i/a}=\sum_{n=0} C_{i/a}^{(n)} (\alpha_s/\pi)^n$
are perturbatively calculable coefficient functions for finding a
parton $i$ from a parton $a$, given in Ref.~\cite{Qiu:2000hf}.  

The non-perturbative function $\tilde{F}^{\rm NP}$ in Eq.~(\ref{qz-W-sol}) has 
the form, 
\begin{eqnarray}
F^{\rm NP}(b,M_Z,x_1,x_2;b_{\rm max}) 
& = & 
\exp\Bigg\{ -\ln\left(\frac{M_Z^2 b_{\rm max}^2}{c^2}\right) \left[
    g_1 \left( (b^2)^\alpha - (b_{\rm max}^2)^\alpha\right) \right. 
\nonumber \\
& + & 
\left. g_2 \left( b^2 - b_{\rm max}^2\right) \right] 
   -\bar{g}_2 \left( b^2 - b_{\rm max}^2\right) \Bigg\}\, .
\label{qz-fnp-m}
\end{eqnarray}
where the explicit logarithmic dependence, $\ln(M_Z^2\,b_{\rm max}^2/c^2)$, 
was derived by solving the Collins-Soper equation \cite{Collins:1984kg}. 
The ${g}_2$ term is a result of adding a general power correction to the 
renormalization group equation while the $\bar{g}_2$ term represents the size 
of the intrinsic transverse momentum of active partons 
\cite{Qiu:2000hf}.  

The coefficients of the two terms proportional to $b^2$ in Eq.~(\ref{qz-fnp-m}) 
can be combined \cite{Zhang:2002yz},
\ben
G_2 = \ln\left(\frac{M^2b_{\rm max}^2}{c^2}\right)\, g_2 + \bar{g}_2\, ,
\een
to sum the dynamical and intrinsic power corrections. 
By requiring the first and second derivatives of $\tilde{W}$ to be continuous 
at $b=b_{\rm max}$, the parameters $\alpha$ and $g_1$ in Eq.~(\ref{qz-fnp-m}) 
can be uniquely fixed, leaving only one parameter, $G_2$, sensitive to the 
power corrections and other nonperturbative effects.
Taking $\bar{g}_2=0.25\pm 0.05$ GeV$^2$ and $g_2=0.01\pm 0.005$ GeV$^2$, 
$G_2^{pp} = 0.324$~GeV$^2$.  Predictions employing Eq.~(\ref{css-gen}) are 
consistent with all $p + \overline p$ and $p+p$ data from the Tevatron and the 
LHC \cite{Qiu:2000hf}. 

The EPS09 NLO parameterization \cite{Eskola:2009uj} is used
to account for the leading-twist nuclear effects on the parton densities in
$p+A$ collisions.  
Following the method proposed in Ref.~\cite{Zhang:2002yz},  
the nuclear-size-enhanced multiple scattering effects are accounted for
by choosing $g_2\to g_2 A^{1/3}$.  Then for $Z^0$ production, $G_2^{p {\rm Pb}} 
= 0.689 {\rm ~GeV}^2$.

In Fig.~\ref{fig:z0}, the predictions for $Z^0$ production are shown. The 
cross section including resummation in Eq.~(\ref{css-gen}) is evaluated 
employing the CTEQ6M parton densities at factorization scale 
$\mu=M_T/2=0.5 \sqrt{M_Z^2+p_T^2}$.  The $Y$ term is calculated at NLO in 
$\alpha_s$ \cite{Qiu:2000hf}.
\bef
\includegraphics[width=0.55\textwidth]{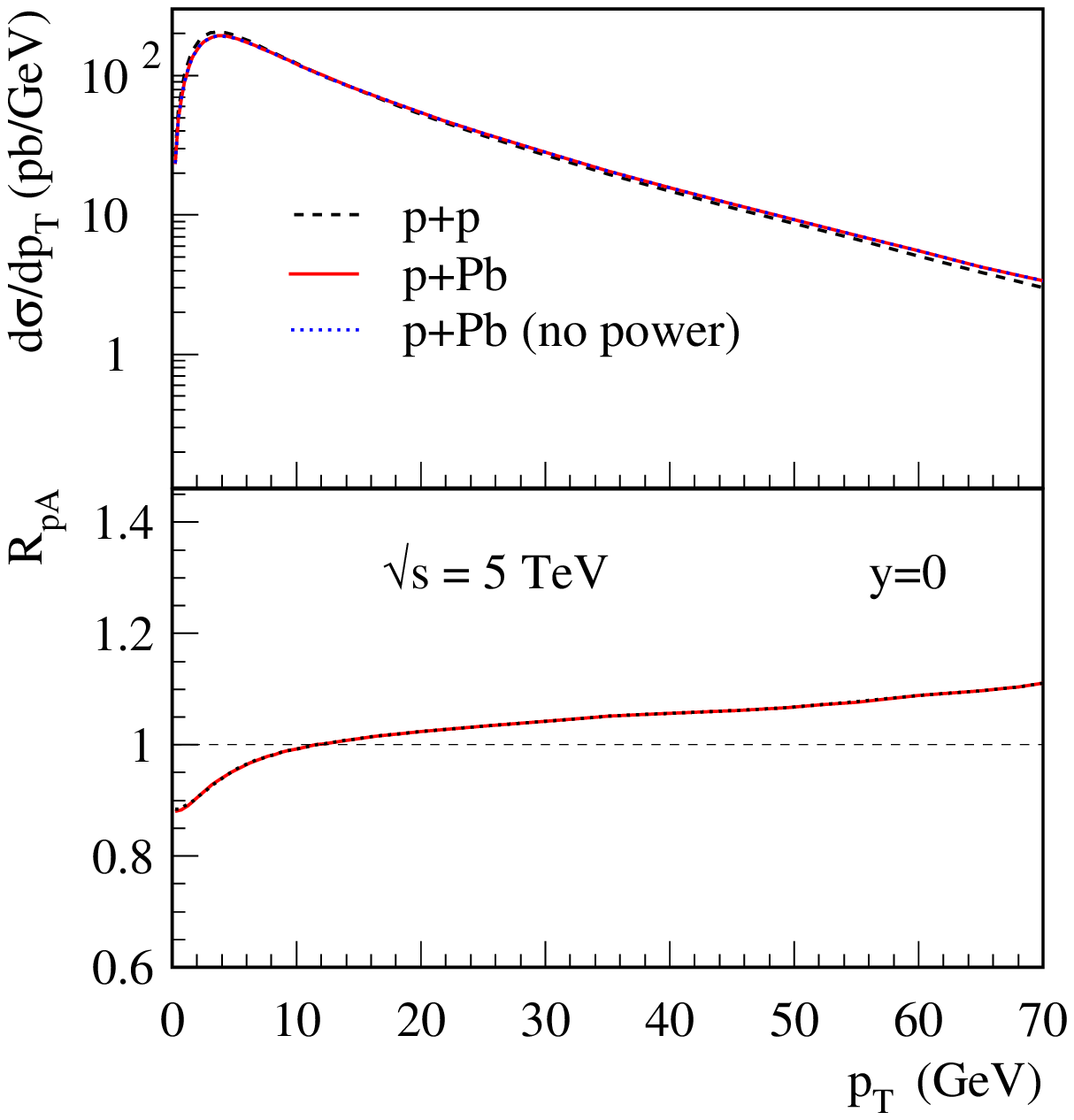}
\caption[]{\label{fig:z0}
$Z^0$ boson production in $p+p$ and $p+$Pb collisions at $\sqrt{s}=5$ TeV 
and $y=0$ \protect\cite{Kang:2012am}.}
\eef
The upper panel shows the $Z^0$ production cross section as a function of $p_T$.
The black dashed curve is the $p+p$ baseline while the red solid curve shows 
the minimum bias $p+$Pb result.  The blue dotted curve is the minimum bias
$p+$Pb result without the $A^{1/3}$ enhancement of $g_2$ so that the 
nuclear-size-enhanced dynamical power corrections from multiple scattering are
absent.   The lower panel presents the nuclear modification factor $R_{pA}$. 

In Fig.~\ref{fig:z0}, the red solid curves are almost indistinguishable from 
the dotted curves.  Thus, power corrections are not important for $Z^0$ 
production at LHC energies.  Therefore, $Z^0$ production in $p+A$ collisions 
is an ideal probe of the modification of the parton densities in nuclei
as well as of the high energy ``isospin'' effect.

The $Z^0$ cross section at $y=0$ is dominated by gluon-initiated subprocesses 
for $p_T > 20$~GeV/$c$.  That is, $R_{pA}$ is an excellent observable to study
nuclear modifications of the gluon distribution, heretofore effectively unknown,
especially at the values of $x$ and $\mu$ probed by $Z^0$ production.  At 
factorization scale $\mu = M_{Z}$, the EPS09 gluon shadowing factor 
\cite{Eskola:2009uj} is less than unity (shadowing) for $x<0.005$ 
and greater than unity (antishadowing) over a sufficiently large range, $0.005 
< x < 0.2$, as shown in Fig.~\ref{fig:eps09}.
\bef
\includegraphics[width=0.48\textwidth]{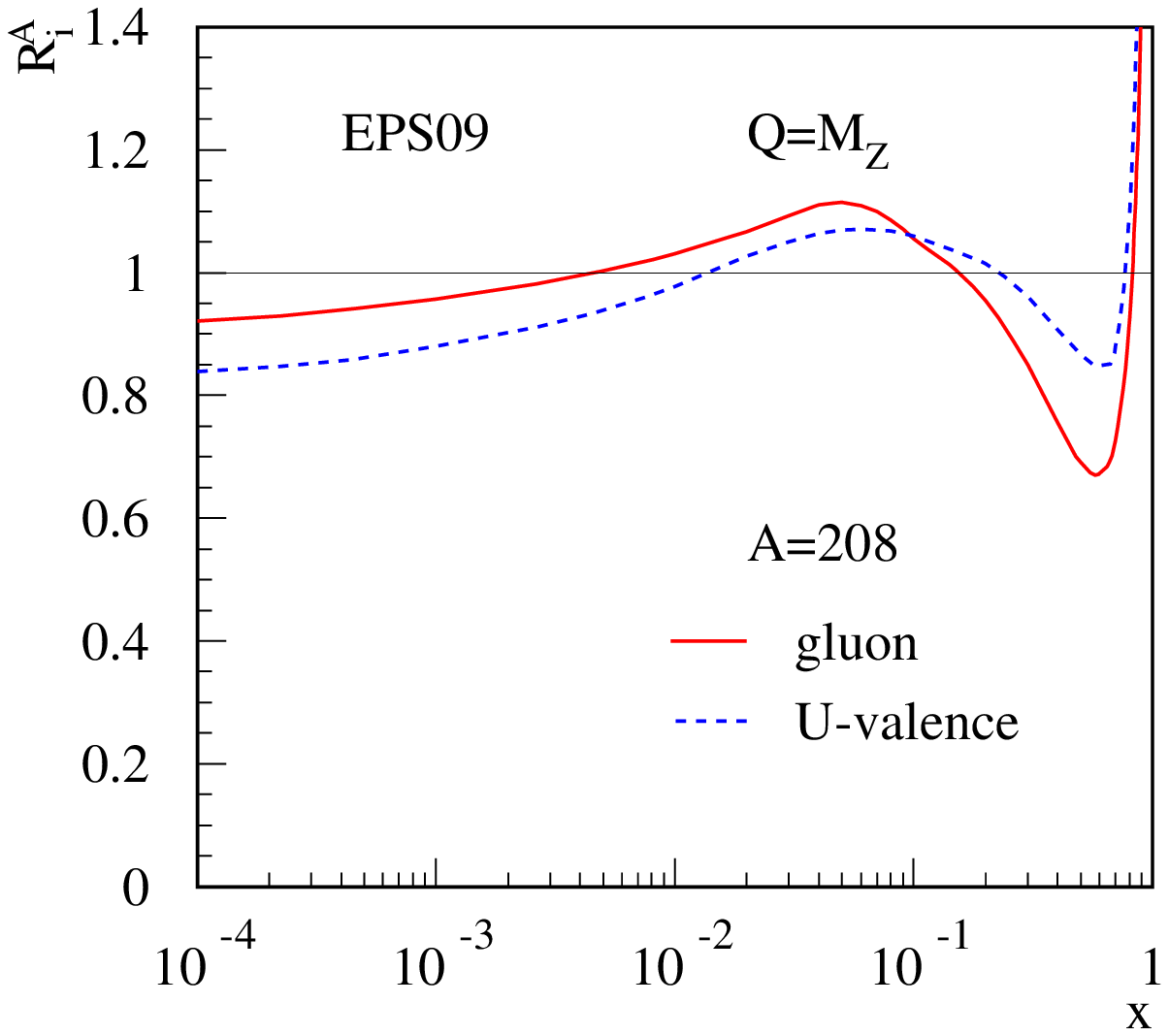}
\caption[]{
\label{fig:eps09}
The EPS09 shadowing ratio at scale $Q=M_Z$: $R_{i}^A = f_{i/A}(x, Q^2)/f_{i/p}(x, 
Q^2)$. The red solid curve shows the gluon ratio while the blue dashed curve
shows the valence $u$-quark ratio.}
\eef

As indicated in the lower half of Fig.~\ref{fig:z0},  the nuclear modification 
factor $R_{pA}$ for $Z^0$ production is suppressed at $p_T < 10$~GeV/$c$ and 
enhanced at high $p_T$, up to $p_T \sim 70$~GeV/$c$.  The low $p_T$ suppression
is a result of low $x$ shadowing in EPS09.  However, the strong enhancement over
such a large $p_T$ range was unexpected.  After a careful examination of 
the kinematics, it was found that, at $y=0$, $x\sim M_T/\sqrt{s} \sim 0.013$, 
already in the EPS09 antishadowing region.  Thus, the clear enhancement of 
$R_{pA}$ in the large $p_T$ region in Fig.~\ref{fig:z0} can be explained by the
broad EPS09 antishadowing region at the $Z^0$ scale.  The large $p_T$ 
enhancement of $R_{pA}$ vanishes for other shadowing parameterizations with
smaller gluon antishadowing.  Therefore, the measurement of $R_{pA}$ for $Z^0$ 
production in $p+$Pb collisions at the LHC provides a clean and unique test of 
gluon antishadowing as proposed in the EPS09 parameterization.  Furthermore, 
$R_{pA}$ is a direct measurement of nuclear gluon distribution since the cross 
section is dominated by gluon-initiated subprocesses for $p_T> 20$~GeV/$c$.

\subsection[Transverse momentum broadening of vector boson 
production]{Transverse momentum broadening of vector boson 
production (Z.-B. Kang and J.-W. Qiu)}
\label{sec:Zbroad}

Finally, transverse momentum broadening of inclusive vector boson production, 
$A(p_1)+B(p_2)\to V[{\rm J/}\psi, \, \Upsilon, \, Z^0](q)+X$ at the LHC is
discussed.  The average squared transverse momentum of vector boson production 
is  
\ben
\langle q_T^2\rangle(y)_{AB}
\equiv 
\int dq_T^2 \, q_T^2\,
     \frac{d\sigma_{AB\to V}}{dy\, dq_T^2}
\left[ \int dq_T^2 \, 
     \frac{d\sigma_{AB\to V}}{dy\, dq_T^2} \right]^{-1}  \, .
\label{avg-qt2}
\een
The transverse momentum broadening in $p+$Pb collisions is defined as
\ben
\Delta\langle q_T^2\rangle_{p\, {\rm Pb}}(y)
\equiv 
\langle q_T^2\rangle(y)_{p\, {\rm Pb}}
- \langle q_T^2\rangle(y)_{pp} \, .
\label{eq:pt-broad}
\een

Following the derivation in Refs.~\cite{Kang:2008us,Xing:2012ii}, the first 
nonvanishing contribution to
the transverse momentum broadening of heavy quarkonium production is 
\ben
\Delta\langle q_T^2\rangle_{\rm HQ}^{\rm CEM}
=\left(\frac{8\pi^2\alpha_s}{N_c^2-1}\, \lambda^2\, A^{1/3}\right) 
\frac{(C_F+C_A)\, \sigma_{q\bar{q}}+2\,C_A\, \sigma_{gg} + \Delta\sigma_{gg}}
     {\sigma_{q\bar{q}}+\sigma_{gg}} 
\label{cem-qt2}
\een
where the superscript ``CEM'' indicates that heavy quarkonium production is 
evaluated in the Color Evaporation Model (CEM).  A similar result was derived 
in the NRQCD approach \cite{Kang:2008us}.   The $\sigma_{q\bar{q}}$ and 
$\sigma_{gg}$ partonic cross sections are contributions from quark-antiquark 
and gluon-gluon subprocesses, respectively \cite{Kang:2008us}.  The 
$\Delta\sigma_{gg}$ term is a small, color-suppressed correction to the $gg$ 
subprocess derived in Ref.~\cite{Xing:2012ii}.  In the region where the $gg$ 
subprocess dominates heavy quarkonium
production, $\sigma_{gg}\gg \sigma_{q\bar{q}}, \, \Delta\sigma_{gg}$, heavy 
quarkonium broadening is further simplified as \cite{Kang:2008us}
\ben
\Delta\langle q_T^2\rangle_{\rm HQ}^{\rm CEM}
\approx
2\, C_A
\left(\frac{8\pi^2\alpha_s}{N_c^2-1}\, \lambda^2\, A^{1/3}\right) 
\, . 
\een

In Fig.~\ref{fig:broadening}, the predictions of transverse momentum broadening
of Drell-Yan type vector boson production in $p+$Pb collisions at the LHC at 
$y=0$ are shown as a function of $N_{\rm coll}$.  To determine the effective 
dependence on $N_{\rm coll}$, the $A^{1/3}$ in Eq.~(\ref{cem-qt2}) is replaced 
by $A^{1/3} N_{\rm coll}(b)/N_{\rm coll}(b_{\rm min \, bias})$.  In $p+$Pb collisions at 
the LHC, a Glauber-model calculation with $\sigma_{\rm NN}^{\rm in}=70$~mb at 
$\sqrt{s}=5$~TeV gives $N_{\rm coll}(b_{\rm min \, bias}) \sim 7$.  
In addition to heavy quarkonium production, the broadening of $W/Z^0$ 
production, calculated using the formalism derived in Ref.~\cite{Kang:2008us}, 
is also shown in Fig.~\ref{fig:broadening}.  The dramatic difference in 
the magnitude of the broadening between heavy quarkonium and $W/Z^0$ production 
in Fig.~\ref{fig:broadening} should be a signature QCD prediction.
\bef
\includegraphics[width=0.55\textwidth]{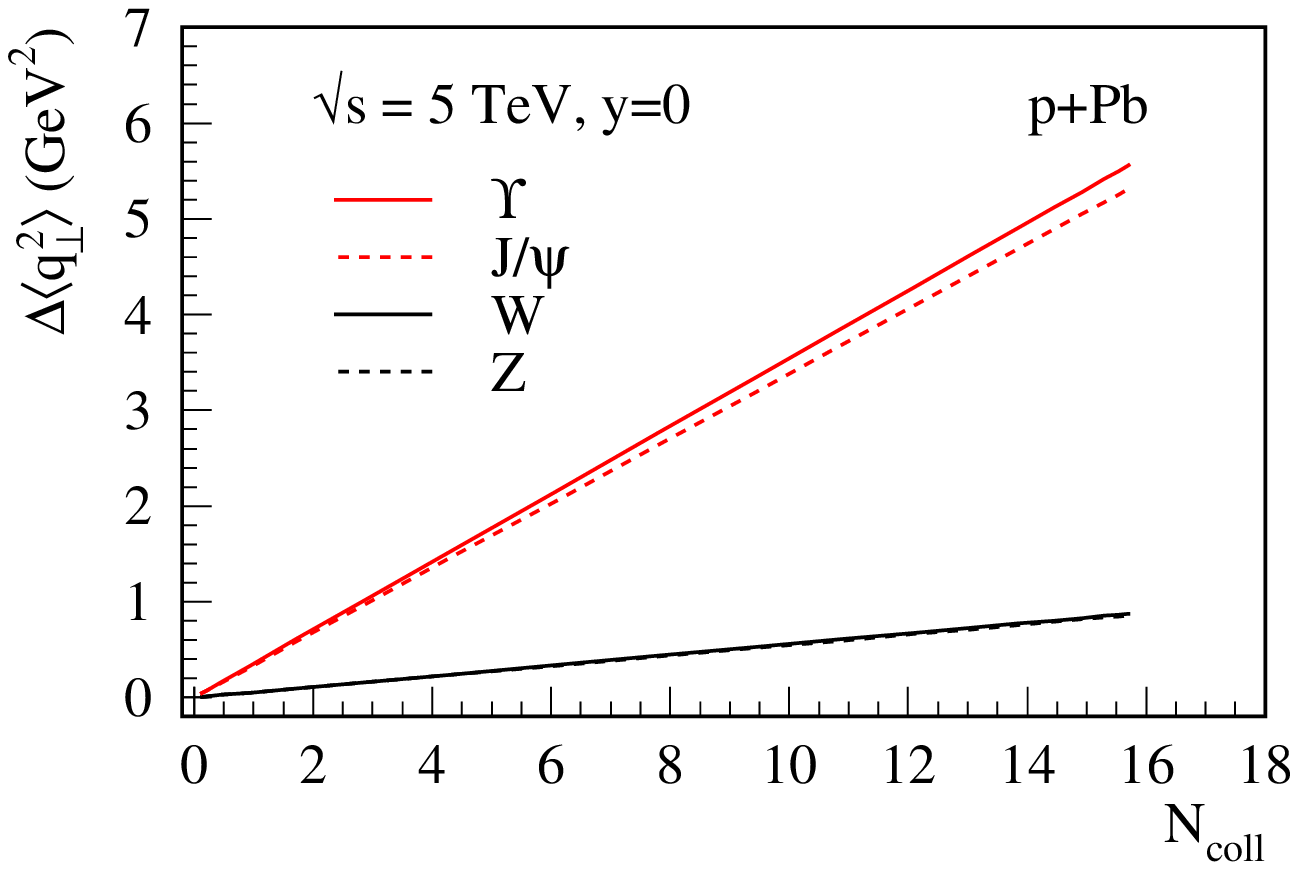}
\caption[]{The transverse momentum broadening of vector boson production in 
$p+$Pb collisions at $y=0$, shown as a function of $N_{\rm coll}$.  The 
$\Upsilon$ (red solid), $J/\psi$ (red dashed), $W^\pm$ (black solid), and 
$Z^0$ (black dashed) results are given \protect\cite{Kang:2012am}.}
\label{fig:broadening}
\eef


\section{Acknowledgments}

The research of J.~L.~Albacete is supported by a fellowship from the 
Th\'{e}orie LHC France initiative funded by  the IN2P3. 
The work of  N.~Armesto was  supported by the European Research Council grant 
HotLHC ERC-2001-StG-279579; by Ministerio de Ciencia e Innovaci\'on of Spain 
grants FPA2008-01177, FPA2009-06867-E and Consolider-Ingenio 2010 CPAN 
CSD2007-00042; by Xunta de Galicia grant PGIDIT10PXIB \-206017PR; and by FEDER.
G.~G.~Barnaf\'oldi was partially supported by the J\'anos Bolyai Research 
Scholarship of the Hungarian Academy of Sciences.
G.~G.~Barnaf\'oldi, M.~Gyulassy, and P.~Levai  also acknowledge Hungarian 
grants OTKA PD73596, NK77816, NK106119, 
NIH TET\_10-1\_2011-0061 and ZA-15/2009. 
J.~Barrette and V.~Topor Pop are supported by the Natural Sciences and 
Engineering Research Council of Canada.
S.~De is grateful to the Department of Atomic Energy of India for financial
support. 
A.~Dumitru is supported by 
the DOE Office of Nuclear Physics through Grant No. DE-FG02-09ER41620 and
by The City University of New York through the PSC-CUNY Research Award
Program, grant 65041-0043.
K. Dusling is supported by the US Department of Energy under DOE Contract No. 
DE-FG02-03ER41260.
K.~J.~Eskola is supported by the Academy of Finland, Project  133005. 
R.~J.~Fries would like to acknowledge
support by NSF CAREER Award PHY-0847538 and by the JET
Collaboration and DOE grant DE-FG02-10ER41682.
H.~Fujii and Y.~Nara are supported in part by Grant-in-Aid for Scientific 
Research (B) 22340064.
F.~Gelis is supported by the Agence Nationale de la Recherche project 
11-BS04-015-01.
M.~Gyulassy is supported by the Division of Nuclear Science, 
U.S. Department of Energy, under Contract No. DE-AC03-76SF00098 and
DE-FG02-93ER-40764 (associated 
with the JET Topical Collaboration Project). 
I.~Helenius is supported by the Magnus Ehrnrooth Foundation.
The work of B.~Z.~Kopeliovich was partially supported by Fondecyt (Chile) 
grant No. 1090291.
The work of K.~Kutak and S.~Sapeta was partially supported by the Foundation 
for Polish Science with
the grant Homing Plus/2010-2/6.
The work of A.~H.~Mueller is supported in part by the US Department of Energy
under contract No. DE-FG02-92ER-40699.
The work of J.~Nemchik was supported by 
grants VZ M\v SMT  6840770039 and LA 08015 (Ministry of Education of the Czech 
Republic).
M.~Petrovici is  supported by the Romanian Authority for Scientific Research, 
CNCS-UEFIS-CDI project number PN-II-ID-2011-3-0368.   
The work of A.~H.~Rezaeian was partially supported by Fondecyt (Chile) grant 
No. 1110781.
R. Venugopalan was supported by US Department of Energy under DOE Contract No.
DE-AC02-98CH10886.
The work of R.~Vogt was performed under the 
auspices of the U.S. Department of Energy by Lawrence Livermore National 
Laboratory under Contract DE-AC52-07NA27344 and within the framework of 
the JET Collaboration.
The work of X.-N.~Wang was performed under the auspices of the
U.S. Department of Energy
under Contract No. DE-AC02-05CH11231, by the National Natural Science 
Foundation of China under grant No. 11221504, and within
the framework of the JET Collaboration.
I.~Vitev is supported by the US Department of Energy, Office of Science, under 
Contract No. DE-AC52-06NA25396 and by the LDRD program at LANL.


\begin{thebibliography}{999}
\bibitem{JET}
http://jet.lbl.gov/

\bibitem{ALICE_dndeta}
  B.~Abelev {\it et al.}  [ALICE Collaboration],
  arXiv:1210.3615 [nucl-ex].

\bibitem{ALICE_rpa}
  B.~Abelev {\it et al.}  [ALICE Collaboration],
  arXiv:1210.4520 [nucl-ex].

\bibitem{Kovchegov:2001sc} 
   Y.~V.~Kovchegov and K.~Tuchin,
  {\it Phys. Rev. D} {\bf 65} (2002) 074026.

\bibitem{JalilianMarian:1997jx}
J. Jalilian-Marian, A. Kovner, A. Leonidov and H. Weigert, {\it Nucl. Phys.} 
B {\bf 504} (1997) 415.

\bibitem{JalilianMarian:1997gr} 
J.~Jalilian-Marian, A.~Kovner, A.~Leonidov and H.~Weigert,
{\it Phys. Rev. D} {\bf 59} (1999) 014014.

\bibitem{Iancu:2000hn}
E. Iancu, A. Leonidov and L. D. McLerran, {\it Nucl. Phys. A} {\bf 692}
(2001) 583.

\bibitem{Ferreiro:2001qy} 
E. Ferreiro, E. Iancu, A. Leonidov and L. D. McLerran, {\it Nucl. Phys.} 
A {\bf 703} (2002) 489.

\bibitem{Dumitru:2005gt}
  A.~Dumitru, A.~Hayashigaki and  J.~Jalilian-Marian,
  {\it Nucl. Phys.}  A {\bf 765} (2006) 464.
  
\bibitem{Altinoluk:2011qy} 
T.~Altinoluk and A.~Kovner, {\it Phys. Rev.}  D {\bf 83} (2011) 105004.

\bibitem{Balitsky:1995ub} 
I.~Balitsky, {\it Nucl. Phys.} B {\bf 463} (1996) 99.

\bibitem{Kovchegov:1999yj} 
Y.~V.~Kovchegov, {\it Phys. Rev. D} {\bf 60} (1999) 034008.

\bibitem{Kovchegov:1999ua}
Y.~V.~Kovchegov, {\it Phys. Rev. D} {\bf 61} (2000) 074018.

\bibitem{Balitsky:2006wa} 
I. Balitsky, {\it Phys. Rev. D} {\bf 75} (2007) 014001.

\bibitem{Albacete:2010sy}
J. L. Albacete, N. Armesto, J.G. Milhano, P. Quiroga Arias and C. A. Salgado, 
{\it Eur. Phys. J. C} {\bf 71} (2011) 1705.

\bibitem{Albacete:2007yr}
  J.~L.~Albacete and Y.~V.~Kovchegov,  {\it Phys. Rev.}  D {\bf 75} (2007) 
125021.

\bibitem{McLerran:1993ni}
L. D. McLerran and R. Venugopalan,
  {\it Phys. Rev. D} {\bf 49} (1994) 2233.

\bibitem{McLerran:1993ka}
L. D. McLerran and R. Venugopalan,
  {\it Phys. Rev. D} {\bf 49} (1994) 3352.

\bibitem{McLerran:1994vd}
L. D. McLerran and R. Venugopalan,
  {\it Phys. Rev. D} {\bf 50} (1994) 2225.

\bibitem{Rezaeian:2012ye} 
  A.~H.~Rezaeian,
  {\it Phys. Lett.} B {\bf 718} (2013) 1058.

\bibitem{Kowalski:2003hm}
  H.~Kowalski and D.~Teaney,
{\it  Phys. Rev. D} {\bf 68} (2003) 114005.

\bibitem{GolecBiernat:1998js}
  K.~J.~Golec-Biernat and M.~Wusthoff,
{\it Phys. Rev. D} {\bf 59} (1998) 014017.

\bibitem{GolecBiernat:1999qd}
  K.~J.~Golec-Biernat and M.~Wusthoff,
{\it Phys. Rev. D} {\bf 60} (1999) 114023.

\bibitem{Bartels:2002cj}
  J.~Bartels, K.~J.~Golec-Biernat and H.~Kowalski,
{\it Phys. Rev. D} {\bf 66} (2002) 014001.

\bibitem{McLerran:1998nk}
  L.~D.~McLerran and R.~Venugopalan,
{\it Phys. Rev. D} {\bf 59} (1999) 094002.

\bibitem{Venugopalan:1999wu}
  R.~Venugopalan,
{\it Acta Phys. Polon. B} {\bf 30} (1999) 3731.

\bibitem{Watt:2007nr} 
 G. Watt and  H. Kowalski, {\it Phys. Rev. D} {\bf 78} (2008) 014016.

\bibitem{Chekanov:2001qu}
  S.~Chekanov {\it et al.}  [ZEUS Collaboration],
{\it Eur. Phys. J. C} {\bf 21} (2001) 443.

\bibitem{Adloff:2000qk}
  C.~Adloff {\it et al.}  [H1 Collaboration],
{\it Eur. Phys. J. C} {\bf 21} (2001) 33.

\bibitem{Chekanov:2002xi}
  S.~Chekanov {\it et al.}  [ZEUS Collaboration],
{\it Eur. Phys. J. C} {\bf 24} (2002) 345.

\bibitem{Aktas:2005xu}
  A.~Aktas {\it et al.}  [H1 Collaboration],
{\it Eur. Phys. J. C} {\bf 46} (2006) 585.

\bibitem{Kowalski:2006hc}
  H.~Kowalski, L.~Motyka and G.~Watt,
{\it Phys. Rev. D} {\bf 74} (2006) 074016.

\bibitem{Rezaeian:2012ji}
  A.~H.~Rezaeian, M.~Siddikov, M.~Van de Klundert and R.~Venugopalan,
  arXiv:1212.2974 [hep-ph].

\bibitem{Tribedy:2010ab}
  P.~Tribedy and R.~Venugopalan,
{\it Nucl. Phys. A} {\bf 850} (2011) 136;
   [Erratum-ibid.\ A {\bf 859} (2011) 185].

\bibitem{Tribedy:2011aa}
  P.~Tribedy and R.~Venugopalan,
{\it Phys. Lett. B} {\bf 710} (2012) 125;
   [Erratum-ibid.\ B {\bf 718} (2013) 1154].

\bibitem{Schenke:2012wb}
  B.~Schenke, P.~Tribedy and R.~Venugopalan,
{\it Phys. Rev. Lett.}  {\bf 108} (2012) 252301.

\bibitem{Schenke:2012hg}
B.~Schenke, P.~Tribedy and R.~Venugopalan,
{\it Phys. Rev. C} {\bf 86} (2012) 034908.
  
\bibitem{Kowalski:2007rw}
  H.~Kowalski, T.~Lappi and R.~Venugopalan,
{\it Phys. Rev. Lett.}  {\bf 100} (2008) 022303.

\bibitem{Blaizot:2004wv}
  J.~P.~Blaizot, F.~Gelis and R.~Venugopalan,
{\it Nucl. Phys. A} {\bf 743} (2004) 57.

\bibitem{Braun:2000bh}
  M.~A.~Braun,
{\it Phys. Lett. B} {\bf 483} (2000) 105.

\bibitem{Gelis:2006tb}
  F.~Gelis, A.~M.~Stasto and R.~Venugopalan,
{\it Eur. Phys. J. C} {\bf 48} (2006) 489.

\bibitem{Kniehl:2000fe} 
B. A. Kniehl, G. Kramer and B. Potter, {\it Nucl. Phys.} B {\bf 582} (2000) 
514.

\bibitem{Wang:1991hta}
  X.~-N.~Wang and M.~Gyulassy,
  {\it Phys. Rev. D} {\bf 44} (1991) 3501.

\bibitem{Gyulassy:1994ew}
  M.~Gyulassy and X.~-N.~Wang,
  {\it Comput. Phys. Commun.}  {\bf 83} (1994) 307
  [arXiv:nucl-th/9502021].

\bibitem{Deng:2010mv}
  W.~-T.~Deng, X.~-N.~Wang and R.~Xu,
  {\it Phys. Rev. C} {\bf 83} (2011) 014915.

\bibitem{Deng:2010xg}
  W.~-T.~Deng, X.~-N.~Wang and R.~Xu,
  {\it Phys. Lett.} B {\bf 701} (2011) 133.  
 
\bibitem{Li:2001xa}
  S.~-y.~Li and X.~-N.~Wang,
  {\it Phys. Lett.} B {\bf 527} (2002) 85.
   
\bibitem{Eskola:2009uj}
  K.~J.~Eskola, H.~Paukkunen and C.~A.~Salgado,
  {\it JHEP} {\bf 0904} (2009) 065.
  
\bibitem{Cronin:1974zm}
  J.~W.~Cronin, H.~J.~Frisch, M.~J.~Shochet, J.~P.~Boymond, R.~Mermod, 
P.~A.~Piroue and R.~L.~Sumner,
  {\it Phys. Rev. D} {\bf 11} (1975) 3105.
 
\bibitem{Xu:2012au}
  R.~Xu, W.~-T.~Deng and X.~-N.~Wang,
  arXiv:1204.1998 [nucl-th].
 
\bibitem{Andersson:1983ia}
  B.~Andersson, G.~Gustafson, G.~Ingelman and T.~Sjostrand,
  {\it Phys. Rept.}  {\bf 97} (1983) 31.

\bibitem{Wang:1991xy}
  X.~-N.~Wang and M.~Gyulassy, {\it Phys. Rev. Lett.}  {\bf 68} (1992) 1480.

\bibitem{ToporPop:2011wk} 
V.~Topor~Pop,~M.~Gyulassy,~J.~Barrette,~and~C.~Gale, 
{\it Phys. Rev. C}  {\bf 84} (2011) 022002.

\bibitem{ToporPop:2010qz} 
V.~Topor~Pop,~M.~Gyulassy,~J.~Barrette,~C.~Gale,~and A. Warburton,
{\it Phys. Rev. C} {\bf 83} (2011) 024902.

\bibitem{Barnafoldi:2011px} 
  G.~G.~Barnafoldi, J.~Barrette, M.~Gyulassy, P.~Levai and V.~Topor Pop,
  {\it Phys. Rev. C} {\bf 85} (2012) 024903.

\bibitem{Pop:2012ug} 
  V.~Topor Pop, M.~Gyulassy, J.~Barrette, C.~Gale and A.~Warburton,
 arXiv:1203.6679 v2 [hep-ph].

\bibitem{Andersson:1986gw}
  B.~Andersson, G.~Gustafson, and B.~Nilsson-Almqvist,
  {\it Nucl. Phys. B} {\bf 281} (1987) 289.

\bibitem{NilssonAlmqvist:1986rx}
 B.~Nilsson-Almqvist and E.~Stenlund,
  {\it Comput. Phys. Commun.}  {\bf 43} (1987) 387.

\bibitem{Bengtsson:1987kr}
  H.~-U.~Bengtsson and T.~Sjostrand,
  {\it Comput. Phys. Commun.}  {\bf 46} (1987) 43.

\bibitem{Levai:2011zz}
P.~Levai,~D.~Berenyi,~A.~Pasztor, and V.~V.~Skokov,
 {\it J. Phys. G} {\bf 38} (2011) 124155.

\bibitem{Duke:1983gd} 
D.~W.~Duke  and J.~F.~Owens, {\it Phys. Rev. D} {\bf 30} (1984) 49.

\bibitem{Gluck:1994uf}
  M.~Gluck, E.~Reya, and A.~Vogt,
  {\it Z. Phys. C} {\bf 67} (1995) 433.

\bibitem{Altarelli:1977zs} 
G.~Altarelli and G.~Parisi, {\it Nucl. Phys. B}
{\bf 126} (1977) 298.

\bibitem{homepage}
Recent and test versions of the AMPT codes are available at 
http://personal.ecu.edu/linz/ampt

\bibitem{Lin:2004en}
Z.~-W.~Lin, C.~M.~Ko, B.~-A.~Li, B.~Zhang and S.~Pal,
{\it Phys. Rev. C} {\bf 72} (2005) 064901.

\bibitem{Xu:2011fi}
J.~Xu and C.~M.~Ko,
{\it Phys. Rev. C} {\bf 83} (2011) 034904.

\bibitem{Kang:2012kc}
  Z.~-B.~Kang, I.~Vitev and H.~Xing,
{\it  Phys. Lett. B} {\bf 718} (2012) 482.

\bibitem{Owens:1986mp}
J. F. Owens, {\it Rev. Mod. Phys.} {\bf 59} (1987) 465.

\bibitem{Pumplin:2002vw}
  J.~Pumplin, D.~R.~Stump, J.~Huston, H.~L.~Lai, P.~M.~Nadolsky and W.~K.~Tung,
  {\it JHEP} {\bf 0207} (2002) 012.

\bibitem{deFlorian:2007aj}
  D.~de Florian, R.~Sassot and M.~Stratmann,
  {\it Phys. Rev. D} {\bf 75} (2007) 114010.

\bibitem{Vitev:2006bi}
  I.~Vitev, J.~T.~Goldman, M.~B.~Johnson and J.~W.~Qiu,
{\it Phys. Rev. D} {\bf 74} (2006) 054010.

\bibitem{Kang:2008wv}
Z.~-B.~Kang, J.~-W.~Qiu and W.~Vogelsang,
{\it  Phys. Rev. D} {\bf 79} (2009) 054007.

\bibitem{Accardi:2002ik}
  A.~Accardi, arXiv:hep-ph/0212148.

\bibitem{Qiu:2003pm} 
  J.~-w.~Qiu and I.~Vitev,
  {\it Phys. Lett.} B {\bf 570} (2003) 161.

\bibitem{Ovanesyan:2011xy} 
  G.~Ovanesyan and I.~Vitev,
  {\it JHEP} {\bf 1106} (2011) 080.

\bibitem{Neufeld:2010dz} 
  R.~B.~Neufeld, I.~Vitev and B.~-W.~Zhang,
  {\it Phys. Lett.} B {\bf 704} (2011) 590.

\bibitem{Vitev:2007ve} 
  I.~Vitev,
  {\it Phys. Rev. C} {\bf 75} (2007) 064906.

\bibitem{Qiu:2004da}
J.~-w.~Qiu and I.~Vitev,
{\it Phys.  Lett. B} {\bf 632} (2006) 507.

\bibitem{Qiu:2004qk} 
  J.~-W.~Qiu and I.~Vitev,
  {\it Phys. Lett.} B {\bf 587} (2004) 52.

\bibitem{Kang:2011bp} 
  Z.~-B.~Kang, I.~Vitev and H.~Xing,
  {\it Phys. Rev. D} {\bf 85} (2012) 054024.

\bibitem{Zhang:2001ce} 
  Y.~Zhang, G.~I.~Fai, G.~Papp, G.~G.~Barnafoldi and P.~Levai,
  {\it Phys. Rev. C} {\bf 65} (2002) 034903.

\bibitem{Papp:2002ub}
    G. Papp, G. G. Barnafoldi, P. Levai, and G. Fai, arXiv:hep-ph/0212249.

\bibitem{Antreasyan:1978cw}
     D. Antreasyan {\it et al.} [Chicago-Princeton Collaboration], {\it Phys. 
Rev. D} {\bf 19} (1979) 764.

\bibitem{Aversa:1988vb} 
F. Aversa, P. Chiappetta, M. Greco, and J. Ph. Guillet, 
{\it Nucl. Phys.} B {\bf 327} (1989) 105.

\bibitem{Aurenche:1998gv}
P. Aurenche, 
M. Fontannaz, J. Ph. Guillet, B. Kniehl, E. Pilon, and M. Werlen,
{\it Eur. Phys. J.}  C {\bf 9} (1999) 107.

\bibitem{Aurenche:1999nz}
P. Aurenche, 
M. Fontannaz, J. Ph. Guillet, B. Kniehl, and M. Werlen,
{\it Eur. Phys. J. C} {\bf 13} (2001) 347.

\bibitem{Wang:1998ww} 
X. N. Wang, {\it Phys. Rev. C} {\bf 61} (2000) 064910.

\bibitem{Wong:1998pq}
C. Y. Wong and H. Wang, {\it Phys. Rev. C} {\bf 58} (1998) 376.

\bibitem{Barnafoldi:2002sj} 
  G.~G.~Barnafoldi, P.~Levai, G.~Papp, G.~I.~Fai and Y.~Zhang,
  {\it Heavy Ion Phys.}  {\bf 18} (2003) 79
  [arXiv:nucl-th/0206006].
 
\bibitem{Barnafoldi:2002xp} 
  G.~G.~Barnafoldi, P.~Levai, G.~Papp, G.~I.~Fai and Y.~Zhang,
  arXiv:nucl-th/0212111.

\bibitem{Martin:2001es}
A. D. Martin, R. G. Roberts, W. J. Stirling, and
R. S. Thorne, {\it Eur. Phys. J. C} {\bf 23} (2002) 73.

\bibitem{Eskola:1998df}
         K. J. Eskola, V. J. Kolhinen and C. A. Salgado,
         {\it Eur. Phys. J. C} {\bf 9}, 61 (1999).

\bibitem{Eskola:2008ca}
K. J. Eskola, H. Paukkunen, and C. A. Salgado, {\it JHEP} {\bf 0807}
(2008) 102.

\bibitem{Hirai:2001np}
         M. Hirai, S. Kumano and M. Miyama, 
         {\it Phys. Rev. D} {\bf 64} (2001) 034003.

\bibitem{Albacete:2012xq}
J.~L.~Albacete, A.~Dumitru, H.~Fujii and Y.~Nara,
arXiv:1209.2001 [hep-ph] (submitted to {\it Nucl. Phys. A}).

\bibitem{Rezaeian:2011ia}
A. H. Rezaeian, {\it Phys. Rev. D} {\bf 85} (2012) 014028.

\bibitem{JalilianMarian:2011dt} 
 J. Jalilian-Marian and A. H. Rezaeian, {\it Phys. Rev. D} {\bf 85} (2012) 
014017. 

\bibitem{Levin:2010dw}
E. Levin and A. H. Rezaeian, {\it Phys. Rev. D} {\bf 82} (2010) 014022.

\bibitem{Levin:2011hr} 
E. Levin and A. H. Rezaeian, {\it Phys. Rev. D} {\bf 83} (2011) 114001.

\bibitem{Levin:2010br} 
  E.~Levin and A.~H.~Rezaeian,
  {\it AIP Conf. Proc.}  {\bf 1350} (2011) 243
  [arXiv:1011.3591 [hep-ph]].

\bibitem{Rezaeian:2011ss} 
A. H. Rezaeian, arXiv:1110.6642 [hep-ph]. 

\bibitem{Levin:2010zy}
E. Levin and A. H. Rezaeian, {\it Phys. Rev. D} {\bf 82} (2010) 054003.

\bibitem{Rezaeian:2012vc} 
A. H. Rezaeian,  arXiv:1208.0026 [hep-ph].

\bibitem{Harris:2012kj} 
  J.~W.~Harris [ALICE Collaboration],
  {\it AIP Conf. Proc.}  {\bf 1422} (2012) 15
  [arXiv:1111.4651 [nucl-ex]].

\bibitem{Kopeliovich:2002yh} 
  B.~Z.~Kopeliovich, J.~Nemchik, A.~Sch\"afer and A.~V.~Tarasov,
  {\it Phys. Rev. Lett.}  {\bf 88} (2002) 232303.
  
\bibitem{Kopeliovich:1999am}
B. Z.~Kopeliovich, A.~Sch\" afer and A. V.~Tarasov, 
{\it Phys. Rev. D} {\bf 62} (2000) 054022.

\bibitem{Cole:2007ru} 
  B.~A.~Cole, G.~G.~Barnafoldi, P.~Levai, G.~Papp and G.~Fai,
  arXiv:hep-ph/0702101.

\bibitem{Adeluyi:2008gj}
  A.~Adeluyi, G.~G.~Barnafoldi, G.~Fai and P.~Levai,
  {\it Phys. Rev. C} {\bf 80} (2009) 014903.

\bibitem{Kharzeev:2003wz}
  D.~Kharzeev, Y.~V.~Kovchegov, and K.~Tuchin,
  {\it Phys. Rev. D} {\bf 68} (2003) 094013.

\bibitem{Levai:2011qm}
P. Levai, {\it Nucl. Phys. A} {\bf 862-863} (2011) 146.

\bibitem{Barnafoldi:2008ec}
G.~G.~Barnafoldi, G.~Fai, P.~Levai, B.~A.~Cole and G.~Papp, {\it 
Indian J. Phys.}  {\bf 84} (2010) 1721.

\bibitem{d'Enterria:2003qs}
D. d'Enterria, arXiv:nucl-ex/0302016. 

\bibitem{Martin:2007bv}
A. D. Martin, W. J. Stirling, R. S. Thorne and G. Watt, {\it Phys. Lett.} B 
{\bf 652} (2007) 292.

\bibitem{Martin:2009iq} 
A. D. Martin, W. J. Stirling, R. S. Thorne and G. Watt, {\it Eur. Phys. J.} 
C {\bf 63} (2009) 189.

\bibitem{Albacete:2010ad} 
J. L. Albacete and A. Dumitru,  arXiv:1011.5161. 

\bibitem{QuirogaArias:2010wh}
P. Quiroga-Arias, J. G. Milhano and U. A. Wiedemann, {\it Phys. Rev. C} 
{\bf 82} (2010) 034903.

\bibitem{Arleo:2011gc}
  F.~Arleo, K.~J.~Eskola, H.~Paukkunen and C.~A.~Salgado, {\it JHEP} 
{\bf 1104} (2011) 055.

\bibitem{Brodsky:1977de}
  S.~J.~Brodsky, J.~F.~Gunion, and J.~H.~Kuhn,
  {\it Phys. Rev. Lett.}  {\bf 39} (1977) 1120.

\bibitem{Adil:2005qn}
  A.~Adil and M.~Gyulassy,
  {\it Phys. Rev.}  C {\bf 72} (2005) 034907.

\bibitem{Helenius:2012wd}
  I.~Helenius, K.~J.~Eskola, H.~Honkanen and C.~A.~Salgado,
  {\it JHEP} {\bf 1207} (2012) 073.

\bibitem{Albino:2008fy}
  S.~Albino, B.~A.~Kniehl and G.~Kramer,
  {\it Nucl. Phys.} B {\bf 803} (2008) 42.

\bibitem{Lourenco:2008sk}
 C. Louren\c{c}o, R. Vogt and H. W\"{o}hri, {\it JHEP} {\bf 0902} (2009) 014. 

\bibitem{McGlinchey:2012bp}
D. C. McGlinchey, A. D. Frawley and R. Vogt, arXiv:1208.2667 [nucl-th].

\bibitem{Nelson:2012bc}
R. E. Nelson, R. Vogt and A. D. Frawley, {\it Phys. Rev. C}, 
in press [arXiv:1210:4610 [hep-ph]].

\bibitem{Vogt:2010aa}
R. Vogt, {\it Phys. Rev. C} {\bf 81} (2010) 044903.

\bibitem{Nadolsky:2008zw} 
  P.~M.~Nadolsky, H.~-L.~Lai, Q.~-H.~Cao, J.~Huston, J.~Pumplin, D.~Stump, 
W.~K.~Tung and C.~-P.~Yuan,
  {\it Phys. Rev. D} {\bf 78} (2008) 013004.

\bibitem{Catani:2002ny} 
  S.~Catani, M.~Fontannaz, J.~P.~Guillet and E.~Pilon,
  {\it JHEP} {\bf 0205} (2002) 028.

\bibitem{Aurenche:2006vj}
  P.~Aurenche, M.~Fontannaz, J.~-P.~Guillet, E.~Pilon and M.~Werlen,
  {\it Phys. Rev. D} {\bf 73} (2006) 094007.

\bibitem{Bourhis:1997yu} 
  L.~Bourhis, M.~Fontannaz and J.~P.~Guillet,
  {\it Eur. Phys. J. C} {\bf 2} (1998) 529.
 
\bibitem{Xing:2012ii} 
  H.~Xing, Z.~-B.~Kang, I.~Vitev and E.~Wang,  
{\it Phys. Rev. D} {\bf 86} (2012) 094010.

\bibitem{Gluck:1992zx}
M. Gluck, E. Reya and A. Vogt, {\it Phys. Rev. D} {\bf 48} (1993) 116, 
[Erratum-ibid D {\bf 51} (1995) 1427]. 

\bibitem{Gelis:2002ki}
  F.~Gelis and J.~Jalilian-Marian,
  {\it Phys. Rev. D} {\bf 66} (2002) 014021.

\bibitem{Baier:2004ti}
  R.~Baier, A.~H.~Mueller and D.~Schiff,
  {\it Nucl. Phys. A} {\bf 741} (2004) 358.

\bibitem{JalilianMarian:2012bd}
  J. Jalilian-Marian and A. H. Rezaeian, {\it Phys. Rev. D} {\bf 86}
(2012) 034016.

\bibitem{Rezaeian:2009it}
A. H. Rezaeian and A. Schaefer, {\it Phys. Rev. D} {\bf 81} (2010) 114032.

\bibitem{Iancu:2003ge}
E. Iancu, K. Itakura and S. Munier, {\it Phys. Lett.} B {\bf 590} (2004) 199.

\bibitem{Rezaeian:2012wa} 
  A.~H.~Rezaeian,
  {\it Phys. Rev. D} {\bf 86} (2012) 094016.

\bibitem{Gelis:2010nm}
F. Gelis, E. Iancu, J. Jalilian-Marian and R. Venugopalan, 
{\it Ann. Rev. Part. Nucl Sci.} {\bf 60} (2010) 463. 

\bibitem{Triantafyllopoulos:2012uv} 
  D.~N.~Triantafyllopoulos,
  arXiv:1209.3183 [hep-ph].

\bibitem{Mueller:2002zm} 
  A.~H.~Mueller and D.~N.~Triantafyllopoulos,
  {\it Nucl. Phys. B} {\bf 640} (2002) 331.

\bibitem{Aad:2010bu}
G.~Aad {\it et al.} [ATLAS Collaboration],
 {\it Phys. Rev. Lett.}  {\bf 105} (2010)  252303.

\bibitem{:2012is}
  G.~Aad {\it et al.}  [ATLAS Collaboration],
  arXiv:1208.1967 [hep-ex].

\bibitem{Chatrchyan:2011sx}
  S.~Chatrchyan {\it et al.}  [CMS Collaboration],
{\it Phys. Rev. C} {\bf 84} (2011) 024906.
    
\bibitem{Chatrchyan:2012nia}
  S.~Chatrchyan {\it et al.}  [CMS Collaboration],
{\it Phys. Lett. B} {\bf 712} (2012) 176.

\bibitem{Chatrchyan:2012gt} 
  S.~Chatrchyan {\it et al.}  [CMS Collaboration],
  arXiv:1205.0206 [nucl-ex].
  
\bibitem{Chatrchyan:2012gw}
  S.~Chatrchyan {\it et al.}  [CMS Collaboration],
  arXiv:1205.5872 [nucl-ex].

\bibitem{Frixione:1995ms}
S.~Frixione, Z.~Kunszt and A.~Signer,
{\it Nucl. Phys. B} {\bf 467} (1996) 399.

\bibitem{Frixione:1997np}
S.~Frixione,
{\it Nucl. Phys. B} {\bf 507} (1997) 295.

\bibitem{Frixione:1997ks}
S.~Frixione and G.~Ridolfi,
{\it Nucl. Phys. B} {\bf 507} (1997) 315.

\bibitem{Eskola:1998iy}
K.~J.~Eskola, V.~J.~Kolhinen and P.~V.~Ruuskanen,
{\it Nucl. Phys. B} {\bf 535} (1998) 351.

\bibitem{Cacciari:2008gp}
  M.~Cacciari, G.~P.~Salam and G.~Soyez,
{\it JHEP} {\bf 0804} (2008) 063.
  
\bibitem{Accardi:2003be}
A.~Accardi {\it et al.},
arXiv:hep-ph/0308248.

\bibitem{Accardi:2003gp}
A.~Accardi {\it et al.}, arXiv:hep-ph/0310274.

\bibitem{pPblumi} M. Lamont at the {\it 111th LHCC Meeting} (CERN, September 
26-27 2012) [{\tt http://indico.cern.ch/conferenceDisplay.py?confId=207964}].

\bibitem{Cacciari:2007fd}
  M.~Cacciari and G.~P.~Salam,
{\it Phys. Lett. B} {\bf 659} (2008) 119.

\bibitem{Kunszt:1992tn}
  Z.~Kunszt and D.~E.~Soper,
{\it Phys. Rev. D} {\bf 46} (1992) 192.

\bibitem{Ellis:1990ek}
  S.~D.~Ellis, Z.~Kunszt and D.~E.~Soper,
{\it Phys. Rev. Lett.}  {\bf 64} (1990) 2121.

\bibitem{Ellis:1992en}
  S.~D.~Ellis, Z.~Kunszt and D.~E.~Soper,
{\it Phys. Rev. Lett.}  {\bf 69} (1992) 1496.

\bibitem{Vitev:2009rd} 
  I.~Vitev and B.-W.~Zhang,
{\it Phys. Rev. Lett.}  {\bf 104} (2010) 132001.

\bibitem{He:2011pd} 
  Y.~He, I.~Vitev and B.-W.~Zhang,
{\it Phys. Lett. B} {\bf 713} (2012) 224.

\bibitem{He:2011sg} 
  Y.~He, B.-W.~Zhang and E.~Wang,
{\it Eur. Phys. J. C} {\bf 72} (2012) 1904.

\bibitem{deFlorian:2012qw}
  D.~de Florian, R.~Sassot, M.~Stratmann and P.~Zurita,
   arXiv:1204.3797 [hep-ph].  

\bibitem{deFlorian:2011fp}
  D.~de Florian, R.~Sassot, P.~Zurita and M.~Stratmann,
{\it Phys. Rev. D} {\bf 85} (2012) 074028.

\bibitem{Hirai:2007sx}
  M.~Hirai, S.~Kumano and T.~-H.~Nagai,
{\it Phys. Rev. C} {\bf 76} (2007) 065207.

\bibitem{Aad:2010ad}
G.~Aad {\it et al.}  [ATLAS Collaboration],
{\it Eur. Phys. J. C} {\bf 71} (2011) 1512.

\bibitem{Ellis:1994dg}
  S.~D.~Ellis and D.~E.~Soper,
{\it Phys. Rev. Lett.} {\bf 74} (1995) 5182.
 
\bibitem{HZW2012} 
  Y.~He, B.-W.~Zhang and E.~Wang,
  in preparation.

\bibitem{Gribov:1984tu}
  L.~V.~Gribov, E.~M.~Levin and M.~G.~Ryskin,
  {\it Phys. Rept.}  {\bf 100} (1983) 1.

\bibitem{Kutak:2012rf}
  K.~Kutak and S.~Sapeta,
  arXiv:1205.5035 [hep-ph].

\bibitem{Deak:2010gk}
  M.~Deak, F.~Hautmann, H.~Jung and K.~Kutak,
  arXiv:1012.6037 [hep-ph].

\bibitem{Albacete:2010pg}
  J.~L.~Albacete and C.~Marquet,
{\it Phys. Rev. Lett.}  {\bf 105} (2010)  162301.

\bibitem{Dumitru:2010iy}
  A.~Dumitru, K.~Dusling, F.~Gelis, J.~Jalilian-Marian, T.~Lappi and
R.~Venugopalan,
{\it Phys. Lett. B} {\bf 697} (2011)  21.

\bibitem{Catani:1990eg}
  S.~Catani, M.~Ciafaloni, F.~Hautmann,
{\it Nucl. Phys. B} {\bf 366} (1991)  1. 

\bibitem{Marquet:2003dm}
  C.~Marquet and R.~B.~Peschanski,
{\it Phys. Lett. B} {\bf 587} (2004) 201.

\bibitem{Deak:2009xt}
  M.~Deak, F.~Hautmann, H.~Jung and K.~Kutak,
{\it JHEP} {\bf 0909} (2009) 121.

\bibitem{Kutak:2003bd}
  K.~Kutak and J.~Kwiecinski,
{\it Eur. Phys. J. C} {\bf 29} (2003) 521.

\bibitem{Kutak:2004ym}
  K.~Kutak and A.~M.~Stasto,
{\it Eur. Phys. J. C} {\bf 41} (2005) 343.

\bibitem{Khachatryan:2010gv} 
  V.~Khachatryan {\it et al.}  [CMS Collaboration],
{\it JHEP} {\bf 1009} (2010) 091.

\bibitem{Dusling:2012iga} 
  K.~Dusling and R.~Venugopalan,
{\it Phys. Rev. Lett.}  {\bf 108} (2012) 262001.

\bibitem{Dusling:2012cg} 
  K.~Dusling and R.~Venugopalan,
  arXiv:1210.3890 [hep-ph].

\bibitem{CMS:2012qk} 
  S.~Chatrchyan {\it et al.}  [CMS Collaboration],
{\it Phys. Lett. B} {\bf 718} (2013) 795.

\bibitem{Dusling:2012wy} 
  K.~Dusling and R.~Venugopalan,
  arXiv:1211.3701 [hep-ph].

\bibitem{Kang:2008us} 
  Z.~-B.~Kang and J.~-W.~Qiu,
  {\it Phys. Rev. D} {\bf 77} (2008) 114027 (2008).

\bibitem{Aad:2011gj}
G. Aad {\it et al.} [ATLAS Collaboration], 
{\it Phys. Lett. B} {\bf 705} (2011) 415.

\bibitem{Aad:2011fp} 
G. Aad {\it et al}, [ATLAS Collaboration], {\it Phys. Rev. D} {\bf 85} (2012)
012005.

\bibitem{Catani:2009sm}
S. Catani, L. Cieri, G. Ferrera, D. de Florian, and M. Grazzini,
{\it Phys. Rev. Lett.} {\bf 103} (2009) 082001.

\bibitem{Collins:1984kg}
J. C. Collins, D. E.~Soper and G.~Sterman, {\it Nucl. Phys. B} {\bf 250}
(1985) 199.

\bibitem{Qiu:2000hf} 
  J.~-W.~Qiu and X.~-f.~Zhang,
{\it Phys. Rev. D} {\bf 63} (2001) 114011.

\bibitem{Zhang:2002yz} 
  X.~-f.~Zhang and G.~I.~Fai,
{\it Phys. Lett. B} {\bf 545} (2002) 91.

\bibitem{Kang:2012am}
Z.-B.~Kang and J.-W.~Qiu, arXiv:1212.6541 [hep-ph].

\end{thebibliography}
\end{document}